\documentstyle[12pt,epsfig,axodraw,graphicx]{article}
\newcommand{\lsim}{\raisebox{-0.13cm}{~\shortstack{$<$ \\[-0.07cm] $\sim$}}~}
\newcommand{\gsim}{\raisebox{-0.13cm}{~\shortstack{$>$ \\[-0.07cm] $\sim$}}~}

\newcommand{\ra}{\rightarrow}
\newcommand{\ee}{e^+e^-}
\newcommand{\s}{\\ \vspace*{-3.5mm} }
\newcommand{\nn}{\noindent}
\newcommand{\non}{\nonumber}

\newcommand{\tb}{\mbox{tan}\beta}
\def\beq{\begin{equation}}
\def\eeq{\end{equation}}

\def\beqn{\begin{eqnarray}}
\def\eeqn{\end{eqnarray}}
\relax
\def\ba{\begin{array}}
\def\ea{\end{array}}

\renewcommand{\theequation}{\thesection.\arabic{equation}}
\def\pb{~{\rm pb}^{-1}}

\def\gv{\rm GeV/c^2}
\def\Ebar{{\rm E}_{\perp}\hspace{-.27cm}/\hspace{+.01cm} }
\begin{document}

\begin{flushright}
\hfill{PM/98--45} \\
\hfill{December 1998} \\
\end{flushright}
\vspace*{.8cm} 

\begin{center}

{\Large\sc \bf The Minimal Supersymmetric Standard Model: }

\vspace*{0.4cm}

{\Large\sc \bf Group Summary Report} 

\vspace*{1.6cm} 

Conveners: \\ 
{\sc A. Djouadi$^1$, S. Rosier-Lees$^2$}

\vspace*{0.6cm}

 Working Group: \\
{\sc 
M.~Bezouh$^1$, 
M.-A.~Bizouard$^3$, 
C.~Boehm$^1$, 
F.~Borzumati$^{1,4}$, 
C.~Briot$^2$, 
J.~Carr$^5$, 
M.~B. Causse$^6$,
F.~Charles$^7$, 
X.~Chereau$^2$, 
P.~Colas$^8$,
L.~Duflot$^3$,
A.~Dupperin$^9$, 
A.~Ealet$^5$, 
H.~El-Mamouni$^9$, 
N.~Ghodbane$^9$, 
F.~Gieres$^9$,  
B.~Gonzalez-Pineiro$^{10}$, 
S.~Gourmelen$^9$,  
G.~Grenier$^9$, 
Ph.~Gris$^{8}$, 
J.-F.~Grivaz$^3$, 
C.~Hebrard$^6$, 
B.~Ille$^9$, 
J.-L.~Kneur$^1$, 
N.~Kostantinidis$^5$, 
J.~Layssac$^1$, 
P.~Lebrun$^9$, 
R.~Ledu$^{11}$, 
M.-C.~Lemaire$^{8}$, 
Ch.~LeMouel$^1$, 
L.~Lugnier$^9$
Y.~Mambrini$^1$, 
J.P.~Martin$^9$,  
G.~Montarou$^6$, 
G.~Moultaka$^1$, 
S.~Muanza$^9$, 
E.~Nuss$^1$,   
E.~Perez$^{8}$, 
F.~M. Renard$^1$,
D.~Reynaud$^1$, 
L.~Serin$^{3}$,
C.~Thevenet$^9$, 
A.~Trabelsi$^{8}$, 
F.~Zach$^9$,} and {\sc D.~Zerwas$^3$.}

\vspace*{1.cm}

{\small

$^1$ LPMT, Universit\'e Montpellier II, F--34095 Montpellier Cedex 5.  \\

$^2$ LAPP, BP 110, F--74941 Annecy le Vieux Cedex. \\

$^3$ LAL, Universit\'e de Paris--Sud, Bat--200, F--91405 Orsay Cedex \\

$^4$ CERN, Theory Division, CH--1211, Geneva, Switzerland. \\

$^5$ CPPM, Universit\'e de Marseille-Luminy, Case 907, F-13288 Marseille Cedex 9. \\  

$^6$ LPC  Clermont, Universit\'e Blaise Pascal, F--63177 Aubiere Cedex. \\ 

$^7$ GRPHE, Universit\'e de Haute Alsace,  Mulhouse \\
 
$^8$ SPP, CEA--Saclay, F--91191 Cedex \\

$^9$ IPNL, Universit\'e Claude Bernard de Lyon, F--69622 Villeurbanne Cedex.\\

$^{10}$ LPNHE, Universit\'es Paris VI et VII, Paris Cedex. 

$^{11}$ CPT, Universit\'e de Marseille-Luminy, Case 907, F-13288 Marseille 
Cedex 9.   \\
}

\vspace*{1cm}

\centerline{\it Report of the MSSM working group for the Workshop 
``GDR--Supersym\'etrie".} 

\newpage

{\bf \large CONTENTS} 
\end{center} 

\vspace*{0.3cm}

{\bf 1. Synopsis} \hfill 4 \\

{\bf 2. The MSSM Spectrum} \hfill 9 \\

\hspace*{0.20cm} 2.1 The MSSM: Definitions and Properties \hfill  \s

\hspace*{0.6cm} 2.1.1 The uMSSM: unconstrained MSSM \hfill \s

\hspace*{0.6cm} 2.1.2 The pMSSM: ``phenomenological" MSSM \hfill  \s 

\hspace*{0.6cm} 2.1.3 mSUGRA: the constrained MSSM  \hfill   \s

\hspace*{0.6cm} 2.1.4 The MSSMi: the intermediate MSSMs \hfill \s

\hspace*{0.20cm} 2.2 Electroweak Symmetry Breaking \hfill  \s

\hspace*{0.6cm} 2.2.1 General features \hfill  \s

\hspace*{0.6cm} 2.2.2 EWSB and model--independent tan$\beta$ bounds \hfill \s 

\hspace*{0.20cm} 2.3 Renormalization Group Evolution \hfill   \s

\hspace*{0.6cm} 2.3.1 The one--loop RGEs \hfill  \s

\hspace*{0.6cm} 2.3.2 Exact solutions for the Yukawa coupling RGEs \hfill  \\

\nn {\bf 3. The Physical Parameters} \hfill 28  \\

\hspace*{0.20cm} 3.1 Particle Masses and Couplings \hfill \s

\hspace*{0.60cm} 3.1.1 Mass matrices and couplings \hfill  \s

\hspace*{0.60cm} 3.1.2 Inverting the MSSM spectrum \hfill \s 

\hspace*{0.20cm} 3.2 The program SUSPECT \hfill  \s

\hspace*{0.6cm} 3.2.1 Introduction \hfill  \s

\hspace*{0.6cm} 3.2.2 The ``phenomenological" MSSM  \s 

\hspace*{0.6cm} 3.2.3 Constrained MSSM \hfill  \s

\hspace*{0.6cm} 3.2.4 Example of input/output files \hfill  \s 

\hspace*{0.6cm} 3.2.5 Discussions and outlook \hfill   \\

{\bf 4. Higgs Boson Production and Decays} \hfill 49  \\

\hspace*{0.20cm} 4.1 MSSM Higgs boson production at the LHC \hfill   \s

\hspace*{0.60cm} 4.1.1 Physical set--up \hfill  \s

\hspace*{0.60cm} 4.1.2 Higgs production in the gluon--fusion mechanism \s 

\hspace*{0.60cm} 4.1.3 Higgs production in association with light stops \s

\hspace*{0.20cm} 4.2 Higgs boson decays into SUSY particles \hfill \s

\hspace*{0.6cm} 4.2.1 SUSY decays of the neutral Higgs bosons \hfill \s

\hspace*{0.6cm} 4.2.2 Decays of the charged Higgs bosons \hfill  \s

\hspace*{0.20cm} 4.3 MSSM Higgs production in $\ee$ collisions \hfill \s

\hspace*{0.6cm} 4.3.1 Production mechanisms \hfill  \s

\hspace*{0.6cm} 4.3.2 The program HPROD \hfill   \s

\hspace*{0.6cm} 4.3.3 Higgs boson production in association with stops 
\hfill   \\

{\bf 5. SUSY Particle Production and Decays} \hfill 73 \\

\hspace*{0.20cm} 5.1 Virtual SUSY effects \hfill  \s

\hspace*{0.20cm} 5.2 Correlated production and decays in 
$\ee \ra  \chi_1^+\chi_1^-$ \hfill  \s

\hspace*{0.20cm} 5.3 Chargino/Neutralino production at hadron colliders 
\hfill  \s

\hspace*{0.6cm} 5.3.1 Theoretical cross sections \hfill  \s

\hspace*{0.6cm} 5.3.2 Searches at the LHC \hfill  \s

\hspace*{0.20cm} 5.4 Two and three--body sfermion decays \hfill  \s

\hspace*{0.20cm} 5.5 Stop and Sbottom searches at at LEP200 \hfill  \s

\hspace*{0.6cm} 5.5.1 Squark production at LEP200 \hfill  \s

\hspace*{0.6cm} 5.5.2 Squark decays \hfill   \\

{\bf 6. Experimental Bounds on SUSY Particle Masses}  \hfill 91 \\

\hspace*{0.20cm} 6.1 Introduction \hfill  \s

\hspace*{0.20cm} 6.2 Scalar particle sector \hfill  \s

\hspace*{0.6cm} 6.2.1 Higgs bosons \hfill  \s

\hspace*{0.6cm} 6.2.2 Scalar leptons \hfill  \s

\hspace*{0.6cm} 6.2.3 Scalar quarks \hfill   \s

\hspace*{0.20cm} 6.3 Gaugino sector \hfill  \s

\hspace*{0.6cm} 6.3.1 Gluinos \hfill  \s

\hspace*{0.6cm} 6.3.2  Charginos and neutralinos  \hfill  \s

\hspace*{0.20cm} 6.4 Summary and bounds on the MSSM parameters  \hfill \\

%{\bf 7. Outlook} \\

{\bf 7. References} \hfill 114

\newpage

\section*{1. Synopsis} 

This report summarizes the activities of the working group on the 
``Minimal Supersymmetric Standard Model" or MSSM for the GDR--Supersym\'etrie. 
It is divided into five main parts: a first part dealing with the general
features of the MSSM spectrum, a second part with the SUSY and Higgs particle
physical properties, and then two sections dealing with the production and 
decay properties of the MSSM Higgs bosons and of the supersymmetric particles, 
and a last part summarizing the experimental limits on these states. A brief 
discussion of the prospects of our working group are given at the end. 
In each section, we first briefly give some known material for completeness 
and to establish the used conventions/notations, and then summarize the
original contributions to the GDR--Supersym\'etrie workshop\footnote{
Since the spectrum of the various contributions is rather wide, this
leads to the almost unavoidable situation of ``un rapport quelque peu 
decousu"...}. For more details, we refer to the original GDR notes [1--18]. 
The report contains the following material: \\

Section 2 gives a survey of the MSSM spectrum. We start the discussion by 
defining what we mean by the MSSM in section 2.1, thus setting the framework 
in which our working group is acting \cite{MSSMrev}. We first introduce the 
unconstrained MSSM (uMSSM) defined by the requirements of minimal gauge group 
and particle content, R--parity conservation and all soft--SUSY 
breaking terms allowed by the symmetry. 
We then briefly summarize the various phenomenological problems 
of the uMSSM which bring us to introduce what we call the ``phenomenological 
MSSM" (pMSSM) which has only a reasonable set of free parameters. We then 
discuss the celebrated minimal Supergravity model (mSUGRA), in which 
unification conditions at the GUT scale are imposed on the various parameters, 
leading only to four continuous and one discrete free parameters. We finally 
discuss the possibility of relaxing some of the assumptions of the mSUGRA 
model, and define several intermediate MSSM scenarii (MSSMi), between mSUGRA 
and the pMSSM. \s

In section 2.2, we focus on the [radiative] Electroweak Symmetry Breaking
mechanism, starting by a general discussion of the two requirements which make 
it taking place in the MSSM. We then summarize some analytical results 
for theoretical bounds on the parameter $\tan \beta$, obtained from the study 
of electroweak symmetry breaking conditions at the one--loop order 
\cite{LMM,LMM1}. The key--point is to use, on top of the usual conditions, 
the positivity of the Higgs boson squared masses which are needed to ensure a 
[at least local] minimum in the Higgs sector. We then show analytically how 
these constraints translate into bounds on $\tan \beta$ which are then 
necessary and fully model--independent bounds. The generic form of the bounds 
are given both in the Supertrace approximation and in a more realistic one, 
namely the top/stop--bottom/sbottom approximation. \s

In section 2.3, we first write down for completeness the Renormalization
Group Equations for the MSSM parameters in the one--loop approximation, 
that we will need for later investigations. In a 
next step, we discuss some exact analytical solutions of the
one--loop renormalization group evolution equations of the Yukawa couplings 
\cite{AM}. Solutions of the one--loop equations for the top and bottom
Yukawa couplings exist in the literature only for some limiting cases. 
Here, we give the exact solutions with no approximation: they are valid 
for any value of top and bottom Yukawa couplings and immediately generalizable 
to the full set of Yukawa couplings. These solutions allow to treat the
problematic large $\tan \beta$ region exactly, and thus are of some 
relevance in the case of bottom--tau Yukawa coupling unification scenarii. \\

Section 3 is devoted to the physical parameters of the MSSM. In section 3.1. 
we first briefly summarize the general features of the chargino/neutralino, 
sfermion mass matrices to set the conventions and the notations for what will 
follow, and discuss our parameterization of the Higgs sector exhibiting
the Higgs boson masses and couplings that we will need later. We then discuss 
a method which allows to derive the MSSM Lagrangian parameters of the gaugino 
sector from physical parameters such as the chargino and neutralino masses 
\cite{KM}. This ``inversion" of the MSSM spectrum is non--trivial, especially 
in the neutralino sector which involves a $4\times4$ matrix to de--diagonalize. 
The algorithm gives for a given $\tan\beta$, the values of the supersymmetric 
Higgs--higgsino parameter $\mu$ and the bino and wino soft--SUSY breaking
mass parameters $M_1$ and $M_2$, in terms of three arbitrary input masses, 
namely either two chargino and one neutralino masses, or two neutralino and 
one chargino masses. Some subtleties like the occurrence of discrete ambiguities
are illustrated and a few remarks on inversion in the sfermion and Higgs 
sectors are made. \s

Section 3.2 describes the Fortran code SUSPECT which calculates the masses
and couplings of the SUSY particles and the MSSM Higgs bosons \cite{SUSPECT}. 
The specific aim of this program is to fix a GDR--common set of parameter 
definitions and conventions, and to have as much as possible flexibility on 
the input/output parameter choice; it is hoped that it may be readily 
usable even with not much prior knowledge on the MSSM. In the present version,
SUSPECT1.1, only the two extreme ``models" pMSSM on the one side and mSUGRA 
on the other side are available. 
We describe the most important subroutines in the case of the pMSSM and 
then discuss the mSUGRA case with the various choices of approximations and 
refinements, paying a special attention on the renormalization group
evolution and the radiative electroweak symmetry breaking. We then briefly 
describe the input and output files and finally collect a list of various 
available options and/or model assumptions, with a clear mention 
of eventual limitations of the present version of the code, and list the
improvements which are planed to be made in the near future.  \\

In section 4, we discuss some aspects of the production of the MSSM Higgs 
bosons at future hadron and $\ee$ colliders, and their possible standard
and SUSY decays. We start in section 4.1 with Higgs boson production at the 
LHC and first make a brief summary in section 4.1.1 of the main production 
mechanisms of the lightest Higgs boson $h$ [in particular in the decoupling 
limit where it is SM--like] and the heavier particles $H,A$ and $H^\pm$, 
as well as the most interesting detection signals. \s

In section 4.1.2 we analyze the effects of stop loops on the main production 
mechanism of the lightest $h$ boson at the LHC, the gluon fusion mechanism 
$gg \ra h$, and on the important decay channel $h \ra \gamma \gamma$ 
\cite{ggh}. We show that if the off--diagonal entry in the $\tilde{t}$ 
mass matrix is large, the lightest stop  can be rather light 
and its couplings to the $h$ boson strongly enhanced; its contributions 
would then interfere destructively with the ones of the top quark, leading 
to a cross section times branching ratio $\sigma( gg \ra h) \times {\rm BR}(h 
\ra \gamma \gamma)$ much smaller than in the SM, even in the decoupling regime. 
This would make the search for the $h$ boson at the LHC much more difficult
than expected. Far from the decoupling limit, the cross section times 
branching ratio is further reduced due to the additional suppression of the 
Higgs couplings to SM fermions and gauge bosons. In the case of 
the heavy $H$ boson, squark loop contributions to the cross section 
$\sigma( gg \ra H)$ can be also large, while they are absent for the $A$ 
boson because of CP--invariance. \s

In section 4.1.3, we discuss the production of the light $h$ particle in 
association with light top--squark pairs at proton colliders, $pp \ra gg + 
q \bar{q} \ra \tilde{t} \tilde{t}h$ \cite{hstop}. The cross section for
this process can  substantially exceed the rate for the SM--like 
associated production with top quarks, especially for large values of the 
off--diagonal entry of the $\tilde{t}$  mass matrix which, as mentioned 
previously, make the lightest stop  much lighter than the  other squarks and 
increase its coupling to the $h$ boson. This process can strongly enhance 
the potential of the LHC to discover the $h$ boson in the $\gamma \gamma +$ 
lepton channel. It would also allow for the possibility of the direct 
determination of the $\tilde{t} \tilde{t} h$ coupling, the largest electroweak 
coupling in the MSSM, opening thus a window to probe directly the trilinear 
part of the soft--SUSY breaking scalar potential. Finally, this reaction could 
be a new channel to search for relatively light top squarks at hadron 
colliders. \s

In section 4.2 we analyze the possible decays into SUSY particles of the
neutral \cite{hdec} and charged \cite{franc} Higgs particles of the MSSM. 
For the light $h$ boson the only SUSY decay allowed by present experimental 
data are the invisible decay into a pair of lightest neutralinos or sneutrinos. 
The decays are possible only in small areas of the parameter space in the 
constrained MSSM; however, relaxing for instance the assumption of a universal
gaugino mass at the grand unification scale, leads  to possibly very light 
neutralinos and the decays into the latter states occurs in a much larger area.
Decays of the heavy neutral $H,A$ bosons into chargino/neutralino pairs and $H$
boson decays into stop pairs can be also dominant in some areas of the 
parameter space.  We then show that the decays of the $H^\pm$ particles into 
chargino/neutralino and slepton pairs are also still allowed and can be 
dominant in some areas of the parameter space;  we also briefly discuss some 
additional charged Higgs boson decay modes present in non--supersymmetric 
two--Higgs doublet models. The SUSY decays should not be overlooked as they 
can strongly suppress the branching ratios of the Higgs boson detection modes, 
and therefore might jeopardize the search for these particles at the LHC. \s

In section 4.3 we briefly summarize the main production mechanisms
of the MSSM Higgs bosons at $\ee$ colliders and describe the Fortran program 
HPROD \cite{hprod} which calculates the production cross sections for SM and
MSSM Higgs particles in $\ee$ collisions. In the SM, it includes the 
bremsstrahlung off the $Z$--boson line and the $WW/ZZ$ fusion processes; some 
higher order production processes, such the production in association with 
$t\bar{t}$ pairs and the Higgs boson pair production in the bremsstrahlung and 
the WW/ZZ fusion processes, are also included. For the MSSM CP--even Higgs 
bosons, it includes the Higgs--strahlung, the associated production with the 
pseudoscalar Higgs boson $A$, and the $WW/ZZ$ fusion processes; for the $H^\pm$
boson it includes the pair production in $\ee$ collisions as well as the top 
quark decay process. The complete radiative corrections in the renormalization 
group improved effective potential approach are incorporated in the program, 
which computes both the running and pole Higgs boson masses. The possibilities 
of having off--shell $Z$ or Higgs boson production in the bremsstrahlung and 
in the pair production processes, as well as initial state radiation, are 
allowed. Future improvements will be listed. \s

Finally, in section 4.3.2, we discuss the associated production of the lightest
$h$ boson with stop pairs in $\ee$ collisions \cite{eehtt}. The final state 
$\tilde{t}_1\tilde{t}_1 h$ can be generated in three ways: $(i)$ the production
of $\ee \ra \tilde{t}_2 \tilde{t}_1$ through $Z$--boson exchange and the 
subsequent decay $\tilde{t}_2 \ra \tilde{t}_1 h$; $(ii)$ the production in the 
continuum in $\ee$ collisions $\ee \ra \tilde{t}_1 \tilde{t}_1 h$ with the main contribution coming from $h$ emission from the $\tilde{t}_1$ lines; and $(iii)$
the production in the $\gamma \gamma$ mode of the $\ee$ collider, $\gamma 
\gamma \ra \tilde{t}_1 \tilde{t}_1 h$. Due to the clean environment of $\ee$ 
colliders, this final state might be easier to be detected than at the LHC if 
kinematically allowed, and would provide a more precise determination of the 
$\tilde{t}_1 \tilde{t}_1 h$ coupling.  \\

In Section 5, we discuss some of the production and decay properties of the 
SUSY particles as well as their virtual effects. We start in section 5.1 by 
discussing some of the virtual effects of charginos, neutralinos and sleptons 
of the first generation at LEP2 energies \cite{GDR1}. In the production of 
lepton pairs in $\ee$ collisions, box diagrams involving neutralino/selectron 
or charginos/sneutrino pairs occur and alter the production cross sections and 
asymmetries; at LEP2 energies the effects can be sizeable and experimentally 
measurable in the process $\ee \ra \mu^+\mu^-$ if the masses of the neutralinos 
and/or charginos are close to the beam energy due to threshold effects. \s

In section 5.2, we analyze the correlated production and decay of a pair
of the lightest charginos \cite{choi}\footnote{This contribution has not been,
strictly speaking, entirely made in the framework of the GDR--Supesrym\'etrie. 
However, one of the authors 
%[who is incidentally a member of the GDR, and
%has spent part of the last six months dealing with the problem] 
could not resist to the temptation of including it in this report since it is 
one of the hot topics of 
the GDR, especially in the working group ``Outils".}. We show that the chargino
polarization and the spin--spin correlations give two observables 
which do not depend on the final decays of the charginos, and hence on the
neutralino sector. Combined with the production cross section and with the 
chargino mass which can be measured via a threshold scan, these two observables
allow a complete determination of the SUSY parameters $\mu, M_2$ and $\tan 
\beta$ [and the sneutrino mass] in a completely model--independent way. With 
the knowledge of the lightest neutralino mass from the energy distribution of 
the final particles in the chargino decays, the parameter $M_1$ can be also 
determined, leading to a full reconstruction of the gaugino sector.  \s

Section 5.3 deals with the production of a chargino/neutralino pair at proton 
colliders, $pp\ra q q \ra \chi^\pm_1 \chi^0_2$. We first discuss the 
analytical expression of the tree--level cross section \cite{clermont}, that 
is calculated with the help of the FeynMSSM package that we briefly describe, 
and compare with the results available in the literature. We then discuss this 
process at the LHC, and propose to use the charge asymmetry in the 
three--leptons plus missing energy channel to determine some MSSM parameters 
\cite{steve}. In particular, it will be shown that 
this asymmetry depends only on the mass of the final state, and can be used to 
measure the sum of the chargino and neutralino masses in a model--independent 
way. \s

Some important three--body decay modes of sfermions \cite{yann} are 
discussed in section 5.4. In particular, we discuss the decays of the 
lightest stop into a $b$--quark, the lightest neutralino and a $W$ or $H^+$ 
boson which can compete with the loop--mediated decay into charm+neutralino
in the case of light stops. 
We also discuss decays of the heavier stop (sbottom) into the lighter one
and a fermion pair through off--shell gauge or Higgs bosons, as well
as decays of squarks into third generation sleptons+leptons (squarks+quarks) 
through a virtual exchange of a neutralino/chargino (gluino). These 
decays can have sizeable branching fractions in some areas of the MSSM
parameter space. \s

In section 5.5, we discuss stop and sbottom squark searches at LEP200
\cite{LEP200}. These squarks have a special place in the SUSY spectra due 
to the strong Yukawa couplings of their partners, the $t$ and $b$ quarks, 
and could have masses accessible at LEP200. After a brief analysis
of the cross section including all important radiative corrections, we
discuss the two main decay modes which are relevant for stop squarks
with masses accessible at LEP200, namely the loop induced flavor changing 
decay into a charm quark and the lightest neutralino, $\tilde{t}_1 \ra c 
\chi_{1}^0$, and the three--body decay $\tilde{t}_1 \rightarrow b l^+ \tilde 
\nu$ through the exchange of an off--shell chargino. Some remarks
will be given on the decays of a light sbottom. \\

Finally, section 6 deals with the limits and constraints on the SUSY particle
and MSSM Higgs boson masses from present experiments \cite{sylvie}, mainly 
from LEP and the Tevatron. At LEP, the searches for SUSY particles concern 
sleptons, stops, sbottoms, charginos and neutralinos, while at the Tevatron 
the main focus is on squarks and gluinos. The lightest CP--even $h$ and the 
CP--odd $A$ bosons are also searched for at LEP, while the charged Higgs bosons
$H^\pm$ are searched for at the Tevatron. Here, we will summarize the 
experimental limits on the production cross sections and masses of these
particles at LEP2 and the Tevatron, paying attention to all the decay channels. 
The most recent results have been used, including preliminary 
results reported at the last summer conference in Vancouver and the last LEPC 
meeting.

\newpage

\setcounter{equation}{0}
\renewcommand{\theequation}{2.\arabic{equation}}

\section*{2. The MSSM Spectrum} 

\subsection*{2.1. Definitions and properties} 

The Minimal Supersymmetric Standard Model (MSSM) is the most economical 
low--energy supersymmetric (SUSY) \cite{wessb} extension of the Standard Model 
(SM); for reviews see Refs.~\cite{R1,HaberKane,R2}. In this 
section, we discuss the basic assumptions which define the model and the 
various constraints which can be imposed on it. This will allow us to set 
our notations and conventions for the rest of the discussion. We will mainly 
focus on the unconstrained MSSM (uMSSM), what we will call the 
phenomenological MSSM (pMSSM) and the constrained minimal Supergravity 
(mSUGRA) model. 

\subsubsection*{2.1.1 The uMSSM: unconstrained MSSM} 

The unconstrained MSSM is defined by the following four basic assumptions: \s

\nn {\it (a) Minimal gauge group:} \s

The MSSM is based on the gauge symmetry ${\rm SU(3)_C \times  SU(2)_L \times 
U(1)_Y}$, i.e. the SM symmetry. SUSY implies then, that the spin--1 gauge
bosons, and their spin--1/2 superpartners the gauginos [bino $\tilde{B}$,
winos $\tilde{W}_{1-3}$ and gluinos $\tilde{G}_{1-8}$] 
are in vector supermultiplets. \s

\nn {\it (b) Minimal particle content:} \s

In the MSSM, there are only three generations of spin--1/2 quarks and leptons
[no right--handed neutrino] as in the SM. The left-- and right--handed 
chiral fields belong to chiral superfields together with their spin--0 SUSY
partners the squarks and sleptons: 
\beq
{\hat{ Q}} \ , \ {\hat{ u}}_{R} \ , \ {\hat{ d}}_{R} \ , \
{\hat{ L}} \ ,  \ {\hat{ l}}_{R} 
\eeq
In addition, two chiral superfields $\hat{H}_d$, $\hat{H}_u$ with respective 
hypercharges $-1$ and $+1$ for the cancellation of chiral anomalies, are needed
\cite{S3,inoue}. Their scalar components: 
\begin{eqnarray} 
H_d= \left( \begin{array}{c} 
               H_d^0 \\
               H_d^{-} \\
               \end{array} \right) \hspace{2cm}
H_u= \left( \begin{array}{c} 
               H_u^{+} \\
               H_u^0 \\
               \end{array} \right) 
\end{eqnarray}
give separately masses to the isospin +1/2 and $-$1/2 fermions. Their spin--1/2
superpartners, the higgsinos, will mix with the winos and the bino, to give
the mass eigenstates, the charginos $\chi_{1,2}^\pm$ and neutralinos
$\chi^0_{1,2,3,4}$. \s

\nn {\it (c) R--parity conservation:} \s

To enforce lepton  and baryon  number conservation, a discrete and 
multiplicative symmetry called R--parity is imposed \cite{S4}. 
It is defined by: 
\beq
R= (-1)^{2s+3B+L} \non 
\eeq
where L and B are the lepton and baryon numbers, and $s$ is the spin
quantum number. The R--parity quantum numbers are then $R=+1$ for
the ordinary particles [fermions, gauge and Higgs bosons], and $R=-1$ 
for their supersymmetric partners. In practice the conservation of
$R$--parity has important consequences: the SUSY particles are 
always produced in pairs, in their decay products there is always
an odd number of SUSY particles, and the lightest SUSY particle (LSP)
is absolutely stable.  \bigskip

The three conditions listed above are sufficient to completely 
determine  a globally supersymmetric Lagrangian. The kinetic part
of the Lagrangian is obtained by generalizing the notion of covariant 
derivative to the SUSY case. The most general superpotential [i.e. a 
globally supersymmetric potential] compatible with gauge invariance,
renormalizability and R--parity conserving is written as: 
\begin{equation}
W=\sum_{i,j=gen} - Y^u_{ij} \, {\hat {u}}_{R}^{i} \hat{H_u}.{\hat{ Q}}^j+
     Y^d_{ij} 
\, {\hat{ d}}_{R}^{i} \hat{H}_d.{\hat{ Q}}^j+
       Y^l_{ij} \,{\hat{ l}}_{R}^{i} \hat{H}_u.{\hat{ L}}^j+
     \mu \hat{H}_u.\hat{H}_d
\end{equation}
The product between SU(2)$_{\rm L}$ doublets reads $H.Q \equiv \epsilon_{a b} 
H^a Q^b$ where $a, b$ are SU(2)$_{\rm L}$ indices and $ \epsilon_{12}=1 = -
\epsilon_{21}$, and $Y^{u,d,l}_{ij}$ denotes the Yukawa couplings among 
generations. The first three terms in the previous expression are nothing
else than a generalization of the Yukawa interaction in the SM, while the
last term is a globally supersymmetric Higgs mass term. 
The supersymmetric part of the tree--level potential $V_{\rm tree}$ is the sum 
of the so--called F-- and D--terms \cite{S5}, where the 
F--terms come from the superpotential 
through derivatives with respect to all scalar fields $\phi_{a}$
\begin{equation}
V_{F}={\sum_{a} |W^{a}|^2} \ \ , \ \ \ W^{a} = \partial{W}/\partial{ \phi_a}
\end{equation}
and the D--terms corresponding to respectively ${\rm U(1)_Y}$, ${\rm SU(2)_L}$, 
and ${\rm SU(3)_C}$ gauge symmetries are given by
\begin{equation}
V_{D}= \frac{1}{2} (D_{1}D_{1}+ D_{2}D_{2}+ D_{3}D_{3} )
\end{equation}
with
\begin{eqnarray}
D_1&=&g_1 \bigg[{\sum_{i=gen}{({\frac{1}{6}} {\tilde{Q}}_i^{\dagger}
{\tilde{Q}}_i -{\frac{1}{2}} {\tilde{L}}_i^{\dagger} {\tilde{L}}_i 
-{\frac{2}{3}} {\tilde{u}}_{R_i}^{\dagger}{\tilde{u}}_{R_i} +
{\frac{1}{3}} {\tilde{d}}_{R_i}^{\dagger}{\tilde{d}}_{R_i}+  
{{\tilde{l}}_{R_i}^{\dagger}{{\tilde{l}}_{R_i} )}}}} 
 + {\frac{1}{2}}H_u^{\dagger} H_u-{\frac{1}{2}}H_d^{\dagger} H_d \bigg]\non \\
D_2&=&g_2 \bigg[{\sum_{i=gen}{({\tilde{Q}}_i^{\dagger} {\frac{\vec{\sigma}}{2}} 
{\tilde{Q}}_i + 
{\tilde{L}}_i^{\dagger} {\frac{\vec{\sigma}}{2}} {\tilde{L}}_i)+ 
H_u^{\dagger} {\frac{\vec{\sigma}}{2}}{H_u}
      +H_d^{\dagger} {\frac{\vec{\sigma}}{2}}{H_d} \bigg] }} \non \\
D_3&=&g_3 \bigg[{\sum_{i=gen}{{\tilde{Q}}_i^{\dagger} {\frac{\vec{\lambda}}{2}} 
{\tilde{Q}}_i- 
{\tilde{u}}_{R_i}^{\dagger} {\frac{\vec{\lambda^{*}}}{2}} {\tilde{u}}_{R_i}-
         {\tilde{d}}_{R_i}^{\dagger}{\frac{\vec{\lambda^{*}}}{2}} 
{{\tilde{d}}_{R_i} }}} \bigg]
\end{eqnarray}
Here the tildes denote the scalar quark and lepton fields, and   
$({\sigma_{k}})_{k=1-3}$ and $( {\lambda}_{k})_{k=1-8}$ the Pauli 
and Gell--Mann matrices; $g_{1,2,3}$ are the three gauge couplings. \s

For completeness we also write down the fermion--scalar sector 
of the supersymmetric part of the Lagrangian, which is needed for
later purpose. 
Following the conventions of Ref.~\cite{wessb} , with two--component spinors, 
this part of the Lagrangian contains on the 
one hand a purely chiral contribution
\begin{equation}
{\cal L}_{\rm chir.}=-{\frac{1}{2}}{\sum_{a,b} { {W^{ab}} \psi_{a} \psi_{b}}}+{\rm h.c.} \ \
 , \ \  {W^{ab}}=\partial^2{W}/\partial{ \phi_a \partial{ \phi_b}}
\end{equation}
where $\psi_{a}$ is the supersymmetric fermionic partner of $\phi_{a}$, and on 
the other hand a mixed chiral--vector contribution coming from the gauged 
matter field kinetic term yielding in component form 
\begin{equation}
{\cal L}_{\rm mix}=-i{\sqrt{2}} {\sum_{A,a}(D_{A})^{a} \psi_{a} \lambda_{A}}
+\ {\rm h.c.} \ \ , 
\ \, (D_{A})^{a} = \partial{D_{A}}/\partial{ \phi_a}
\end{equation}
The index {\small A} denotes the gauge group with which  the gaugino $\lambda_{A}$  is associated [there is of course also a contribution from the usual gauged kinetic 
term for the Higgs which we do not write here]. \\

\nn {\it (d) Soft--SUSY breaking:} \s

To break Supersymmetry, while preventing the reappearance of the quadratic
divergences [soft--breaking] we add to the supersymmetric Lagrangian a set
of terms which explicitly but softly break SUSY \cite{S6}: 
\begin{itemize} 

\item[$\bullet$] Mass terms for the gluinos, winos and binos:
\beq
- {\cal L}_{\rm gaugino}=\frac{1}{2} \left[ M_1 \tilde{B}  
\tilde{B}+M_2 \sum_{a=1}^3 \tilde{W}^a \tilde{W}_a +
M_3 \sum_{a=1}^8 \tilde{G}^a \tilde{G}_a  \ + \ {\rm h.c.} 
\right]
\eeq

\item[$\bullet$] Mass terms for the scalar fermions: 
\beq
-{\cal L}_{\rm sfermions} = 
{\sum_{i=gen} m^2_{{\tilde {Q}},i} {\tilde{Q}}_i^{\dagger}{\tilde{Q}}_i+
m^2_{{\tilde{ L}},i} {\tilde{L}}_i^{\dagger} {\tilde{L}}_i +
         m^2_{ {\tilde{u}},i} |{\tilde{u}}_{R_i}|^2+m^2_{ {\tilde{d}},i} 
|{\tilde{d}}_{R_i}|^2+  m^2_{{\tilde{l}},i} | {\tilde{l}}_{R_i}|^2}   
\eeq
\item[$\bullet$] Mass and bilinear terms for the Higgs bosons: 
\beq
-{\cal L}_{\rm Higgs} = m^2_{H_u} H_u^{\dagger} H_u+m^2_{H_d}  H_d^{\dagger} 
H_d + B \mu (H_u.H_d + {\rm h.c.} ) 
\eeq
\item[$\bullet$] Trilinear couplings between sfermions and Higgs bosons 
\beq
-{\cal L}_{\rm tril.}= 
{\sum_{i,j=gen} { \left[ A^u_{ij} Y^u_{ij}  {\tilde{u}}_{R_i} H_u. 
{\tilde{Q}}_j+
A^d_{ij} Y^d_{ij}  {\tilde{d}}_{R_i} H_d.{\tilde{Q}}_j
+A^l_{ij} Y^l_{ij} {\tilde{l}}_{R_i} H_u.{\tilde{L}}_j\ + \ {\rm h.c.} 
\right] }}
\eeq
\end{itemize} 
The soft--SUSY breaking scalar potential, which will be discussed later
in detail, is the sum of the three last terms:
\beq
V_{\rm soft} = -{\cal L}_{\rm sfermions} -{\cal L}_{\rm Higgs}-
{\cal L}_{\rm tril.}
\eeq
Up to now, no constraint is applied to this Lagrangian, although for generic
values of the parameters, it might lead to severe phenomenological problems, 
such as flavor changing neutral currents [FCNC], unacceptable amount 
of additional CP--violation, color and charge breaking minima, etc... The MSSM 
defined by the four hypotheses $(a)$--$(d)$ above, will be called the 
unconstrained MSSM or in short the uMSSM. 

\subsubsection*{2.1.2 The pMSSM: phenomenological MSSM}

The uMSSM contains a huge number of free parameters, which are mainly 
coming  from the scalar potential $V_{\rm soft}$. Indeed, the sfermions 
masses are in principle 
$3 \times 3$ hermitian matrices in generation space, with complex matrix
elements, leading to $5 \times 6 \times 2$ arbitrary parameters. The 
matrices $A^u$ and $A^d$ for the trilinear couplings on the other hand 
are arbitrary $3\times 3$ complex matrices in generation space leading
to $2 \times 9 \times 2 = 36$ parameters [the $A^l$ matrices are diagonal
and the matrix elements are real, since in the SM the neutrinos are 
massless\footnote{A signal for neutrino oscillations, thus implying  
neutrino masses, has been very recently reported by Super--Kamiokande 
\cite{kamio}. However, these neutrino masses are so tiny that they have 
no effect on the discussion given here.} 
and there is a separate conservation of the $e, \mu$ and $\tau$ lepton 
numbers, leading to a much smaller number of parameters]. 
Thus, if we allow for intergenerational mixing and complex phases, 
the soft--SUSY breaking terms will introduce a huge number of unknown
parameters, 105 parameters \cite{S7} in addition to the 19 parameters  
of the SM! This feature of course will make any phenomenological analysis 
a daunting task,  if possible at all, and in addition induces severe 
phenomenological problems as mentioned above. One definitely needs 
to reduce the number of free parameters  to be able to use the model 
in a reasonable and somewhat predictive way. \s

There are, fortunately, several phenomenological constraints which make
some assumptions reasonably justified to constrain the uMSSM. 
These assumptions will be discussed in the report of the ``Saveurs" 
group \cite{S8} to which we refer for details. Here we will simply 
and briefly mention them. \\

\nn {\it (a) No new source of CP--violation} \s

New sources of CP--violations are constrained by the experimental
limits on the electron and neutron electric moments and in the $K$
system [e.g. $\epsilon'/\epsilon]$ which are extremely tight. [Of course, 
since the phases at hand in the uMSSM are numerous, a kind of fine tuning 
can be made which will allow for cancelling contributions in the various 
quantities]. Assuming that all phases in the soft--SUSY breaking potential 
are zero eliminates all new sources of CP--violation and 
leads to a drastic reduction of the number of parameters. \\

\nn {\it (b) No Flavor Changing neutral currents} \s

The non--diagonal terms in the sfermion mass matrices and in the trilinear
coupling matrices, can induce large violations of FCNC which are severely 
constrained by present experimental data. Constraints have then to be
imposed to suppress operators which lead to these large effects. 
These constraints amount  to a severe limitation of 
the pattern of the sfermion mass matrices: either they are close to the 
unit matrix in flavor space, or they are almost proportional to the
corresponding fermion masses [flavor universality and  flavor alignment
respectively]. We will assume here that both the matrices for the sfermion 
masses
and for the trilinear couplings are diagonal, which also leads to a drastic 
reduction of the number of new parameters. \\

\nn {\it (c) First and Second Generation Universality}  \s

Experimental data, e.g. from $K^0$--$\bar{K}^0$ mixing, severely limit the 
splitting between the masses of the first and second generation squarks, unless 
squarks are significantly heavier than 1 TeV. One can assume therefore that 
the soft--SUSY breaking scalar masses are the same for the first and second 
generations. There is no experimental constraint on the third generation masses 
[note in addition that in this sector significant mass 
splitting between the mass eigenstates can be generated by the off--diagonal 
matrix elements, as will be discussed later]. Furthermore, one can assume also 
that $A^u, A^d$ and $A^l$ are the same for the two generations. In fact,
since they are always proportional to the fermion masses, these trilinear 
couplings are important only in the case of the third generation; one 
can therefore set the ones of the first two generations to zero without
any phenomenological consequence in this context.  \s

In addition, some parameters in the Higgs sector can be related to SM 
parameters, see below. Thus, making the three assumptions $(a)$--$(c$)  
will lead to the following 19 input parameters only: \s

\nn \hspace*{2cm} $\tan \beta$: the ratio of the vev of the two--Higgs doublet 
fields.\\
 \hspace*{2cm} $M_A$: the mass of the pseudoscalar Higgs boson \\
 \hspace*{2cm} $\mu$: the Higgs--higgsino mass parameter \\
 \hspace*{2cm} $M_1, M_2, M_3$: the bino, wino and gluino mass parameters. \\
 \hspace*{2cm} $m_{\tilde{q}}, m_{\tilde{u}_R}, m_{\tilde{d}_R}, 
               m_{\tilde{l}}, m_{\tilde{e}_R}$: first/second generation
 sfermion masses\\ 
  \hspace*{2cm} $m_{\tilde{Q}}, m_{\tilde{t}_R}, m_{\tilde{b}_R}, 
               m_{\tilde{L}}, m_{\tilde{\tau}_R}$: third generation
 sfermion masses\\
  \hspace*{2cm} $A_t, A_b, A_\tau$: third generation trilinear couplings. \s

Note that the remaining three parameters $m^2_{H_u}, m^2_{H_d}$ and $B$
are determined through the electroweak symmetry breaking conditions {\sl and} 
the value of $M_A$; alternatively, one can use directly the Higgs mass 
relations, which are equivalent to the electroweak symmetry breaking 
conditions, only when supplemented with an extra relation [see section 2.2 
for a further discussion of these issues.] \s

Such a model, with this relatively moderate number of parameters [especially
that, in general, only a small subset appears when one looks at a given sector
of the model] has much more predictability and is much easier to be discussed
phenomenologically, compared to the uMSSM. We will refer to the MSSM with the 
set of 19 free input parameters given above as the ``phenomenological" MSSM or 
pMSSM\footnote{Note, however, that at the time being the program SUSPECT which 
will be discussed later does not use in the option pMSSM, exactly the 
same set of input parameters as proposed above. The underlying physical 
assumptions are nevertheless identical [see section 3.2].}. 
 
\subsubsection*{2.1.3 mSUGRA: the constrained MSSM} 

All the phenomenological problems of the unconstrained MSSM discussed 
previously are solved at once if one assumes that the MSSM parameters obey 
a set of boundary conditions at the Unification scale.  These assumptions 
are natural [but not compulsory, see Ref.~\cite{nonuniversal} e.g.] 
in scenarii where the SUSY--breaking occurs in a hidden sector which
communicates with the visible sector only through gravitational 
interactions. These unification  and universality hypotheses are 
as follows \cite{S6}: 
\begin{itemize} 
\item[$\bullet$]  Gauge coupling unification: 
\beq
\alpha_{1} (M_U) = \alpha_2 (M_U) = \alpha_3 (M_U) \equiv \alpha_U
\eeq
with $\alpha_i=g_i^2/4\pi$. Strictly speaking, this is not an assumption 
since these relations are verified given the experimental results from LEP1 
\cite{LEPunif}. In fact, one can view these relations as fixing the Grand 
Unification scale $M_U$. 

\item[$\bullet$]  Unification of the gaugino masses: 
\beq
M_1 (M_U)=M_2(M_U)=M_3(M_U) \equiv m_{1/2}
\eeq
Since the gaugino masses and the gauge couplings are governed by the same 
renormalization group equations, the former at the electroweak scale
are given by:
\beq
M_i = \frac{\alpha_i(M_Z)}{\alpha_U} m_{1/2} \ \longrightarrow 
\ M_3(M_Z)=\frac{\alpha_3(M_Z)} {\alpha_2(M_Z)} M_2(M_Z)
          =\frac{\alpha_3(M_Z)} {\alpha_1(M_Z)} M_1(M_Z)
\eeq
For instance, one has the well--known relation: $M_1=\frac{5}{3} 
{\rm tg}^2\theta_W M_2\sim \frac{1}{2}M_2$. 

\item[$\bullet$] Universal scalar [sfermion and Higgs boson] masses
\beq
M_{\tilde{Q}} =M_{\tilde{u}_R} =M_{\tilde{d}_R} =M_{\tilde{L}} 
= M_{\tilde{l}_R} =M_{H_u} =M_{H_d} \equiv  m_0
\eeq

\item[$\bullet$] Universal trilinear couplings: 
\beq
A_u (M_U) = A_d (M_U) = A_l (M_U) \equiv  A_0
\eeq
\end{itemize} 

Besides the three parameters $m_{1/2}, m_0$ and $A_0$ the
supersymmetric sector is described at the GUT scale by the bilinear
coupling $B$ and the Higgs--higgsino mass parameter $\mu$. However, 
one has to require that electroweak symmetry breaking  takes
place. This results in two necessary minimization
conditions of the Higgs potential [see next section for details].
% at the low--energy scale in the tree 
%approximation, they are given by 
%\begin{eqnarray}
%{1\over 2}M_Z^2={{m_{H_d}^2-m_{H_u}^2\tan ^2\beta }
%\over {\tan ^2\beta -1}}-\mu ^2 \ , \ 
%B\mu ={1\over 2}(m_{H_u}^2+m_{H_d}^2+2\mu ^2)\sin 2\beta \;
%\end{eqnarray}
%For given values of the parameters $\tan \beta, m_{H_u}$ and $m_{H_d}$ 
The first minimization equation can be solved for $|\mu|$; 
the second equation can then be solved for $B$. 
Therefore, in this model, we will have only four continuous 
and one discrete free parameters:
\beqn
\tan \beta \ , \ m_{1/2} \ , \ m_0 \ , \ A_0 \ , \ \ {\rm sign}(\mu)
\eeqn
This model is clearly appealing and suitable for 
thorough phenomenological and experimental scrutinity. This constrained
model, is usually referred to as the minimal Supergravity model, or mSUGRA. 
In addition, one can also require the unification of the 
top, bottom and tau Yukawa couplings 
at the GUT scale \cite{btau}. This would lead to a further constraint if 
the ``fixed point" solutions are chosen. Depending on whether one 
includes or not the top Yukawa couplings, the parameter $\tb$ should be
either small $\tan \beta \sim 1.5$, or large $\tb \sim 50$ \cite{btaup}. 
The values taken by the parameter $A_0$ happen to be also constrained in 
this case; a situation which further reduces the number of free parameters. 
 
\subsubsection*{2.1.4 The MSSMi: the intermediate MSSMs} 

mSUGRA is a well defined model of which the possible phenomenological 
consequences and experimental signatures have been widely studied in the 
literature. However, it should not be considered as THE definite model, in the 
absence of a truly fundamental description of SUSY--breaking. Indeed, some 
of the assumptions inherent to the model might turn out not to
be correct. In fact, in many models some of the universality conditions 
of mSUGRA are naturally violated; see e.g. 
Ref.~\cite{nonuniversal,SO10} for a discussion. \s

To be on the safe side from the experimental point of view 
it is therefore wiser to depart from this model, and to study the 
phenomenological implications of relaxing some defining assumptions.
However, it is desirable to limit the number of extra free parameters, 
in order to retain a reasonable amount of predictability 
when attempting detailed investigations 
of possible signals of SUSY. Therefore, it is more interesting to relax 
only one [or a few] assumption at a time and study the phenomenological 
implications. Of course, since there are many possible directions, 
this would lead to several intermediate MSSM's between mSUGRA
and pMSSM, denoted here by MSSMi's [with i an integer 
and finite, although possibly large, number]. Some of these MSSMi's are 
similar to the Minimal Reasonable Model discussed 
recently \cite{MRSSM}. 
A partial list of possible MSSMi's can be as follows
[many other possibilities are of course possible including the relaxation
of two assumptions at a time and  the introduction of an amount of 
CP--violation]:  \\

\nn {\it (1) MSSM1: mSUGRA with no sfermion and Higgs boson 
mass unification}: \s

The Higgs sector of the MSSM can be studied in a relatively model
independent way, since at the tree--level only two input parameters
are needed: $\tb$ and one of the Higgs boson masses. Although the 
squark mass parameters and
the trilinear couplings of the third generations, as well as the parameter
$\mu$ enter through radiative corrections, one can study their impact in 
a thorough manner without invoking any strong assumption [c.f. the LEP
analyses]. The mSUGRA model is therefore too restrictive, and to have more
freedom, one can 
decouple the Higgs sector for the squark sector by relaxing the equality
of the sfermion and Higgs boson masses in eq.~(2.18):
\beq
m_{\tilde{Q}} =m_{\tilde{u}_R} =m_{\tilde{d}_R} =m_{\tilde{L}}
=m_{\tilde{e}_R} \neq M_{H_u} =M_{H_d}
\eeq
Different sfermion and Higgs boson masses are in fact suggested by some 
SUSY--GUT theories e.g. based on SO(10) \cite{SO10}. 
One would then have an additional 
parameter since one of the Higgs masses for instance will remain free. \\

\nn {\it (2) MSSM2: mSUGRA without sfermion mass unification}: \s

Scenarii with light stops are appealing in several respects; for instance,
a light stop with a mass of the order of 100 GeV might trigger Baryogenesis
at the electroweak scale \cite{baryogen}. However, it is rather difficult, 
with the scalar
mass unification assumption to have a light stop while the other squarks 
remain rather heavy [without e.g. large $A_t$ values] . One can then disconnect 
the third generation from the 
first two ones [where the experimental constraints on FCNC are most stringent]
and allow for non universal scalar masses in the third generation squarks
[and similarly for sleptons] 
\beq
m_{\tilde{Q}} \neq m_{\tilde{t}_R} \neq m_{\tilde{b}_R} \neq 
m_{\tilde{L}} \neq m_{\tilde{\tau}_R}  \neq  m_{\tilde{f}} 
\eeq
with $m_{\tilde{f}}$ the (common) mass parameter of the first/second 
generation sfermions. \\

\nn {\it (3) MSSM3: mSUGRA with no gaugino mass unification}: \s

The chargino/neutralino sector depends only on three parameters: 
$\mu, M_2$ and $\tb$ if gaugino unification is assumed. This allows to 
make thorough experimental analyses 
which led to important results, such that the mass of the lightest
neutralino should be larger than $\sim 40$ GeV [see section 6]. Furthermore, 
searches for charginos at LEP2 and gluinos at the Tevatron are connected since
the gaugino masses are related. One can go one step downwards  in
model--dependence and relax the gaugino mass unification: 
\beq
M_1 (M_U) \neq M_2(M_U) \neq M_3(M_U)
\eeq
This would make the connection between the chargino, neutralino  and the 
gluino sectors less strong and will e.g. leave open the possibility 
of very light neutralinos. \\

\nn {\it (4) MSSM0: pMSSM with partial mass and coupling unification} \s

This model, with 7 free parameters, is the most used in phenomenological 
analyses: 
\beqn
\tb \ \ , \ \ M_A \ \ , \ \ \mu \hspace*{5cm} \non \\
M_1 (M_U) = M_2(M_U)= M_3(M_U)=m_{1/2}  \ , 
\ A_t=A_b=A_\tau=A_0 
\hspace*{1.5cm} \non \\
m_{\tilde{Q}} = m_{\tilde{t}_R} = m_{\tilde{b}_R} =
m_{\tilde{q}} = m_{\tilde{u}_R} = m_{\tilde{d}_R} \ \ , \ 
m_{\tilde{L}} = m_{\tilde{\tau}_R} = m_{\tilde{l}} 
=m_{\tilde{l}} = m_{\tilde{e}_R}
\eeqn

\subsection*{2.2 Electroweak Symmetry Breaking}

\subsubsection*{2.2.1 General features}

We turn now to the discussion of the electroweak symmetry breaking (EWSB). 
Using the notations introduced in the previous section, the Higgs potential 
takes the form   
\begin{eqnarray} 
V_{\rm Higgs}&=& V_{\rm tree} +V_1 \non \\
V_{\rm tree} &=& 
(m^2_{H_d}+\mu ^2) H_d^{\dagger} H_d + (m^2_{H_u}+\mu ^2) H_u^{\dagger} H_u + B \mu (H_u.H_d +
{\rm h.c.}) \nonumber \\
 &&+ \frac{g^2}{8}  ( H_d^{\dagger} H_d - H_u^{\dagger} H_u)^2 + \frac{g_2^2}{2} 
(H_d^{\dagger}  H_u) (H_u^{\dagger}  H_d) \label{vhiggs}
\end{eqnarray}
where $g^2 \equiv g_1^2 + g_2^2$; in some cases we will also use the 
shorthand notations:
\beq
m_1^2=m^2_{H_d}+\mu ^2 \ , \   
m_2^2=m^2_{H_u}+\mu ^2  \ , m_3^2 =- B \mu 
\eeq
$V_1$ contains the higher order corrections to this potential. Here we 
consider solely the one--loop corrections which have the well--known form
in the $\overline{\rm MS}$ scheme \cite{CW}
\begin{equation}
V_1= \frac{\hbar}{64 \pi^2} {\rm Str}[ M^4 ({\rm Log} \frac{M^2}{\mu_R^2} -
3/2) ] \label{EP}
\end{equation}
where $\mu_R$ denotes the renormalization scale, $M^2$ the field dependent 
squared mass matrix of the scalar or vector or fermion fields, and Str$[...] 
\equiv \sum_{\rm spin} (-1)^{2 s} (2 s + 1) (...)_s $, where the 
sum runs over gauge boson, fermion and scalar contributions.\s

Even if no model assumption apart from minimal Supersymmetry is made, i.e. 
no unification of the gauge couplings, no universality of the soft--SUSY
breaking terms, no Yukawa coupling unification, etc... it is clearly important 
to still require the electroweak symmetry breaking to take place. The usual 
necessary condition for EWSB is obtained from 
eq.~(\ref{vhiggs}) as  
\begin{eqnarray}
\frac{1}{2} M_Z^2 = \frac{\overline{m}_1^2 - \overline{m}_2^2 \tan^2 \beta}
{\tan^2 \beta -1 }  &,&
\sin 2\beta= \frac{-2 \overline{m}_3^2}{\overline{m}_1^2 + \overline{m}_2^2}
\label{EWSBcond} 
\end{eqnarray}
with  $M_Z^2=g^2/4 ( v_u^2 + v_d^2) $ and $\tan \beta= v_u/v_d$, where we 
assumed 
\begin{eqnarray} 
<H_d> \ = \ \frac{1}{\sqrt{2}}\left( \begin{array}{c} 
               v_d \\
               0 \\
               \end{array} \right) \hspace{2cm}
<H_u>\ = \ \frac{1}{\sqrt{2}}\left( \begin{array}{c} 
               0 \\
               v_u \\
               \end{array} \right)
\label{neutrdir} 
\end{eqnarray}
Eqs.~(\ref{EWSBcond}) contain two complementary and necessary requirements,
for {\sl i)} the breaking of the electroweak symmetry, 
{\sl ii)} the Z mass value to be reproduced correctly through this breaking.
These two conditions are however generally not sufficient to ensure 
electroweak symmetry breaking. For one thing, beyond tree--level they only 
express the existence of a stationary point not necessarily a global 
(nor even local) minimum of the effective potential. In a phenomenological 
analysis one then usually
checks numerically for the globality of the minimum [see next section for
an analytical approach]. The other reason is the possible existence of
color or charge breaking minima which can be either lower than the electroweak
minimum or sufficiently stable to become dangerous from a cosmological point
of view [we will have, however, nothing to say about these minima in the present report].
\s

Thus it should be clear that even in the unconstrained MSSM one should at least
impose the constraints of eq.~(\ref{EWSBcond}). 
These equations correlate not only
the parameters of the Higgs sector, but actually all the other parameters
of the model when the radiative corrections to $V_{\rm Higgs}$ are taken 
into account. Then one has typically 
\begin{eqnarray}
\overline{m}^2_1 &=& m^2_{H_d}+\mu^2 +\mbox{rad. corr.}\, (v_u, v_d) 
\label{mbar1} \non \\
\overline{m}^2_2 &=& m^2_{H_u}+\mu^2 + \mbox{rad. corr.}\, (v_u, v_d)
 \label{mbar2} \non \\
\overline{m}^2_3 &=& - B \mu  \label{mbar3}      
\end{eqnarray}
In this case eqs.~(\ref{EWSBcond}) are no more quadratic in 
$\tan \beta$ and linear in $\sin 2 \beta$ [not even polynomial 
in these variables anymore]
so that one usually resorts to numerical methods in solving
these equations. At this level we should perhaps restate a question of
terminology. What is usually called {\sl radiative} electroweak symmetry
breaking is the fact that eqs.~(\ref{EWSBcond}) are satisfied at a given
scale [presumably the electroweak scale] through the running of the quantities
which enter these equations, down from a high scale where some initial
conditions were assumed. Relaxing the radiative breaking requirement simply
means that one no more insists on starting from a high scale and specifying
initial conditions, but requires directly the electroweak symmetry breaking.
This would be a fully model--independent but still a physically consistent
approach. Of course one could eventually run the masses and couplings
up to a high scale to assess their consistency with any model assumption
at that scale.\s

Finally one can also implement eqs.~(\ref{EWSBcond}) indirectly. For the sake
of illustration we give here a tree--level example: the usual tree--level
Higgs boson mass relations
\begin{eqnarray}
M^2_{h, H}&=&\frac{1}{2}(M_Z^2 + M_{A}^2 \mp \sqrt{(M_Z^2 + M_{A}^2)^2 
- 4 M_Z^2 M_{A}^2 \cos^2 2 \beta}\;\; ) \non \\
M^2_{H^\pm} &=& M_{A}^2 + M_W^2 \label{hneut}
\end{eqnarray}
which are a consequence of the special form of the tree--level part of the 
Higgs potential eq.~(\ref{vhiggs}) and of eqs.~(\ref{EWSBcond}). The four free
parameters $ m_1^2, m_2^3, m_3^2$ and $\tan \beta$ reduce to two free
parameters due to the constraints of eq.~(\ref{EWSBcond}), usually chosen as 
$M_A$ and $\tan \beta$ in eqs.~(\ref{hneut}). \s

However one can show explicitly that eqs.~(\ref{hneut}) do 
not necessarily imply electroweak symmetry breaking
i.e. using expressions with non--zero masses is not sufficient to 
ensure vacuum stability.
Such mass relations can be realized and still have
eqs.~(\ref{EWSBcond}) violated. Only if one imposes on top of these mass
relations $M_A$ to be given by
\beq
M_{A}^2 = m_1^2 + m_2^2 
\eeq 
that EWSB is assured. Also this result is established at tree--level and it is 
not clear whether it still holds when loop corrections are taken into account.
Thus it is always safer to check explicitly eqs.~(\ref{EWSBcond}) whenever 
possible. 

\subsubsection*{2.2.2 EWSB and model--independent $\tan \beta$ bounds}

In this section we report on some analytical results for model independent 
theoretical bounds on $\tan \beta$ obtained from the study of electroweak 
symmetry breaking conditions to one--loop order \cite{LMM, LMM1}. 
The point is to use, on top of eqs.~(\ref{EWSBcond}),
the positivity of the Higgs boson squared masses  which are needed to ensure a, 
at least local, minimum in the Higgs sector. We then study analytically how
these constraints translate into bounds on $\tan \beta$
which are then necessary and fully model--independent bounds on this parameter.
We recall here that the positivity of the squared Higgs boson masses is 
automatically satisfied as a consequence  of eq.~(\ref{EWSBcond}) 
at the tree--level [or tree--level renormalization group improved] effective 
potential.
However, this property is not expected to be
generic when the one--loop corrections are taken into account 
[beyond the ones included in the runnings]. This was explicitly shown
in Ref.~\cite{LMM} in a specific approximation.
Since the full one--loop effective potential
has a complicated form we relied on two different approximations and showed
that they both lead to qualitatively similar results.
The first of these approximations consists in absorbing {\sl all}
logarithms of the one--loop effective potential in the runnings of the
tree--level quantities thus casting $V_{\rm Higgs}$ 
in the form, see eqs.~(\ref{vhiggs},\ref{EP}),
\begin{equation}
 V_{\rm Higgs}= \overline{V}_{\rm tree}(\mu_R^2) + \frac{\hbar}{64 \pi^2} 
\left( -\frac{3}{2}\right) {\rm Str} M^4\label{Poteff}
\end{equation}

~From now on we refer to this approximation as the Supertrace approximation.
Here $\overline{V}_{\rm tree}(\mu_R^2)$ is obtained from $V_{\rm tree}$ by 
replacing all the tree--level quantities by their running counterparts.
This would be fully justified in a model with just one mass scale and would
mean that we have resummed to all orders the leading logarithms in the
$\overline{\rm MS}$ scheme. Note that the residual one--loop correction in
eq.~(\ref{Poteff}) is scheme dependent but should be consistently kept 
as a residual correction [and not
reabsorbed in the runnings as it is sometimes suggested] since it would 
otherwise jeopardize the resummation procedure; see for instance 
Ref.~\cite{bando}.
Of course the MSSM has many different mass scales and the above approximation
is therefore very rough. It has however the merit of allowing a full analytical proof
of the existence of bounds on $\tan \beta$ free from any phenomenological 
assumption. This is significant in the sense that our approximation with
one mass scale tends to increase the symmetry of $V_{\rm Higgs}$ so that if new
bounds appear because of the difference in structure between tree--level and
one--loop, then these would hardly disappear in more realistic, and less
symmetric, approximations. \s

The analysis will not be carried further here,
the interested reader is referred to Ref.~\cite{LMM} for full details. 
Hereafter we only give the generic form of the $\tan \beta$ bounds
and then discuss briefly how some of these bounds can arise also in a more 
realistic approximation, namely the top/stop-bottom/sbottom approximation. \s

In the Supertrace approximation, the bounds on $\tan \beta$ read:
\begin{eqnarray}
\mbox{\, if \,} \tan \beta > 1 \mbox{:}\;\;\;
&\tan \beta_{-} \leq 
\tan \beta \leq
{\rm Min}(\tan \beta_{+}, -X_{m_1}^2/ X_{m_3}^2 )& 
\end{eqnarray}
where 
\begin{eqnarray}
\tan \beta_{-}={\rm Min}( T_{+}, m_t/m_b) \ \ 
\mbox{and} \;\;\;\tan \beta_{+}={\rm Max}( T_{+}, m_t/m_b) 
\end{eqnarray}
\begin{eqnarray}
\mbox{\, if \,} \tan \beta < 1  \mbox{:} \;\;\; &{\rm Max}(T_{-}, 
-X_{m_3}^2/X_{m_2}^2 ) \leq \tan\beta <1 & 
\end{eqnarray}
where 
\begin{equation}
T_{\pm} = \frac{1}{2 X_{m_3}^2} 
\left[ -X_{m_1}^2 - X_{m_2}^2 \mp \sqrt{(X_{m_1}^2 + X_{m_2}^2)^2 -
4  X_{m_3}^4}  \right] 
\end{equation}
The $X_{m_i}^2$  are generalizations of the $\overline{m}_i^2$'s
which include residual one--loop corrections from eq.~(\ref{Poteff}).
Note here that these bounds are slightly improved with respect to the 
ones in Ref.~\cite{LMM} due to the presence of $-X_{m_1}^2/X_{m_3}^2$ 
and $-X_{m_3}^2/ X_{m_2}^2$. Since we choose by convention $\tan \beta > 
0$, the positivity of $M_A^2$ translates into $X_{m_3}^2 < 0$, while 
the signs of $X_{m_1}^2$ and $X_{m_2}^2$ are not fixed. \\

We come now to the top/stop--bottom/sbottom approximation and give here a little
more details about our approach.
In this approximation, the one--loop correction 
to the Higgs effective potential is approximated by  
\begin{eqnarray}
\label{potet}
V_1&=& {\frac{6 \hbar}{64 \pi^2}} \bigg[ \sum_{i=1,2}{m_{\tilde{t}_{i}}^4 
\bigg( {\rm Log}[{\frac{m_{\tilde{t}_{i}}^2}{Q^2}}]}-{\frac{3}{2}}\bigg)
  -2 m^4_{t}\bigg( {\rm Log}[{\frac{m^2_{t}}{Q^2}}]-{\frac{3}{2}}\bigg) 
\bigg]  \nonumber \\
&&+{\frac{6 \hbar}{64 \pi^2}} \bigg[ \sum_{i=1,2}{m_{\tilde{b}_{i}}^4
\bigg( {\rm Log}[{\frac{m_{\tilde{b}_{i}}^2}{Q^2}}]}-{\frac{3}{2}} \bigg)
         -2 m^4_{b} ({\rm Log}[{\frac{m^2_{t}}{Q^2}}]-{\frac{3}{2}}
\bigg) \bigg] 
\end{eqnarray}
Taking a linear combination of the two stationarity conditions of
$V_{\rm Higgs}$ at the 
electroweak minimum, and defining $u\equiv v_u v_d$ and $t\equiv \tan \beta$, 
one finds
\begin{equation}
\label{TT0}
{\cal{T}}_0\equiv {\cal{K}}_2(u) u^2 + {\cal{K}}_1(u)  u +{\cal{K}}_0(u) t =0
\end{equation}
where
\begin{equation}
\label{K2}
{\cal{K}}_2(u)= {\frac{\kappa_0}{96}} (t^2-1)^2 \bigg[
{\frac{ (3 g_2^2-5 g_1^2)^2}
{(m_{\tilde{t}_{1}}^0)^2-(m_{\tilde{t}_{2}}^0)^2}}
{\rm  Log}[\frac{(m_{\tilde{t}_{1}}^0)^2}{(m_{\tilde{t}_{2}}^0)^2}]+ 
{\frac{ (3 g_2^2- g_1^2)^2}{(m_{\tilde{b}_{1}}^0)^2-(m_{\tilde{b}_{2}}^0)^2}}
{\rm   Log} [\frac{(m_{\tilde{b}_{1}}^0)^2}{(m_{\tilde{b}_{2}}^0)^2}] \bigg] 
\end{equation}
\begin{equation}
\label{K1}
{\cal{K}}_1(u)={\cal{K}}_1^{(0)}(u)+{\cal{K}}_1^{(1)}(u)+{\cal{K}}_1^{(2)}(u)
\end{equation}
with 
\begin{eqnarray}
{\cal{K}}_1^{(0)}(u)&=&   \kappa_0 t \bigg[ 6 Y_t^4 t^2 {\rm Log}
[{\frac{(m_{\tilde{t}_{1}}^0)^2 (m_{\tilde{t}_{2}}^0)^2}{(m_{t}^0)^4}}]+
 6 Y_b^4 {\rm Log} [{\frac{(m_{\tilde{b}_{1}}^0)^2 (m_{\tilde{b}_{2}}^0)^2}
{(m_{b}^0)^4}}] \bigg]     \nonumber \\ && \nonumber \\
{\cal{K}}_1^{(1)}(u)&=& -{\frac{3 g^2}{4}}  (t^2-1) 
\kappa_0 t \bigg[ Y_t^2 ({\rm Log}[{\frac{(m_{\tilde{t}_{1}}^0)^2}{Q^2}}]+
{\rm Log} [{\frac{(m_{\tilde{t}_{2}}^0)^2}{Q^2}}]-2) \nonumber \\
&& \hspace*{2cm} -Y_b^2 ({\rm Log} [{\frac{(m_{\tilde{b}_{1}}^0)^2}{Q^2}}]+
{\rm Log}[{\frac{(m_{\tilde{b}_{2}}^0)^2}{Q^2}}]-2)  \bigg] \\ \nonumber && \\
{\cal{K}}_1^{(2)}(u)&=& \kappa_0 Y_t^2 
{\frac{{\rm Log} [(m_{\tilde{t}_{1}}^0)^2/(m_{\tilde{t}_{2}}^0)^2]}
{(m_{\tilde{t}_{1}}^0)^2-(m_{\tilde{t}_{2}}^0)^2}} 
\bigg[ 12 t Y_t^2(A_t t-\mu)^2 -\frac{1}{2}(3g_2^2-5 g_1^2) (t^2-1) t \non \\ 
&& \times (m^2_{{\tilde{Q}}_3}-m^2_{{\tilde{u}}_3}) 
-\frac{3}{4} [g^2 (t^2-1)- 8 Y_t^2 t^2] (A_t t-\mu) (A_t-\mu t) \bigg]   
\nonumber \\
&&+ \kappa_0 Y_b^2 {\frac{{\rm Log} [(m_{\tilde{b}_{1}}^0)^2/
(m_{\tilde{b}_{2}}^0)^2]}{(m_{\tilde{b}_{1}}^0)^2-(m_{\tilde{b}_{2}}^0)^2}}
\bigg[ 6 Y_b^2 [(A_b t-\mu) (A_b-\mu t)+ 2t (A_b-\mu t)^2] \nonumber \\
&&+\frac{3}{2}(g_2^2-g_1^2) (t^2-1) t (m^2_{{\tilde{Q}}_3}-m^2_{{\tilde{d}}_3})
+\frac{ 3}{4} g^2 (t^2-1)(A_b t-\mu) (A_b-\mu t) \bigg]  \non 
\end{eqnarray}
The last term in ${\cal{T}}_0$ takes the following form:
\begin{eqnarray}
\label{K0}
{\cal{K}}_0(u)&=&  \frac{1}{2} (1+t^2) \; \Delta_{A}^0(u)  +
t \tilde{{\cal{K}}} _0(u)
\end {eqnarray}
where:
\begin{eqnarray}
\label{delta0}
\Delta_{A}^0 (u) \equiv  -2 \kappa_0 \mu 
\bigg\{ 6 A_t Y_t^2 \bigg[ \bigg( {\rm Log} [\frac{(m_{\tilde{t}_{1}}^0)^2
(m_{\tilde{t}_{2}}^0)^2} {(Q^2)^2} ] - 2\bigg) 
+ \frac{ {\rm Log}[\frac{(m_{\tilde{t}_{1}}^0)^2}{(m_{\tilde{t}_{2}}^0)^2}]}
{(m_{\tilde{t}_{1}}^0)^2-(m_{\tilde{t}_{2}}^0)^2} 
(m^2_{{\tilde{Q}}_3}+m^2_{{\tilde{u}}_3}) \bigg] \nonumber \\
 + 6 A_b Y_b^2 \bigg[ \bigg( {\rm Log}[\frac{(m_{\tilde{b}_{1}}^0)^2
(m_{\tilde{b}_{2}}^0)^2}{(Q^2)^2}]-2 \bigg)
+ \frac{{\rm Log}[\frac{(m_{\tilde{b}_{1}}^0)^2}{(m_{\tilde{b}_{2}}^0)^2}]}
{(m_{\tilde{b}_{1}}^0)^2-(m_{\tilde{b}_{2}}^0)^2}
(m^2_{{\tilde{Q}}_3}+m^2_{{\tilde{d}}_3}) \bigg] \bigg\}+2 m_3^2 
\hspace*{0.5cm}
\end{eqnarray}
and 
\begin{eqnarray}
\label{KK0}
\tilde{{\cal{K}}} _0 (u)&\equiv&  m_1^2+m_2^2+6 \kappa_0  \bigg\{ 
Y_t^2 \bigg[ \bigg( {\rm Log} [{\frac{(m_{\tilde{t}_{1}}^0)^2}{Q^2}}]
+{\rm Log}[{\frac{(m_{\tilde{t}_{2}}^0)^2}{Q^2}}]-2 \bigg) 
(A_t^2+\mu^2+m^2_{{\tilde{Q}}_3}+m^2_{{\tilde{u}}_3}) \nonumber \\
&&+{\frac{{\rm Log} [(m_{\tilde{t}_{1}}^0)^2/(m_{\tilde{t}_{2}}^0)^2]}{
(m_{\tilde{t}_{1}}^0)^2-(m_{\tilde{t}_{2}}^0)^2}} 
\big( (A_t^2+\mu^2) (m^2_{{\tilde{Q}}_3}+m^2_{{\tilde{u}}_3})+(m^2_{
{\tilde{Q}}_3}-m^2_{{\tilde{u}}_3})^2 \big) \bigg] \nonumber \\
&&+ Y_b^2 \bigg[ \bigg( {\rm Log} [{\frac{(m_{\tilde{b}_{1}}^0)^2}{Q^2}}]+
{\rm Log} [{\frac{(m_{\tilde{b}_{2}}^0)^2}{Q^2}}]-2 \bigg) 
( A_b^2+\mu^2+m^2_{{\tilde{Q}}_3}+m^2_{{\tilde{d}}_3})  \nonumber \\
&&+ \frac{{\rm Log}[(m_{\tilde{b}_{1}}^0)^2/(m_{\tilde{b}_{2}}^0)^2]}
{(m_{\tilde{b}_{1}}^0)^2-(m_{\tilde{b}_{2}}^0)^2} 
\bigg( (A_b^2+\mu^2) (m^2_{{\tilde{Q}}_3}+m^2_{{\tilde{d}}_3})+
(m^2_{{\tilde{Q}}_3}-m^2_{{\tilde{d}}_3})^2 \bigg) \bigg] \bigg\} 
\end{eqnarray}
[with $\kappa_0\equiv \hbar/(64 \pi^2)$.]\s

In deriving the above formulae, the dependence on the Higgs fields
in the logarithms was fully taken into account, together with the 
convention $m_{\tilde{t}_1}^2 \geq  m_{\tilde{t}_2}^2, m_{\tilde{b}_1}^2 
\geq  m_{\tilde{b}_2}^2 $. The key point now is  to note that on one hand 
\begin{equation}
{\cal{K}}_2 (u)\geq 0
\label{ineqK2}
\end{equation}
and on the other hand
\begin{equation}
{\cal{K}}_1 (u) t \geq 0
\label{ineqK1}
\end{equation}
The first inequality is obvious from eq.~(\ref{K2}), while the second requires
some mild phenomenological assumptions to overcome the analytical complexity
of eq.~(\ref{K1}) Let us give here just two examples
of such assumptions, leaving a more detailed study to Ref.~\cite{LMM1}. 
$(i)$ If the mixing between the stop eigenstates is weak, 
$ A_t \tan \beta - \mu \ll 1$, and neglecting the gauge contributions
to ${\cal{K}}_1$, one easily sees that ${\cal{K}}_1$ is dominated by
${\cal{K}}_1^{(0)}$ which behaves like $m_t^4$. Since in this limit 
the stops are almost degenerate and heavier than the top [assuming
all squared soft masses to be positive], then eq.~(\ref{ineqK1}) is
readily verified. $(ii)$ A second example of mild phenomenological 
assumption, is to take the heaviest stop  mass larger than $\sim 
360$ GeV, and $\tan \beta$ small $(\gsim 1$). Combined with 
the experimental lower bound on the lightest stop [$ \sim 80$ GeV,
see section 6], one again obtains eq.~(\ref{ineqK1}). \s

\begin{equation}
{\cal{K}}_0 t \leq 0
\end{equation}
The positivity of $M_A^2$ is not automatic beyond the tree--level, and
should be imposed explicitly] one retrieves a generalization of the 
$T_{\pm}$ bounds of the previous approximation. With the convention $\tan 
\beta \geq 0$,
which we did not need to impose in the previous discussion, the generalized
$T_{\pm}$ bounds read:
\begin{eqnarray}
\label{nTpm}
&&{\rm for} \ \ \tan \beta \ge 1 \ \ , \ \  \tan \beta \ge T_+ \\
&&{\rm for} \ \ 0 \le \tan \beta \le 1 \ \ , \ \ \tan \beta \le T_{-} 
\end{eqnarray}
with ($T_{+} \ge T_{-}$):
\begin{equation}
T_{\pm}= {\frac{-\tilde{{\cal{K}}} _0\pm {\sqrt{(\tilde{{\cal{K}}} _0)^2-  ({\Delta_{A}^0})^2}}}{ {\Delta_{A}^0} }} 
\end{equation}   
In summary, we have determined analytically and in a model--independent 
context, calculable bounds on $\tan \beta$. These bounds are nothing but
partial necessary constraints coming from the requirement of electroweak
symmetry breaking in the MSSM. Since these {\sl necessary} bounds  
are calculable in terms of the parameters of the MSSM, they can be used to 
delineate domains which are theoretically inconsistent with EWSB.
This would be a valuable guide in the standard procedure of the
numerical check of electroweak symmetry breaking, where one can avoid
from the start inconsistent domains in the input parameter space.
Furthermore, we emphasize that the above bounds have in principle a wider
applicability, and are more quantitative, than the usual bounds
$ 1 \lsim \tan \beta \lsim m_t/m_b$. \s
 
We show in Fig.~1 the sensitivity of $T_{+}$ to the mixing parameter
$\tilde{A}_t \equiv A_t \tan \beta - \mu$ in the stop sector. 
This illustrates the amount of exclusion 
of $\tan \beta$ depending on the approximation used. \\

\noindent
\begin{figure}[htb]
\vspace*{-.01cm}
\begin{center}
\mbox{ 
\psfig{figure=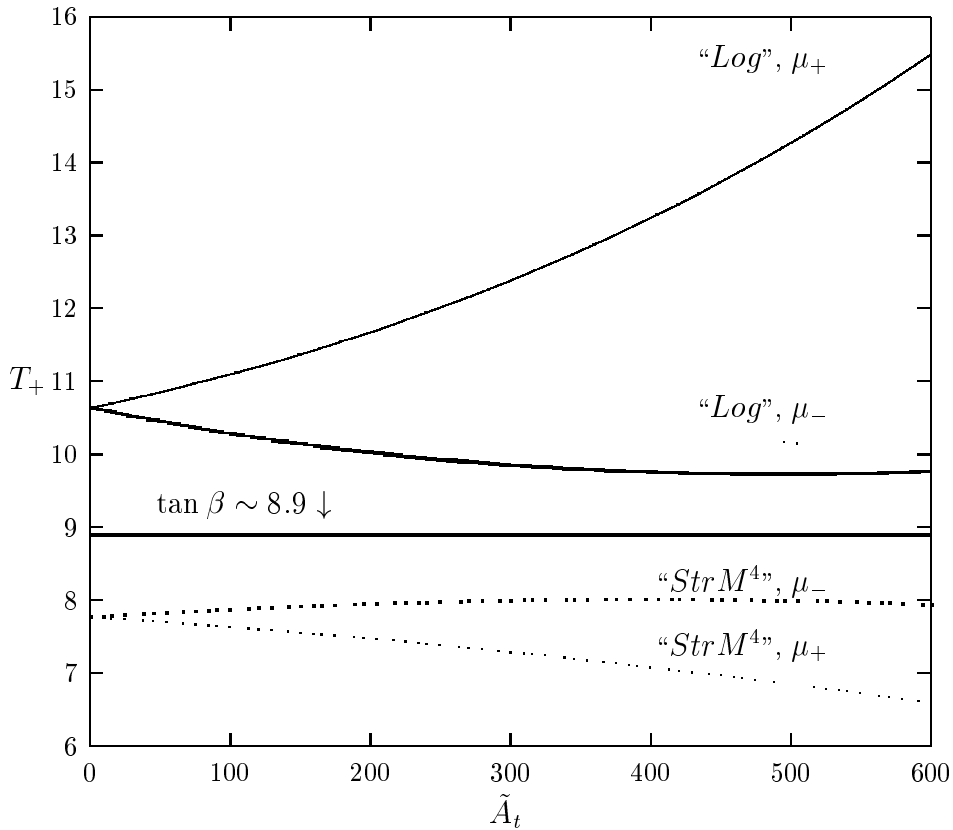}}
\end{center}
\vspace*{-13.3cm}
\caption[]{The $T_{+}$ model--independent lower bound on $\tan \beta$
in the top/stop--bottom/sbottom approximation and in the Supertrace 
approximation. Here: $M_A^2=m_1^2 + m_2^2=(300\; {\rm GeV})^2$, 
$m_3^2 = - (100\; {\rm GeV})^2$, $Q = m_t=175$ GeV, $M_{\rm SUSY}= 630$ GeV, 
$\mu_\pm= \pm 100$ GeV and $ Y_t=0.9$  } 
\end{figure}

\newpage

\subsection*{2.3 Renormalization Group Evolution} 

In the minimal SUGRA model, the MSSM parameters [couplings and masses] 
are defined at the Unification scale with some unification conditions,
and then are evolved down to the electroweak scale through Renormalization
Group Equations (RGE). The RGE evolution is thus an important ingredient
of mSUGRA, and more generally in any theory incorporating GUT unification. 
In the RGE's of the MSSM, different levels of approximations are available 
as will be discussed later. In this section, we will write for completeness 
the RGE's for the masses and couplings in the one--loop approximation, that
we will need later when we will discuss the program SUSPECT. \s

In a next step, we will discuss some exact analytical solutions of the
one--loop RG evolution equations of the Yukawa couplings. Solutions to 
these equations exist in the literature since many years \cite{solutions} 
in the the limit where the top Yukawa coupling is assumed to dominate all 
the others. This limiting solution is relevant for small $\tan \beta$ values 
and remain numerically useful for $\tan \beta$ values 
up to 10 or so. Later on, various attempts were made to obtain general 
solutions, but still relying on some approximations, such as 
neglecting the U(1) coupling as compared to those of SU(3) and SU(2). 
However, even in this case, the solutions given where actually implicit
\cite{FL} and not directly generalizable to include
more than two Yukawa couplings ($Y_t, Y_b$). Below, we will give the
exact solutions with no approximation whatsoever. They are valid for any
value of $Y_t$ and $Y_b$ and immediately generalizable to the full set
of Yukawa couplings.    

\subsubsection*{2.3.1 The one--loop RGE's} 

In the following, we list the RGE's for the MSSM parameters in the 
one--loop approximation. Although two--loop evolution equations for 
some parameters [such as the gauge and the Yukawa couplings] are 
available, for many purposes it is a rather good approximation to 
use only the one--loop equations [especially that it is much faster
when implemented in numerical programs]. The list that we give below
is ordered as in the program SUSPECT which will be discussed in section 3.2. 
%Defining the evolution parameter $t$ by ??, 
%\begin{equation}
%t={\rm Log}(M_U^2/Q^2) \label{tdef}
%\end{equation}
\s

\nn $\bullet$ Gauge Couplings [$g_i$ with $i=1,2,3$ and $n_g$ the generation 
number]: 
\beqn 
\label{coupjauge}
{\frac {dg_i}{dt}}={\frac {1}{32 \pi^2} b_i g_i^3 } \ \ {\rm with} \ \ 
b_1=-1-{\frac {10}{3}} n_{g} \ , \ b_2=5-2 n_{g} \ , \ b_3=9-2 n_{g}
\eeqn 
\nn $\bullet$ Yukawa Couplings [$i=1,2,3$ generations]: 
\begin{eqnarray}
 {\frac {dY_{u}^i}{dt}}&=&-{\frac {Y_{u}^i}{32 \pi^2}} \bigg[ 
{\sum_{k} 3 (Y_u^k)^2+(Y_d^i)^2} + 3 (Y_u^i)^2 - \bigg({\frac {13}{9}} 
g_1^2+3 g_2^2+\frac {16}{3}g_3^2 \bigg) \bigg] \label{yukup} \\
{\frac {dY_{d}^i}{dt}}&=&-{\frac {Y_{d}^i}{32 \pi^2}} \bigg[
\sum_{k}\{3 (Y_d^k)^2 + (Y_l^k)^2\} +(Y_u^i)^2 + 3 (Y_d^i)^2   
-\bigg({ \frac {7}{9}} g_1^2+3 g_2^2+\frac {16}{3}g_3^2 \bigg) \bigg]
\label{yukdown} \\
{\frac {dY_{l}^i}{dt}}&=&-{\frac {Y_{l}^i}{32 \pi^2}} \bigg[ 
\sum_{k}\{ (Y_l^k)^2+3 (Y_d^k)^2\} +3 (Y_l^i)^2-3( g_1^2+ g_2^2) \bigg]
\label{yuklep} 
\end{eqnarray}
\nn $\bullet$ The $\mu$ term and the vacuum expectation values $v_1$, 
$v_2$ [$k=1,2,3$]: 
\begin{eqnarray} 
 {\frac {d\mu}{dt}}&=&-{\frac {\mu}{32 \pi^2}} \bigg[ 
\sum_{k}3 \{(Y_u^k)^2+3 (Y_d^k)^2+(Y_l^k)^2 \}-( g_1^2+ 3 g_2^2) \bigg] 
\\ \label{eqv1}
{\frac {dv_1}{dt}}&=&{\frac {v_1}{32 \pi^2}} \bigg[ 
\sum_{k} \{3(Y_d^k)^2+(Y_l^k)^2 \}-{\frac {3}{4}} \bigg( {\frac {1}{3}} 
g_1^2+ g_2^2 \bigg) \bigg]  \\ \label{eqv2}
{\frac {dv_2}{dt}}&=&{\frac {v_2}{32 \pi^2}} \bigg[ 
{\sum_{k}3 (Y_u^k)^2}-{\frac {3}{4}} \bigg( {\frac {1}{3}} g_1^2+ g_2^2
\bigg) \bigg] 
\end{eqnarray}
\nn $\bullet$ The scalar Higgs masses and the parameter $B$ [$k=1,2,3$]: 
\begin{eqnarray}
\label{eqmH1}
{\frac {d m^2_{H_1}}{dt}}&=&-{\frac {1}{16 \pi^2}} \bigg[
\sum_{k} \{ 3 (Y_d^k)^2 P_{\tilde{d}}^{k}+(Y_l^k)^2 P_{\tilde{l}}^{k} \}-
{\frac {1}{2}}g_1^2 {\rm Tr} (Y m^2) -( g_1^2 M_1^2+3 g_2^2 M_2^2) \bigg] 
\\ 
\label{eqmH2}
{\frac {d m^2_{H_2}}{dt}}&=&-{\frac {1}{16 \pi^2}} \bigg[ 
{\sum_{k}3 (Y_u^k)^2 P_{\tilde{u}}^{k}}+{\frac {1}{2}}g_1^2 {\rm Tr}
(Y m^2)-(g_1^2 M_1^2+3 g_2^2 M_2^2) \bigg]   \\ \label{eqB}
{\frac {d B}{dt}}&=&-{\frac {1}{16 \pi^2}} \bigg[ 
\sum_{k} \{ 3 A_u^k (Y_u^k)^2+3 A_d^k (Y_d^k)^2+ A_d^k (Y_l^k)^2 \} -
( g_1^2 M_1+3 g_2^2 M_2) \bigg]
\end{eqnarray}
\nn $\bullet$ The trilinear $A$ couplings $[i,k=1,2,3$]: 
\begin{eqnarray}
{\frac {d A_u^i}{dt}}&=&-{\frac {1}{32 \pi^2}} \bigg[ 
6 A_u^i (Y_u^i)^2 + 2 A_d^i (Y_d^i)^2+6{\sum_{k}A_u^k (Y_u^k)^2} \non \\
&& \hspace*{1.4cm} 
-\bigg( {\frac {26}{9}} g_1^2 M_1+6 g_2^2 M_2+{\frac {32}{3}}g_3^2 M_3 
\bigg) \bigg]   \\
 {\frac {d A_d^i}{dt}}&=&-{\frac {1}{32 \pi^2}} \bigg[ 6 A_d^i (Y_d^i)^2 
+ 2 A_u^i (Y_u^i)^2 +2 \sum_{k}\{ A_l^k (Y_l^k)^2+3 A_d^k (Y_d^k)^2\} \non \\ 
&& \hspace*{1.4cm} 
-\bigg( {\frac {14}{9}} g_1^2 M_1+6 g_2^2 M_2+{\frac {32}{3}}g_3^2 M_3 
\bigg) \bigg] \\ \label{eqAl}
 {\frac {d A_l^i}{dt}}&=&-{\frac {1}{32 \pi^2}} \bigg[ 
6 A_l^i (Y_l^i)^2 +2 \sum_{k} \{A_l^k (Y_l^k)^2+3 A_d^k (Y_d^k)^2 \} 
-6( g_1^2 M_1+ g_2^2 M_2) \bigg] 
\end{eqnarray}
\nn $\bullet$ The scalar fermion masses [with $
P^k_{\tilde{u},\tilde{d},\tilde{l}} \equiv 
 m^2_{H_2, H_1,H_2}
+m^2_{ {\tilde{Q}}_{k},{\tilde{Q}}_{k},{\tilde{L}}_{k} }
+m^2_{\tilde{u}_k,\tilde{d}_k,\tilde{l}_k}
+(A^k_{u,d,l})^2$] 
\begin{eqnarray}
 {\frac {d m^2_{{\tilde{l}}_{R_i}}}{dt}}&=&-{\frac {1}{16 \pi^2}} \bigg[
2 (Y_l^i)^2 P_{\tilde{l}}^{i}+g_1^2 {\rm Tr}(Y m^2)-4 g_1^2 M_1^2
\bigg] \\
 {\frac {d m^2_{{\tilde{L}}_i}}{dt}}&=&-{\frac {1}{16 \pi^2}} \bigg[
(Y_l^i)^2 P_{\tilde{l}}^{i}  
 -{\frac {1}{2}}g_1^2 {\rm Tr} (Y m^2)-( g_1^2 M_1^2+3 g_2^2 M_2^2) \bigg] \\
 {\frac {d m^2_{{\tilde{d}}_{R_i}}}{dt}}&=&-{\frac {1}{16 \pi^2}} \bigg[
2 (Y_d^i)^2 P_{\tilde{d}}^{i} 
+{\frac {1}{3}}g_1^2 {\rm Tr} (Y m^2)- \bigg( 
{\frac {4}{9}} g_1^2 M_1^2+{\frac {16}{3}} g_3^2 M_3^2 \bigg) \bigg]  \\
 {\frac {d m^2_{{\tilde{u}}_{R_i}}}{dt}}&=&-{\frac {1}{16 \pi^2}} \bigg[
2 (Y_u^i)^2 P_{\tilde{u}}^{i} 
  -{\frac {2}{3}}g_1^2 {\rm Tr} (Y m^2)- \bigg(
{\frac {16}{9}} g_1^2 M_1^2+{\frac {16}{3}} g_3^2 M_3^2 \bigg)  \bigg] \\
\label{eqQi}
 {\frac {d m^2_{{\tilde{Q}}_i}}{dt}}&=&-{\frac {1}{16 \pi^2}} \bigg[
(Y_u^i)^2 P_{\tilde{u}}^{i}  
                    +(Y_d^i)^2 P_{\tilde{d}}^{i} +{\frac {1}{6}}g_1^2 
{\rm Tr} (Y m^2) \non \\
&& \hspace*{1.4cm} - \bigg(
{\frac {1}{9}} g_1^2 M_1^2+3 g_2^2 M_2^2+{\frac {16}{3}} g_3^2 M_3^2
\bigg) \bigg]   
\end{eqnarray}
\nn $\bullet$ The gaugino masses [$i=1,2,3$ and the $b_i$ are given above]
\begin{equation}
\label{jaugino}
{\frac {dM_i}{dt}}={\frac {1}{16 \pi^2} M_i b_i g_i^2 } \label{rgs}
\end{equation}
A few remarks are in order, here: \s

$(i)$ The evolution parameter $t$ is defined by $t={\rm Log}(M_U^2/Q^2)$;
this is different from the one used in the RGE's of the program SUSPECT, 
where $t= 1/2{\rm Log}(Q^2/M_U^2)$. $(ii)$ Tr$(Ym^2)$ is the  isospin
pondered sum of the squared soft masses of the scalar fermions; in the 
case of universal soft masses, the trace vanishes at any scale due to 
anomaly cancellation. $(iii)$ The RGE's for the gaugino masses and the 
gauge couplings are related and from eqs.~(\ref{coupjauge}) and (\ref{rgs}), 
one can easily  see that $d/dt(M_i/g_i^2)=0$. 

\subsubsection*{2.3.2 Exact solutions for the Yukawa coupling RGE's} 

We are interested here in eqs.~(\ref{yukup}--\ref{yuklep}). They have the nice
feature of being completely decoupled from the rest of the system, especially
from the gauge couplings whose running is determined a priori via 
eq.~(\ref{coupjauge}). [This is no more true at two--loop order where the 
gauge and Yukawa equations become interwound.] When all Yukawa couplings 
except $Y_t$ are neglected, eqs.~(\ref{yukdown}) and (\ref{yuklep}) become 
trivial while eq.~(\ref{yukup}) becomes of the Bernoulli type in the variable 
$y_t\equiv Y_t^2$ 
\begin{equation}
\frac{d}{dt} y_t = f_1(t) y_t + b y_t^2
\end{equation}
where 
\begin{eqnarray}
f_1(t)= \frac{1}{16 \pi^2}( \frac{16}{3} g_3^2 + 3 g_2^2 + 
\frac{13}{9} g_1^2 ) \ \ , \ \  b = -\frac{6}{16 \pi^2}  \label{f1} 
\end{eqnarray}
and is easily solved to give \cite{diffeq,solutions} 
\beqn
y_t(t)=\frac{y^0 E(t)}{1 -b y^0 \int_0^t E(t') dt'} \label{ytsol0}
\eeqn 
where 
\beqn  
E(t)= e^{\int_0^t f_1(t') dt'}
 \ \ {\rm and} \ \ y^0= Y_t^2(t=0) 
\eeqn
In the more general case where both $Y_t$ and $Y_b$ are kept in the game, but 
neglecting all other Yukawa couplings, eqs.~(\ref{yukup},\ref{yukdown}) 
become after the change of variable, $y_t\equiv Y_t^2, y_b\equiv Y_b^2$
\begin{eqnarray}
\frac{d}{dt} y_t&=& f_1(t) y_t + a y_b y_t + b y_t^2 \non \\
\frac{d}{dt} y_b&=& f_2(t) y_b + a y_b y_t + b y_b^2 \label{ytb} 
\end{eqnarray}
where $f_1(t)$ and $b$ are given in eqs.~(\ref{f1}) and 
\begin{eqnarray}
f_2(t)= \frac{1}{16 \pi^2}( \frac{16}{3} g_3^2 + 3 g_2^2 + 
\frac{7}{9} g_1^2 ) \ \ , \ \  a= -\frac{1}{16 \pi^2} 
\end{eqnarray}
As far as we know, the system eqs.~(\ref{ytb}) is not treated in standard
text books, and although it looks at first sight simple, we could not find 
a systematic way of relating it to a standard 
form\footnote{The situation would be much simpler if $f_1(t) = f_2(t)$,  i.e. when 
neglecting $g_1$. In this case the equations can be solved in quadrature 
after some change of variables, leading though only to implicit solutions 
involving some hypergeometric functions \cite{FL}}. It is also relatively easy 
to solve the system up to first order in $Y_b$ in the region $Y_t \gg Y_b$ 
\cite{AM}. This is already an improvement of the known solutions with $Y_b 
\sim 0$. It extends the numerical validity much further than $\tan \beta 
\simeq 10$. More importantly, this approximate solution gave us a hint of 
the structure of the {\sl exact} solution which was then found by sheer 
guess \cite{AM}:
\begin{eqnarray}
&&y_t=\frac{y_t^0 E_{12}(t)}{1 - b y_t^0 \int_0^t E_{12}(t') dt'} \label{ytsol} \\
&& \nonumber \\
&&y_b=\frac{y_b^0 E_{21}(t)}{1 - b y_b^0 \int_0^t E_{21}(t') dt'} \label{ybsol}
\end{eqnarray}
where
\begin{eqnarray}
&&E_{12}(t)= \frac{ E_1(t)}{ (1 - b y_b^0 \int_0^t E_{21}(t') dt')^{a/b}} 
\label{e12} \\
&& \nonumber \\
&&E_{21}(t)= \frac{ E_2(t)}{ (1 - b y_t^0 \int_0^t E_{12}(t') dt')^{a/b} } 
\label{e21}  \\
&& \nonumber \\
&& E_i= e^{\int_0^t f_i(t') dt'} \;\;\; i=1,2 \\
\nonumber
\end{eqnarray}
and $y_t^0\equiv Y_t^2(t=0), y_b^0\equiv Y_b^2(t=0)$ are arbitrary 
initial conditions.
The solutions eqs.~(\ref{ytsol},\ref{ybsol}) are exact for any value of 
$\tan \beta$.
They resemble formally eq.~(\ref{ytsol0}) of which they are a generalization.
One should note however the important difference, namely that due to 
eqs.~(\ref{e12},\ref{e21}), the general solutions for $y_t$ and $y_b$
are actually continued integrated fractions.
Indeed eqs.~(\ref{e12},\ref{e21}) give an implicit definition
of $E_{12}(t)$ and $E_{21}(t)$, the first being defined in terms of the
second and vice-versa.  One could still write $E_{12}(t)$
in an explicit though infinite series form:
\begin{equation}
E_{12}(t)=\frac{E_{1}(t)}{( 1- b y_b^0 \large{\int_0^t}
           \frac{ E_{2}(t_1) dt_1}{( 1- b y_t^0 \int_0^{t_1}
            \frac{ E_{1}(t_2) dt_2}{( 1- b y_b^0 \int_0^{t_2}
             \frac{ E_{2}(t_3) dt_3}{( 1- b y_t^0 \int_0^{t_3} \cdots
              )^{a/b}}  )^{a/b}} )^{a/b}} )^{a/b}}
\end{equation}
and similarly for  $E_{21}(t)$ with the substitution $1 \leftrightarrow 2,
y_t^0 \leftrightarrow y_b^0$. We will see later on that both forms for
$E_{i j}$ are useful. In any case, we should stress here that the solutions
for $y_t, y_b$ are themselves explicit. \s

What do we gain from these exact analytical solutions? \s

$(i)$ First of all one can prove rigorously the convergence of the infinite 
series
$E_{i j}(t)$ and determine explicitly the convergence criteria \cite{AM}.
This implies that, for practical purposes, keeping only the first iteration
of the $E_{i j}(t)$ series is numerically a very good approximation
[within the convergence region]. We give in Table 1 a numerical comparison
versus a Runge-Kutta method. \s

$(ii)$ \
The large $\tan \beta$ region is treated exactly and one can follow
precisely the various features of the running of the Yukawa couplings
in this regime. \s

$(iii)$ 
The fact that the coefficients of the quadratic parts of eqs.~(\ref{ytb})
are equal and that there are only two non-zero Yukawa couplings
is actually unessential in finding the general solutions in the present
form. This form generalizes straightforwardly for an exact solution
of eqs.~(\ref{yukup}--\ref{yuklep}) which will be given elsewhere \cite{AM},
and would thus be of relevance in the case of bottom--tau Yukawa 
coupling unification scenarios \cite{btau}. \s

$(iv)$ 
Finally, one can control analytically the acceptable regions for
the initial conditions $y_b^0, y_t^0$. This is related to the necessity
of avoiding Landau poles and more generally of requiring
$y_t, y_b$ to remain positive throughout the evolution,
being the squares of $Y_t, Y_b$. This is of relevance if one wants
to run the quantities between two low-energy scales choosing
some initial conditions at one of these scales without referring
explicitly to the initial values at the unification scale $M_U$. 
[Such a possibility is being implemented in the fortran code SUSPECT,
see section 3.2]. \bigskip

%\vspace*{9cm}

\begin{table}[htb]
%\stretchline{1.5}
\begin{center}
\begin{tabular}{|c|c|c|c|c|c|c|}
\hline\hline
&&&&&& \\
$\tan \beta$ &$Y_b(t=0)$ & $Y_t(t=0)$ & $Y_b(t)$ &  $Y_b(t)$ & $Y_t(t)$&  
$Y_t(t)$ \\
&& & ``exact"&  Runge-Kutta & ``exact"&  Runge-Kutta \\ 
&&&&&& \\ \hline\hline
&&&&&& \\
2&0.0387453 &1.13007 & 0.0145059&0.0145050 &0.775788 &  0.775974 
              \\  &&&&&& \\ \hline 
&&&&&& \\
10&0.174138 &1.01581 & 0.0630978 &0.0631052 &0.54263 & 0.542743          
      \\  &&&&&& \\ \hline
&&&&&& \\
50&0.866544 &1.01097 & 0.435682&0.439526 &0.585453 &    0.590258         
    \\  &&&&&& \\ \hline \hline
\end{tabular}
\end{center}
\caption{Numerical comparison of the exact one--loop solution, truncated 
to the first iteration, with the Runge--Kutta RG evolution, obtained with 
only a one step evolution over 10 orders of magnitude.}
\end{table}

\newpage

\setcounter{equation}{0}
\renewcommand{\theequation}{3.\arabic{equation}}

\section*{3. The Physical Parameters} 

\subsection*{3.1 Particle masses and couplings} 

\subsubsection*{3.1.1 Mass matrices and couplings} 

In this section, we discuss the general features of the chargino/neutralino,  
sfermion and Higgs boson sectors to set the conventions and the notations 
which will be used further on. \\

\nn {\bf a) The chargino/neutralino sector} \\

\nn The general chargino mass matrix depends on the parameters $M_2, \mu$ and
$\tb$ \cite{HaberKane,Haber}
\begin{eqnarray}
{\cal M}_C = \left[ \begin{array}{cc} M_2 & \sqrt{2}M_W \sin \beta
\\ \sqrt{2}M_W \cos \beta & \mu \end{array} \right]
\end{eqnarray}
is diagonalized by two real matrices $U$ and $V$, 
\begin{eqnarray}
U^* {\cal M}_C V^{-1} \ \ \ra \ \ U={\cal O}_- \ {\rm and} \ \ V = 
\left\{
\begin{array}{cc} {\cal O}_+ \ \ \ & {\rm if \ det}{\cal M}_C \geq 0  \\
            \sigma_3  {\cal O}_+ \ \ \ & {\rm if \ det}{\cal M}_C <0  
\end{array}
\right. 
\end{eqnarray}
where the $\sigma_3$ matrix and the  ${\cal O}_\pm$ 
matrices are given by [with the appropriate signs depending on 
the values of $M_2$, $\mu$, and $\tan\beta$] 
\begin{eqnarray}
{\sigma_3} = \left[ \begin{array}{cc} + 1 & 0
\\ 0 & - 1 \end{array} \right] \ \ , \ \ 
{\cal O}_\pm = \left[ \begin{array}{cc} \cos \theta_\pm & \sin \theta_\pm
\\ -\sin \theta_\pm & \cos \theta_\pm \end{array} \right] 
\end{eqnarray}
with
\begin{eqnarray}
\tan 2 \theta_- &=&  \frac{ 2\sqrt{2}M_W(M_2 \cos \beta
+\mu \sin \beta)}{ M_2^2-\mu^2-2M_W^2 \cos \beta} \non \\ 
\tan 2 \theta_+  &=&  \frac{ 2\sqrt{2}M_W(M_2 \sin \beta
+\mu \cos \beta)}{M_2^2-\mu^2 +2M_W^2 \cos \beta} 
\end{eqnarray}
This leads to the two chargino masses,  
\begin{eqnarray}
m_{\chi_{1,2}^+} = && \frac{1}{\sqrt{2}} \left[ M_2^2+\mu^2+2M_W^2
\right. \\ &&  \left. 
\mp \left\{ (M_2^2-\mu^2)^2+4 M_W^4 \cos^2 2\beta+4M_W^2 (M^2_2+\mu^2
+2M_2\mu \sin 2\beta) \right\}^{\frac{1}{2}} \right]^{\frac{1}{2}} \non 
\label{MCexp}
\end{eqnarray}
In the limit $|\mu| \gg M_2, M_Z$, the masses of the two charginos reduce to
$[\epsilon_\mu \equiv {\rm sign}(\mu)$]
\begin{eqnarray}
m_{\chi_{1}^+}  & \simeq &  M_2 - \frac{M_W^2}{\mu^2} 
\left( M_2 +\mu \sin 2 \beta
\right) \non \\
m_{\chi_{2}^+}  & \simeq & |\mu| + 
\frac{M_W^2}{\mu^2} \epsilon_\mu \left( M_2 \sin 
2 \beta +\mu \right)  
\end{eqnarray}
For $|\mu| \ra \infty$, the lightest chargino corresponds to a pure wino state 
with $m_{\chi_{1}^+} \simeq M_2$, while the heavier chargino corresponds to a 
pure higgsino state with $m_{\chi_{2}^+} = |\mu|$.  

\bigskip 

In the case of the neutralinos, the four--dimensional neutralino mass matrix 
\cite{HaberKane} depends on the same two mass parameters $\mu$ and $M_2$, 
if the GUT 
relation $M_1=\frac{5}{3} \tan^2 \theta_W$ $ M_2 \simeq \frac{1}{2} M_2$ 
is used. In the $(-i\tilde{B}, -i\tilde{W}_3, \tilde{H}^0_1,$ 
$\tilde{H}^0_2)$ basis, it has the form  [$s_W^2=1-c_W^2 \equiv \sin^2 
\theta_W$]
\begin{eqnarray}
{\cal M}_N = \left[ \begin{array}{cccc}
M_1 & 0 & -M_Z s_W \cos\beta & M_Z  s_W \sin\beta \\
0   & M_2 & M_Z c_W \cos\beta & -M_Z  c_W \sin\beta \\
-M_Z s_W \cos\beta & M_Z  c_W \cos\beta & 0 & -\mu \\
M_Z s_W \sin \beta & -M_Z  c_W \sin\beta & -\mu & 0
\end{array} \right]
\end{eqnarray}
It can be diagonalized analytically \cite{egypte} by a single real matrix $Z$;
the [positive] masses of the neutralino states $m_{\chi_i^0}$ have complicated
expressions which will not be given here. In the limit of large $|\mu|$ 
values, the masses of the neutralino states however simplify to 
\begin{eqnarray}
m_{\chi_{1}^0} &\simeq& M_1 - \frac{M_Z^2}{\mu^2} \left( M_1 +\mu \sin 2 \beta
\right) s_W^2 \non \\
m_{\chi_{2}^0} &\simeq& M_2 - \frac{M_Z^2}{\mu^2} \left( M_2 +\mu \sin 2 \beta
\right) c_W^2 \non \\
m_{\chi_{3}^0} &\simeq& |\mu| + \frac{1}{2}\frac{M_Z^2}{\mu^2} \epsilon_\mu 
(1-\sin 2\beta) \left( \mu + M_2 s_W^2+M_1 c_W^2 \right) \non \\
m_{\chi_{4}^0} &\simeq& |\mu| + \frac{1}{2}\frac{M_Z^2}{\mu^2} \epsilon_\mu 
(1+\sin 2\beta) \left( \mu - M_2 s_W^2 - M_1 c_W^2 \right) 
\end{eqnarray}
Again, for $|\mu| \ra \infty$, two neutralinos are pure gaugino states 
with masses $m_{\chi_{1}^0} \simeq M_1$ , $m_{\chi_{2}^0} =M_2$, while
the two others are pure higgsino states, with masses 
$m_{\chi_{3}^0} \simeq m_{\chi_{4}^0} \simeq |\mu|$. \\

\nn {\bf b) The sfermion sector} \\

Assuming a universal scalar mass $m_0$ and gaugino mass $m_{1/2}$ at 
the GUT scale, one obtains relatively simple expressions for the 
left-- and right--handed sfermion masses when performing the RGE 
evolution to the weak scale at one--loop order, if the
the Yukawa couplings in the RGE's are neglected [for third generation 
squarks this is a poor approximation since these couplings can be large; 
in this case numerical analyses are needed as will be discussed later]. 
One has: 
\begin{eqnarray}
m_{\tilde{f}_{L,R}}^2 = m_0^2 + \sum_{i=1}^{3} F_i (f) 
m^2_{1/ 2} \pm ( I^3_f - e_{f} s_W^2 ) M_Z^2 
\cos 2 \beta
\end{eqnarray}
$I_f^3$ and $e_{{f}}$ are the weak isospin and the 
electric charge of the sfermion and $F_i$ are the RGE 
coefficients for the three gauge couplings at 
the scale $Q \sim M_Z$, given by
\begin{eqnarray}
F_i = \frac{c_i(f)}{b_i} \left[1- \left( 1 - \frac{\alpha_U}{4\pi} 
b_i {\rm Log} \frac{Q^2}{M_U^2} \right)^{-2} \right] 
\end{eqnarray}
The coefficients $b_i$, assuming that all the MSSM particle spectrum 
contributes to the evolution from $Q$ to the GUT scale $M_G$, are given by:
$b_1=33/5, b_2=1, b_3=-3$. 
The coefficients $c(\tilde{f})=(c_1,c_2,c_3) (\tilde{f})$ depend on the 
hypercharge and color of the sfermions  
\begin{eqnarray}
c(\tilde{L}) = \left( \begin{array}{c} {3 \over 10} \\ {3 \over 2}
\\ 0 \end{array} \right) , 
c(\tilde{l}_R)= \left( \begin{array}{c} {6 \over 5} \\ 0 \\ 0 
\end{array} \right) , 
c(\tilde{Q})= \left( \begin{array}{c} {1 \over 30} \\ {3 \over 2} \\ 
{8 \over 3} \end{array} \right) , 
c(\tilde{u}_R)= \left( \begin{array}{c} {8 \over 15} \\ 0 \\ {8 \over 3} 
\end{array} \right) , 
c(\tilde{d}_R)= \left( \begin{array}{c} {2 \over 15} \\ 0 \\ {8 \over 3}
 \end{array} \right) \nonumber
\end{eqnarray}
With the input gauge coupling constants at the scale of the $Z$ boson mass
$\alpha_1 (M_Z) \simeq 0.01, \alpha_2 (M_Z) \simeq 0.033$ and $\alpha_3 (M_Z) 
\simeq 0.118$, one obtains $M_U \sim 1.9 \times 10^{16}$ GeV for the GUT scale 
and $\alpha_U = 0.041$ for the coupling constant $\alpha_U$. Using these
values, one obtains for the left-- and right--handed sfermion masses
%\begin{eqnarray}
%m^2_{\tilde{f}} = m_0^2 + a_{\tilde{f}}m^2_{1/2} +d_{\tilde{f}}M_Z^2 
%\cos2\beta
%\end{eqnarray} 
%$(a_{\tilde{u}_L}, d_{\tilde{u}_L})=(6.28,0.35) ,
%(a_{\tilde{d}_L}, d_{\tilde{d}_L})=(6.28,-0.42) ,
%(a_{\tilde{u}_R}, d_{\tilde{u}_R})=(5.87,0.16) ,
%(a_{\tilde{d}_R}, d_{\tilde{d}_R})=(5.82,-0.08) ,
%(a_{\tilde{\nu}_L}, d_{\tilde{\nu}_L})=(0.52,0.5),
%(a_{\tilde{e}_L}, d_{\tilde{e}_L})=(0.52,-0.27) ,
%(a_{\tilde{e}_R}, d_{\tilde{e}_R})=(0.15,-0.23)$. 
\begin{eqnarray}
m^2_{\tilde{u}_L} &=& m_0^2 +6.28 m^2_{1/2} +0.35 M_Z^2 \cos2\beta \non \\
m^2_{\tilde{d}_L} &=& m_0^2 +6.28 m^2_{1/2} -0.42 M_Z^2 \cos2\beta \non \\
m^2_{\tilde{u}_R} &=& m_0^2 +5.87 m^2_{1/2} +0.16 M_Z^2 \cos2\beta \non \\
m^2_{\tilde{d}_R} &=& m_0^2 +5.82 m^2_{1/2} -0.08 M_Z^2 \cos2\beta \non \\
m^2_{\tilde{\nu}_L} &=& m_0^2 +0.52 m^2_{1/2} +0.50 M_Z^2 \cos2\beta \non \\
m^2_{\tilde{e}_L} &=& m_0^2 +0.52 m^2_{1/2} -0.27 M_Z^2 \cos2\beta \non \\
m^2_{\tilde{e}_R} &=& m_0^2 +0.15 m^2_{1/2} -0.23 M_Z^2 \cos2\beta 
\label{smass}
\end{eqnarray} 
In the case of the third generation sparticles, left-- and right--handed 
sfermions will mix \cite{qmix}; for a given sfermion $\tilde{f} = \tilde{t}, 
\tilde{b}$ and $\tilde{\tau }$, the mass matrices which determine the mixing 
are given by
\begin{eqnarray}
 M_{\tilde{f}}^2 \ = \ 
\left[ \begin{array}{cc} m_{\tilde{f}_L}^2 + m_f^2 & m_f (A_f - \mu r_f) 
\\ m_f (A_f - \mu r_f)  & m_{\tilde{f}_R}^2 + m_f^2 \end{array} \right]
\end{eqnarray}
where the sfermion masses $m_{\tilde{f}_{L,R}}$ are given above, $m_f$
are the masses of the partner fermions and $r_{b} = r_\tau =1/r_t= \tb$. 
These matrices are diagonalized by orthogonal matrices; the mixing angles
$\theta_f$ and the squark eigenstate masses are given by 
\begin{eqnarray}
\sin 2\theta_f = \frac{2 m_f (A_f -\mu r_f)} { m_{\tilde{f}_1}^2
-m_{\tilde{f}_2}^2 } \ \ , \ \ 
\cos 2\theta_f = \frac{m_{\tilde{f}_L}^2 -m_{\tilde{f}_R}^2} 
{m_{\tilde{f}_1}^2 -m_{\tilde{f}_2}^2 } \hspace*{0.8cm}  \\
m_{\tilde{f}_{1,2}}^2 = m_f^2 + \frac{1}{2} \left[ 
m_{ \tilde{f}_L}^2 + m_{\tilde{f}_R}^2 \mp \sqrt{
(m_{\tilde{f}_L}^2 - m_{\tilde{f}_R}^2)^2 + 4m_f^2 (A_f -\mu r_f)^2 } 
\right] 
\end{eqnarray}
Due to the large value of $m_t$, the mixing is particularly strong in the stop 
sector. This generates a large splitting between the masses of the two stop 
eigenstates, possibly leading to a lightest top squark much lighter than the 
other squarks and even the top quark. \\

\nn {\bf c) The Higgs sector} \\

We come now to a description of our parameterization of the MSSM Higgs 
sector \cite{R2}. Besides the four masses, $M_h, M_H, M_A$ and $M_{H^\pm}$, 
the Higgs sector is described at the tree level by two additional parameters, 
$\tb$ and a mixing angle $\alpha$ in the CP--even Higgs sector. Due to SUSY 
constraints as discussed before, only two of them are independent and the 
two inputs are in general taken to be $\tb$ and $M_A$. Radiative corrections, 
the leading part of which grow as the fourth power of the top mass and 
logarithmically with the common squark mass \cite{radcor,mh}, change 
significantly the relations between masses and couplings and shift the mass 
of the lightest $h$ boson upwards. These radiative corrections are  very 
important and should therefore be included in any analysis. Here we will,
however, only discuss the leading part of this correction which in the 
simplest case can be parameterized in terms of the quantity \cite{radcor}
\beqn
\epsilon = \frac{3 G_{F}}{\sqrt{2}\pi^2} \frac{m_t^4}{\sin^2\beta}
{\rm Log}\left( 1+\frac{m_{\tilde{q}}^2}{m_t^2} \right) \ .
\eeqn
The CP--even Higgs boson masses are then given in terms of the
pseudoscalar mass $M_A$ and $\tb$, and the charged Higgs boson mass 
in terms of $M_A$, are given by
\begin{small}
\beqn
M_{h,H}^2=\frac{1}{2}\Bigg[ M_A^2+M_Z^2+\epsilon 
\mp \sqrt{(M_A^2+M_Z^2+\epsilon)^2- 4M_A^2 M_Z^2 \cos^2 2\beta
 - 4\epsilon( M_A^2 \sin^2 \beta + M_Z^2 \cos^2 \beta)} \Bigg] \non \\
M_{H^\pm}^2=M_A+M_W^2 \hspace*{6cm} 
\eeqn 
\end{small}
The mixing angle $\alpha$ reads in terms of $M_A$ and $\tb$
\beqn
\tan 2\alpha &=& \tan 2\beta
 \frac{M_A^2 + M_Z^2}{M_A^2 - M_Z^2 +\epsilon/\cos 2\beta} \ , \hspace*{1cm}
- \frac{\pi}{2} \leq \alpha \leq 0 \ .
\eeqn
Once $\tb$ and $M_A$ are chosen and the leading radiative correction is
included in $\alpha$, all the couplings of the Higgs bosons to fermions, 
gauge bosons and Higgs bosons are fixed.  However, in the trilinear MSSM 
Higgs boson couplings, there are also large radiative corrections which are
not entirely mapped into the angle $\alpha$, but the leading part can also 
be expressed in terms of the leading correction $\epsilon$ \cite{radcor1}. 

\bigskip

\begin{center}
\begin{tabular}{|c||c|c|c||c|} \hline
& & & & \s
$\hspace{1cm} \Phi \hspace{1cm} $ &$ g_{ \Phi \bar{u} u} $ & $
g_{\Phi \bar{d} d} $ & $ g_{\Phi VV} $ & $g_{\Phi VA} $\\
& & & & \\ \hline \hline
& & & & \\ 
$h$  & \ $\; \cos\alpha/\sin\beta       \; $ \ & \ $ \; -\sin\alpha/
\cos\beta \; $ & $ \sin(\beta-\alpha) $ & $ \cos(\beta-\alpha)$ \ \\
$H$  & \       $\; \sin\alpha/\sin\beta \; $ \ & \ $ \; \cos\alpha/
\cos\beta \; $ & $ \cos(\beta-\alpha) $ & $- \sin(\beta-\alpha)$ \\
$A$  & \ $\; 1/ \tb \; $        \ & \ $ \; \tb \; $ & $0$ & $0$ \\[0.3cm] 
\hline
\end{tabular}
\end{center}

\vspace*{2mm}

\nn Table 2: Higgs boson couplings in the MSSM to fermions 
and gauge bosons relative to the SM Higgs  couplings, and coupling to Higgs 
and gauge bosons. 

\bigskip

The couplings of the charged Higgs boson  to down (up) type fermions are 
(inversely) proportional to $\tb$, as in the case of the pseudoscalar 
Higgs boson. For the CP--even Higgs bosons, the couplings to 
down (up) type fermions are enhanced (suppressed) compared to the SM Higgs 
couplings [$\tb>1$]; the couplings to gauge bosons are suppressed by 
$\sin/\cos(\beta-\alpha)$ factors. $A$ has no tree level 
couplings to gauge bosons. Note also that while the couplings of the $h$ 
and $H$ bosons to $ZA$ and $W^+H^-$ pairs are proportional to $\cos$ and 
$\sin(\beta-\alpha)$ respectively, the $W^+H^-A$ coupling is not suppressed 
by these factors. The couplings of the MSSM neutral Higgs bosons to fermions 
and gauge bosons [normalized to the SM Higgs coupling $g_{H_{\rm SM}ff} = 
(\sqrt{2} G_F)^{1/2} m_f$ and $g_{H_{\rm SM}VV} = 2(\sqrt{2} G_F)^{1/2} 
M_V^2$] and to gauge and Higgs bosons [normalized to $(\sqrt{2} G_F)^{1/2} 
/M_Z (p_\Phi + p_A)_\mu$ with $p_\Phi$ and $p_A$ the Higgs bosons 4--momenta] 
are given in Table 2. \s

Let us turn now to the $h$ boson couplings to stop squarks which will be a 
very important ingredient for the discussion of the next section 4. 
Normalized to $2M_Z^2(\sqrt{2}G_F)^{1/2}$, they are given in the decoupling 
limit $M_A \gg M_Z$, by 
\begin{eqnarray}
g_{h \tilde{t}_1 \tilde{t}_1 } &=& - \cos 2\beta \left[ 
\frac{1}{2} \cos^2 \theta_{\tilde{t}} - \frac{2}{3} s^2_W \cos 2
\theta_{\tilde{t}} \right] 
- \frac{m_t^2}{M_Z^2} + \frac{1}{2} \sin 2\theta_{\tilde{t}} 
\frac{m_t \tilde{A}_t } {M_Z^2} \nonumber \\
g_{h \tilde{t}_2 \tilde{t}_2 } &=& - \cos 2\beta \left[ 
\frac{1}{2} \sin^2 \theta_{\tilde{t}} - \frac{2}{3} s_W^2 \cos 2
\theta_{\tilde{t}} \right] 
- \frac{m_t^2}{M_Z^2} - \frac{1}{2} \sin 2\theta_{\tilde{t}} 
\frac{m_t\tilde{A}_t } {M_Z^2} 
\label{ghtt}
\end{eqnarray} 
and involve components which are proportional to $\tilde{A}_t = A_t -\mu/
\tan \beta$. For large values of the parameter $\tilde{A}_t$ which 
incidentally make the $\tilde{t}$ mixing angle  maximal, $|\sin 2 \theta_{
\tilde{t}}| \simeq 1$, the last components can strongly  enhance the 
$g_{h\tilde{t} \tilde{t}}$ couplings and make them larger than the top 
quark coupling of the $h$ boson, $g_{htt} \propto m_t/M_Z$. The couplings
of the heavy $H$ boson to stops involve also components which can be large;
in the case of the lightest stops, the coupling reads in the decoupling 
limit:
\begin{eqnarray}
g_{H \tilde{t}_1 \tilde{t}_1 } = \sin 2\beta \left[ 
\frac{1}{2} \cos^2 \theta_{\tilde{t}} - \frac{2}{3} s^2_W \cos 2
\theta_{\tilde{t}} \right] - \frac{m_t^2}{M_Z^2} \frac{1}{\tb}
+ \frac{1}{2} \sin 2\theta_{\tilde{t}} \frac{m_t}{M_Z^2} (\frac{A_t}{\tb} 
+\mu)
\end{eqnarray}
For large $\tb$ values, the $m_t^2$ and the $A_t$ components are 
suppressed; only the component proportional to $\mu$ is untouched. The 
pseudoscalar $A$ couples only to $\tilde{t}_1 \tilde{t}_2$ pairs because 
of CP--invariance, the coupling is given by:
\begin{eqnarray}
g_{A \tilde{t}_1 \tilde{t}_2 } &=&  \frac{1}{2} \frac{m_t}{M_Z^2} 
(A_t/\tb -\mu) 
\end{eqnarray}
In the maximal mixing case, $|\sin2\theta_{\tilde{t}}| \simeq 1$, this is also 
the main component of the $H$ boson coupling to $\tilde{t}_1 \tilde{t}_2$ 
pairs except that the sign of $\mu$ is reversed. \s

Finally, let us exhibit the couplings of the $h$ boson to sneutrinos and the
lightest neutralinos that we will need in section 4.2. For sneutrinos, the 
couplings is just given in eq.~(\ref{ghtt}) and setting the fermion mass and 
the mixing angle to zero:
\beqn
g_{h \tilde{\nu} \tilde{\nu}} =  - \frac{1}{2} \cos 2\beta 
\label{ghsnu}
\eeqn
For the normalized couplings of the $h$ boson to the LSP, one has 
\beqn
g_{h \chi_1^0 \chi_1^0} = (Z_{12}-\tan\theta_W Z_{11})(\sin \beta Z_{14} 
-\cos \beta Z_{13}) \label{gchi}
\eeqn 
with $Z$ is the matrix diagonalizing the neutralino mass matrix discussed 
above. 

\subsubsection*{3.1.2 Inverting the chargino/neutralino spectrum}

Once a few super--partners will be discovered at LEP/LHC, the next immediate 
task would be to reconstruct from the measured parameters, as precisely as 
possible, the structure of the soft--SUSY breaking Lagrangian. Although the 
relationship between physical parameters [mass eigenvalues, mixing angles and 
physical couplings] and e.g. the phenomenological MSSM Lagrangian is 
well established~\cite{R1,HaberKane,R2}, it would be useful to invert such a 
relationship, namely to derive Lagrangian parameters directly from physical 
parameters. This is however especially non--trivial in the neutralino 
sector, which involves the $4 \times 4$ mass matrix shown above to 
``de--diagonalize". \s

Here, we illustrate a relatively simple scheme for such an analytic inversion
\cite{KM} for most of the Lagrangian parameters of the phenomenological 
MSSM, in terms of a minimal set of appropriately chosen physical input 
parameters. In the pure gaugino sector, the algorithm gives for a given  
$\tan\beta$ the values of the $\mu$, $M_1$ and $M_2$ parameters in terms of  
three arbitrary input masses, chosen indifferently  among four, namely
either two chargino and one neutralino masses or two neutralino and one 
chargino masses.

\smallskip

Note that, in a more standard approach [i.e. from the Lagrangian to the 
physical parameters], one may obtain a similar information, e.g. by some 
systematic scanning of the whole parameter space. However, the
advantage of having relatively simple [and thus fast] analytical 
expressions should be obvious, since in practice a complete scanning of the 
unconstrained MSSM parameters would be rather tedious and probably not 
particularly illuminating [for a recent detailed analysis, although in 
the more constrained mSUGRA scenario,  see Ref.~\cite{Denegri} for instance]. 
In addition, there are some subtleties in such  a reconstruction, like the 
occurrence of possible discrete ambiguities which are most clearly seen via 
explicit inversion, as we shall illustrate. \s

Let us now sketch our general procedure to reconstruct the gaugino sector 
parameters\footnote{Note that we restrict ourselves to real--valued parameters,
but do not necessarily assume universality of gaugino masses; without loss of 
generality, one thus can choose $M_2$ to be always positive, while the signs 
of $M_1$ and $\mu$ remain arbitrary, as a result of the phase 
re-parameterization freedom of MSSM~\cite{phaseconv}.}; for more details we 
refer to Ref.~\cite{KM}. First, one should fix a specific choice of 
input/output parameters and a simple starting point is to assume that 
$\tan\beta$ is an input i.e. that it has been extracted from 
elsewhere prior to gaugino reconstruction [see e.g. section 5.2]. 
Then, we consider two basic algorithms or ``scenarii", that
we call $S_1$ and $S_2$. \s

{\bf Scenario $S_1$}: here, the input masses are assumed to be the two 
chargino masses $m_{\chi^\pm_1}$ and $m_{\chi^\pm_2}$ and one (but any) 
neutralino mass $\pm m_{\chi_i^0}$. Then, the algorithm $S_1$ gives the 
parameters: $\mu$, $M_2$, $M_1$, plus the values of the three other 
neutralino masses $m_{\chi_j^0}, (j \neq i)$. 
More precisely, when assuming that $\tan\beta$ and the two chargino masses 
are given, the basic equations giving $\mu$ and $M_2$ are simply obtained by 
inverting the chargino mass expressions eqs.~(\ref{MCexp}), obtaining 
\beqn
\label{invmu2}
\mu^2 = \frac{1}{2}(m^2_{\chi_1^\pm}+m^2_{\chi_2^\pm}-2M^2_W 
\pm [ (m^2_{\chi_1^\pm}+m^2_{\chi_2^\pm}-2M^2_W)^2 -4(M^2_W \sin 2\beta 
\pm m_{\chi_1^\pm} m_{\chi_2^\pm})^2 ]^{1/2} ) \non \\
M_2  =  [m^2_{\chi_1^\pm}+m^2_{\chi_2^\pm}-2M^2_W -\mu^2 ]^{1/2} 
\hspace*{5cm}
\eeqn 
with the sign of $\mu$ determined from
\beqn
\label{sgnmu}
M_2 \;\mu = M^2_W \sin 2\beta \pm m_{\chi_1^\pm} m_{\chi_2^\pm}
\eeqn
Note that in eq.~(\ref{invmu2}), the $\pm$ in front of the square root simply 
reflects the spurious $\mu^2 \leftrightarrow M^2_2$ ambiguity, while the $\pm$ 
inside the square root or in eq.~ (\ref{sgnmu}) corresponds to a true 
ambiguity, i.e. when the expression under the root is positive (or zero) for 
both sign choice there are two possible solutions for ($\mu$, $M_2$). 
The occurrence of this twofold ambiguity crucially depends, obviously, on the 
mass values $m_{\chi_1^\pm}$, $m_{\chi_2^\pm}$ and $\tan\beta$, as will be 
illustrated. It is relatively easy to determine in which parameter domain one 
has either no solution, a unique or twofold solution.\s 

Concerning the neutralino mass inversion, we note first that since we restrict 
ourselves to the case where $M_1, M_2$ and $\mu$ are all real--valued, the 
neutralino mass matrix is symmetric and can be diagonalized through a 
similarity transformation, i.e.
\begin{equation}
P M_N P^{-1} = M_N^{\rm diagonal}
\end{equation}
Now, rather than an analytically cumbersome inversion of the mass eigenvalue 
expressions, a simpler procedure is to start from the four basic invariants 
\begin{eqnarray}
\label{fourinv}
{\rm Tr} M \ , \ 
\frac{1}{2}({\rm Tr}M)^2 - \frac{1}{2} {\rm Tr} M^2 \ , \ 
\frac{1}{6} ({\rm Tr} M)^3 - \frac{1}{2}{\rm Tr} M \, {\rm Tr} M^2
+ \frac{1}{3} {\rm Tr} M^3 \ , \ {\rm Det} M 
\end{eqnarray}
under similarity transformations. These invariants contain the complete 
information on the relationship between the mass eigenvalues and the initial 
parameters in the neutralino mass matrix, but do not favor any particular 
set of parameters. Thus, the system may be solved in many different ways 
depending on the choice of input/output one is interested in. In fact, the 
necessary and sufficient conditions for the existence of solutions to 
eq.~(\ref{fourinv}) take the form 
\begin{eqnarray}
&&  P_{2 3}^2  
+  (\mu^2 + M_Z^2 - M_1 M_2 + (M_1 + M_2) S_{2 3} - S_{2 3}^2) P_{2 3} 
\nonumber \\
&& + \mu M_Z^2 (c_W^2 M_1 + s_W^2 M_2 ) \sin 2 \beta -\mu^2 M_1 M_2 =0
\label{condition1} 
\end{eqnarray}
\begin{eqnarray}
&&  (M_1 + M_2 - S_{2 3}) P_{2 3}^2+  (\mu^2 (M_1 + M_2)  + 
            M_Z^2 (c_W^2 M_1 + s_W^2 M_2  - \mu \sin 2 \beta)) P_{2 3} 
\nonumber \\
&&+ \mu ( M_Z^2 (c_W^2 M_1 + s_W^2 M_2 ) \sin 2 \beta -\mu M_1 M_2) 
    S_{2 3}=0 
\label{condition2} 
\end{eqnarray}
where $S_{2 3}= \tilde{m}_{\chi^0_2} + \tilde{m}_{\chi^0_3}$ and
$ P_{2 3}= \tilde{m}_{\chi^0_2} \tilde{m}_{\chi^0_3}$ [with similar
equations for all possible combinations of two neutralino masses, 
($\tilde{m}_{\chi^0_i}$, $\tilde{m}_{\chi^0_j}$)]. These equations
constitute our basics to invert the neutralino sector. For instance, 
in scenario $S_1$ we can extract $M_1$ and the three physical masses 
$m_{\chi^0_1}$, $m_{\chi^0_3}$, $m_{\chi^0_4}$ as functions of the mass
$m_{\chi^0_2}$, from any one of eqs.~(\ref{condition1}) or 
(\ref{condition2}). Note that the mass $m_{\chi^0_2}$ plays the role of 
any neutralino mass to be given as input, i.e. there will be a relabeling 
of neutralino states depending on the values of the other parameters. \s

{\bf Scenario $S_2$:} here, we assume that $\mu$ [alternatively 
$M_2$] is an input parameter together with two neutralino masses, say 
$m_{\chi^0_2}$ and $m_{\chi^0_3}$. Then, the quadratic system 
eqs.~(\ref{condition1}) and (\ref{condition2}) gives $M_2$ and $M_1$. 
The key point is that it is relatively simple to merge
these two basic algorithms, $S_1$ and $S_2$, to also obtain 
$\mu$, $M_2$, $M_1$ consistently from  $m_{\chi^\pm_1}$,
$m_{\chi^0_2}$, and $m_{\chi^0_3}$; for instance choosing an 
arbitrary initial guess value for $m_{\chi^\pm_2}$ [alternatively $M_2$],
one simply has to use $S_1$ followed by $S_2$, eventually iterating 
until a consistent, i.e. convergent, set of values is obtained. In most 
practical cases, convergence is very fast after 2 or 3 iterations. This 
peculiar decomposition, 
with this choice of input/output masses is deliberately
chosen as the one giving the most algebraically tractable
inversion in the
gaugino sector. It does not imply however, a strong 
particular prejudice on the chronology of discovery of the gauginos.
The most likely situation where one would presumably first discover
two neutralinos and only one chargino, is precisely
tractable from  the combined $S_1$ + $S_2$ 
algorithm as explained above. The price to pay however, is that
scenario $S_1 + S_2$ [with only one chargino mass input $m_{\chi^\pm_1}$, 
and without further model assumption]  
potentially gives more ambiguities than $S_1$ alone. 
The upshot is that up to four (at most)
distinct solutions for ($\mu$, $M_1$, $M_2$) 
can occur for some $m_{\chi^\pm_1}$, $m_{\chi^0_2}$,
$m_{\chi^0_3}$ input choices once all constraints are taken into account 
[including in particular our necessary sign convention $M_2 > 0$]. \s

Let us illustrate with some representative plots the results
of the inversion in the gaugino sector according to the algorithms
$S_1$ and $S_2$. As it turns out, a number 
of general and interesting properties of the inversion can directly 
be seen irrespective of the precise values of the other parameters 
that have to be fixed, like $\tan\beta$ typically. \s

\nn $(i)$ {\it Two charginos plus one neutralino input} \s
 
We first discuss the basic algorithm $S_1$ in Fig.~2, where to exhibit
as much as possible the dependence on the physical inputs, we fixed only one 
chargino mass, say $m_{\chi^\pm_1}$, and varied the other mass 
$m_{\chi^\pm_2}$.  Fig.~2 exhibits characteristics that are quite generic; 
there are three distinct zones as regards the existence, uniqueness, or 
possible ambiguities on the parameters $\mu$, $M_2$, $M_1$: \s

\begin{figure}[htb]
\epsfxsize=100mm
\epsfysize=100mm
\begin{center}
\vspace*{-0.5cm}
\hspace*{1.3cm}
\epsffile{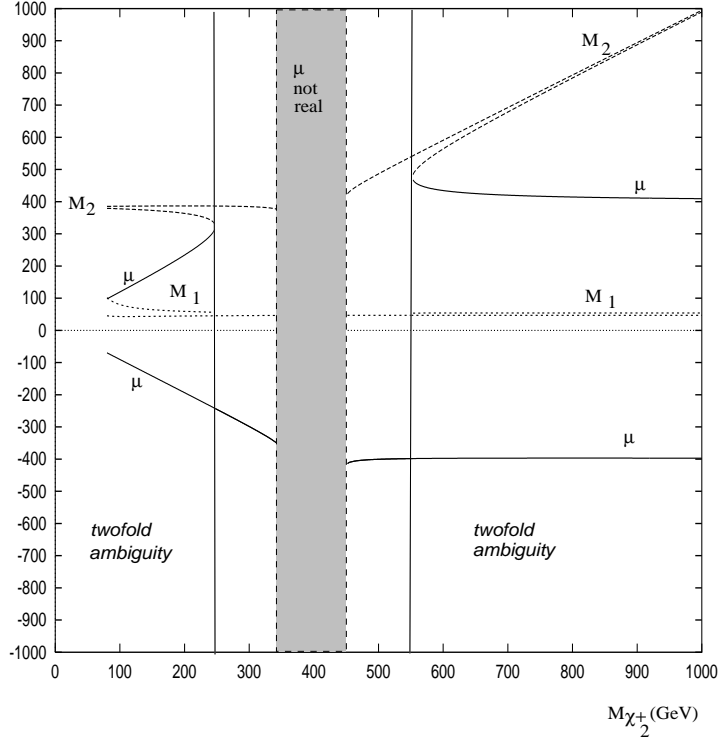}
\vspace*{-0.4cm}
\caption{\label{fig2} The parameters $\mu$, $M_2$ and $M_1$ as functions of  
$m_{\chi^\pm_2}$ for fixed $m_{\chi^\pm_1}= 400$ GeV, $m_{\chi^\pm_2}= 50$
GeV and $\tan\beta = 2$.}
\vspace*{-0.7cm}
\end{center}
\end{figure}

{\it a)} The grey shaded region corresponds to $m_{\chi^\pm_1}^2 +
m_{\chi^\pm_2}^2 <  2 M_W^2$ where there are no solutions for real 
$\mu$ that has to be rejected according to our basic 
assumptions\footnote{Of course, more generally, one 
could be interested in complex $\mu$ solutions. However, 
the present algorithm is not entirely consistent if $\mu$ and
$M_2$ are assumed complex so that in the present context
a complex $\mu$ solution of eq.~(\ref{invmu2}) has to be
rejected.}; if one takes a smaller or larger $m_{\chi^\pm_1}$
value, this region around $m_{\chi^\pm_1}$ will be simply displaced 
accordingly.  ${\it b)}$ In the left and right border zones are the 
regions of twofold ambiguities on $\mu$, $M_2$  as indicated. 
{\it c)} Finally the two bands in between zones {\it a)} and {\it b)} 
correspond to the region  where eqs.~(\ref{invmu2}) give a 
unique solution for $\mu$ and $M_2$; note that those bands are 
narrower when $\tan\beta$ is increasing [$\tan\beta =2$ in Fig.~2], 
irrespective of the $m_{\chi^\pm_1}$ values, becoming e.g. only a few GeV 
wide for $\tan\beta > 35$. \s

Note also that $\mu$ and $M_2$ are rather insensitive to $\tan\beta$, apart 
from the discontinuous change occurring for one of the solution at the border 
between zones {\it b)} and {\it c)}. One can also see from Fig.~2 the 
relatively simple shape of $\mu$ and $M_2$ as function of $m_{\chi^\pm_2}$, 
with an almost constant or linear dependence apart in some narrow regions. 
This is easily understood since from eqs.~(\ref{invmu2}),  one 
obtains $\mu (M_2) \simeq m_{\chi^\pm_1} (m_{\chi^\pm_2})$ for $m_{\chi^\pm_2} 
\ll m_{\chi^\pm_1}$ or  $m_{\chi^\pm_2} \gg m_{\chi^\pm_1}$. 
In Fig.~2 we also plot $M_1$ for the corresponding values of $\mu$ and
$M_2$ and for fixed $m_{\chi^0_2}$ [the almost constant behavior of $M_1$ 
in this plot, apart from small $m_{\chi^\pm_2} \simeq $ 100 GeV, is not 
completely obvious but is explained in details in Ref.~\cite{KM}]. \s

Thus, the information from the plots in Fig. 2 is that, apart from some small 
regions, for a very wide range of $\vert m_{\chi^\pm_1} -m_{\chi^\pm_2}\vert$ 
the dependence of $\mu$, $M_2$ [and even $M_1$ to some extent] upon the latter 
mass difference is strongly correlated. It is straightforward to obtain some 
resulting values of the parameters $\mu$, $M_2$ and $M_1$ at the GUT scale,
when a RG evolution of these parameters  is applied after the inversion 
algorithm $S_1$. The behavior of each parameter as a function 
of input masses remains essentially the same apart from a systematic shift 
due to the RG evolution. A comparison with SUSY--GUT model assumptions is then 
possible at this level. \s

\nn $(ii)$ {\it One chargino plus two neutralinos input} \s

Next we illustrate the probably more phenomenologically relevant combined 
scenario $S_1$ plus $S_2$, namely where $m_{\chi^\pm_1}$,  $m_{\chi^0_2}$ 
and $m_{\chi^0_3}$ are given as input. As expected, Fig.~3 reflects the more 
involved inversion when combining algorithm $S_1$ [with unknown $m_{\chi^\pm_2
}$] and $S_2$ due to the possible occurrence of a larger number of 
distinct solutions for 
($\mu$, $M_1$, $M_2$) in this case. However, apart from relatively 
messy--looking but narrow
zones where twofold solutions occur for this particular input 
choice, for a wide range of the $m_{\chi^2_2}$ values 
the solution is unique at least for these input values. 
The shaded regions again corresponds to a  zone where one (at least) 
of the output parameters ($\mu$, $M_1$, $M_2$) 
becomes complex-valued (that is, for any possible solution). 
In Fig.~3 we only show 
on purpose a range of values such that all masses are relatively
small, while for larger $\vert m_{\chi^2_2}\vert $ the dependence of
$\mu$, $M_2$, $M_1$ upon the latter 
becomes simpler and almost linear,  in 
accordance with the behavior in the 
previous figure for scenario $S_1$ alone. 
Also, the dependence of the scenario $S_1 +S_2$ upon
$\tan\beta$ variations is relatively mild. In contrast,
varying $m_{\chi^\pm_1}$ and/or  $m_{\chi^0_3}$ input values
for plots similar to   
Fig.~3 has more drastic effects since in particular, 
the number of distinct solutions crucially 
depend on those inputs. More
precisely, when varying those two input masses, some of the 
branches in Fig.~3 may disappear or  on the opposite, 
extra branches may appear although the behavior of a given 
unaffected branch as a function of $m_{\chi^2_2}$, 
remains essentially the same.

\begin{figure}[htb]
\label{figs2}
\vspace*{-.8cm}
\epsfxsize=100mm
\epsfysize=100mm
\begin{center}
\hspace*{1.3cm}
\epsffile{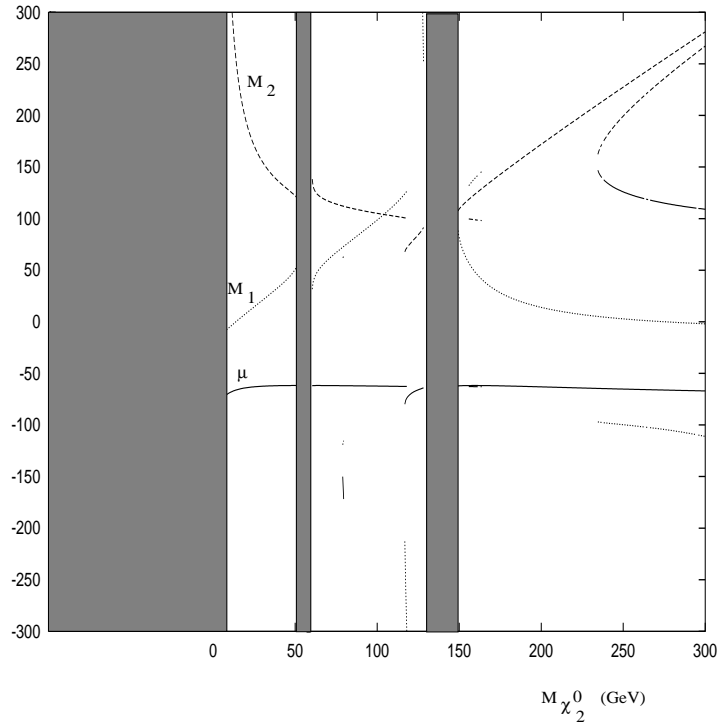}
\vspace*{-.8cm}
\end{center}
\caption{The parameters $\mu$, $M_2$ and $M_1$ as a function of  
$m_{\chi^0_2}$ for fixed $m_{\chi^0_3}= -100$ GeV and $m_{\chi^\pm_1} 
= 80$ GeV for $\tan\beta = 2$.}
\vspace*{-0.3cm}
\end{figure}

Finally, let us make a remark on the other MSSM parameters inversion. 
In parallel to the reconstruction of the gaugino sector soft--breaking 
parameters from the physical masses, it is natural to attempt such a 
reconstruction for the remaining part of the soft--breaking Lagrangian. 
In contrast to the gaugino sector however, the de--diagonalization of 
the sfermion sector and the Higgs sector does not present much analytical 
difficulties, provided of course that  one knows a sufficient number of 
physical masses and/or couplings. Here, we briefly mention that such an 
inversion is indeed possible and we refer for more details and illustrations 
to Ref.~\cite{KM}. We should only note that, in contrast with the 
gaugino sector inversion where the only difficulty was due to algebraically 
non--trivial de--diagonalization, for a realistic inversion of the Higgs 
parameters care should be taken with the correct account of one--loop 
corrections to the scalar potential and Higgs masses. 

\newpage

\subsection*{3.2 The program SUSPECT\footnote{During the last GDR--SUSY 
meeting in Montpellier, some people complained about the former name of 
the fortran code: MSSMSPEC seemed to be difficult to pronounce by some of our
(presumably non Slavic) colleagues, probably due to a local cluster
of consonants. We propose a change of name
to SUSYSPECT, or to make short SUSPECT [since every code, is, a priori!]. 
%(the additional T is however, only moderately welcome by some of us!). 
%It is needless to say that we believe that the program does not deserves 
%this name.
\\ }}

\subsubsection*{3.2.1 Introduction} 

It is a well--known fact that the proliferation of Supersymmetry breaking 
terms in the unconstrained MSSM Lagrangian makes the Lagrangian--to--physical 
parameters [i.e. particles masses and couplings, etc.] relationship a rather 
tedious task to derive in an exhaustive manner. Although several systematic 
routines doing this work with different levels of refinement are available, it 
turned out to be highly desirable to develop our own tool within the GDR 
workshop,  with the specific aim among other things, to fix a GDR--common 
set of parameter definitions and conventions once for all, and to have as 
much as possible flexibility on the input/output parameter choice. \s

Such a program should necessarily include the possibility of renormalization
group evolution of the relevant parameters, due to the all importance of the 
high/low energy scales interplay in most modern SUSY scenarii, with GUT models 
as a typical but not only example. Besides, it unavoidably poses numerous 
questions on the relevance of various possible ``options" [or more precisely 
model assumptions] to be available, as well as the choice of an adequate level
of approximation at different stages, such as typically, in the effective 
potential related to radiative corrections to Higgs boson  masses. \s

This brought us to develop the Fortran code SUSPECT which calculates the
masses and couplings of the SUSY particles and the Higgs bosons of the MSSM.
The code contains one source file [SUSPECT.f] and one input file [SUSPECT.in]; 
any choice and option is driven from this input file where one can change data 
almost at will. All results, including comments when useful, 
are to be found in the output file [SUSPECT.out] which is 
created at any run of the program. It is hoped that the code may be readily 
usable even with not much prior knowledge on the MSSM. \s

The code is as yet at a rather 
preliminary stage [version 1.1]. As mentioned previously, 
its main purpose at present is to propose some conventions, definitions, 
and possible flexibility choices which are largely open to 
discussions/suggestions. 
In the present version, only two extreme ``models" are 
readily available: the 
``phenomenological MSSM" [but with the possibility of RG evolution
to arbitrary scales] and the mSUGRA model; 
many ``intermediate" models 
can be very easily included [suggestions are particularly welcome here].
For instance the pMSSM with unification of the gaugino masses at the
GUT scale, or with a common scalar mass are straightforward to implement 
by just setting constraints ``by hand" in the input file. 
The output for the MSSM mass spectrum has been compared with
several other existing similar codes [see the ``tools for SUSY"
working group report \cite{outils} for more details on those comparisons],
and is in reasonably good agreement. Up to a few percent 
differences in the output
is to be sometimes expected, however,  due to different 
approximations used in different codes. \s

The rest of this subsection is organized as follows: we first describe 
the most important subroutines in the case of the ``phenomenological" MSSM
in particular those which compute the chargino/neutralino, sfermion and Higgs 
boson masses and couplings. We will then discuss the case of the constrained
MSSM with the various choices of approximations and refinements, paying
a special attention on the RG evolution and the radiative electroweak 
symmetry breaking. Next, a description of the input and output files to be 
driven by the user is given. In doing so we shall try to be as close as 
possible to the definitions and conventions given in the previous subsections. 
We  finally collect a list of various available [or not yet] options and/or 
model assumptions, with a clear mention of eventual limitations of the 
present version of the code. 

\subsubsection*{3.2.2 The ``phenomenological" MSSM} 

The purely phenomenological MSSM, that is, with the set of input parameters 
as defined in section 2.1, and without any RG evolution, 
is implemented by enforcing in the SUSPECT.in 
input file 
the choice ICHOICE(1)=0. Another alternative of interest
is to have 
the full set of parameters as defined by the phenomenological model, 
but with the further possibility to evolve those parameters
to --or from-- an arbitrary scale prior to calculating the physical spectrum. 
This latter
case is readily
implemented and enforced  by the choice ICHOICE(1)=1. \s

There are three main subroutines evaluating all possible SUSY 
physical masses, respectively 
chargino/neutralinos, Higgs boson and sfermion masses, 
for given input Lagrangian parameters: 
\begin{verbatim}
       CALL GAUGINO(MU,M1,M2,M3,BETA,ALPHA,GMC,GMN,XMN)
       CALL SUSYCP(TGBETA)  
       CALL SFERMION(MSQ,MTR,MBR,MSL,MTAUR,AL,AU,AD,MU,
     .               GMST,GMSB,GMSL,GMSU,GMSD,GMSE,GMSN)
\end{verbatim}
%(NB the indicated calling order is important, at least for the first call, 
%since e.g. the Higgs mass calculated by SUSYCP uses some of the output from 
%GAUGINO, as we will see).\\
For the couplings, only a subset is implemented at present: the couplings of 
the MSSM Higgs bosons to standard and SUSY particles [the angles $\alpha$
and $\beta \equiv {\rm atan} (\tan\beta)$ 
of the MSSM Higgs sector are called by a common]. Note that the
couplings are not readily 
default output in the SUSPECT1.1 version, 
but are 
straightforwardly available, 
from various common blocks,  as specified below. 
The remaining couplings [sfermion/ino couplings to gauge bosons and 
fermion--sfermion--ino] will be included soon. \\

\nn In the subroutine {\bf GAUGINO}, the input parameters are $\mu$,
the gaugino masses $M_1,M_2,M_3$ [which can be made related], the mixing 
angles $\beta$ and $\alpha$ in the Higgs sector [the former is an input 
while the latter is calculated in the subroutine SUSYCP as explained below].
Signs and other relevant conventions are consistent with the ones discussed 
in section 3.1. The output is: \s

\nn \hspace*{2cm} GMC(1), GMC(2): ordered values of the chargino masses\\
    \hspace*{2cm} GMN(1)--GMN(4): ordered absolute values of the neutralino
masses \\
    \hspace*{2cm} XMN(1)--XMN(4): neutralino masses including the eventual 
negative signs. \\

\nn The neutralino mass matrix is diagonalized analytically.
The negative signs in  XMN(1)--XMN(4) are relevant for instance for the 
couplings to Higgs bosons; if needed, these couplings are available via a 
common defined in the subroutine:  \s

\nn \hspace*{2cm} AC1(2,2),AC2(2,2),AC3(2,2): chargino--$h,H,A$ couplings\\
\hspace*{2cm} AN1(4,4),AN2(4,4),AN3(4,4): neutralino--$h,H,A$ couplings\\ 
\hspace*{2cm} ACNL(2,4),ACNR(2,4): chargino--neutralino--$H^\pm$ couplings. \\

The Higgs boson spectrum and couplings are calculated by the subroutine
{\bf SUSYCP}. It uses as two important input parameters: $\tan\beta$ and the 
mass of the pseudoscalar Higgs boson $M_A$ which may be either given as 
direct input or calculated from the two soft--SUSY breaking scalar 
masses $M_{H_u}$ and $M_{H_d}$ and $\mu$ as in eq.~(\ref{EWSBcond}). In 
addition, SUSYCP uses as input the third generation soft scalar masses and 
trilinear couplings as well as $\mu$ which enter
in the radiative corrections. In the case where the pole Higgs boson
masses are calculated it needs the parameter $M_2$, as well as the 
masses of stops and sbottoms and their couplings to Higgs bosons [which
are calculated similarly as in the subroutine SFERMION].  
SUSYCP calls the subroutine SUBH \cite{CQW}, which calculates the 
renormalization group improved values of the Higgs boson masses and the 
couplings $\lambda$ in the scalar potential\footnote{Note that compared to
the original version \cite{CQW}, in the one used in the program some 
bugs [e.g. in the calculation of the $\tilde{b}$ contributions to 
the running Higgs boson masses] have been corrected and some improvements
[such as using the standard parameters given in the input file and 
calculating the running quark masses and $\alpha_s$ using the 
subroutines RUNM and ALPHAS, as well as calculating the Higgs couplings
to squarks as in SFERMION, see below] have been made. Furthermore, the 
function $F0$ which calculates the loop functions is given explicitly.}. 
The SUSYCP output gives the SM and $A,h,H,H^\pm$ Higgs boson masses and the 
running $A$ mass [the SM Higgs boson mass is in fact the input $M_A$] \s

\nn \hspace*{4cm} AMSM,AMA,AML,AMH,AMHC,AMAR \s

\nn Depending on the flag 
ICHOICE(10) one has either the running (=0) or pole (=1) Higgs boson
masses masses [in the former case AMAR$\equiv$AMA]. The output also 
provides via common blocks the 
self--couplings $\lambda_{1-7}$ in the scalar potential, and the Higgs 
couplings to fermions, gauge bosons and the self--couplings [the notation 
being obvious, with T,B,L standing for $t,b,\tau$ and L,H,A for $h,H,A$;
GLPM=$\sin(\beta-\alpha)$ and GHPM=$\cos(\beta-\alpha)$] as well as the 
running angles $\beta$ and $\alpha$: \s

\nn COMMON/HMASS/AMSM,AMA,AML,AMH,AMCH,AMAR \\
COMMON/HSELF/LA1,LA2,LA3,LA4,LA5,LA6,LA7 \\
COMMON/COUP/GAT,GAB,GLT,GLB,GHT,GHB,GZAH,GZAL,GHHH,GLLL,\\
\hspace*{3.4cm} GHLL,GLHH,GHAA,GLAA,GLVV,GHVV,GLPM,GHPM,B,A \\

\nn In the subroutine {\bf SFERMION}, the input parameters are the soft--SUSY
breaking mass parameters for left--handed and right--handed squarks and 
sleptons, the trilinear couplings $A_t, A_b$ and $A_\tau$ for the third 
generation and the parameter $\mu$. The output are 2--vectors: \s

\hspace*{2cm}   GMST,GMSB,GMSL,GMSU,GMSD,GMSE,GMSN \s

\nn for the masses of the two $\tilde{t}, \tilde{b}, \tilde{\tau}$ eigenstates, 
their first/second generation partners as well as the mass of the 
$\tilde{\nu}$'s. The D--terms are calculated by the subroutine and full 
mixing in the third generation is implemented. Again, the couplings to Higgs 
bosons are not explicitly written but are available if desired from a common 
defined in SFERMION: \s

\nn \hspace*{2cm} GLEE(2,2),GLTT(2,2),GLBB(2,2): $h$ couplings to $\tilde{\tau},
\tilde{t}, \tilde{b}$\\ \hspace*{2cm}
GHEE(2,2),GHTT(2,2),GHBB(2,2): $H$ couplings to $\tilde{\tau},
\tilde{t}, \tilde{b}$\\ \hspace*{2cm}
GAEE,GATT,GABB: 
$A$ couplings to $\tilde{\tau}_1 \tilde{\tau}_2$,  
$\tilde{t}_1 \tilde{t}_2$,   $\tilde{b}_1 \tilde{b}_2$\\ \hspace*{2cm}
GCEN(2,2),GCTB(2,2): $H^\pm$ couplings to $\tilde{\tau}\tilde{\nu}$,
$\tilde{t} \tilde{b}$. \s

The conventions for the signs of $A_f, \mu$ as well as the mixing
terms are as introduced previously. [The sfermion masses are running
masses, since they are obtained from the Lagrangian parameters of the
scalar sector; note indeed that  
for stops and sbottoms the running quark partner masses 
are implemented in the mass matrices]. \s

Of course, the three subroutines need in addition 
some standard input parameters.
Apart from ``electroweak" coupling constant inputs 
[such as the Fermi constant
or the QED coupling defined at zero--momentum] and some particles masses 
[such as the W,Z gauge boson and fermion masses] which are precisely
known and which are included in the source code, there are other 
less precisely known parameters which are  very important
[especially in the constrained MSSM, since e.g. the GUT and SUSY thresholds 
are quite sensitive to the values of some of them]. These constants
[defined at the low scale, where they are experimentally measured, 
currently $M_Z$] are: \s

\nn \hspace*{1cm} ALFINV: the inverse QED constant $1/\alpha(M_Z)$ in the 
$\overline{\rm MS}$  scheme (=127.9) \\
 \hspace*{1cm}  SW2: $\sin^2 \theta_W (M_Z)$ in the $\overline{\rm MS}$ 
scheme (=0.2315 for $m_t=175$ GeV); \\
 \hspace*{1cm}  ALPHAS:  the value of $\alpha_S(M_Z)$ at the $M_Z$ scale 
(=0.119 ); \\
 \hspace*{1cm}  MT,MB,MC: the $t,b,c$ pole masses (=175, 4.7, 1.42 GeV) \s 

\nn The reference values for those parameters as indicated in parenthesis 
above correspond to central values quoted by the most recent 
data~\cite{data}; they may accordingly be varied within the known error 
bounds.

\subsubsection*{3.2.3 Constrained MSSM} 

The case of mSUGRA is dealt with by the flag ICHOICE(1)=10 in SUSPECT.in. 
In the mSUGRA choice, all the SUSY parameters are determined in terms of 
four arbitrary and one discrete parameters which are given at a ``high" scale 
which, if unification of the gauge couplings is imposed, corresponds to the 
GUT scale $M_U$. These parameters, as discussed in section 2.1,  are: 
\begin{verbatim}
                 TGBETA(MZ), m0(GUT), mHALF(GUT), A0(GUT), sign(MU)
\end{verbatim}
Other parameters also enter the game as ``starting guess" values, 
and must be specified in some cases:
\begin{verbatim}
                      MU, B, SUSYM, EHIGH, ELOW
\end{verbatim}
Typically, the actual values of $\mu$ and $B$ are determined consistently 
from EWSB, but the latter almost unavoidably needs some iteration 
procedure, and
thus some starting ``seed" $\mu, B$  values are necessary.  
If the unification of the gauge couplings scenario is not 
chosen two additional inputs must be given: SUSYM corresponding to an 
initial guess of the 
SUSY threshold [a single universal one to simplify], 
and EHIGH corresponding to the high energy end of the RG
evolution [if gauge unification is enforced, both SUSYM and EHIGH
 can be safely set to arbitrary values, as 
more adequate values will be then calculated]. 
Finally, one needs
to specify ELOW which corresponds to the lower--energy end of the
RG evolution equations; this is always necessary\footnote{For 
all mass and coupling calculations performed by the routines 
GAUGINO, SUSYCP and SFERMION described above, a scale is 
implicitly involved since the input Lagrangian parameters are 
implicitly defined at some scale, 
the latter being eventually the 
one resulting from an RG evolution; see the discussion below.} and important 
as it fixes
the final scale at which all physical masses will be computed. Moreover,
ELOW should be below or at most equal to the EWSB scale
[see the discussion in section 2.2]. 
Thus ELOW should be roughly of the order of the 
electroweak scale, with some flexibility, say 
$M_Z  < {\rm ELOW}  < 1000$ GeV. In the mSUGRA case 
all other parameters are then determined by 
GUT universality and RG evolution. \s

In addition, independently of the model choice driven by ICHOICE(1), 
there are several possibilities for the RG 
evolution equations and for electroweak 
symmetry breaking [which are 
however not always all relevant, 
depending on the choice of models]: \s
\begin{itemize} 
\item ICHOICE(2): elaboration level of the RG evolution equations. \\
=11 (21): one (two)--loop evolution with simple [single SUSY--threshold 
scale] threshold effects. \\
=12 (22): one (two)--loop evolution, with more realistic thresholds
[i.e. step functions for each particle specy thresholds]. 
% \\ Other possible options could be easily implemented later. 

\item ICHOICE(3): Gauge coupling and Yukawa coupling unification. \\
= 0: no unification \\
= 1: only gauge coupling unification [then approximate GUT and simple
SUSY--threshold scales are automatically determined]. \\
= 2: gauge and bottom--tau Yukawa coupling unification 
 
\item ICHOICE(4): Accuracy of RG evolution. \\
= 0: fast but less accurate RG evolution. \\
= 1:  accurate evolution but rather slow. 

\item ICHOICE(5): for electroweak symmetry breaking. \\
=0:  no radiative EWSB 
[electroweak breaking is of course there and determines
consistent values of $B$ and $\mu$ at the EWSB scale; 
the difference between ``radiative" and ``no radiative" EWSB 
was explained in section 2.2]. \\ 
=1: radiative EWSB is implemented, i.e. 
$B$ and $\mu$ are fixed at the EW
scale and by other parameters to be consistent with 
RG evolution from a given (generally EHIGH $\simeq M_{U}$)
scale. 
[NB: ICHOICE(5)=1 is automatically enforced for the mSUGRA case ICHOICE(1)=10]
\item ICHOICE(6)--ICHOICE(9) are reserved for later use.

\item ICHOICE(10): =0 (1) computes running (pole) Higgs boson masses\footnote{
Physical masses are, strictly speaking, defined as the pole masses:
$M^{\rm pole} \equiv  m(M^{\rm pole})+$ radiative 
corrections, where $m(Q)$ is the corresponding running mass, whose evolution 
with energy scale $Q$ is entirely dictated by the RG; the radiative correction
parts are all non--RG finite loop corrections. However, apart from the exception
of the Higgs boson and gluino masses, 
for most other MSSM particles the non--RG corrections are 
assumed to be negligible with respect to the present level of approximation 
and are 
not included in SUSPECT.}.
\end{itemize} 

Note that we anticipated several possible steps between the pMSSM and mSUGRA 
and therefore left the possibilities ICHOICE(6--9) reserved for later use. 
Besides the physical spectrum calculation for input Lagrangian parameters 
as described above, the two other main tasks performed by the code are the RG 
evolution of the parameters, and the consistent implementation  of 
EWSB, which we discuss is some more details now. \s

\nn {\it (i)  Renormalization Group evolution}: \s

ODEINT is a subroutine 
returning a set of masses and coupling parameters
at a specified scale $Q_{\rm out}$, when given at an initial scale 
$Q_{\rm in}$. It is based on a Runge--Kutta 
numerical algorithm solving differential equations
by Numerical Recipes~\cite{NR}. RG evolution is trivially
reversible, so that the choice of $Q_{\rm in}$ 
and $Q_{\rm out}$ is in principle quite flexible; it 
is also possible in particular to use this evolution back and forth if 
needed.
\begin{verbatim}
       CALL ODEINT(y,n,x1,x2,eps,h1,1.d-5,nok,nbad,deriv1,rkqc)
\end{verbatim}
The input are the values $y(n)$ of all evolving couplings
and masses relevant to the phenomenological MSSM,
given at the scale $Q_{in} =e^{x_1}$, the latter being also given as 
input; see section 2.3 for the precise assignment of $y(n)$
components in terms of MSSM couplings and masses. [Quite obviously, 
when specific  constraints on  the $y(n)$ are implemented at some
scale, like in the mSUGRA model typically, it simply 
implies specific initial
conditions for those. But the 
evolution of each mass or coupling parameters is driven by 
its own specific beta function and we therefore 
should keep the latter as model--independent as possible]. 
The output are the values of $y(n)$ at the scale $Q_{\rm out} =e^{x_2}$ to 
be specified. \s

According to the previous discussion ODEINT also needs the beta functions: 
$\beta(y) (\equiv d(y)/d\ln Q$), for all the relevant $y(n)$, which are 
provided by the subroutines: 
\begin{verbatim}
                    subroutine DERIV1(x,y,dydx)
                    subroutine DERIV2(x,y,dydx)
\end{verbatim}
for the one and two-loop RG beta functions respectively,  including 
threshold effects. \\
 
\nn {\it (ii) Electroweak Symmetry Breaking}: \s

There is a set of subroutines [in fact essentially one] which
evaluates 
the one--loop contribution to the 
effective potential, whose detailed shape study
settles the occurrence [or not] of the ${\rm SU(2)\times U(1)}$ 
spontaneous symmetry breaking, irrespective  of particular model
building assumptions.  What is
however, a model assumption chosen as option flag, is whether
the EWSB is radiative or not; namely whether the relevant 
parameters of the Higgs potential at the EWSB scale are consistent 
or not with their respective values chosen at a higher scale, 
given their specific RG evolution 
properties [see again section 2.2 for a
 more complete discussion on EWSB].       
\begin{verbatim}
       call VLOOP2(Q2,mt2,mst22,mst12,mb2,msb22,msb12,
     .        mtau2,mstau22,mstau12,-rmu,dVdvd2,dVdvu2)
\end{verbatim}
The inputs are: \s

\nn \hspace*{2cm} Q2: the scale $Q^2$ ($\geq $ ELOW) at which EWSB is supposed 
to happen; \\
\hspace*{2cm} mt2, mst22, mst12: the $t, \tilde{t}_1, \tilde{t}_2$ masses; \\ 
\hspace*{2cm} mb2, msb22, msb12: the $b, \tilde{b}_1, \tilde{b}_2$ masses; \\
\hspace*{2cm} mtau2, mstau22, mstau12: the $\tau, \tilde{\tau}_1, \tilde{\tau}_2$ \\ 
\hspace*{2cm} $-$RMU: the parameter $\mu$ \\

The output are dVdvd2, dVdvu2, which are [up to some appropriate overall 
constants] the derivatives $\partial(V_{\rm eff})/
\partial(v^2_d)$, $\partial(V_{\rm eff})/\partial(v^2_u)$ respectively, the
basic quantities entering Higgs mass corrections and EWSB consistency
conditions. As already mentioned,  our present algorithm calculates values 
of the parameters $B$ and $\mu$ consistent with EWSB by iterating the two 
expressions of $\mu^2$ and $B$, eqs.~(\ref{EWSBcond}) of section 2.2. 
The iteration is non trivial because in eqs.~(\ref{EWSBcond}):
\beq 
\tilde M^2_{H_d} \equiv M^2_{H_d} +\partial(V_{\rm eff})/\partial(v^2_d)\ \ 
{\rm and} \ \ \tilde M^2_{H_u} \equiv M^2_{H_u} +\partial(V_{\rm eff})/
\partial(v^2_u)
\eeq
involve other MSSM particle masses and couplings, which themselves
depend on $\mu$ [in most practical cases, however, the algorithm 
converges very fast, after 2 or 3 iterations]. Besides eqs.~(\ref{EWSBcond}),
there are other constraints to fulfill in order to assert that one reached a 
local minimum of the potential, as discussed in section 2.2. 
This is  simply checked numerically in SUSPECT. In contrast, there 
is no consistency
checks about the possible occurrence of charge and color breaking [CCB] 
minima in the present version 1.1.  
Note, however,  that a more refined different 
algorithm, partly based on analytical studies 
of the complete potential at its
minimum~\cite{LMM}, will be implemented later. 

\subsubsection*{3.2.4 Example of input/output files} 

The following example of SUSPECT.in input file, corresponds to the mSUGRA 
so--called ``SNOWMASS" point 1:
\begin{verbatim}
1) choice of different options/models, etc flags:
ICHOICE(1) = 10
ICHOICE(2) = 21
ICHOICE(3) = 12
ICHOICE(4) = 0
ICHOICE(5) = 1
ICHOICE(6) = 0
ICHOICE(7) = 0
ICHOICE(8) = 0
ICHOICE(9) = 0
ICHOICE(10)= 1

2) Initialize SM parameters: 
ALFINV   =127.9d0
SW2      =.2315d0
ALPHAS   =.12d0
MT       =175.d0
MB       = 4.7d0
MC       =1.42d0

3) Initialize SUSY parameters 
SUSYM    = 200.d0
EHIGH(GUT=2.d16
ELOW     = 433.d0 
(here SUGRA case with universality): 
m0       = 400.d0
m1/2     = 200.d0
A0       = 0.d0
TGBETA   = 2.d0
SGNMU0   = -1.d0

4) pMSSM case (at HIGH scale!) 
MHU2     = .5D4
MHD2     = 5.d5
M1       = 200.D0
M2       = 200.D0
M3       = 200.D0
MSL      = 1.D2
MTAUR    = 1.D2
MSQ      = 1.d2
MTR      = 1.d2
MBR      = 1.D2
MEL      = 1.D2
MER      = 1.D2
MUQ      = 1.D2
MUR      = 1.D2
MDR      = 1.D2
AL       = 200.D0
AU       = 200.D0
AD       = 200.D0
\end{verbatim}
The corresponding  SUSPECT.out output file is shown below [the comments 
are self--explanatory]:
\begin{verbatim}
 HIGH(GUT), LOW (Final) and SUSY-threshold     SCALES:
  0.274650E+17    433.000        175.000    
 EVOLVE FIRST TIME from MZ to HIGH(GUT) scale 
 gauge cpl^2 1,2,3 at HIGH(GUT) scale  0.529596 0.526736 0.517345    
 RUNNING mtau,mb,mtop at HIGH scale    1.58621  1.18050  76.0442    
 EVOLVE DOWN to Low-energy
 
__________________________________________________________________
            RESULTING values at Low-energy scale: 
 tan(beta) at Low scale:    1.93532    
 mtau,mb,mtop at Low scale:    1.76582        2.69798        151.422    
 Atau,Ab,Atop at Low scale:   -135.228       -633.479       -368.151    
 m(phi_u)^2, m(phi_d)^2   -137416.        179919.    
 m_tauR, m_L, m_bR, m_tR, m_Q: 
   407.225      424.685      607.580      399.425      530.837    
 rel. % err. in B, mu:   0.280874E-03    0.338679E-03
 (indicate departure from consistent radiative  EW breaking)
 loop-level EW breaking + stability tests:
 m1^2*m2^2-B^2*MU^2, m1^2+m2^2 +/-2*B*MU:
 -0.767807E+09   994235.       100948.    
 FINAL SOFT-BREAKING PARAMETERS: 
 mu(Q_EWB), B(Q_EWB):   -504.916       -442.295    
 gauginos 1,2,3 masses:    83.3875        165.443        524.285    

 __________________________________________________________________
            FINAL PHYSICAL PARAMETER RESULTS: 
    tgbeta       M_A        M_h       M_H        M_H+  
      2.000   735.902    79.555   742.020   743.895

    chi_+^1    chi_+^2    chi_0^1    chi_0^2    chi_0^3    chi_0^4 
   172.336   515.130    85.597   172.466   506.938   516.171

   stop_1     stop_2     sup_1      sup_2   
   429.182   557.382   621.448   608.878

   sbottom_1  sbottom_2    sdown_1    sdown_2 
   532.761   607.879   624.448   608.114

    stau_1     stau_2    selec_1    selec_2      sneut 
   408.361   426.429   426.354   408.756   421.948
 __________________________________________________________________
\end{verbatim}

\subsubsection*{3.2.5 Discussions and outlook} 

In summary, the main core of the code SUSPECT gives the physical masses and 
couplings of the SUSY particles and the MSSM Higgs bosons as functions of 
input Lagrangian parameters, taking into account renormalization group 
evolution if needed, and consistent electroweak symmetry breaking. In 
addition, there are a number of options, which might be useful for many 
practical purposes, that we briefly describe now.\\

INVERTOR: \s

\nn This is a subroutine [based on the discussion in section 3.1.2],  
which will be included soon in the next version, 
as an option in SUSPECT and whose purpose is to 
determine the inverted spectrum relationship 
[i.e. recovering the Lagrangian parameter values directly from  
physical masses and/or couplings]. 
The algorithm in its present form essentially deals with the non--trivial
inversion in the gaugino parameter sector, where the input can be either
two charginos and one neutralino, or two neutralinos and one chargino
physical masses. The output are the Lagrangian gaugino parameters,
$\mu$, $M_1$, $M_2$ [the $M_3$ to $M_{\rm gluino}$ relation obviously does not
need inversion]. \\

RGEXACT: \s

\nn This is a subroutine also to be included in the next
version as an option in SUSPECT
which implements an exact RG evolution solver [limited to one--loop
approximation]. We refer to section 2.3 for a more detailed 
discussion of the procedure. Let us simply mention here, that the exact 
solutions of the relevant SUSY RG equations which have been recently derived 
for arbitrary values of $\tan\beta$, should not only be useful to improve the 
general RG evolution algorithm, but more importantly, should provide 
a better control on some non--trivial issues of the RG evolution, such as the 
possible occurrence of Landau poles in the Yukawa couplings typically. \\

Finally we emphasize that the code is largely in a developing stage, and
many other extensions of the present algorithms and options
are foreseen, as well as interfaces with other numerical  codes for 
SUSY\footnote{For instance, the three subroutines GAUGINO, SFERMION and SUSYCP
are also used in the program HDECAY which calculates the branching ratios 
of the MSSM Higgs bosons and that SUSYCP is used in HPROD [see next section]
for the Higgs 
production cross sections in $\ee$ collisions: an interface is therefore 
easily possible and will be done soon.}.  Some of the present limitations 
and approximations of the version SUSPECT1.1 are: \\

\nn {\it (i) Model choice}: \s

\nn Only two extreme ``models", phenomenological MSSM 
on one side and mSUGRA with all universal terms on the other side, are 
treated at present. Many ``intermediate" models can be enforced by hand by
choosing appropriately the input parameters and will be very easily 
included later. \\

\nn {\it (ii) Scale choice}: \s

\nn The input scale at which to define parameters is 
not at present completely flexible: in the phenomenological MSSM, 
all SUSY parameters 
should be entered at the same scale [the latter scale is however 
arbitrary, provided it is chosen between 1 TeV and say $10^{18}$ GeV].
An important exception is  $\tan\beta$, which should be given at the 
$Z$--boson scale and of course the SM  parameters such as $\alpha_s$, etc..  
[see the example of input file].  \\

\nn {\it (iii) Approximations in specific calculations}: \s

\nn There are approximations in the RG evolution, for instance an option on 
one--loop versus two--loop evolution, where in the latter only the gauge 
and Yukawa couplings truly involve two--loop contributions. Also, there are 
approximations in the contributions to the one--loop corrections to the 
scalar potential [relevant for the radiative EWSB consistency conditions]: 
that is, only the third generation fermion and sfermion contributions
are included. This implies, strictly speaking, some inconsistency
with the RG evolution since in the latter much more contributions
are taken into account. The inconsistency however, should not be very 
significant numerically. A more complete and consistent treatment of the 
one--loop scalar effective potential will be  available in the next 
version of the program.   \\

\nn {\it (iv) Threshold effects}: \s

These are not yet fully realistic at the
present moment: only  a single, average SUSY threshold scale can be chosen. 
Inclusion of more realistic thresholds [i.e. for each particle species] will 
be soon available. \\

\nn {\it (v) Electroweak symmetry breaking}: \s

As explained previously, the 
EWSB conditions are consistently implemented numerically by iteration on the
parameters $\mu$ and $B$, and the occurrence of a local minimum is
checked numerically. There is however, no check about the possible
occurrence of inconsistent charge and color breaking minima in the
present version. These should be implemented in the next version,
together with a complete and more analytical treatment 
of the EWSB conditions. \\

\nn {\it (vi) Known bugs}: \s

Finally to be fair one cannot omit
mentioning that one serious bug that has been identified in this
1.1 version, is not yet completely understood. The
observation is that, for some choices of the
low scale input parameters, and in particular for 
a relatively low value of $\tan\beta$, $\lsim 1.3$--$1.4$, there is an
uncontrollable growth of the Yukawa/gauge couplings, when running the RGE
in the upwards direction. This seemingly ``Landau pole" effects is not yet
under control and we hope to solve this problem for the next version [the
exact RG solver RGEXACT will certainly help us to resolve the issue]. \\

The next version SUSPECT1.2 should therefore be hopefully more complete.

\newpage

\setcounter{equation}{0}
\renewcommand{\theequation}{4.\arabic{equation}}

\section*{4. Higgs Boson Production and Decay}

One of the main motivations of supersymmetric theories, as discussed in 
section 2, is the fact that they provide an elegant way to break the 
electroweak symmetry and to stabilize the huge hierarchy between the GUT 
and Fermi scales \cite{R1,HaberKane}. The probing of the Higgs sector of the 
MSSM \cite{R2} is thus of utmost importance. The search for the CP--even Higgs 
bosons $h$ and $H$, the pseudoscalar Higgs boson $A$ and the charged Higgs 
particles  $H^\pm$ of the MSSM is therefore one of the main entries in the 
LEP2 \cite{LEP2}, Tevatron \cite{TEV}, LHC \cite{LHCex,LHCth} and future 
$\ee$ colliders \cite{LC} agendas. \s

In the theoretically well motivated models, such as the mSUGRA scenario
discussed previously, the MSSM Higgs sector is in the so called decoupling 
regime \cite{decoup} for most of the SUSY parameter space allowed by present 
data  constraints \cite{data}: the heavy CP--even, the CP--odd  and the 
charged Higgs bosons are rather heavy and almost degenerate in mass, while 
the lightest neutral CP--even Higgs  particle reaches its maximal allowed 
mass value $ M_h \lsim $ 80--130 GeV \cite{radcor,mh} depending on the SUSY 
parameters. In this scenario, the $h$ boson  has almost the same properties 
as the SM Higgs boson and would be the sole Higgs particle accessible at 
the next generation of colliders. \s

In this section, we  discuss the production of the MSSM Higgs bosons
at the next generation of  colliders, mainly at the LHC and a future $\ee$ 
linear collider as well as their decays modes, in particular the SUSY channels. 
The new material presented in this section is based on 
Refs.~\cite{ggh,hstop,hdec,franc,hprod}. 

\subsection*{4.1 MSSM Higgs boson production at the LHC}

\subsubsection*{4.1.1 Physical set--up} 

At the LHC, the most promising channel \cite{LHCex,LHCth} for detecting the
lightest $h$ boson is the rare decay into two photons, $h \rightarrow \gamma 
\gamma$, with the Higgs particle dominantly produced  via the top quark loop 
mediated gluon--gluon fusion mechanism \cite{R7,R7c}
\beqn
gg  \rightarrow h
\eeqn
In the decoupling regime, the two LHC collaborations expect to detect the 
narrow $\gamma \gamma$ peak in the entire Higgs mass range, 80 $\lsim M_h
\lsim 130$ GeV, with an integrated luminosity $\int {\cal L} \sim 300$ 
fb$^{-1}$ corresponding to three years of LHC running \cite{LHCex};
see Fig.~4. \s

Two other channels can be used to detect the $h$ particle in this mass 
range:  the production in association with a $W$ boson \cite{R8} or
in association with top quark pairs \cite{R9}
\beqn
pp \ \ra \ hW \ \ \ {\rm and} \ \ \ pp \ra \bar{t}t h
\eeqn
with the $h$ boson decaying into 2 photons and the $t$ quarks into $b$ quarks 
and $W$ bosons [for the latter process, the Higgs boson detection with $h \ra 
b\bar{b}$ final states looks also promising; see Ref.~\cite{bbh} for instance].
Although the cross sections are smaller  compared to the $gg \ra h$ case, 
the background cross sections are also  small if one requires a lepton from 
the decaying $W$ bosons as an additional  tag, leading to a significant signal.
Furthermore, the cross section $\sigma( pp \ra \bar{t}th)$ is directly 
proportional to the top--Higgs Yukawa coupling,  the largest electroweak 
coupling in the SM; this process would therefore allow the measurement of 
this parameter, and the experimental test of a fundamental prediction of 
the Higgs mechanism: the Higgs couplings to fermions and gauge bosons are 
proportional to the particle masses. \s

The additional vector boson fusion mechanisms \cite{wwfus}, 
\beqn 
pp \ra W^* W^* /Z^* Z^*  \ra qq h
\eeqn
is less interesting to detect the $h$ boson than in the SM, but can be 
useful to test the $h$ particle properties [for instance to measure the 
$hWW$ and $hZZ$ couplings]. \s 

The heavy CP--even $H$ and CP--odd $A$ bosons can be searched for 
at the LHC through their decays modes into $\tau^+ \tau^-$ pairs 
with the Higgs bosons produced in the $gg$ fusion mechanism or in 
association with $b\bar{b}$ pairs:
\beqn
gg \ra H/A \ \ \ {\rm and} \ \ \ gg, q \bar{q} \ra b\bar{b} \; + \; H/A  
\eeqn
This needs large values $\tb \gsim 5$ for the Higgs boson masses $M_{H,A}
\gsim 300$ GeV, to enhance the cross sections of the above processes 
and the $\tau^+ \tau^-$ decay branching ratios, which in the case where 
only standard modes are allowed reach the asymptotic value of $\sim 10\%$
\cite{Hbrs}.  
The decays of the $H$ and $A$ bosons into muon pairs, $H/A \ra \mu^+ 
\mu^-$, give a rather clean signal and can be used despite of the very small
branching ratios, $\sim 4.10^{-4}$ in the asymptotic limit.  For lower values 
$\tb \lsim 3$ [most of which will be covered by the upgrade of LEP to 
$\sqrt{s}=200$ GeV] and not too large $M_{H,A}$ values, the decays $H/A 
\ra t\bar{t}$ can be also used. The decays $H \ra hh \ra b\bar{b} \gamma \gamma$
and $A \ra Zh \ra l^+l^- b \bar{b}$ can also be used in a tiny area of the
MSSM parameter space; see Fig.~4. \s

At the  LHC, the only way to detect the charged Higgs boson is when $M_{H^\pm} 
< m_t-m_b$ and the $H^\pm$ particles can be then produced in top quark decays, 
\beqn
pp \ \ra \ t\bar{t}  \ \ {\rm with} \ \  t \ra H^+ b 
\eeqn
The usual signature is to look at a breakdown of lepton universality by 
selecting the decay  $H^+ \ra \tau^+ \nu_\tau$, which together with the decay 
$H^+ \ra c\bar{s}$ for small $\tb$ values, is considered as the main decay 
mode [see however the discussion in section 4.2]; Fig.~4. \s

In the next subsections we will discuss the effects of supersymmetric 
particles in the production of the lightest $h$ boson at the LHC: first 
the contributions of light stops in the production of the $h$ boson in the 
$gg \ra h$ mechanism which can significantly alter the expected production 
rate \cite{ggh} and then the associated production of the $h$ boson with 
$\tilde{t} $--squark pairs for which the cross section can be rather large, 
exceeding the one for the SM--like process $pp \ra t\bar{t}h$ \cite{hstop}. \s

One of the most important ingredients of these discussions is that stops can 
alter significantly the phenomenology of the MSSM Higgs bosons\footnote{This 
might also be the case of the sbottoms for large values of $\tb$ and the 
parameters $\mu$ and $A$; however this will not be discussed here and we will 
assume that the mixing is zero in this sector.}. The reason is two--fold: 
$(i)$ as discussed in section 3.1, the current eigenstates, $\tilde{t}_L$ 
and $\tilde{t}_R$, mix to give the mass eigenstates $\tilde{t}_1$ and 
$\tilde{t}_2$; the mixing angle $\theta_{\tilde{t}}$ is proportional to 
$m_t \tilde{A}_t$, with $\tilde{A}=A_t - \mu/\tan \beta$, and can be very
large, leading to a scalar top squark $\tilde{t}_1$ much  lighter than the
$t$--quark and all other scalar quarks; $(ii)$ the couplings of the top 
squarks to the neutral Higgs bosons $h$ given in section 3.1, involve 
components which are proportional to $\tilde{A}_t$ and for large values of 
this parameter, the $g_{h\tilde{t}_1 \tilde{t}_1 }$ coupling can be strongly 
enhanced.\s

The strong $h$ couplings\footnote{The couplings of the neutral heavy Higgs 
bosons $H,A$ to $\tilde{t}$--squarks and of the charged Higgs bosons $H^\pm$ to $\tilde{t} \tilde{b}$--squarks involve also large components as discussed 
in section 3.1. For small $\tb$ values, the decays of these 
heavy Higgs bosons to stop squarks if kinematically allowed can be therefore 
enhanced by these strong couplings, as will be discussed in the next section.}
to stops would result in a stop contribution to the $h gg$ and $h\gamma \gamma$
vertices that is comparable or even larger than the top contribution, altering 
significantly the rate for the process $gg \ra h \ra \gamma \gamma$.
It also leads to a possibly substantial cross section for the production 
of the $h$ boson in association with stops, $pp \ra \tilde{t} \tilde{t} h$. 

\begin{figure}[hbtp]
\vspace*{-5.0cm}
\hspace*{-.5cm}
\epsfxsize=17cm \epsfysize=22cm \epsfbox{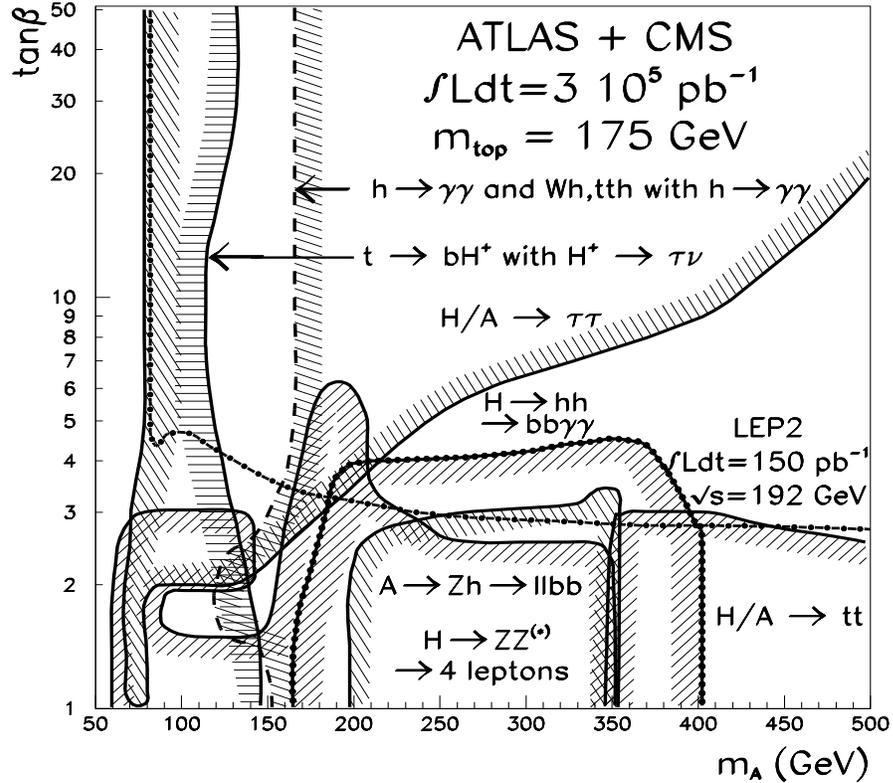}
\vspace*{-5.5cm}
\caption[]{MSSM parameter space with the contours of the Higgs bosons visible 
at the LHC with a luminosity $\int {\cal L} dt = 300$ fb$^{-1}$ and combining 
the experimental data of ATLAS and CMS; from Ref.~\cite{LHCcont}.}
\end{figure}

\newpage

\subsubsection*{4.1.2 Higgs production in the gluon fusion mechanism}

In the SM, the Higgs--gluon--gluon vertex is mediated by heavy 
[mainly top and to a lesser extent bottom] quark loops, while the rare 
decay into two photons is mediated by $W$--boson and heavy fermion 
loops, with the $W$--boson contribution being largely dominating. In the 
MSSM however, additional contributions are provided by SUSY particles: 
$\tilde{q}$ loops in the case of the $hgg$ vertex, and $H^\pm, \tilde{f}$
and $\chi^\pm$ loops in the case of the $h \ra \gamma \gamma$ decay. 
In the latter case \cite{gamma}, the contributions of  $H^\pm$ bosons, 
sleptons and the scalar partners of the light quarks are in general small 
given the experimental bounds on the masses of these particles \cite{data}. 
Only the contributions of relatively light $\tilde{t}$ squarks [and to a 
lesser extent $\tilde{b}$ for large $\tan \beta$ values and $\chi_1^\pm$ 
for masses close to 100 GeV, which could contribute at the 10\% level] can 
alter significantly the loop induced $hgg$ and $h \gamma \gamma$ vertices. \s

The expressions of the partial width for the decay $h \ra gg$, can be found in 
Ref.~\cite{R7c}; the cross section $\sigma(gg \ra h$) is directly proportional 
to the decay width $\Gamma(h \ra gg)$. The latter cross section is affected by 
large QCD radiative corrections \cite{R7c}; however they are practically the 
same for quark and squark loops, and if only deviations compared to the 
standard case are considered, they drop out in the ratios. The partial width 
for the decay $h \ra \gamma \gamma$ can be found e.g. in Ref.~\cite{gamma}; 
the QCD corrections are small and can be neglected. The $\gamma \gamma$ and 
$gg$ decay widths of the $h$ boson are evaluated numerically with the help of 
an adapted version of the program HDECAY \cite{HDECAY}. \s

Figs.~5 and 6 show, as a function of $\tilde{A}_t$ for $\tb=2.5$, the 
deviations from their SM values of the partial decay widths of the $h$ boson 
into two photons and two gluons as well as their product which gives the 
cross section times branching ratio 
$\sigma(gg \ra h \ra \gamma \gamma)$. The quantities $R$ are defined as 
the partial widths including the SUSY loop contributions [all charged SUSY 
particles for $h \ra \gamma \gamma$ and squark loops for $h \ra gg$] 
normalized to the partial decay widths without the SUSY contributions, 
which in the decoupling limit correspond to the SM contributions: 
$R=\Gamma_{\rm MSSM}/\Gamma_{\rm SM}$. \s

\begin{figure}[htb]
\vspace*{-.5cm}
\mbox{\psfig{figure=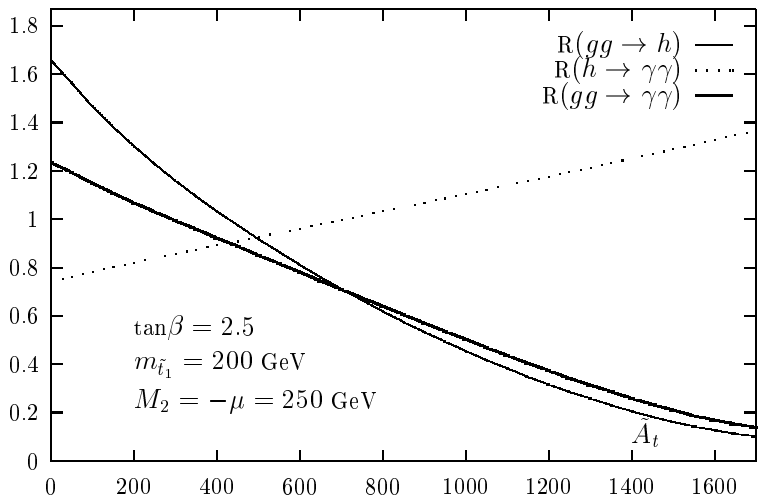,width=14.5cm}}
\vspace*{-13.2cm}
\caption[]{SUSY loop effects on R$(gg \ra h)$, R$(h \ra \gamma \gamma)$ and
R$(gg \ra \gamma \gamma)$ as a function of $\tilde{A}_{t}$ for $\tb=2.5$ and 
$m_{\tilde{t}_1}=200$ GeV, $M_2=-\mu=250$ GeV.}
\end{figure}
\begin{figure}[htb]
\vspace*{-.5cm}
\mbox{\psfig{figure=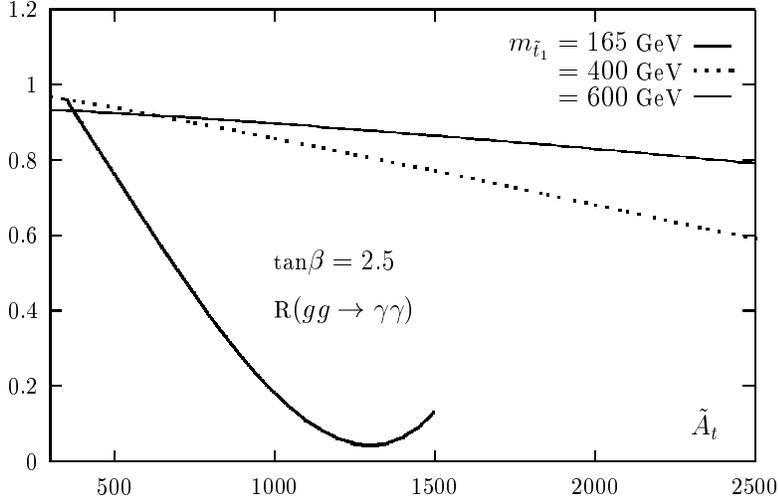,width=14.5cm}}
\vspace*{-13.2cm}
\caption[]{SUSY loop effects on R$(gg \ra \gamma \gamma)$ as a function of 
$\tilde{A}_{t}$ for $\tb=2.5$ and $m_{\tilde{t}_1}=165,400$ and 600 GeV with 
$M_2=-\mu=500$ GeV for $m_{\tilde{t}_1}\geq 400$ GeV.} 
\end{figure}

In Fig.~5, the stop mass is set to $m_{\tilde{t}_1}=200$ GeV. For small values 
of $\tilde{A}_t$ there is no mixing in the stop sector and the dominant 
component of the $h\tilde{t} \tilde{t}$ couplings, eq.~(\ref{ghtt}),  is the 
one proportional to $m_t^2/M_Z^2$ [here, both $\tilde{t}_1$ 
and $\tilde{t}_2$ contribute since their masses and couplings to $h$ are almost 
the same]. The sign of this component, compared to the $ht\bar{t}$ coupling, is 
such that the $t$ and $\tilde{t}$ contributions interfere constructively in the 
$hgg$ and $h\gamma \gamma$ amplitudes. This leads to an enhancement of the $h 
\ra gg$ decay width up to $60\%$ in the MSSM. However, the $h \ra \gamma 
\gamma$ decay 
width is dominated by the $W$ amplitude which interferes destructively with the 
$t$ and $\tilde{t}$ amplitudes, and the $\tilde{t}$ contributions reduce the 
$h \ra \gamma \gamma$ decay width by an amount up to $-20\%$. The product 
R($gg \ra \gamma \gamma$) in the MSSM is then enhanced by a factor $\sim 1.2$ 
in this case. \s

With increasing $\tilde{A}_{t}$, the two components of $g_{h\tilde{t}_1 
\tilde{t}_1}$ [which have opposite sign since $\sin2\theta_{\tilde{t}}
\propto m_t \tilde{A}_t$ in eq.~(\ref{ghtt})] interfere destructively and 
partly cancel each other, resulting in a rather small stop contribution. 
For larger values of $\tilde{A}_{t}$, the second component of $g_{h\tilde{t}_1 
\tilde{t}_1}$ becomes the most important one, and the $\tilde{t}_1$ loop 
contribution [$\tilde{t}_2$ is too heavy to contribute] interferes 
destructively with the one of the $t$--quark. This leads to an enhancement of 
R$(h \ra \gamma \gamma)$ and a reduction of R$(gg \ra h)$. However, the 
reduction of the latter is much stronger than the enhancement of the former 
[recall that the $W$ contribution in the $h \ra \gamma \gamma $ decay is much 
larger than the $t$ contribution] and the product R($gg \ra \gamma \gamma$) 
decreases with increasing
$\tilde{A}_t$. For $\tilde{A}_t$ values of about 1.5 TeV, the signal for 
$gg\ra h \ra \gamma \gamma$ in the MSSM is smaller by a factor of 
$\sim 5$ compared to the SM case\footnote{Note that despite of the large
$(\tilde{t},\tilde{b}$) mass splitting generated by large $\tilde{A}_t$ values, 
the contributions of the isodoublet to high--precision observables stay below 
the experimentally acceptable level \cite{drho}.}. \s

Fig.~6 shows the deviation  R$(gg \ra \gamma \gamma)$ with the same parameters
as in Fig.~5 but with different $\tilde{t}_1$ masses, $m_{\tilde{t}_1}=
165,400$ and 600 GeV. For larger masses, the top squark contribution 
$\propto 1/ m_{\tilde{t}_1}^2$, will be smaller than in the previous case. 
In the no--mixing case, the enhancement (reduction) of the $hgg (h\gamma 
\gamma)$ 
amplitude is only of the order of 10\% for $m_{\tilde{t}_1}\simeq 400$ GeV,
and leads to an almost constant cross section times branching ratio for the 
$gg \ra h \ra \gamma \gamma$ process compared to the SM case. Again the stop 
contribution vanishes for some intermediate value of $\tilde{A}_t$, and then 
increases again in absolute value for larger $\tilde{A}_t$. However, 
for $m_{\tilde{t}_1}\simeq 400$ GeV, the effect is less striking compared to 
the 
case of $m_{\tilde{t}_1}=200$ GeV, since here $\sigma(gg \ra h) \times {\rm BR}
(h \ra \gamma \gamma$) drops by less than a factor of 2, even for extreme 
values 
of $\tilde{A}_t \sim 2.5$ TeV. As expected, the effect of the top squark loops 
will become less important if the $\tilde{t}_1$ mass is increased further to 
600 
GeV for instance. In contrast, if the stop mass is reduced to $m_{\tilde{t}_1}
\simeq 165$ GeV, the drop in  R$(gg \ra \gamma \gamma)$ will be even more 
important: for  $\tilde{A}_t \sim 1.5$ TeV, the $gg \ra \gamma \gamma$ cross 
section times branching ratio including stop loops is an order of magnitude 
smaller than in the SM. For $\tilde{A}_t \sim 1.3$ TeV, the $\tilde{t}$ 
amplitude almost 
cancels completely the $t/b$ quark amplitudes; the non--zero value
of R$(gg \ra \gamma \gamma)$ is then due to the imaginary part of the 
$b$-quark contribution. \s

One should recall that $M_h$ varies with $\tilde{A}_t$, and no constraint on 
$M_h$ has been set in Figs.~5--6. Requiring $M_h \gsim 90$ GeV, the 
lower range $\tilde{A}_t \lsim 350$ GeV and the upper ranges $\tilde{A}_t 
\gsim 1.5 (2.3)$ TeV for $m_{\tilde{t}_1}=200 (400)$ GeV for instance, are 
ruled out. This means that the scenario where R$(gg \ra \gamma \gamma) >1$, 
which occurs only for $\tilde{A}_t \lsim 300$ GeV for $m_{\tilde{t}_1}=200$ GeV
is ruled out for $M_h \gsim 90$ GeV. Therefore, the rate for the $gg \ra \gamma 
\gamma$ process in the MSSM will always be smaller than in the SM case, making
more delicate the search for the $h$ boson at the LHC with this process. \s

For large values of $\tb$, $\tb \gg1$, the mixing in the
sbottom sector can also be very large, leading to $m_{\tilde{b}_1}$ 
possibly rather small, and a large $ g_{h \tilde{b}_1 \tilde{b}_1 }$ 
coupling which can also generate large $\tilde{b}_1$ loop contributions 
to the $hgg$ and $h\gamma \gamma$ vertices. Indeed, for $\tb \sim 50$
and $m_{\tilde{b}_1}=200$ GeV, the deviations of the R$(h \ra gg)$  
and thus R$(gg \ra h \ra \gamma  \gamma$) observables from unity are 
substantial for large values of $|\mu|$. For instance, for $|\mu| 
\simeq 1$ TeV the $gg \ra \gamma \gamma$ cross section in the MSSM 
can be suppressed compared to the SM case by a factor of 5. When
$m_{\tilde{b}_1}$ is increased (reduced) the effect becomes 
less (more) striking. \s

In the case where the decoupling limit is not yet reached, the $hWW$ 
and $htt$ couplings are smaller than in the SM, and both the $gg \ra h$ cross 
section and $h \ra \gamma \gamma$ widths are suppressed compared to the SM 
case, even in the absence of the squark loops. Including light $\tilde{t}$ 
contributions will further decrease the amplitudes for large $\tilde{A}_t$. For 
large $\tb$ values, the $hgg$ amplitude can be enhanced by the $b$--loop 
contribution, but the $h \ra \gamma \gamma$ branching ratio is strongly 
suppressed due to the absence of the $W$--loop and the increase of the total 
decay width $\propto m_b^2 \tan^2\beta$. \s

Finally, in the case of the heavy $H$ boson, squark loop 
contributions to the cross section $gg \ra H$ can be even larger since because 
of the larger value of $M_H$, more room will be left for the $\tilde{t}$ 
[and $\tilde{b}$] squarks before they decouple form the $Hgg$ amplitude. 
In addition, for $M_H$ values above the squark pair threshold, the
decays $H \ra \tilde{t}_1 \tilde{t}_1$ or $H \ra \tilde{b}_1 \tilde{b}_1$
will be kinematically allowed and could have large branching ratios, 
therefore suppressing the other decay modes including the $H \ra \tau^+ \tau^-$ 
channel. For the pseudoscalar $A$ boson, however, squark loops 
will not have drastic effects on $\sigma (gg \ra A)$: because of 
CP--invariance, the $A$ boson couples only to $\tilde{t}_1 \tilde{t}_2$ 
or  $\tilde{b}_1 \tilde{b}_2$ 
pairs while the gluon coupling to different squarks is absent; the $Agg$
amplitude cannot be built at lowest order by scalar quark loops. 

\subsubsection*{4.1.3 Higgs production in association with light stops} 

If one of the stop squarks is light and its coupling to the $h$ boson is 
enhanced, an additional process might provide a new important source for 
Higgs particles: the associated production with $\tilde{t}$ states, 
\beq
pp \ra gg + q \bar{q} \ra \tilde{t} \tilde{t}h
\eeq 
At lowest order, i.e. at ${\cal O}( G_F \alpha_s^2)$, the process is  
initiated by 12 Feynman diagrams: 10 diagrams  for the $gg$ mechanism 
[including those  with the quartic gluon--squark interaction and the 
three--gluon vertex] once the various  possibilities for emitting the Higgs 
boson from the squark lines and the crossing of the two gluons are added
and 2 diagrams for the $q\bar{q}$  annihilation process; some generic diagrams 
are shown in Fig.~7. Due to the larger gluon luminosity at high energies, the 
contribution of the $gg$--fusion diagrams is much larger  than the 
contribution of the $q\bar{q}$ annihilation diagrams at the LHC. 

\noindent
\begin{figure}[htb]
\begin{center}
\mbox{ 
\psfig{figure=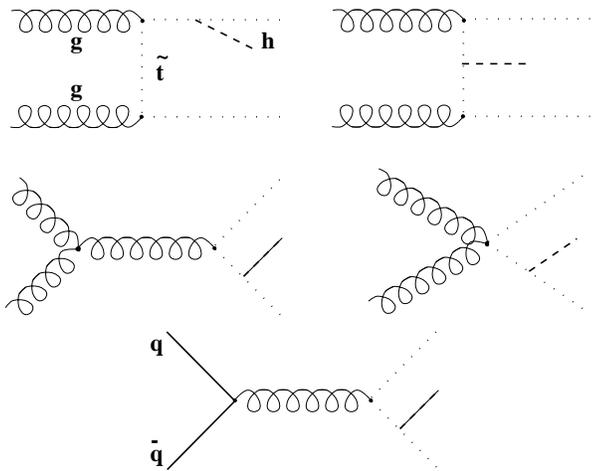,width=5.5cm,angle=-90}}
\end{center}
\caption[]{Generic Feynman diagrams for the production of the $h$ boson 
in association with top squarks via $gg$ fusion and $q\bar{q}$ 
annihilation.}
\end{figure}

In Fig.~8, the $pp \ra \tilde{t}_1 \tilde{t}_1h$ cross section [in pb] is 
displayed as a function of $m_{\tilde{t}_1}$ for $\tb=2$, in the case of 
no--mixing [$A_t=200, \mu=400$ GeV], moderate mixing [$A_t=500$ and $\mu=100$ 
GeV] and large mixing 
[$A_t=1.5$ TeV and $\mu=100$ GeV]. We have used $m_{\tilde{t}_L}= 
m_{\tilde{t}_R} \equiv m_{\tilde{q}}$ as is approximately the case in GUT  
scenarios and for illustration, $\tb=2$ and $30$.
Note for comparison, that the cross section for the standard--like $pp \ra 
\bar{t} t h$ process is of the order of 0.6 pb for $M_h 
\simeq 100$ GeV \cite{kunszt}; $m_t=175$ GeV, and the CTEQ4 parameterizations 
of the structure functions \cite{CTEQ} are chosen. \s

If there is no mixing in the stop sector, $\tilde{t}_1$  and
$\tilde{t}_2$ have the same mass and approximately the same couplings to the 
$h$ boson since the $m_t^2/M_Z^2$ components are dominant. The cross  section,
which should be then multiplied by a factor of two to take into account  both
squarks, is comparable to the $\sigma(pp \ra t\bar{t} h)$ in the  low
mass range $m_{\tilde{t}}\lsim  200$ GeV\footnote{If the  
$\tilde{t}$ masses are related to the masses of the light quark partners,  
$m_{\tilde{q}}$, the range for which the cross section is rather large is
therefore ruled out by the experimental constraints on $m_{\tilde{q}}$ 
\cite{data}.}. For intermediate values of $\tilde{A}_t$ the two components of the  
$h \tilde{t}_1 \tilde{t}_1$ coupling interfere destructively and partly 
cancel each other, resulting in a rather small cross section, unless
$m_{\tilde{t}_1} \sim {\cal O}(100)$ GeV. 
In the large mixing case $\tilde{A}_t \sim 1.5$ TeV $\sigma(pp \ra \tilde{t}_1 
\tilde{t}_1 h)$ can be very large. It is above the rate for the standard 
process $pp \ra \bar{t}th$ for values of $m_{\tilde{t}_1}$ smaller than 220 GeV.
If  $\tilde{t}_1$ is lighter than the top quark, the $\tilde{t}_1 \tilde{t}_1 
h$ cross section significantly exceeds the one for $\bar{t}th$ final states.
For instance, for $m_{\tilde{t}_1}=140$ GeV corresponding to $M_h\sim 76$ GeV, 
$\sigma( pp \ra \tilde{t}_1 \tilde{t}_1 h)$ is an order of magnitude larger 
than $\sigma(pp \ra t\bar{t}h)$. \s

\begin{figure}[htb]
\begin{center} 
\vspace*{-.6cm}
\mbox{\psfig{figure=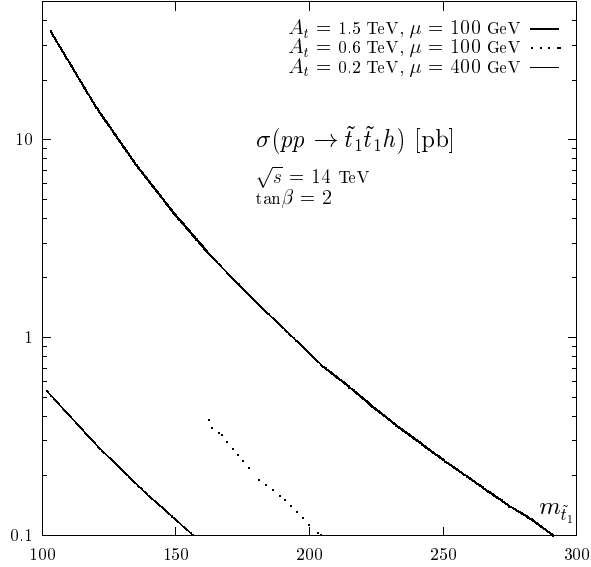,width=11cm}}
\vspace*{-7.3cm}
\caption[]{The production cross section $\sigma (pp \ra \tilde{t}_1 
\tilde{t}_1 h$) [in pb] as a function of the $\tilde{t}_1$ mass and three 
sets of $A_t$ and $\mu$ values and $\tb$ is fixed to $\tb=2$.}
\end{center}
\end{figure}

In Fig.~9, we fix the lightest top squark mass to $m_{\tilde{t}_1} =165$ GeV 
$ \sim m_{t}^{\rm \overline{MS}}$ and display the $pp \ra gg+q\bar{q} \ra 
\tilde{t}_1 \tilde{t}_1h$ cross section as a function of $\tilde{A}_t$. 
For comparison, the $*$ and $\bullet$ give the standard--like  $pp \ra 
\bar{t}th$ cross section for $M_h=100$ GeV and $\tb=2$ and 30, respectively. 
For $\tb=30$ the cross section is somewhat smaller than for $\tb=2$, a mere 
consequence of the increase of the $h$ boson mass with $\tb$ \cite{mh}.  As 
can be seen again, the production cross  section is substantial for the 
no--mixing case, rather small for intermediate mixing [becoming negligible for 
$\tilde{A}_t$ values between 200 and 400 GeV], and then becomes very large exceeding 
the reference cross section for values of $\tilde{A}_t$  above $\sim 1$ TeV. 
For instance, for the inputs of Fig.~8, $\sigma(pp \ra
\tilde{t}_1 \tilde{t}_1 h)$ exceeds $\sigma(pp \ra t\bar{t}h)$ in the SM for the
same Higgs boson mass when $\tilde{A}_t \gsim 1(1.05)$ TeV for $\tb=2(30)$. \s

\begin{figure}[htb]
\begin{center} 
\vspace*{-.6cm}
\mbox{\psfig{figure=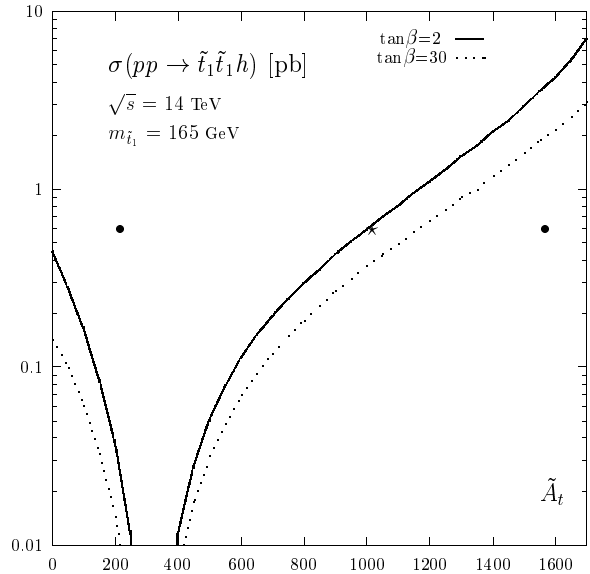,width=11cm}}
\vspace*{-7.3cm}
\caption[]{The production cross section $\sigma (pp \ra \tilde{t}_1 
\tilde{t}_1 h$) [in pb] as a function as a function of $\tilde{A}_t$ for
 fixed $m_{\tilde{t}_1}=165$ GeV and for $\tb=2,30$ (b).}
\end{center}
\end{figure}

For the signal, in most of the parameter space, the stop decay is $\tilde{t}_1 
\ra b\chi^+$, if  $m_{\tilde{t}_1} <m_t+m_{\chi_1^0}$  where $\chi_1^0$ is the 
LSP, or $\tilde{t}_1 \ra t \chi_1^0$, in the opposite case [see section 5.4]. 
In the interesting region where  $\sigma(pp \ra \tilde{t}_1  \tilde{t}_1 h)$ 
is large, i.e. for light $\tilde{t}_1$, the decay  $\tilde{t}_1 \ra b \chi^+$ 
is dominant, unless  $m_{\tilde{t}_1} -m_{\chi_1^+}$ is very small, in which 
case the loop induced  decay, $\tilde{t}_1 \ra c \chi^0_1$, can become 
competitive.  Assuming that sleptons are heavier than the chargino, $\chi_1^+$ 
will mainly decay into $bW^+ + \ {\rm missing \ energy}$ leading to
$\tilde{t}_1 \ra bW^+$ final states. This is the same topology as 
the decay, $t \ra bW^+$, except that in the case of the $\tilde{t}$ there is a 
large amount of missing energy. If sleptons are also relatively 
light, charginos decays will also lead to $l \nu \chi_1^0$ final states. 
The only difference between the final states generated by the $\tilde{t}
\tilde{t}h$ and $t\bar{t}h$ processes, will be due to the softer energy 
spectrum of the charged leptons coming from the chargino decay in the former
case, because of the energy carried by the invisible LSP. \s

The Higgs boson can be tagged through its $h \ra \gamma \gamma$ decay mode.
In the decoupling limit, and for light top squarks and large $\tilde{A}_t$
values, the branching ratio for this mode can be substantially enhanced
compared to the SM Higgs boson  as discussed in the previous 
subsection. Therefore, $\gamma \gamma$+ charged 
lepton events can be  more copious than in the SM, and the 
contributions of the $pp \ra \tilde{t} \tilde{t} h$ process to these events 
can render the detection of the $h$ boson much easier than with the process 
$pp \ra t \bar{t}h$ alone. \s

Although a  detailed Monte--Carlo analysis will be required to  assess the 
importance of this signal and to
optimize the cuts needed not to dilute the contribution of the $\tilde{t}
\tilde{t}h$ final states, it is clear that in a substantial area of the
MSSM parameter space, the contribution of the top squark to the $\gamma \gamma
l^\pm$ signal can significantly enhance the potential of the LHC to discover
the lightest MSSM Higgs boson in this channel. 
This would be a new and very 
interesting means to search for top squarks at the LHC, which due to the large 
QCD background from $\bar{t} t$ production, are otherwise difficult to detect 
in other channels. Last but not least, and as welcome bonus, this process 
would  allow to measure the $h\tilde{t} \tilde{t}$ coupling, the potentially
largest electroweak coupling in the MSSM, opening thus a window to probe 
directly the trilinear part of the soft--SUSY breaking scalar potential.

\subsection*{4.2 Higgs boson decays into SUSY particles} 

\subsubsection*{4.2.1 SUSY decays of the neutral Higgs bosons}

As discussed previously, in mSUGRA scenarii the MSSM Higgs bosons $H,A,H^\pm$
are rather heavy and degenerate in mass, while the lightest $h$ boson is 
in the decoupling regime. We have seen in the previous section 4.1 that the 
search of the 
$h$ boson at the LHC relies heavily on the rare $h \ra \gamma \gamma$ decay 
mode, while the almost only possibility of detecting the heavy $H,A$ bosons 
for masses beyond 300 GeV and large enough $\tb$ values, $\tb \gsim 5$, would 
be to look for the decays into tau pairs, $H/A \ra \tau^+ \tau^-$. With  high 
enough luminosity, these signals will be visible at the LHC [see Fig.~4]. \s

However, in these analyses, it is always assumed that the heavy $H/A$ bosons 
decay only into standard particles, and that the SUSY decay modes are shut. 
But for such large values of $M_{H,A}$, at least 
the decays into the lightest neutralinos and charginos, and possibly into 
to light $\tilde{t}$ and $\tilde{b}$, can be kinematically allowed. These 
modes could have large decays widths, and thus could suppress the $H/A \ra
\tau^+ \tau^-$ branching ratios drastically. For the $h$ boson, because 
of its small mass, only a little room is left for decays into SUSY 
particles by present experimental data \cite{data}. However, the 
possibility of $h$ decays into neutralinos is not yet completely ruled out,
especially if one relaxes the gaugino mass unification; decays into sneutrinos
are also still possible. When these
invisible decays occur, they can be dominant, hence reducing the probability 
of the $h \ra \gamma \gamma$ decay to occur. These SUSY decays should therefore
not be overlooked as they might jeopardize the detection of the Higgs 
particles at the LHC. \\

\nn {\bf a) Invisible decays of the h boson} \\ 

\noindent 
Despite the lower bound of $91\,$GeV on the mass of the lightest chargino 
$\chi_1^+$ and the constraints from $\chi_0^1 \chi_0^2$ searches at LEP2
\cite{data}, the decay of the lightest $h$ boson into a pair of
lightest neutralinos is still kinematically possible. Even in the constrained
MSSM with a common gaugino mass at the GUT scale, leading to the 
well--know relation between the wino and bino masses $M_1=\frac{5}{3} 
{\rm tg}^2\theta_W M_2\sim \frac{1}{2}M_2$, the lower bound on the LSP 
mass is only $m_{\chi_0^1} \gsim 40$ GeV \cite{data}. Since the upper 
bound on the lightest $h$ boson in the MSSM is $M_h \sim 130$ \cite{mh}, there 
is still room for the invisible decay $h \ra \chi_0^1 \chi_0^1$ to occur. 
\s

Although high values of $\tb$ are required to be closer to the
upper bound of $M_h$, the $hb\bar{b}$ coupling 
is SM--like if the $h$ boson is in the decoupling regime: in 
this case the $h\chi_0^1 \chi_0^1$ coupling can be much larger
then the $hb\bar{b}$ Yukawa coupling 
and the decay of the $h$ boson into the lightest neutralinos can be 
dominant, resulting in a much smaller BR($h\ra \gamma \gamma$) 
than in the SM. Far from the decoupling limit, 
the coupling $g_{hbb} \sim \tan \beta$ is strongly enhanced for 
$\tb \gsim 3$, while the $h$ boson couplings to $W$ bosons and top 
quarks [which provide the main contributions to the $h \gamma \gamma$ 
loop vertex] are suppressed. This again will result in a strong 
suppression of BR($h \ra \gamma \gamma$). \s 

The partial width for the decay $h \ra \chi_0^1 \chi_0^1$ is given by
[$\beta_\chi^2=  1 - 4m^2_{\chi_1^0}/M_h^2$ and the coupling $g_{h \chi \chi}$
is given in eq.~(\ref{gchi}) of section 3.1]

\beqn 
\Gamma (h \ra \chi_0^1 \chi_0^1)= \frac{G_F M_W^2 M_h}{2 \sqrt{2} \pi}
\ g^2_{h \chi \chi} \ \beta^3_{\chi} 
\eeqn 
The decay
is important only for moderate values of $M_2$ and $\mu$ [with a preference 
for $\mu>0$] since the $h$ boson prefers to couple to neutralinos which
are a mixture of gauginos and higgsinos. In this range, the decay $h \ra 
\chi_0^1 \chi_0^1$ is dominant if $M_h$ is above the $2m_{\chi_0^1}$ threshold;
close to the threshold, the decay width is strongly suppressed by the 
$\beta^3_\chi$ factor. \s

As an illustration of this possibility we show in Fig.~10 the fraction
BR$(h \ra \gamma \gamma)$ as a function of $\mu$ for two values of $\tb=
2,30$. We choose $M_A =1$ TeV [to be in the decoupling regime] and  the
``maximal mixing" scenario $A_t=\sqrt{6} m_{\tilde{q}}$ with the 
common squark mass parameter fixed to 1 TeV [to maximize the $h$
boson mass]; this gives $M_h \simeq 126$ GeV for $\tb=30$ and 
$ M_h \simeq 106$ GeV for $\tb=2$ [the variation with $\mu$ is almost 
negligible]. 
Focusing first on the $\tan \beta=30$ and $M_2=140$ GeV case, for  $|\mu| 
\gsim  200$ GeV the channel $h \ra \chi_1^0 \chi_1^0$ is kinematically 
closed and the $\gamma \gamma$ branching ratio is SM--like, BR $\simeq 
2.3 \times 10^{- 3}$. In the range $110 \lsim |\mu| \lsim 200$ GeV, the LSP 
is lighter that $M_h/2$ while the chargino is still heavier than 91 GeV, 
the decay $h \ra \chi_1^0 \chi_1^0$ is thus allowed to occur and 
suppresses the $\gamma \gamma$ branching ratio. The suppression is stronger 
with decreasing $|\mu|$ since on the one hand the phase-space becomes more 
favorable, and on the other hand the LSP tends to be an equal mixture of 
higgsino and gaugino. The maximum drop of BR($h \ra \gamma \gamma$) is a factor 
of three and two for $\mu>0$ and $\mu<0$ respectively. For values $|\mu| 
\lsim  110$ GeV, $m_{\chi_1^\pm}$ exceeds its experimentally allowed lower 
bound. \s

\begin{figure}[htb]
\vspace*{-.5cm}
\mbox{\psfig{figure=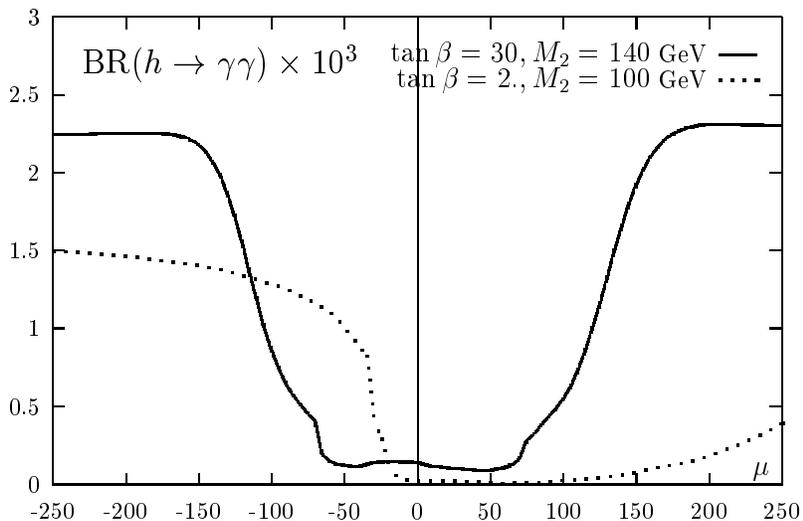,width=15cm}}
\vspace*{-13.5cm}
\caption[]{Branching ratios in units of $10^{-3}$ for the decays $h \ra 
\gamma \gamma$ as a function of $\mu$ for $\tb=2(30)$ and $M_2=100(140)$ GeV.} 
\end{figure}

In the case $\tb=2$ and  $M_2=100$ GeV, the only experimentally allowed
region is $|\mu| \gsim 110$ GeV with $\mu<0$, since elsewhere the chargino is 
heavier than 91 GeV. In this $|\mu|$ range, the decay $h \ra \chi_1^0 \chi_1^0$
is kinematically allowed, but the branching ratio is very small, less than 
0.5\%. This is due to the fact that in this area $\chi_0^1$ is a pure bino 
state and its couplings to the $h$ boson are strongly suppressed. What makes
the $h \ra \gamma \gamma$ branching ratio drop by almost a factor two
compared to the previous case is first, the smaller mass of the $h$ boson
[the decay width grows with the third power of the Higgs mass] but also 
because of the contribution of the chargino loops to the $h \ra \gamma 
\gamma$ decay which interfere destructively for $\mu<0$ with the dominant 
contribution due $W$ loops [the reduction is nevertheless very mild, at 
most 15\% in this case]. \s

If the constraint on the unification of the gaugino masses at the GUT scale
is relaxed, there is practically no lower bound on the LSP mass. Indeed,
for relatively large values of the Higgs--higgsino mass parameter $\mu$, 
the lightest chargino $\chi_1^+$ and the next--to--lightest neutralino
$\chi_2^0$ are wino--like with a mass $\sim M_2$ while the lightest
neutralino is bino--like with a mass $\sim M_1$. Since $M_1$ is a free
parameter, it can be as small as possible leading to a possibly very
light LSP. The decay $h \ra \chi_0^1 \chi_0^1$ will then have more room
to occur. \s

\begin{figure}[htb]
\vspace*{-.5cm}
\mbox{\psfig{figure=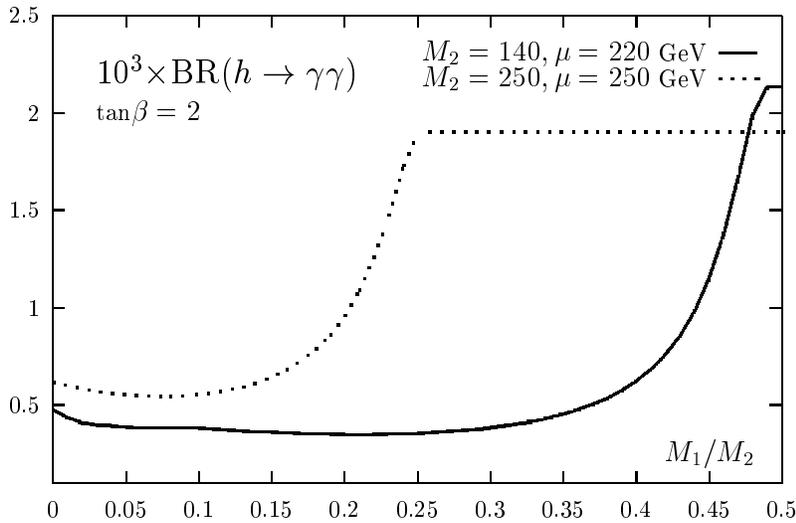,width=15cm}}
\vspace*{-13.5cm}
\caption[]{Branching ratios for the decays $h \ra \gamma \gamma$ in units
of $10^{-3}$ as a function
of $M_1/M_2$ for $\tb=2$ and two sets of $M_2,\mu$ values.} 
\end{figure}

The branching ratio for the decay $h \ra \chi_1^0 \chi_1^0$ can be rather 
large thus suppressing the $\gamma \gamma$ branching ratio. This is exemplified 
in Fig.~11, where BR($h\ra \gamma \gamma$) is shown as a function of the 
ratio $M_1/M_2$ for $\tb=2$ and for two sets of $M_2$ and $\mu$ values;
$M_2=140$ GeV, $\mu=220$ GeV leading  to $m_{\chi_1^+}\simeq 96$ GeV, and 
$M_2=\mu=250$ GeV leading to $m_{\chi_1^+}\simeq 175$ GeV; the remaining 
inputs are as in the previous figure. 
In the first scenario [solid lines], when the LSP is very small BR$(h \ra 
\gamma \gamma$) drops to the level of $5.10^{-4}$, a strong reduction compared
to the expected rate $\sim 2.10^{-3}$. With increasing $M_1/M_2$ and hence
with increasing LSP mass, it stays almost constant until the $2m_{\chi_1^0}$ 
threshold is reached for $M_1 \sim M_2/2$ and the branching ratio recovers its
standard value. \s

In the second scenario [dotted lines], 
BR($h\ra \gamma \gamma$) starts at the same level 
as previously, but increases more
rapidly and reaches approximately the standard value for $M_1 \sim M_2/4$ 
which corresponds to the kinematical limit for the decay $h \ra \chi_1^0 
\chi_1^0$. When the LSP decay is shut, the difference between the $\gamma
\gamma$ branching ratios in the two scenarios is due to a constructively 
interfering chargino loop contribution [the sign of the $\chi_1^\pm$ 
contribution goes with the sign of $\mu$] in the case where $m_{\chi_1^+}
\simeq 96$ GeV and which enhances the $\gamma \gamma$ decay width by 20\% 
or so. 
This picture is expected not to be altered significantly for larger values of 
$\tb$ if the $h$ boson is in the decoupling regime as discussed previously 
[in fact for large $\tb$ values and for some moderate values of the parameters 
$M_2$ and $\mu$, even the decays into the lightest and the next to lightest 
neutralinos is possible]. \bigskip

Another kinematically still 
possible SUSY mode for the lightest $h$ boson is the 
decay into sneutrinos. Indeed, the experimental lower bound on the $\tilde{
\nu}$ masses is still rather low, $m_{\tilde{\nu}} \gsim 45$ GeV
\cite{data}, leaving
some room for the decay $h \ra \tilde{\nu} \tilde{\nu}$ to occur. However, 
because of SU(2)$_{\rm L}$ invariance, the sneutrino and the left--handed 
charged slepton masses are related and one should avoid being into conflict
with the experimental bounds on the $\tilde{l}_L$ mass which are stronger, 
$m_{\tilde{l}_L} \gsim 70$ GeV. However, even in this case one can obtain 
a rather light sneutrino since a splitting 
between the $\tilde{\nu}$ and $\tilde{l}_L$ masses can be generated by the 
D--terms. Indeed, recalling eqs.~(\ref{smass}) and denoting the common scalar 
mass by $\tilde{m}$, one has:
\beqn
m_{\tilde{\nu}}^2 \simeq \tilde{m}^2 +0.50 M_Z^2 \cos2\beta \ \ , \ \ 
m_{\tilde{l}_L}^2 \simeq \tilde{m}^2 -0.27 M_Z^2 \cos2\beta
\eeqn
For small values of $\tilde{m}$, the slepton masses are governed by the 
D--terms, and for large values of $\tb$, $\cos 2\beta \ra -1$ and the 
D--terms become maximal. Since they tend to increase $m_{\tilde{l}_L}$
and decrease $m_{\tilde{\nu}}$, relatively low masses for sneutrinos can be 
kept while still having rather heavy left--handed\footnote{The D--terms
for right--handed charged sleptons are approximately the same as for
the left--handed ones and tend also to decrease the mass. However,
in GUT scenarii such as mSUGRA, the $\tilde{l}_R$ tends to be lighter 
than the sneutrinos for reasonable values of the gaugino mass $m_{1/2}$; 
see eqs.~(\ref{smass}). In this case, the decay $h \ra \tilde{\nu} 
\tilde{\nu}$ is forbidden because of the experimental bound $m_{\tilde{l}_R} 
\gsim 70$ GeV.} sleptons [note however, that the $\tilde{\nu}$ should not be
lighter than the lightest neutralino which is expected to be the LSP]. \s

In the decoupling limit, the $h$ boson coupling to sneutrinos is also 
proportional to $\cos 2\beta$ [eq.~(\ref{ghsnu}) of section 3.1], and for large 
$\tb$ values it becomes also maximal. And since it is a  ``gauge" coupling, 
it is much larger than the $hb\bar{b}$ Yukawa coupling, and the decay $h \ra 
\tilde{\nu} \tilde{\nu}$ is always largely dominating once it is kinematically 
allowed. The partial decay width for the decay, summing over the three 
sneutrinos, is given by
\beqn 
\Gamma (h \ra \tilde{\nu} \tilde{\nu})= \frac{3 G_F M_Z^4}{8 \sqrt{2} \pi M_h}
\beta_{\tilde{\nu}} \ \ , \ \ \beta_{\tilde{\nu}} = \left[ 1 - \frac{4 
m^2_{\tilde{\nu}}}{M_h^2} \right]^{1/2} 
\eeqn 
Modulo the velocity factor $\beta_{\tilde{\nu}}$, 
the partial width is larger than the otherwise dominant
$b\bar{b}$ decay width by a huge factor: $M_Z^4/(2 m_b^2 M_h^2) \sim  
230$ for $M_h=130$ GeV. Thus, if the  $h \ra \tilde{\nu} \tilde{\nu}$ decay
mode is allowed, all the branching ratios for the other decay channels
including the $h \ra \gamma \gamma$ mode, will be suppressed by two
orders of magnitude! Since
the sneutrinos will decay invisibly in this mass range [$m_{\tilde{\nu}} < 
m_{\chi_1^\pm}$ and the only possible mode is $\tilde{\nu} \ra 
\nu \chi_1^0$],  the $h$ boson would be then also very difficult to detect at 
the LHC\footnote{At $e^+ e^-$ colliders missing mass techniques allow for an 
easy detection in the process $e^+ e^- \ra hZ$.}. \bigskip

\nn {\bf b) H/A decays into SUSY particles} 

\bigskip

If the CP--even and the CP-odd Higgs bosons $H$ and $A$ are heavy, $M_{H,A}
\gsim 300$ GeV, at least the decays into the lightest neutralinos and possibly 
charginos should be kinematically allowed. If the couplings to $b\bar{b}$
and $\tau^+ \tau^-$ pairs [which together with $t\bar{t}$ states account 
for most of the total decay width in the absence of SUSY modes] are not 
strongly enhanced, and hence for not too large values of $\tan \beta$, these 
decays might be dominant and suppress drastically the branching ratios for the 
$H/A \ra \tau^+ \tau^-$ signals. This is exemplified in Fig.~12 where BR$(H/A 
\ra \tau^+\tau^-)$ are plotted as a function of the $H/A$ masses for three 
values $\tb=5,10$ and 30. The choice $M_2=-\mu=200$ GeV has been made leading 
to $m_{\chi_1^0} \sim 90$ GeV and $m_{\chi_1^+} \sim 160$ GeV [with a small 
variation with $\tb$]. \s

\begin{figure}[htb]
\vspace*{-.5cm}
\mbox{\psfig{figure=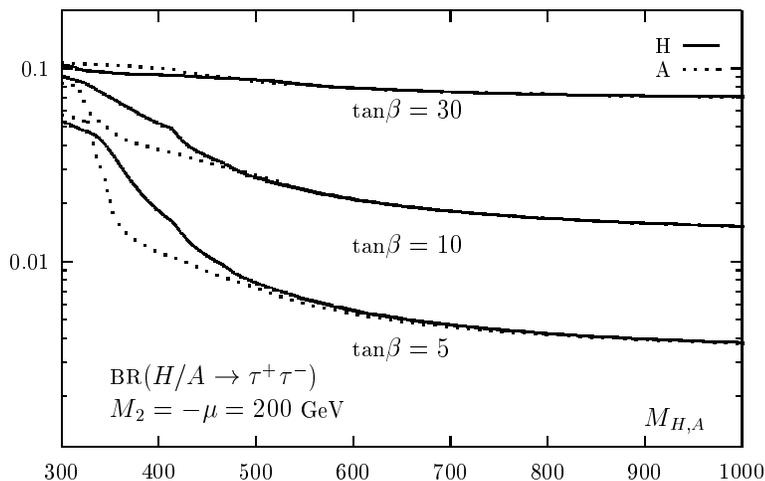,width=14cm}}
\vspace*{-12.6cm}
\caption[]{Branching ratios for the decays $H/A \ra \tau^+ \tau^-$ as a function
of $M_{H,A}$ for $\tb=5,10$ and 30 and for the values $M_2=-\mu=200$ GeV.} 
\end{figure}

The branching ratios for $H$ and $A$ decays are almost the same except for 
small values of $\tb$ and relatively small Higgs masses: in this case, the 
decoupling limit is not yet reached and additional [and different] decay 
modes occur for the $H$ and $A$ bosons; see section 4.1. For $\tb=5$
the $H/A$ couplings to down--type fermions are not very strongly enhanced
and the decays into charginos and neutralinos have large branching ratios: 
they decrease BR$(H/A \ra \tau^+ \tau^-)$ from the standard $\sim 10\%$ 
value for small Higgs masses [where only a few SUSY channels are open and 
some are suppressed by phase space] to less than 0.4\% for very heavy Higgs 
boson masses $M_{H,A} \sim 1$ TeV [here most of the neutralino/chargino
channels are open and they are not suppressed by phase space], thus a 
reduction by more than a factor of 20 compared to the branching ratio without 
the SUSY decays. For $\tb=10$, the couplings to $b$--quarks and $\tau$--leptons
are more enhanced and BR$(H/A \ra \tau^+\tau^-)$ are larger by slightly more 
than a factor two compared to the previous case. For even larger values of 
$\tb$, $\tb=30$, the decays into charginos and neutralinos are not dominating 
anymore, and the branching ratios for the $H/A$ decays into tau pairs are 
suppressed only slightly, less than a factor of two. The branching ratios for 
the decays into $\mu^+ \mu^-$ can be obtained from Fig.~12 by rescaling the 
numbers by $m_\mu^2/m_\tau^2$. 

In the preceding discussion, the decays of the $H$ and $A$ bosons into 
sfermions were assumed to be shut. However, as discussed previously, at 
least one of the stops can be rather light and its couplings  to the Higgs 
bosons enhanced; the decays of the heavier Higgs bosons $H,A$ to stop squarks 
might be therefore kinematically accessible and could dominate over the 
standard decays, and even over the decays into charginos and 
neutralinos\footnote{The decay widths of the $H$ bosons into the light fermion
partners are proportional to $ G_F M_W^4 \sin^2 2\beta /M_{H}$ for $M_H \gg 
m_{\tilde{f}}$. They are thus suppressed by the heavy $H$ mass and cannot 
compete with the decays into fermions [$t, b, \tau$ and possibly $\chi$ states]
for which the widths grow as $M_H$. The pseudoscalar $A$ boson cannot decay 
into the partners of light fermions, if the fermion mass is neglected.}.  
Indeed, in the decays of the $H$ boson into stops, the partial widths up to 
mixing angle factors are proportional  to $G_F m_t^4/ (M_H \tan^2\beta)$ 
or/and $G_F m_t^2 (\mu -A_t/\tb)^2 /M_H$; for small $\tb$ values and  not too 
large $M_H$ and for intermediate $\tb$ values and for large $\mu$ and $A_t$, 
the widths for the decays $H \ra \tilde{t} \tilde{t}$ can be very large and 
can compete with, and even dominate over, the other [standard and SUSY]
decay channels. The branching ratios for the $H$ decays into $\tau$ pairs 
would be then further suppressed. \s

%\begin{figure}[htb]
%\vspace*{-.5cm}
%\mbox{\psfig{figure=f3b.eps,width=13cm}}
%\vspace*{-11.8cm}
%\caption[]{Branching ratios for the decays $H \ra \tilde{t}_1 \tilde{t}_1$ 
%as a function of $M_{H}$ for $\tb=2.5,5,10$ and 30 and for the values 
%$M_2=\mu=m_{\tilde{f}}/2=250$ GeV and $A_t=1.5$ TeV.}  
%\vspace*{-4mm}
%\end{figure}

\begin{figure}[htb]
\hspace*{-2.cm}
\vspace*{-1.cm}
\mbox{\psfig{figure=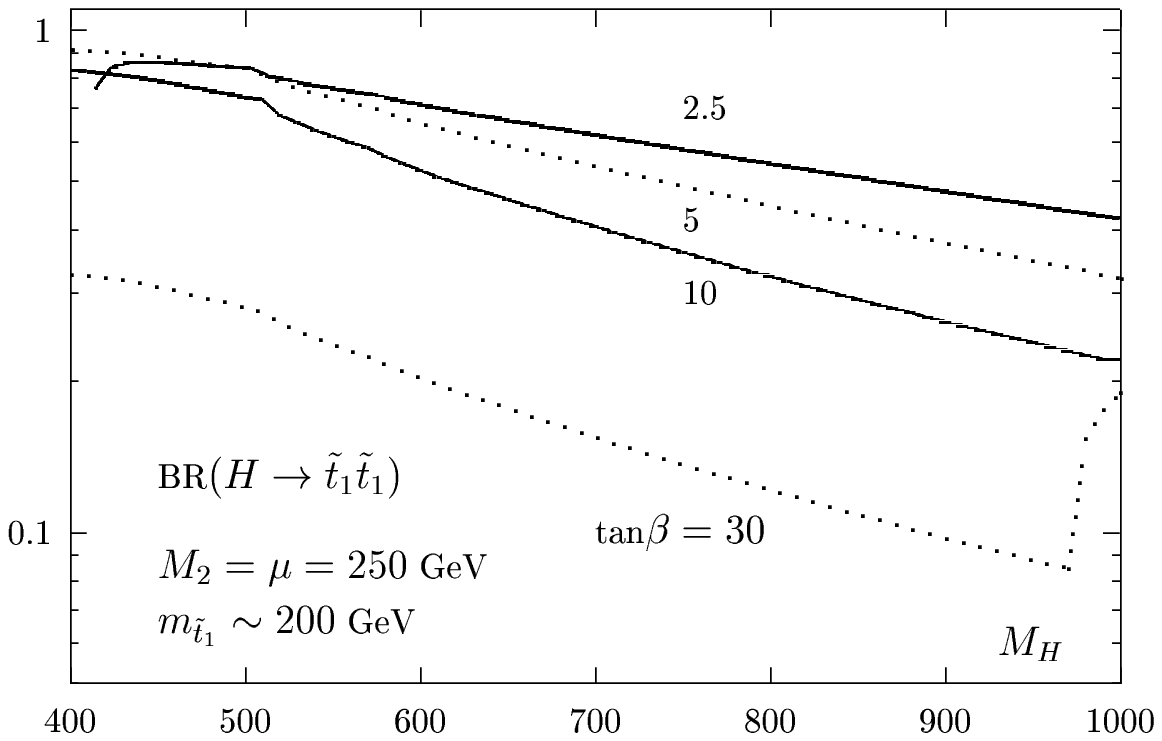,width=11cm}, \hspace*{-3.1cm} \psfig{figure=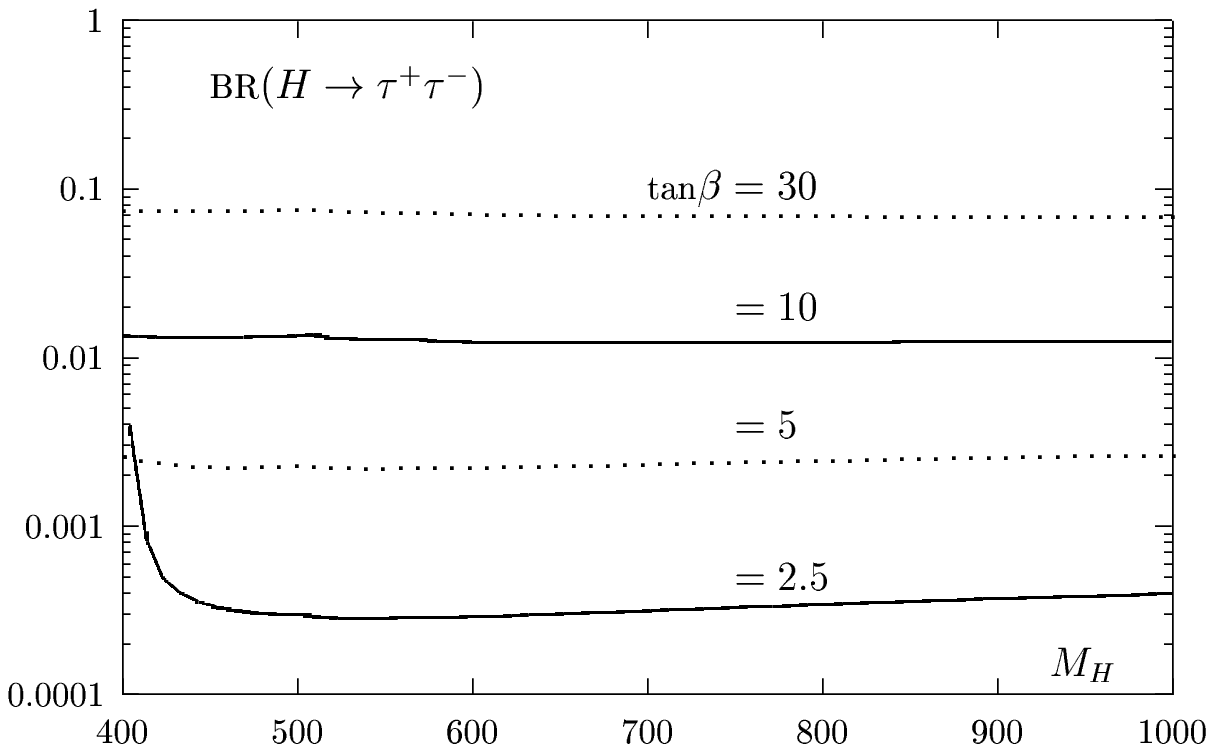,width=11cm}}
\vspace*{-9.4cm}
\caption[]{Branching fractions for $H \ra \tilde{t}_1 \tilde{t}_1$ 
[left] and $H \ra \tau^+ \tau^-$ [right] as a function of $M_{H}$ for 
$\tb=2.5,5,10,30$ and $M_2=\mu=m_{\tilde{f}}/2=250$ GeV and $A_t=1.5$ TeV.}
\vspace*{-3mm}
\end{figure}

This is illustrated in Fig.~13, where BR$(H\ra \tilde{t}_1\tilde{t}_1$)
is shown as a function of $M_H$ for $\tb=2.5,5,10,30$ and  $m_{\tilde{t}_1} 
\simeq 200$ GeV [for $\tb=2.5$ this is achieved by setting $m_{\tilde{f}_L}
=m_{\tilde{f}_R} = 500$ GeV and $A_t =1.5$ TeV]; $M_2=\mu=250$ GeV. 
As expected BR$(H \ra \tilde{t}_1 \tilde{t}_1$) decreases with increasing $\tb$ 
values and increasing $M_H$. However, it is still at the level of $\sim 50\%$ 
for $\tb=5$ and $M_H=1$ TeV. For $\tb=30$, the channel $H \ra \tilde{t}_1 
\tilde{t}_2$ opens up for $M_H \sim 900$ GeV, and the curve shows the sum 
BR$(H \ra \tilde{t}_1 \tilde{t}_1+ \tilde{t}_1 \tilde{t}_2$). But for this 
large $\tb$ value, the branching ratio barely exceeds the level of $20\%$ in 
contrast to lower $\tb$ values where it can reach almost unity for small $M_H$.
For larger $M_H$, the decays into charginos and neutralinos become more 
important and will dominate; so BR$(H \ra \tau^+ \tau^-)$ is reduced anyway. 
For the $A$ boson the only important decay into sfermions is $A\ra \tilde{t}_1
\tilde{t}_2$ [and maybe $\tilde{b}_1 \tilde{b}_2$ for $\tb \gg 1$]. Thus 
both stops must be light for the decay to be allowed by kinematics. This 
happens only in a small area of the parameter space, unless all squarks are 
relatively light. For instance, in the scenario of Fig.~13, $m_{\tilde{t}_2} 
\sim 700$ GeV and the decays $A,H \ra \tilde{t}_1 \tilde{t}_2$ occur only 
for masses close to 1 TeV.

\subsubsection*{4.2.2 Decays of the charged bosons} 
%into $\chi^\pm \chi^0$ and $\tilde{\tau} \tilde{\nu}$ states} 

Charged Higgs bosons are searched for at the Tevatron and will be 
searched for at the LHC in the mass range $M_{H^\pm} < m_t -m_b$, {\rm i.e.} 
when produced by a decaying $t$--quark. Some experimental limits 
from Tevatron [a $\tb$--$M_{H^\pm}$ exclusion contour] are already 
available \cite{H+TEV}. However, the detection techniques rely on the 
assumption that no 
$H^+$ decay mode other than $c \bar{s}$ and $\tau^+ \nu_\tau$ is kinematically 
significant,  and in particular that no decay into SUSY particles is 
possible. However, decays into the lightest charginos and neutralinos as well 
as decays into sleptons \cite{franc} 
are still allowed by present experimental data;  
for instance, the decay channel $H^+ \to \chi_1^+ \chi_1^0$ is certainly 
possible for $M_{H^\pm} \gsim 130\,$GeV, despite the lower bound of $m_{\chi_1^+} 
\geq 91$ GeV \cite{data}. \s

In Fig.~14a, the relative branching ratio for the decay $H^+ \to \chi_1^+ 
\chi_1^0$ is shown as a function of $M_{H^+}$, for $\tb=2$, $M_2=110\,$GeV, 
$|\mu|=500\,$GeV, and all remaining soft masses large enough to forbid other 
supersymmetric decay modes. For these values of parameters, one has 
$m_{\chi_1^+} =96.5\,$GeV and $m_{\chi_1^0}\simeq 50\,$GeV. Also shown is the 
case where the assumption of gaugino mass  universality is relaxed: keeping 
fixed all other parameters to the previous values, $M_1$ is set to $20\,$GeV, 
and the lightest neutralino mass becomes $m_{\chi_1^0} \simeq 16\,$GeV. In 
this case, the decay channel $\chi^+_1 \chi^0_1$  opens already at $\sim 115
\,$GeV. For small values of $\tb$, whenever kinematically allowed, the mode 
$\chi_1^+ \chi_1^0$ is the dominant one. For large values of $\tb$, however, 
only the channel $H^+ \to \tau^+ \nu_\tau$ survives. \s

\begin{figure}[htb]
\hspace*{-2.cm}
\vspace*{-1.cm}
\mbox{\psfig{figure=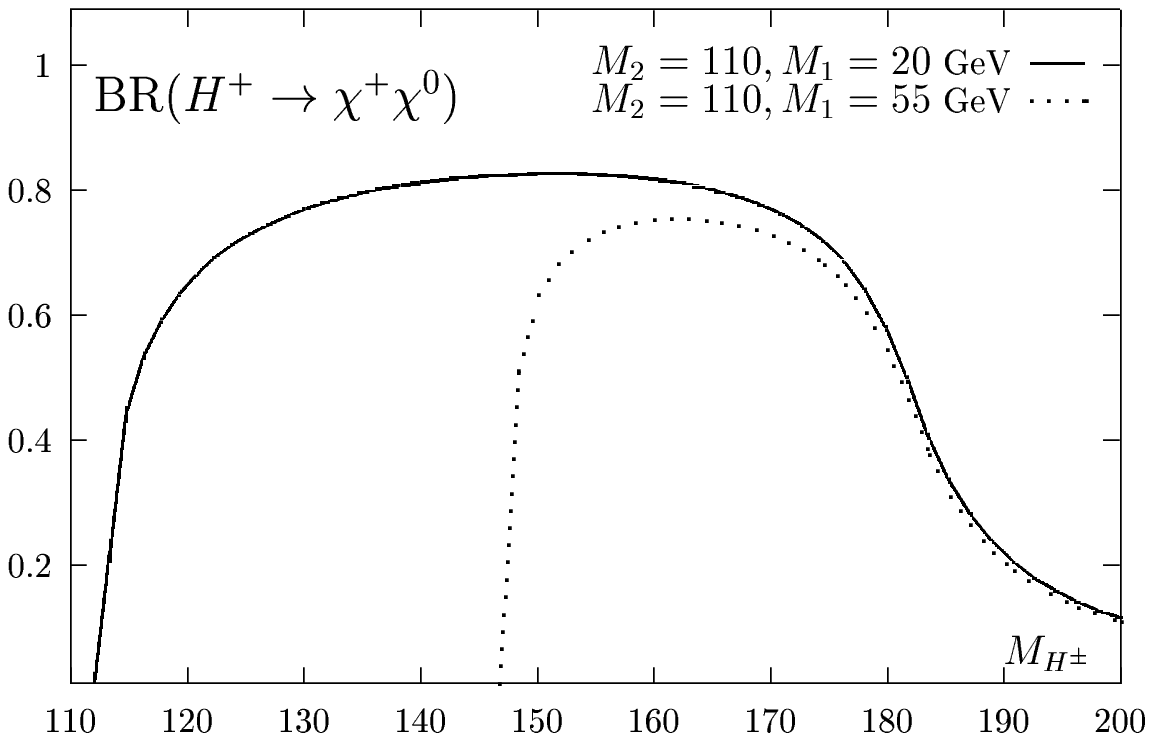,width=11cm}, \hspace*{-3.1cm} \psfig{figure=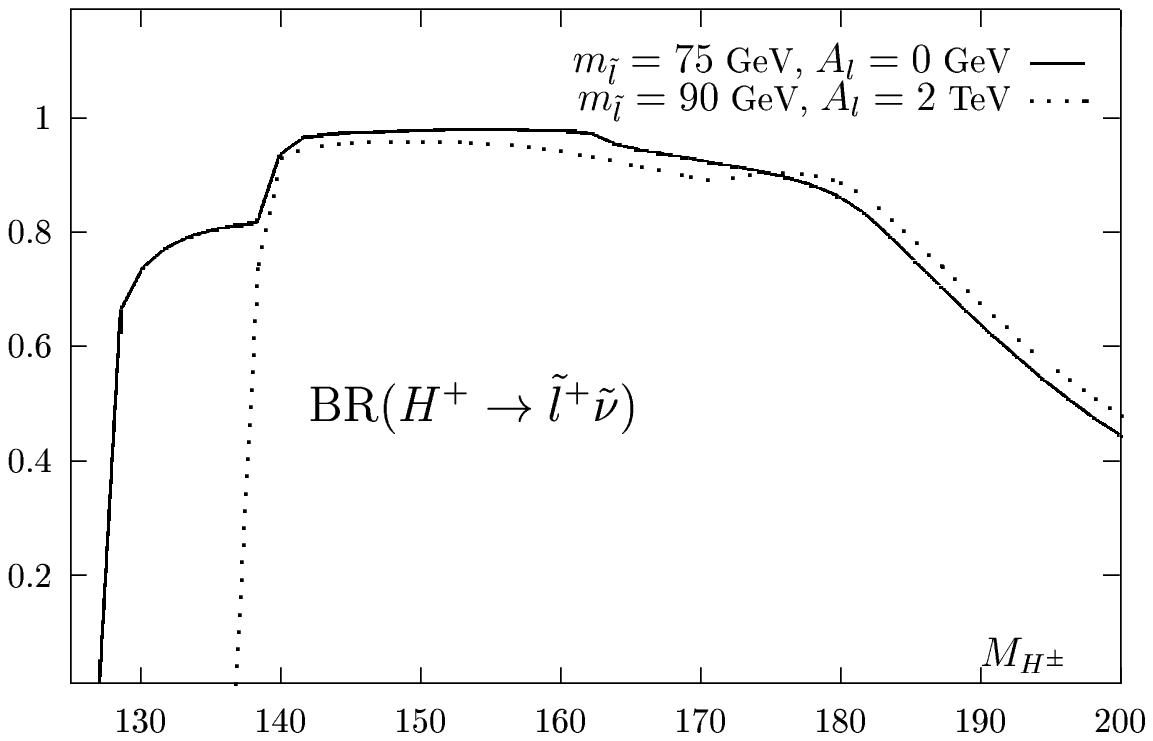,width=11cm}}
\vspace*{-9.4cm}
\caption[]{Branching fractions for the decays $H^+ \to \chi_1^+ 
\chi_1^0$ (left) and $H^+ \to \tilde{l}^+ \tilde{\nu}$ (right) as a function 
of $M_{H^\pm}$, for $\tb=2$, and some $M_2, M_1, \mu$ and $A$ values.}
\vspace*{-3mm}
\end{figure}

The previous discussion applies only to supersymmetric models with heavy 
sfermion masses. The existing lower bounds on slepton masses, however, are 
still rather modest: $\sim 65$ GeV for $\tilde{\tau}$'s and $\gsim\,45$ GeV 
for $\tilde{\nu}$'s \cite{data}. The decay $H^+ \to \tilde{\tau}^+ \tilde{
\nu}_\tau$ is therefore kinematically possible for $M_{H^\pm} \gsim 110$ GeV.  
Fig.~14b shows the relative branching ratio for $\tb=2$, $M_2=120\,$GeV,  
$M_1\sim M_2/2$ and two choices of parameters in the slepton mass matrices: 
{\it (a)} $m_{\tilde{l}_L}=m_{\tilde{l}_R}=m_{\tilde{l}}=75\,$GeV,
$\mu=500\,$GeV, $A_\tau=0$ and {\it (b)} $m_{\tilde{l}}=90\,$GeV,
$\mu=-500\,$GeV, $A_\tau=1\,$TeV. This leads to a slepton spectrum:
$(a)$ $m_{\tilde{\nu}}=56$ GeV, 
$m_{\tilde{e}} \sim m_{\tilde{\mu}}=83$ GeV and the two $\tilde{\tau}$
masses are 10 GeV below and above this value, and $(b)$ 
$m_{\tilde{\nu}}=75\,$GeV, 
$m_{\tilde{e}} \sim m_{\tilde{\mu}}=97\,$GeV and
$m_{\tilde{\tau}_1}=63\,$GeV, and $m_{\tilde{\tau}_2} =121\,$GeV. 

The prominence of the $ H^+ \ra \tilde{\tau}^+ \tilde{\nu}_\tau$ mode 
above threshold is explained by the fact that the $H^+\tilde{\nu}_L^* 
\tilde{l}_L$ coupling $ \propto g/\sqrt{2} M_W \sin 2 \beta$ is very large 
compared to the $H^+{\nu}_L^* {\tau}_R$ coupling $ g/\sqrt{2} (m_\tau/M_W)$.  
Due to the $\sin2 \beta$ dependence, this term quickly dies
off for increasing $\tan \beta$. In this case, however, there
exists other directions of parameter space where 
this decay mode has still a branching ratio close to unity. 
When $A_\tau \sim \mu \tb$, in fact, the left--right mixing
in the slepton mass matrix tends to vanish, but the 
coupling of the Lagrangian term 
$H^+\tilde{\nu}_L^* \tilde{\tau}_R$:
 $-g/\sqrt{2} (m_\tau/M_W) (\mu + A_\tau \tb)$   
acquires a $1/\cos^2 \beta $ dependence, which increases 
with increasing $\tb$. For instance,  $\tb = 10$, 
$A_\tau =2\,$TeV, $\mu\sim 200\,$GeV and $m_l\sim 80\,$GeV,
give a branching ratio above $90\%$ for the decay channel
$H^+ \to \tilde{\tau}^+ \tilde{\nu}_\tau$, when kinematically 
accessible. Such a decay mode produces a final $\tau^+$ plus 
missing energy as the $\tau^+ \nu_\tau$ mode, but with a 
much softer $\tau^+$ than that searched for at hadron colliders. \s

Finally, let us make a few remarks on other possible decay of $H^\pm$
in non--SUSY two--Higgs doublet models (2HDM). Although the 
decay $b \to s \gamma $ excludes masses $M_{H^\pm}$ up to $\sim 165\,$GeV, 
irrespectively of $\tan \beta$~\cite{BSG}, direct searches still allow 
for the possibility  $M_{H^\pm} < M_W$. The $H^\pm$  states are
searched for at LEP2 and the Tevatron, relying again on the $c\bar{s}$ and  
$\tau \nu$ signals [see Ref.~\cite{franc} for details and references]. \s

$(i)$ In 2HDMs, there is no lower bound on the pseudoscalar mass $M_A$ coming 
from LEP: since the mixing angle $\alpha$ is in this case a free parameter, 
it can be chosen in such a way that the coupling $ZhA$ vanishes and the process
$Z \to hA$ does not occur. This makes that the LEP2 bound $M_A>75$ GeV 
\cite{data} does not hold, leaving open the possibility of a very light 
$A$ boson\footnote{Two other production mechanisms are possible at LEP1, 
since they require very large statistics: the fermion loop--mediated decay 
$Z \to A \gamma$ and the radiation off the $b \bar b$ and $\tau^+ \tau^-$ 
lines; however the rates are small especially for moderate values of $\tb$. 
One remains therefore with the rather modest bound from the decay $\Upsilon 
\to A \gamma$ which has been searched for by the Crystal Ball 
Collaboration~\cite{CRISTAL}, $M_A > 5\,$GeV.}. [In this case the 
cross section for the process $e^+ e^- \to Z^* \to hZ$, which is complementary,
is not suppressed compared to the SM Higgs boson, and the bound $M_h 
>88.6\,$GeV~\cite{data} applies here]. Now since $g_{H^+ W^-A}$ is a 
``gauge" coupling, it is clear that the decay $H^+ \to AW^+$ can be
rather important in a 2HDM. Indeed, this remains true even for an off--shell 
$W$--boson, in spite of the virtuality and the additional weak coupling 
suppressions. Even for masses $M_{H^\pm} \sim 55$ GeV, {\rm i.e.} roughly the 
value excluded at LEP2, the branching ratio can be still at the level of 50\%
for small enough $\tb$ and $M_A$ values. For heavier $H^\pm$ bosons that can 
be searched for at the Tevatron, the decay $H^\pm  \ra A W^{(*)}$ can be by far
the dominant one. \s
 
$(ii)$ For charged Higgs boson masses above $\sim 140$ GeV, even if the
decays into $AW^*$ bosons are suppressed [for instance the $A$ boson is 
too heavy], the $H^+ \to \tau \nu_\tau$ and $c \bar{s}$ channels are still 
not the dominant ones for small $\tb$ values. Indeed, for $\tb \sim 1$ 
the three--body decay mode $H^+ \to \bar{b}t^* \to \bar{b} b W^+$ is 
already the dominant decay mode: despite the  virtuality of the top
quark and the fact that the process is of higher--order, the $H^\pm$
coupling to $tb$ quarks is much larger than the $\tau \nu$ and $cs$ 
couplings, leading to a large compensation. For instance, for $M_{H^\pm}=140$ 
GeV and $\tb=1$, the branching ratio is already at the level of 50\%. 
Note that this decay mode [as well as the three--body decay mode $H^\pm \ra 
hW^{*} \ra h f\bar{f}$] can also be important in the MSSM for large enough
$H^\pm$ masses [but still below $m_t-m_b$] and small $\tb$ values.

\subsection*{4.3. MSSM Higgs production in $e^+e^-$ collisions}  

\subsubsection*{4.3.1 Production mechanisms} 

The main production mechanisms of the neutral MSSM Higgs bosons at $\ee$ 
colliders are the Higgs--strahlung process  and pair 
production \cite{strah,pair}, as well as the $WW$ and $ZZ$ fusion 
processes \cite{wwfus}: 
\begin{eqnarray}
(a) \ \ {\rm Higgs\mbox{-}strahlung} \hspace{1cm} \ee & 
           \ra &  (Z) \ra Z+h/H 
\hspace{5cm} \non \\
(b) \ \ {\rm pair \ production} \hspace{1cm} \ee & \ra & (Z) \ra A+h/H  
\non \\
(c) \ \ {\rm fusion \ processes} \hspace{0.8cm} \ \ee & \ra &  \bar{\nu} 
\nu \ (WW) \ra \bar{\nu} \nu \ + h/H \hspace{3.5cm} \non \\
\ee & \ra &  \ee (ZZ) \ra \ee + h/H  
\end{eqnarray}
The CP--odd Higgs boson $A$ cannot be produced in the
Higgs--strahlung and fusion processes to leading order since it does not
couple to vector boson  pairs. \s
 
The charged Higgs particle \cite{pair} can be pair produced through virtual 
photon and $Z$ boson exchange in $\ee$ collisions, and also 
%at $\gamma \gamma$ 
%colliders [with Compton back--scattering \cite{laser} of laser photons] and 
in top-quark decays for masses below $m_t-m_b \sim 170$ GeV: 
\begin{eqnarray}
(d) \ \ {\rm charged \ Higgs } \hspace{0.8cm} \ \ee & \ra &  \ (\gamma , 
Z^* ) \ \ra \ H^+ H^- \hspace*{3.93cm} \nonumber \\
% \gamma \gamma & \ra & H^+ H^- \non \\ 
\ee & \ra & t \bar{t} \ {\rm  with} \ t \ra H^+b 
\end{eqnarray}

The production cross sections for the neutral Higgs bosons are suppressed by 
mixing angle factors compared to the SM Higgs production,
\begin{eqnarray}
\sigma(\ee \ra Zh) \ , \ \sigma(VV \ra h) \ , \ \sigma(\ee \ra AH) \ \ 
\sim \ \sin^2(\beta-\alpha) \non \\
\sigma(\ee \ra ZH) \ , \ \sigma(VV \ra H) \ , \ \sigma(\ee \ra Ah) \ \ 
\sim \ \cos^2(\beta-\alpha) 
\end{eqnarray}
while the cross section for the $H^\pm$ particles does not depend 
on any parameter other than $M_{H^\pm}$ when pair produced in $\ee$
%and $\gamma \gamma$ 
collisions; a $\tb$ dependence is however
present for $H^\pm$ production in top decays, due to the branching 
ratio BR$(t \ra H^+b)$.  \s

Modulo the mixing factors, the cross section for the 
fusion process, $\ee \ra \bar{\nu}_e \nu_e \Phi$ with $\Phi=h,H$ is 
enhanced at high energies since it scales like $M_W^{-2}\log s/M_\Phi^2$; 
the cross section for $\ee \ra e^+e^-\Phi$ is $\sim 16 \cos^2\theta_W$ 
i.e. one order of magnitude smaller. The cross sections for the 
Higgs--strahlung and pair production processes, $\ee \ra \Phi Z, \Phi A$
and $H^+ H^-$ scale like $1/s$ and hence are smaller at high energies. 
Close to the decoupling limit, the factor $\cos(\beta-\alpha) \ra M_Z^2
\sin2\beta/(2M_A^2)$ vanishes, and the only relevant processes for the 
production of the heavy states [when kinematically allowed] are 
the pair production of $A$ and $H$ and the charged Higgs pair production: 
$\ee \ra HA$ and $\ee \ra H^+H^-$. The cross section for the fusion process, 
$\ee \ra \nu \bar{\nu}H$, which in principle is enhanced
at high energies, is only relevant in the mass range of a few hundred GeV 
and for small $\tb$ values where the factor $\cos^2(\beta-\alpha)$ is not 
prohibitively small. For the lightest $h$ boson, the only remaining 
production processes are the bremsstrahlung and vector boson fusion 
mechanisms: $\ee \ra hZ$ and $\ee \ra \nu \bar{\nu}/\ee +h$, exactly like for 
the SM Higgs bosons. % the cross sections are rather large. \\

The cross sections for processes the 
$(a)$--$(c)$ and $(d)$ are shown in Figs.~15
and 16 as functions of the Higgs boson masses for $\tb=2.5$ at a c.m. energy 
$\sqrt{s}=500$ GeV. They are obtained with the help of the program HPROD to
be discussed later. \s

There are several additional processes for the production of the neutral
Higgs particles, and in particular of the lightest $h$ boson, that we will
take as an example. Although these processes are of higher order, the 
production rates can be substantial for high--luminosities, and they might
open a window to the determination of several fundamental properties of
the Higgs particles. \s

$(i)$ The associated production with $t\bar{t}$ pairs \cite{htop} which allows
to measure directly the $t \bar{t}$--Higgs Yukawa coupling, the strongest 
coupling in the standard sector:  
\begin{eqnarray}
\ee  \ \ra  \ (\gamma , Z^* ) \ \ra \ t\bar{t} \Phi 
\end{eqnarray} 

$(ii)$ The double Higgs boson production, either in the bremsstrahlung or in 
the fusion processes, which allows to determine the trilinear Higgs boson 
couplings, and reconstruct the all important scalar potential 
\cite{hhZ1,hhZ2}
\beqn
\ee \ra Zhh \ \ {\rm and} \ \ WW/ZZ \ra hh
\eeqn 

$(iii)$ The production of the Higgs bosons in association with a photon
\cite{hgamma1,hgamma2}, which allows to measure the $h\gamma \gamma$ and 
$hZ\gamma$ 
loop induced vertices and probe the effects of heavy particles which couple 
to the Higgs boson %[this process is similar to the resonant s--channel Higgs 
% boson in $\gamma \gamma$ colliders, $\gamma \gamma \ra h$]  
\beqn
\ee \ra h \gamma
\eeqn
In addition, if stop squarks are light, the production of the $h$ boson with
stop pairs, $\ee  \ra \tilde{t} \tilde{t} h$ \cite{hstop2},  which allows to 
measure the $\tilde{t} \tilde{t} h$ couplings, the largest electroweak 
couplings in the MSSM, opening a window to the probing of trilinear couplings. 
\s

Finally, $\ee$ linear colliders can be made to run in the $\gamma \gamma$ 
mode with the high energy photons coming from Compton back--scattering 
\cite{laser} of laser photons. This lead to the production of the Higgs 
bosons as $s$--channel resonances: $\gamma \gamma \ra h,H,A$, allowing to
measure the Higgs--photon couplings [similarly to the process $\ee \ra 
\gamma$+Higgs]. Furthermore charged Higgs bosons can be pair--produced in this 
mode. 

\subsubsection*{4.3.2 The program HPROD} 

There are many generators which deal with the production of the MSSM Higgs 
bosons at $\ee$ colliders \cite{outils}. Here, we briefly describe the program 
HPROD \cite{hprod} which calculates the cross sections of the SM and MSSM 
Higgs particles at $\ee$ machines. It includes: \s

$(i)$ In the SM: the strahlung and the $WW/ZZ$ fusion processes
as well as the higher order processes $\ee \ra ttH^0, H^0H^0Z$ and $VV \ra 
H^0H^0$ in the longitudinal vector boson approximation. The associated production 
with a photon will be included soon. \s

$(ii)$ In the MSSM it includes all the processes $(a)$--$(d)$
above: the  strahlung and the pair production as well as the $WW/ZZ$ 
fusion processes for neutral Higgs production, and the $\ee$ % \gamma \gamma$
and top decay processes for the charged Higgs boson. The higher
order processes will be included soon. \s

$(iii)$ In the MSSM, the complete radiative corrections in the effective 
potential approach with full mixing in the $\tilde{t}, \tilde{b}$ sectors; 
it uses the RG improved values of the masses and couplings, 
and the relevant leading next--to--leading--order corrections are also 
implemented. Both the pole and the running Higgs boson masses are calculated.\s

$(iv)$ Off--shell $Z$ and Higgs bosons in the bremsstrahlung and the  Higgs
pair production processes [the latter only in the region of parameter space 
where they are relevant: close to the production threshold and for sizeable
Higgs boson decay widths].  The Higgs boson total widths [without the decays 
into SUSY particles] are included in the routines. \s

$(v)$ The initial state radiative corrections for the bremsstrahlung, pair 
production processes and the $WW/ZZ$ fusion processes. The effect of 
beamstralung for the TESLA machine by making an interface with the 
program CIRCEE \cite{circee} will be implemented soon.  \s
 
The basic input parameters, fermion and gauge boson masses and their
total widths, coupling constants and in the MSSM, soft SUSY--breaking
parameters can be chosen from an input file. In this file several
flags allow to switch on/off or change some options [{\it e.g.} choose
a particular Higgs boson, include/exclude initial state radiation, 
use running or pole Higgs masses]. The parameters in the input file
are: \s

$\sqrt{s}$ the c.m. energy of the collider; IHIGGS: an integer which chooses 
the Higgs boson [0,1,2,3 for $H^0,h,H,H^\pm $ and 4 for all MSSM Higgs bosons]; 
$\tb$ and $M_A$: the two basic parameters in the MSSM Higgs sector; $m_b,m_t$ 
the bottom and top masses; $\alpha_s(M_Z), G_F, \alpha(0)$: the strong, 
Fermi and QED coupling constants; 
$M_Z,M_W,\Gamma_Z$: the $W,Z$ boson masses and the $Z$ total width;  
$M_2,\mu, m_{\tilde{Q}_L},m_{\tilde{U}_R}, m_{\tilde{D}_R}$:  the SUSY 
breaking mass parameters; $A_b,A_t$:  SUSY breaking trilinear coupling for 
stops and sbottoms; ISHELZ/ISHELH: integers which for 1(0) include (exclude) 
off-shell $Z$/Higgs bosons in the bremsstrahlung/pair production processes; 
IPOLE: an integer which for 0(1) calculates 
running (pole) MSSM Higgs masses;  ISR: an integer which for 1(0) excludes 
(includes) initial state radiation; IHIGH: an integer which for 0(1) includes
(excludes) the higher order processes for the SM Higgs boson. 
 An example of input file for Higgs boson production in the MSSM 
is shown below. \s

\begin{verbatim}
  SQRS     = 500.D0      ALPH  = 137.036D0     MDR      = 1000.D0  
  IHIGGS   = 4           GF    = 1.1664D-5     AU       = 2400.D0
  TGBET    = 2.5D0       GAMZ  = 2.489D0       AD       = 2400.D0
  MABEG    = 50.D0       MZ    = 91.187D0      IPOLE    = 0
  MAEND    = 250.D0      MW    = 80.33D0       ISHELZ   = 0
  NMA      = 5           MU    = 500.D0        ISHELH   = 0
  MB       = 4.62D0      M2    = 500.D0        ISR      = 1
  MT       = 175.D0      MSQ   = 1000.D0       IHIGH    = 1
  ALS(MZ)  = 0.118       MUR   = 1000.D0
\end{verbatim}

\nn Tab.~3: Example of input file for the MSSM Higgs production at 
$\sqrt{s}=500$ GeV. Figs.~15 and 16 are obtained with the inputs given in 
this file. \newpage

\begin{figure}[htb]
\hspace*{-2.5cm}
\mbox{\psfig{figure=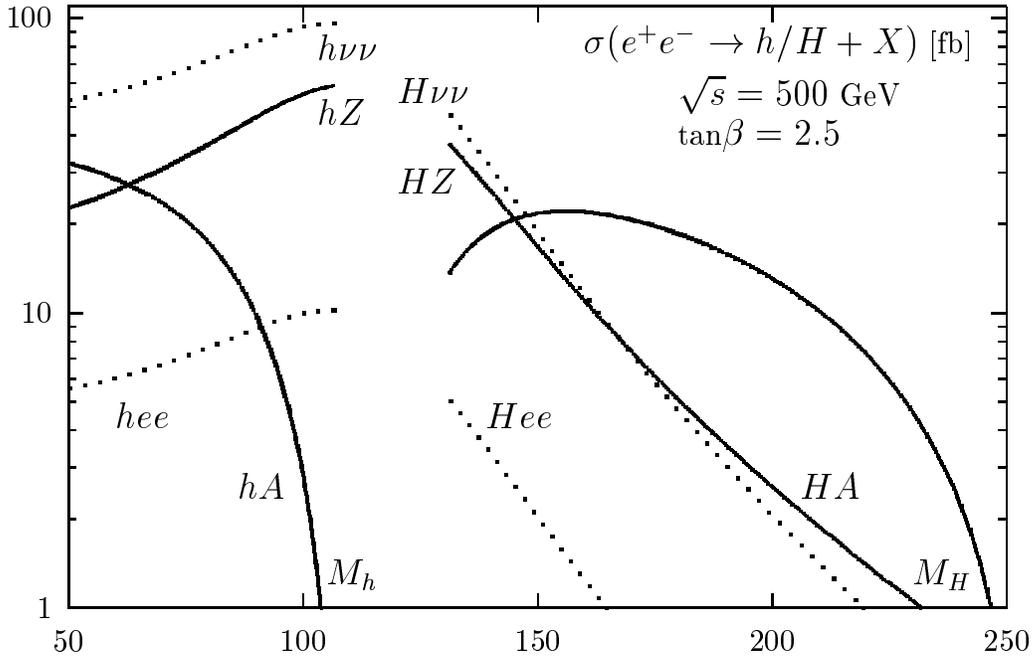,width=19.3cm}}
\vspace*{-18.cm}
\caption[]{Production cross sections for the neutral MSSM Higgs bosons
in $\ee$ collisions as a function of the $h$ and $H$ masses at a c.m. energy 
$\sqrt{s}=500$ GeV and for $\tb=2.5$.}
\vspace*{-1.cm}
\end{figure}
\begin{figure}[htb]
\hspace*{-2.5cm}
\mbox{\psfig{figure=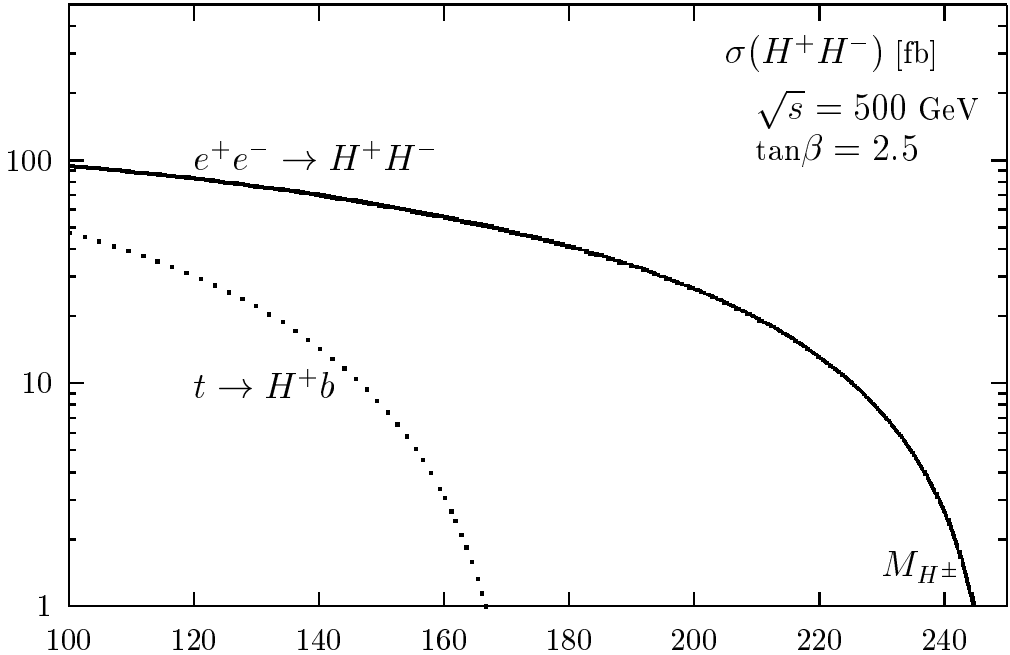,width=19.3cm}}
\vspace*{-18.cm}
\caption[]{Production cross sections for the charged Higgs boson 
in $\ee$ collisions as a function of the $H^\pm$ mass for a c.m. 
energy $\sqrt{s}=500$ GeV and for the decay $t \ra H^+b, \tb=2.5$.} 
\vspace*{-7cm}
\end{figure}

\newpage

The program is written in FORTRAN and has been tested on several
machines. All the necessary subroutines [e.g. for integration] are 
included. The program is lengthy but rather fast [especially 
if some options, e.g. ISHELH, are switched off]. The results for the
many production cross sections in femtobarns are written into 
several output files [with headers indicating the processes and giving 
the inputs]:
\begin{eqnarray}
{\rm zps.out}:  && M_{H^0} \ , \ \sigma(H^0 Z) \ , \ \sigma (H^0 \nu 
\bar{\nu} ) \ , \ \sigma (H^0 \ee ) \non \\
{\rm zpt.out}:  && M_{H^0} \ , \ \sigma(WW \ra H^0 H^0) \ , \ \sigma (ZZ \ra
H^0 H^0) \ , \ \sigma (H^0 H^0 Z) \ , \ \sigma(t\bar{t}H^0)
\non \\
{\rm zpl.out}:  && M_A \ , \  M_{h} \ , \ \sigma(hZ) \ , \ \sigma(h \nu 
\bar{\nu}) \ , \ \sigma(h\ee )  \ , \ \sigma(hA) \non \\
{\rm zph.out}:  && M_A \ , \  M_{H} \ , \ \sigma(HZ) \ , \ \sigma(H \nu 
\bar{\nu})\ , \ \sigma(H\ee) \ , \ \sigma(HA) \non \\
{\rm zpc.out}:  && M_A \ , \  M_{H^\pm} \ , \ \sigma(\ee \ra H^+ H^-) \ , 
\ \sigma(t\bar{t} \ra H^+ H^-) \non 
\end{eqnarray}
\begin{verbatim} 
  MA             Mh          hZ       hnunu       he+e-      hA
 50.0000      48.0416      22.09      52.00      5.483      32.98
 100.000      79.7747      37.30      72.05      7.658      17.48
 150.000      97.2542      52.94      91.32      9.741      4.564
 200.000      103.804      57.82      95.56      10.21      1.019
 250.000      106.497      58.98      95.78      10.23     0.2473
\end{verbatim}
\centerline{Tab.~4a: The output file zpl.out for the input file in Tab.~3.}
\smallskip
\begin{verbatim}
 50.0000      131.191      37.36      46.92      5.032      13.66
 100.000      143.719      22.15      25.60      2.750      20.38
 150.000      173.380      6.506      6.176     0.6653      20.11
 200.000      215.038      1.625      1.164     0.1257      8.726
 250.000      261.096     0.4601     0.2384     0.2575E-01  0.000
\end{verbatim}
\centerline{Tab.~4b: The output file zph.out for the input file in Tab.~3.}
\smallskip
\begin{verbatim}
  MA           MH+       ee->H+H-     tt->H+H-
 50.0000      92.1414      98.79      53.73
 100.000      126.452      78.94      24.77
 150.000      168.790      49.36      0.5294
 200.000      214.453      16.70      0.000
 250.000      261.706      0.000      0.000
\end{verbatim}
\centerline{Tab.~4c: The output file zpc.out for the input file in Tab.~3.}

\bigskip

In a future version of the code, we will implement the remaining higher order
processes [discussed previously] of the SM and for the $h$ boson in the MSSM,
and include the $\gamma \gamma$ option of the collider. 
We also plan to make an interface with the programs SUSPECT [section 3.2] 
and HDECAY  [which calculates the total decay widths in a more precise way]. 
Beamstralung will also be incorporated by making and interface with the program
CIRCEE. The Higgs boson production at hadron machines will be included in the 
next--to--leading version of the code. 

\newpage

\subsubsection*{4.3.2 Higgs boson production in association with light stops} 

As previously discussed [section 3.1 and above], if the off--diagonal entries 
in the $\tilde{t}$  
mass matrix is large, the eigenstate $\tilde{t}_1$ can be rather light and at 
the same time its couplings to the Higgs bosons strongly enhanced. The $h 
\tilde{t}_1 \tilde{t}_1$ coupling would be then the potentially largest 
coupling in the electroweak sector of the MSSM, and its measurement would 
allow to pin down some of the soft--SUSY breaking parameters of the MSSM 
Lagrangian, in particular the trilinear coupling $A_t$. This coupling can be 
measured in the process $pp \ra \tilde{t}_1 \tilde{t}_1h$ at the LHC [see 
section 4.1.3]. 
However, this coupling can be best measured in the associated production 
process $\ee \ra \tilde{t}_1 \tilde{t}_1 h$ \cite{eehtt}. This is first 
due to the fact that the cleaner environment of high--energy electron-positron 
colliders allows for a more efficient search of this complicated final state, 
ans also because of the very high--luminosity expected at such machines [for 
instance for the DESY--machine TESLA: $\int {\cal L} \sim 500$ fb$^{-1}$ in 
a year], which
compensates for the fact that the process is of higher order in perturbation
theory and has a small cross section in priciple. \s

At future linear $\ee$ colliders, the final state $\tilde{t}_1 \tilde{t}_1 h$
may be generated in three ways: \\

\nn {\bf a) Two--body production and decay}: \s

If the center of mass energy of the $\ee$ collider is high enough, one first 
produces a mixed pairs of top squarks, $\ee \ra \tilde{t}_1 \tilde{t}_2$,  
through the exchange of a virtual $Z$--boson, and then makes the heaviest top 
squark decay into the lightest stop and the Higgs boson, $\tilde{t}_2 \ra 
\tilde{t}_1h$, if the splitting between the two stops is larger than the $h$ 
boson mass. In principle, if phase--space allowed, the cross section for the 
two--body production process times the branching ratio for the two--body decay 
should be large enough for the final state to be copiously produced. \s

However, the $Z\tilde{t}_1\tilde{t}_2$ coupling is proportional to 
$\sin 2 \theta_t$ while the $h\tilde{t}_1 \tilde{t}_2$ coupling is
proportional to $\cos 2 \theta_t$; since in a large part of the MSSM parameter
space [as discussed in section 4.1.3] $\sin 2 \theta_t$ is either small 
[no--mixing case] or close to one [maximal mixing case], the cross 
section times branching bratio, which is then proportional to $\sin 4 
\theta_q$, is always 
rather small. [In addition, the decay width $\tilde{t}_2 \ra h\tilde{t}_1$ is 
in general much smaller than the $\tilde{t}_2$ decay widths into chargino and
neutralinos]. 
Nevertheless, there are limited regions of the MSSM parameter 
space where the rate for this process is visible for the high luminosities
$\int {\cal L} \sim 500$ fb$^{-1}$ expected at the linear colliders. \\
   
\nn {\bf b) Production in the continuum in e$^+$e$^-$ collisions}: \s 

There are three types of Feynman diagrams leading to $\tilde{t}_1\tilde{t}_1 h$
finale states in $\ee$ collisions: Higgs boson emission for the $\tilde{t}_1$ 
lines which proceeds
through s--channel photon and $Z$--boson exchange, Higgs boson emission from 
the $\tilde{t}_2$ lines which proceeds through $Z$--boson exchange, and Higgs
boson emission from a virtual $Z$ boson which then splits into $\tilde{t}_1
\tilde{t}_1$ pairs. When the $g_{h \tilde{t}_1 \tilde{t}_1}$ coupling is large, 
the two last types of Feynman diagrams give negligible contributions since
the virtuality of $\tilde{t}_2$ is large and the $Z\tilde{t}_1\tilde{t}_1$
coupling not enhanced. The Dalitz plot density of the process [in terms of the 
scaled energies of the two stops] can be then written in a very simple form. \s

As an illustration, the total cross section for the process $\ee \ra h 
\tilde{t}_1 \tilde{t}_1$
is shown in Fig.~17 as a function of the lightest stop mass for a c.m. 
energy $\sqrt{s}=500$ GeV, for the values $\tilde{A}_t=1.5$ TeV and $\tb=3,30$.
One can see that below masses of the order of $m_{\tilde{t}_1} \sim 160$ GeV,
the cross section is larger than 1 fb, leading to $\sim 500$ events with the 
expected luminosity quoted above. The final state topologies have been
discussed in section 4.1.3; however, here the main decay mode of the $h$ boson,
$h \ra b\bar{b}$, can be used [this calls for good micro--vertex detectors]. 
With the relatively large sample of events, this final state should be 
experimentally possible to be detected even after efficiencies have been
included. Higher c.m. energies would allow to probe larger stop masses. 
The cross section for this process has also been calculated in 
Ref.~\cite{annecy} using the package SUSY--GRACE;  it 
has been shown that combined with the $\ee \ra \tilde{t}_1 \tilde{t}_1$ 
production cross section, the SUSY parameters of the stop sector can be 
determined. 

\begin{figure}[htb]
\vspace*{-.5cm}
\mbox{\psfig{figure=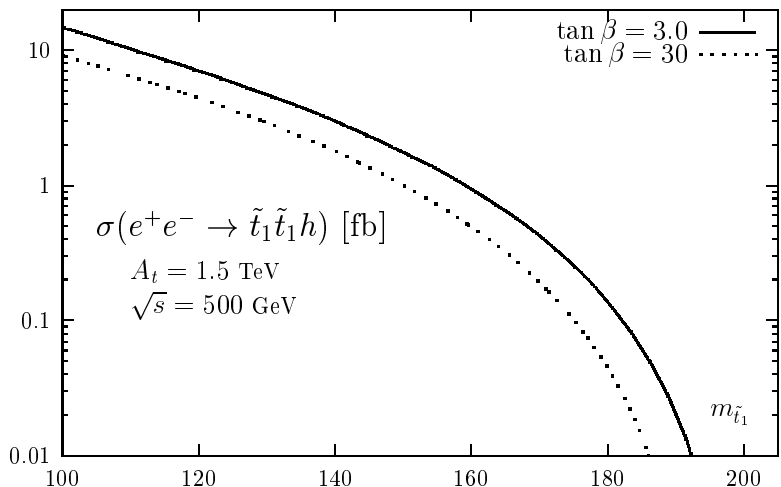,width=14.5cm}}
\vspace*{-13.2cm}
\caption[]{The cross section for the process $\ee \ra h \tilde{t}_1 \tilde{t}_1$
as a function of the lightest stop mass for a c.m. energy $\sqrt{s}=500$ GeV
and for the values $\tilde{A}_t=1.5$ TeV and $\tb=3,30$.}
\end{figure}

\nn {\bf c) Production in the continuum in $\gamma \gamma$ collisions}: \s

In the $\gamma \gamma$ option of future high--energy $\ee$ linear colliders,
with the high energy photons coming from Compton back--scattering 
of laser beams, 
the final state $\tilde{t}_1 \tilde{t}_1 h$ can be generated by emitting the
$h$ boson from the stop lines in the process $\gamma \gamma \ra \tilde{t}_1 
\tilde{t}_1$. In Ref.~\cite{eehtt}, the cross section for this reaction has 
been calculated, and the analytical expression has been found to be slightly 
more complicated than the one obtained in the $\ee$ mode. \s

However, because the c.m. energy of the $\gamma \gamma$
collider is expected to be only $\sim 80\%$ of the one of the original $\ee$
machine, the process is less phase--space favored. In addition, the collected 
luminosity is expected to be somewhat smaller than the one of the $\ee$ mode,
leading to a smaller number of events. Nevertheless, the cross section for
the $\tilde{t}_1 \tilde{t}_1h$ final state is of the same order as in the 
$\ee$ mode for c.m. energies not too close to the kinematical threshold, and 
the process might be useful to obtain complementary information since it does 
not involve the $Z$--boson and $\tilde{t}_2$ exchanges.

\newpage

\setcounter{equation}{0}
\renewcommand{\theequation}{5.\arabic{equation}}

\section*{5. SUSY Particle Production and Decays} 

\subsection*{5.1. Virtual SUSY effects} 
 
Virtual effects of supersymmetry, for example through gauge boson 
self--energy or gauge boson--fermion--fermion vertex corrections, are
known to be invisibly small at the $Z$ resonance unless the masses of the 
SUSY particles are very small, a situation which is by now excluded or
unlikely \cite{data}. It has been noticed \cite{Hollik} that this property 
does not apply 
for energies beyond the $Z$ peak due to the raise of box diagrams. 
At LEP1 energies, the box contributions are strongly suppressed as they do 
not benefit of the enhancement due to the $Z$ resonance. Beyond this energy,
one can expect them to become relatively important. This is what happens in 
the standard electroweak case with the box diagrams due to $WW$ formation 
and neutrino exchanges, where their effect reaches the percent level 
in the cross section for $e^+e^-\to \mu^+\mu^-$ at $\sqrt{s} \sim 200$ GeV. 
A second source of local box enhancement is the threshold effect
appearing around $s=(m_i+m_j)^2$ where $m_i$ and $m_j$ are the masses of
the particles formed in the intermediate state. In the aforementioned 
$WW$ box diagrams the contribution peaks at 1.2$\%$ around 161 GeV. \s

With this standard situation in mind, one can look at the SUSY box 
contributions due to charginos or neutralinos and sleptons for the production
of light fermion pairs \cite{GDR1}. The case of  
$\chi^{0}_i$--$\chi^{0}_j$--slepton
boxes is particularly favored for several reasons: the usual
gaugino--fermion--sfermion couplings have full 
electroweak strength, and the box contribution is not mixed with
other significant virtual corrections [self--energy, vertex corrections]
involving neutralino pairs since the gaugino components are decoupled 
from gauge bosons and their higgsino components are decoupled from light 
fermions,  
so it should be easier to single out this contribution. In order to 
disentangle such contributions from other effects affecting the $Z$ boson tail, 
it is convenient to use the so--called ``$Z$--peak subtraction method" 
\cite{Zsub} which consists in using as inputs the measured $Z$--parameters 
[mass, partial widths and asymmetries] and to describe the additional 
contributions to the observables beyond the $Z$--peak in terms of four 
functions constructed with a subtraction at $s=M^2_Z$. In this way, one
eliminates all known or unknown effects contributing at the $Z$--peak and 
to the $Z$ boson tail, which do not have a strong $s$--dependence. \s

A first numerical exploration, assuming that neutralinos as well as the left-- 
and right--handed sleptons are degenerate, has been made \cite{GDR1}. 
This choice allows to greatly simplify the computation as in this case one 
can use sum rules for reducing the lengthy expressions in the box amplitude. 
For simplicity neutralinos were also taken as pure wino and bino
states. Results are shown in Fig.~18 for the $e^+e^- \to \mu^+\mu^-$ 
cross section and forward--backward asymmetry, for a common neutralino 
mass $m_{\chi^0}=100$ GeV and a slepton mass $m_{\tilde{e}}=60$ GeV.
As expected, a peak associated to the threshold effect appears at 
$s=4m_{\chi^0}^2$, which reaches the  level of 1.5\% in the cross section
and 0.7\% in the FB asymmetry. The half--width is of about 15 GeV. \s

The level reached by these contributions is very encouraging since such a 
precision for the cross section is certainly possible at LEP2 \cite{LEPII}; 
in the case of the FB asymmetry the observability is more
difficult.  Increasing the $\tilde{e}$ mass to 110 GeV, the peak cross 
section however decreases roughly by a factor of two. In the case
of chargino boxes, assuming again $m_{\chi^+}=100$ GeV and $m_{\tilde{\nu}}=
60$ GeV, one obtains a similar effect on the cross section and on the
forward--backward asymmetry [in this case, the sign is opposite]. 
However, in contrast to the neutralino case, there now exist other important 
chargino effects in gauge boson self--energies and vertex corrections; 
cancellations among these various contributions occur and depend on the 
masses and couplings of the SUSY particles. \s

These results are promising since from a general point of view they suggest 
that virtual SUSY particle effects may be observable at LEP2. 
A complete study for both neutralino and chargino contributions at LEP2, 
including all possible effects within the MSSM and varying the parameters 
in their allowed domain, is under way.  

\begin{figure}[htb]
\vspace*{-1.5cm}
\hspace*{3.5cm}\mbox{\psfig{figure=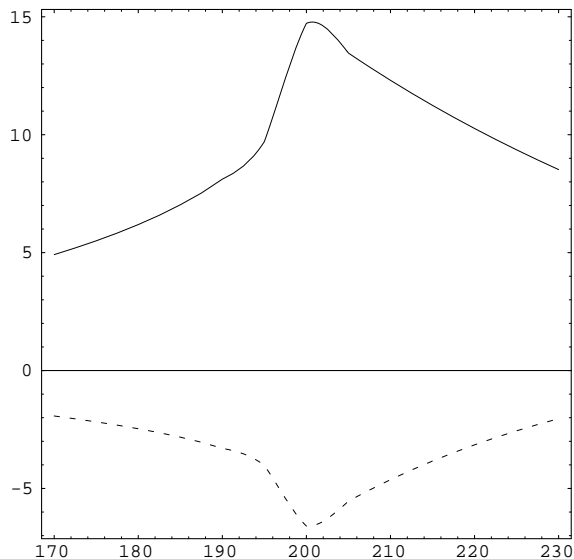,width=7.5cm}}
\vspace*{-1.5cm}
\caption[]{Neutralino box effects on muon pair production at 
LEP2 energies. Relative effect in permille on the cross 
section (solid) and on the forward-backward asymmetry (dashed)
as a function of the c.m. energy.}
\vspace*{-.4cm}
\end{figure}

\subsection*{5.2 Correlated production and decays in $\ee \ra \chi_1^+ 
\chi_1^-$} 

Once charginos will have been discovered, the experimental analysis of 
their properties, production and decay mechanisms, should reveal the basic 
structure of the underlying low--energy supersymmetric theory. This needs
precise experimental measurement, which are more likely to be performed 
in the clean environment of $\ee$ colliders. In the case of the chargino
sector, three parameters are needed to describe it completely: 
$\tan \beta, \mu$ and $M_2$. One needs therefore several experimental 
observables to pin down these parameters. \s

In $e^+e^-$ collisions, charginos are produced 
either in diagonal or in mixed pairs \cite{cpro}. Since the second 
chargino is generally expected to be significantly heavier than the first 
state, at LEP2 and potentially even in the first phase of  
$e^+e^-$ linear colliders, the lightest chargino $\tilde{\chi}_1^\pm$ may 
be, for some time, the only state that can be studied in  detail.
It is therefore wiser to focus on the diagonal pair production of the lightest 
chargino $e^+e^- \rightarrow \chi^+_1 \chi^-_1$. 
The production process is generated by $s$--channel $\gamma$ and $Z$
exchanges, and $t$--channel $\tilde{\nu}$ exchange. It will 
depend on the two angles $\phi_L$ and $\phi_R$ [more precisely on 
$\cos 2\phi_L$ and $\cos 2\phi_R$] which define the 
chargino--chargino--$Z$  and the electron--sneutrino--chargino 
vertices, the chargino mass $m_{\chi_1^+}$ and the sneutrino mass 
$m_{\tilde{\nu}}$. Thus at least four measurements are needed to determine
these parameters. \s

However, the situation is complicated by the decays of the charginos.
Indeed, the charginos will decay into the lightest neutralino and light
fermion pairs \cite{cpro} and since the two neutralinos are stable and escape 
undetected, it is not possible to make a complete reconstruction of the 
events; in particular, one cannot measure the $\chi^\pm_1$ production angle. 
Furthermore, the chargino decays occur through $W$ boson and sfermion 
[and also a marginal contribution from charged Higgs boson] exchanges, and 
the knowledge of the sfermion masses and couplings is in principle 
also required. \s

Recently it has been shown \cite{choi} that even in this situation the 
underlying SUSY parameters can be extracted from the mass
$m_{\chi^\pm_1}$, the total production cross section, and the 
measurement of the polarization with which the charginos are produced.
The $\chi^\pm$ polarization vectors and $\chi$--$\chi$
spin--spin correlation tensors can be determined from the decay distributions 
of the charginos. Beam polarization is helpful but not necessarily required. 
No detailed information on the decay dynamics, nor on the structure of the 
neutralino, is needed to carry out the spin analysis. 
There are already several analyses dealing with polarized chargino production 
[see Ref.~\cite{ccor} or the contribution of Katsanevas et al. \cite{kats} 
for instance]; Ref.~\cite{choi} however attempts to explore analytically the 
event characteristics as will be summarized below. \s

Since the $\chi^\pm_1$ lifetime is very small, only the correlated
production and decay can be observed experimentally
\vspace*{-0.7cm}
\begin{center}
\begin{picture}(300,100)(0,0)
\Text(130,80)[]{$e^+e^-\rightarrow {\chi}^-_1 (p,n) {\chi}^+_1
(\bar{p}, \bar{n})$}
\Line(145,70)(145,40)
\Line(155,70)(155,60)
\Text(155,61)[l]{$\rightarrow\tilde{\chi}^0_1+(f_1\bar{f}_2)$}
\Text(145,41)[l]{$\longrightarrow\hskip 0.15cm\tilde{\chi}^0_1+(\bar{f}_3f_4)$}
\end{picture}
\end{center}
\vskip -1.4cm
where $n$ and $\bar{n}$ are the spin 4--vectors and $p, \bar{p}$ the momenta 
of the charginos. In covariant language the final state distributions are 
found by combining the polarized cross section
\begin{eqnarray}
{\rm d}\sigma=\langle{\rm d}\sigma\rangle\frac{1}{4}
       \left[1-{\cal P}^\mu n_\mu-\bar{\cal P}^\mu \bar{n}_\mu
              +{\cal Q}^{\mu\nu} n_\mu\bar{n}_\nu\right]
\end{eqnarray}
with the polarized distributions for the decays into a neutralinos and
light fermion pairs
\begin{eqnarray}
{\rm d}\Gamma=\langle{\rm d}\Gamma\rangle
                \left[1-{\cal P}^{'\mu}n_\mu\right] \ \ && \ \ 
{\rm d}\bar{\Gamma}=\langle{\rm d}\bar{\Gamma}\rangle
                \left[1-\bar{\cal P}^{'\mu}\bar{n}_\mu\right] 
\end{eqnarray}
Inserting the completeness relations $\sum n_\mu n_\nu=-g_{\mu\nu}+ p_\mu 
p_\nu / m^2_{\tilde{\chi}^-_1}= \eta_{\mu \nu}$ etc, the overall event 
topology can then be calculated from the formula
\begin{eqnarray}
{\rm d}\sigma_{\rm final}&=&\langle{\rm d}\sigma\rangle 
   \langle{\rm d}\Gamma\rangle\langle{\rm d}\bar{\Gamma}\rangle\frac{1}{4}
   \bigg[1+\eta_{\mu\alpha}{\cal P}^\mu{\cal P}^{'\alpha}
      +\bar{\eta}_{\nu\beta}\bar{\cal P}^\nu\bar{\cal P}^{'\beta}
      +\eta_{\mu\alpha}\bar{\eta}_{\nu\beta}
          {\cal Q}^{\mu\nu}{\cal P}^{'\alpha}\bar{\cal P}^{'\beta}\bigg]
\end{eqnarray}
In these expressions ${\cal P}({\cal P}')$ and $\overline{\cal P}
(\overline{\cal P}')$ are the polarization vectors of the produced 
(decaying) $\chi^-, \chi^+$ states, while ${\cal Q}_{\mu \nu}$ is the 
chargino spin--spin correlation matrix. 

In the case of CP--invariance [the CP non--invariant case has been discussed
in Ref.~\cite{nonCP}], 
the overall topology is determined by seven independent functions: 
the unpolarized cross section  $\sigma_{\rm unpol}$, the two
components of the polarization
vector of one of the charginos, and four correlation functions. 
Except for  $\sigma_{\rm unpol}$, these
observables will depend on the final state, and hence on the decays
of the charginos into neutralinos and fermion pairs. However, it was 
shown \cite{choi} that it is possible to construct two observables, denoted 
by ${\cal P}^2/{\cal Q}$ and ${\cal P}^2/{\cal Y}$, which do not depend on 
the chargino decays and hence,  reflect unambiguously the properties of the 
chargino system, being not affected by the neutralinos. 
The energy dependence of the ratios ${\cal P}^2/{\cal Q}$ and
${\cal P}^2/{\cal Y}$ is shown in Fig.~19 for $\tan \beta=2$ and for the
$(M_2,\mu)$ parameters chosen as: gaugino region ($81,-215)$ GeV, 
higgsino region $(215, -810)$ GeV and mixed region $(92, -93)$ GeV,
which lead to a chargino mass $m_{\chi^\pm_1} \sim 95$ GeV.
The two ratios are sensitive to the couplings at sufficiently large 
c.m. energies. 

\begin{center}
\vspace*{-.4cm}
\begin{figure}[htb]
\hbox to\textwidth{\hss\epsfig{file=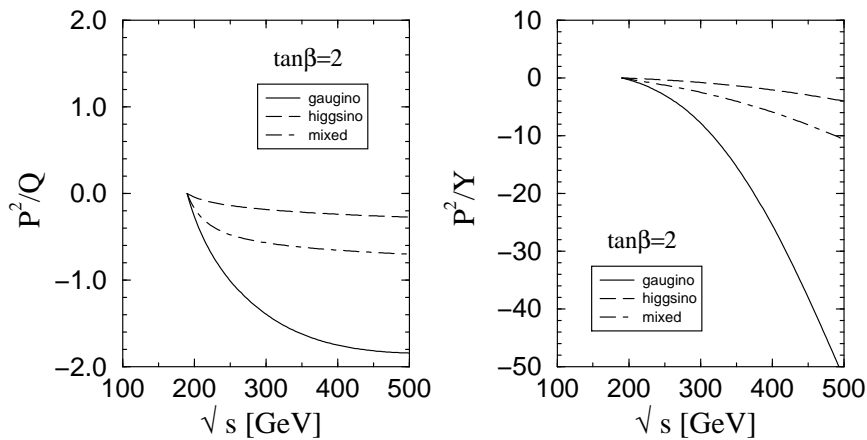,height=6cm}\hss}
\caption[]{The  energy dependence of the ratios ${\cal P}^2/{\cal Q}$ 
             and ${\cal P}^2/{\cal Y}$ for $\tan \beta=2$.}
\vspace*{-1cm}
\end{figure}
\end{center} 

Therefore, in addition to the chargino mass $m_{{\chi}^\pm_1}$ which
can be measured very precisely near the threshold where the production cross 
section $\sigma(e^+e^-\rightarrow {\chi}^+_1 {\chi}^-_1)$ rises 
sharply with the velocity, and the $\tilde{\nu}$ mass which can be 
determined with the energy variation of the cross section [in case
where the determination of $m_{\tilde{\nu}}$ is poor, longitudinal
beam polarization might be helpful to eliminate the $t$--channel
$\tilde{\nu}$ exchange], we have two observables which allow to
determine the two mixing angles $\cos2\phi_L$ and $\cos2\phi_R$. A 
representative example of determining these two parameters is 
shown in Fig.~20 at a c.m. energy $\sqrt{s}=500$ GeV for the ``measured" 
values: $m_{{\chi}^\pm_1}=95 \ {\rm GeV}, \sigma(e^+e^-\rightarrow
{\chi}^+_1 {\chi}^-_1)=0.38\ {\rm pb}, {\cal P}^2/{\cal Q}
=-0.25, {\cal P}^2/{\cal Y}= -5.0$. 
The three contour lines meet at a single point $(\cos 2\phi_L,\cos 2\phi_R)
=(-0.60,-0.49)$ for $m_{\tilde{\nu}}=250$ GeV. \s

\begin{center}
\vspace*{-1.cm}
\begin{figure}[htb]
\hbox to\textwidth{\hss\epsfig{file=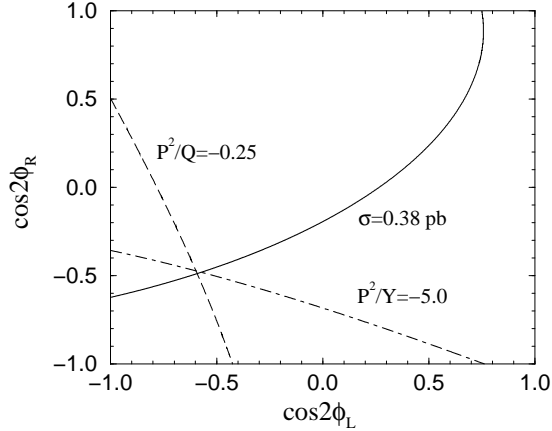,height=7cm}\hss}
\caption[]{Contours for the ``measured values" of the total cross section, 
${\cal P}^2/{\cal Q}$ and ${\cal P}^2/{\cal Y}$ for $m_{{\chi}^\pm_1}=95$ 
             GeV [$m_{\tilde{\nu}}=250$ GeV].}
\vspace*{-0.5cm}
\end{figure}
\end{center}

Once $\cos2\phi_L,\cos2\phi_R$ and $m_{{\chi}^\pm_1}$ are known, the 
SUSY parameters $\tan \beta, M_2, \mu$ can be then determined 
with at most a two--fold discrete ambiguity. Introducing the two 
triangular quantities $ p=\cot(\phi_R-\phi_L)$ and $q=\cot(\phi_R+\phi_L)$
and solving the set 
\beq
p^2+q^2= 2\frac{s_L^2+s_R^2}{(c_L-c_R)^2} \ , \ 
p^2-q^2= \frac{4 s_Ls_R}{(c_L-c_R)^2} \ , \ 
pq = \frac{c_L+c_R}{c_L-c_R} 
\eeq
with $c_{L,R}=\cos 2\phi_{L,R}$ and $s_{L,R}=\sin 2\phi_{L,R}$, the solutions 
$(p,q)$ in point (1) and point (2) of Fig.~21 are found for $\sin 2\phi_L \sin 
2\phi_R >0$ or $<0$, and a second set is found by reversing 
the signs of the solutions pairwise. From this, one obtains for $\tan \beta$:
\begin{eqnarray}
\cos 2\phi_R > \cos 2\phi_L\ \ : \ \ 
\tan\beta = \frac{p^2-q^2\pm 2\sqrt{\chi^2(p^2+q^2+2-\chi^2)}}{
       (\sqrt{1+p^2}-\sqrt{1+q^2})^2-2\chi^2} 
      \; \Rightarrow \; \tan\beta\geq 1 
\end{eqnarray}
where $\chi^2 =m^2_{{\chi}^\pm_1}/M^2_W$. 
If the denominator is positive, there are either up to two solutions
for $\tan\beta >+1$ in point (1) and none in point (2), or
at most one in point (1) and at most one in point (2). The possible
solutions can be counted in an analogous way if the denominator is
negative; the r\^{o}les of point (1) and point (2) are just 
interchanged. The same counting is also valid in the second case
$\cos 2\phi_R < \cos 2\phi_L: \tan \beta \ra 1/\tan \beta$.  
Thus, only a two--fold ambiguity is inferred from all solutions in 
point (1) and point (2).\s

The gaugino and higgsino mass parameters are given in terms of $p,q$ 
by: 
\begin{eqnarray}
\left( \begin{array}{c} M_2 \\ \mu \end{array} \right) =
\frac{M_W}{\sqrt{2}}\bigg[(p \pm q)\sin\beta-(p \mp q)\cos\beta\bigg]
\end{eqnarray}
The parameters $M_2$, $\mu$ are uniquely fixed if $\tan\beta$ is chosen 
properly in point (1) and/or point (2). Since $\tan\beta$ is invariant
under pairwise reflection of the signs in $(p,q)$, the definition 
$M_2 > 0$ can be exploited to remove this additional ambiguity. 
Returning to the ``experimental values" of mass, cross section and spin
correlations introduced above, the following  parameters are  
extracted for point (2) $[\tan\beta; M_2,\mu] =[1.0; 53~{\rm GeV}, 
 -52~{\rm GeV}]$ or $[1.4; 240~{\rm GeV}, 137~{\rm GeV}]$. 
Point (1) gives negative values for $\tan\beta$ so that the solution
derived from the ``experimental values" is unique in this case. \s

Thus,  the fundamental SUSY parameters $\tan\beta;M_2,\mu$
can be derived from the observables $m_{ {\chi}^\pm_1}$
and $\cos 2\phi_R$, $\cos 2\phi_L$ up to at most a two--fold ambiguity,
by considering the production of the lightest chargino pair and 
using the partial information on the chargino polarizations. 
Moreover, from the energy 
distribution of the final particles in the decay of the chargino, the mass 
of the lightest neutralino can also be measured; this allows to determine the 
parameter $M_1$ so that also the neutralino mass matrix can be reconstructed
in a model--independent way. 

\begin{center}
\begin{figure}[htb]
\hbox to\textwidth{\hss\epsfig{file=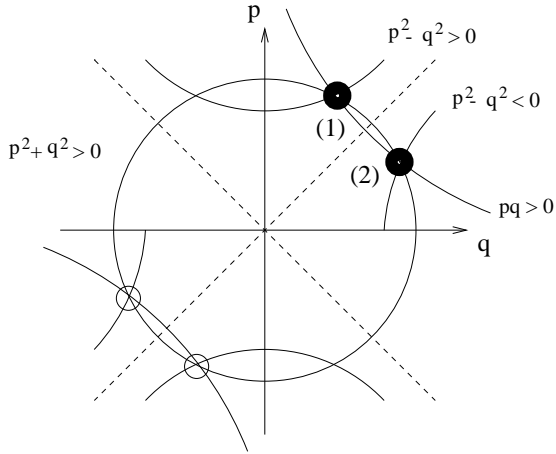,height=6cm}\hss}
\caption[]{Determination of $(p,q)$ from $p^2+q^2$, $pq$ and $p^2-q^2$ 
for $pq>0$.}
\end{figure}
\end{center}

\subsection*{5.3 Chargino/neutralino production at hadron colliders} 

\subsubsection*{5.3.1 Theoretical cross sections} 

Because they are strongly interacting, squarks and gluinos are the most
copiously produced SUSY particles at hadron colliders, and the processes
$pp \ra \tilde{g} \tilde{g}, \tilde{q} \tilde{q}, \tilde{q}\tilde{g}$
have been extensively discussed in the literature; for a review 
see Ref.~\cite{HaberKane}. However, the charginos and neutralinos might 
be the lightest SUSY particles, and would be the first
to be kinematically accessible. In addition, even if the existence of 
supersymmetry would be established by the detection of squarks and gluinos
at the LHC, 
the direct production of charginos and neutralinos would allow various tests 
of the model, and  is therefore very interesting to investigate. \s

Except for squark and gluino production, the next potentially largest cross 
section for SUSY particles at the LHC is the one for the production of the 
lightest chargino and the next--to--lightest neutralino:
\beq
pp \ \ra \ qq \ \ra \ \chi_2^0 \ \chi_1^\pm
\eeq
Although it is an electroweak process, the production rates are large enough 
to allow for a copious number of events at the LHC, even at a low luminosity 
${\cal L} =10$ fb$^{-1}$. The process is of the Drell--Yan type, and proceeds 
via $s$--channel $W$ boson exchange and $t$--channel squark exchange. 
In this section we revisit the tree--level calculation for the
partonic cross section of this process, and this for two  
reasons: $(i)$ this is a good warming--up exercise using the 
semi--automatic program FeynMSSM that we are developing for SUSY 
computations, which we will describe briefly later on, and which we plan
to use, in the future, to calculate the one--loop QCD and 
electroweak radiative corrections to this process; $(ii)$ 
compare our independent calculation with two existing and 
disagreeing results in the literature \cite{Baer,Dawson} to settle the issue. 
By doing so we also took the opportunity to compare our notations 
[mainly based on Ref.~\cite{Haber} and subsequent papers] with the 
notations of the previous references [which, of course, are also distinct 
from each other...]. For more details the reader is referred to 
Ref.~\cite{clermont}. \s 

The differential cross--section for the subprocess has been given in both
Refs.~\cite{Baer} and \cite{Dawson}. However the only place where the 
result of Ref.~\cite{Baer} can be found in an analytical form is in the
program ISAJET. When this expression is extracted from the code and 
compared with the one given in Ref.~\cite{Dawson}, one finds a discrepancy. 
The formula from ISAJET reads
\begin{eqnarray}
d\sigma /dt&=&1/(16\pi s^2)[{\cal A}_s (U+T)/{\cal D}_s+{\cal A}_s''
(U-T)/{\cal D}_s \nonumber \\
& &+{\cal A}_s' \ (m_{\chi_1^+}  m_{\chi_2^0})s/{\cal D}_s
+{\cal A}_u \ U/
{\cal D}_u^2+{\cal A}_t T/{{\cal D}_t'}^2  \nonumber \\
& &+{\cal A}_{st} \ (s-M_W^2)T/({\cal D}_s{\cal D}_t') \nonumber \\
& &+{\cal A}_{st}' \ m_{\chi_1^{+}} \ m_{\chi_2^0} s(s-M^2_W)/
({\cal D}_s{\cal D}_t')+{\cal A}_{su} \ U(s-M^2_W)/({\cal D}_s{\cal D}_u) 
\nonumber \\
& &+{\cal A}_{su}' \  m_{\chi_1^{+}} \ m_{\chi_2^0} s (s-M^2_W)/
({\cal D}_s{\cal D}_u)+{\cal A}_{tu} \ s \ m_{\chi_1^{+}} \ 
m_{\chi_2^0}/({\cal D}_u {\cal D}_t')] 
\label{dsigdt}
\end{eqnarray}
where 
\begin{eqnarray}
U=(m^2_{\chi_1^{+}}-u)(m^2_{\chi_2^0}-u)\ ,\ 
T=(m^2_{\chi_1^{+}}-t)(m^2_{\chi_2^0}-t) 
\end{eqnarray}
$m_{\chi_1^{+}}$ and $m_{\chi_2^0}$ are respectively the chargino and neutralino
masses and $u,\ t$ and $s$ the Mandelstam variables.  ${\cal D}_s$ and  
${\cal D}_u$ are given by
\begin{eqnarray}
{\cal D}_s&=&(s-M^2_W)^2+M^2_W\Gamma^2_W\ ,\ {\cal D}_u=(u-m^2_{{\tilde{u}_L}})\ ,\\
{\cal D}_u'&=&(u-m^2_{\tilde{d}_L})\ ,\ {\cal D}_t'=(t-m^2_{\tilde{d}_L}) .
\end{eqnarray}
where $m_{\tilde{d}_L} , m_{{\tilde{u}_L}}$ are the left--handed
down and up squark masses respectively.  The factors ${\cal A}^{(','')}$ 
encapsulate the information on the chargino--neutralino couplings to $W$ 
bosons and sfermions. Eq.~(\ref{dsigdt}) agrees with the corresponding one in 
Ref.~\cite{Dawson} apart from the term ${\cal A}_s''$ which is missing
in the latter. We repeated the calculation in a completely independent way
and found full agreement with eq.~(\ref{dsigdt}) thus re--ensuring the 
validity of the expression used by ISAJET. \s

This calculation has been performed with the program FeynMSSM, which is 
based on the packages FeynArts [which generates the full set of diagrams 
automatically] and FeynCalc [which calculates the amplitudes, and reduces
them to the standard forms] developed \cite{Mertig} for a semi--automatic 
calculation of amplitudes in the Standard Model. FeynMSSM, incorporates 
the MSSM Feynman rules [following the notations of Ref.~\cite{Haber}] and 
thus generates the amplitudes for any SUSY process in principle. The program 
is running under Mathematica2.0, and has 
been checked for processes with up to three particles in the final state, 
as well as loop diagrams [up to two loops] which is in fact the main 
purpose of the program. When FeynMSSM will be fully operational, the 
hope is that it would be very efficient and useful when dealing with 
higher--order processes, either multi--loop or multi--final state, in 
the MSSM. \s

\subsubsection*{5.3.2 Searches at the LHC}

Let us now turn to the discussion of this process at the LHC. 
To suppress the huge QCD backgrounds, one needs to look at signatures
with isolated leptons in the final state, and the many leptons there are 
the smaller will be the backgrounds. The one--lepton channel suffers from 
the large number of singly produced $W$ bosons, while the two--lepton 
channel suffers from single $Z$ boson and $WW$ pair production. One should 
then look at the signal with three leptons plus missing energy. The
leptons of the signals will come from the decays of the chargino into a 
$W$ boson and the LSP, and the decay of the next--to--lightest neutralino into
a $Z$ boson and the LSP. The branching ratios are sizeable in a large area
of the MSSM parameter space, despite of the small branching ratios of the 
gauge boson [especially the $Z$ boson] decays into charged leptons. The main 
background reactions are due to $WZ$ and $ZZ$ production, as well as the 
semi--leptonic decays of heavy quarks and the $Zb\bar{b}$ background.
There are also backgrounds from the production of other SUSY particles, and 
in particular squark and gluino production [which then decay, among other
final states, into the studied neutralino and chargino via cascades] and 
associated chargino/neutralino production with a squark or a gluino; there 
are also backgrounds from the pair production of charginos and neutralinos. \s

The signal can be distinguished from 
the background events, mainly due to the lower hadronic activity and not from 
the amount of missing energy; applying a set of standard [ATLAS] cuts 
\cite{LHCex}, one can obtain a $5\sigma$ signal at the low luminosity LHC 
for gluino masses smaller than $\sim 500$ GeV; see Ref.~\cite{thesis} for 
details. Furthermore, the mass difference between the produced neutralino 
and the LSP, $m_{\chi_2^0} - m_{\chi_1^0}$, can be measured precisely from
the invariant mass distribution the lepton pair coming from $Z$ decays. \s

In this section, we propose to use the charge asymmetry
\beq
A = (N_+ - N_-)/(N_+ + N_-)
\eeq
where $N_+(N_-)$ is the number of positively (negatively) charged leptons, 
as a new observable in LHC analyses in processes where charged particle final 
states are produced \cite{steve}. What is exploited is the fact that, contrary 
to LEP and the Tevatron, the LHC is an asymmetric machine as far as the electric
charge is concerned. Indeed, at the LHC we expect to produce more $W^+$
than $W^-$ bosons, since the proton contains as valence quarks  two $u$
and one $d$ quarks and the process $u\bar{d} \ra W^+$  is more favored
than $\bar{u}d \ra W^-$. \s

It has been noticed in Ref.~\cite{steve} that the charge asymmetry increases
linearly with increasing mass of the formed final state. It can be therefore
used for indirect measurement of particle masses, based only on the structure
functions. This has been checked with PYTHIA 6.114 for the events $W \ra
e \nu_e$ or $\mu \nu_\mu$ using the structure functions CTEQ4 \cite{CTEQ}, 
and varying artificially the $W$ boson mass. This exercise has been repeated 
for the process $pp \ra WZ \ra 2\; {\rm jets}+e\nu_e$ and $2\; {\rm jets}+ 
\mu \nu_\mu$, which exhibits the same behavior. To check that this effect 
was not due to the particular choice of the structure functions, we used a 
different set of parton densities and the result was the same, except of course
for the different slope of the charge asymmetry as a function of the mass.  \s

These results are illustrated in Fig.~22, where the charge asymmetry in 
percent is plotted against the mass of the final $W$ or $WZ$--system at
the LHC with $\sqrt{s}=14$ TeV, using the CTEQ4M parton densities. The
two dark points at $M\simeq 80$ and $171$ GeV correspond to the asymmetries
of the $W$ and $WZ$ systems, which are respectively 19\% and 21\%.  
The brighter circles and crosses correspond respectively to the 
asymmetries of the $W$ and $WZ$ systems when the effective mass 
is varied [in the latter case, the ratio $M_W/M_Z$ is kept constant]. 
One notices that these asymmetries are quite substantial, $\sim 15$  
to $35$\%,  and can be therefore experimentally measurable. \s

\begin{figure}[htb]
\vspace*{-0.05cm}
\epsfxsize=160mm
\epsfysize=160mm
\vspace*{-1cm} 
\begin{center}
\epsffile{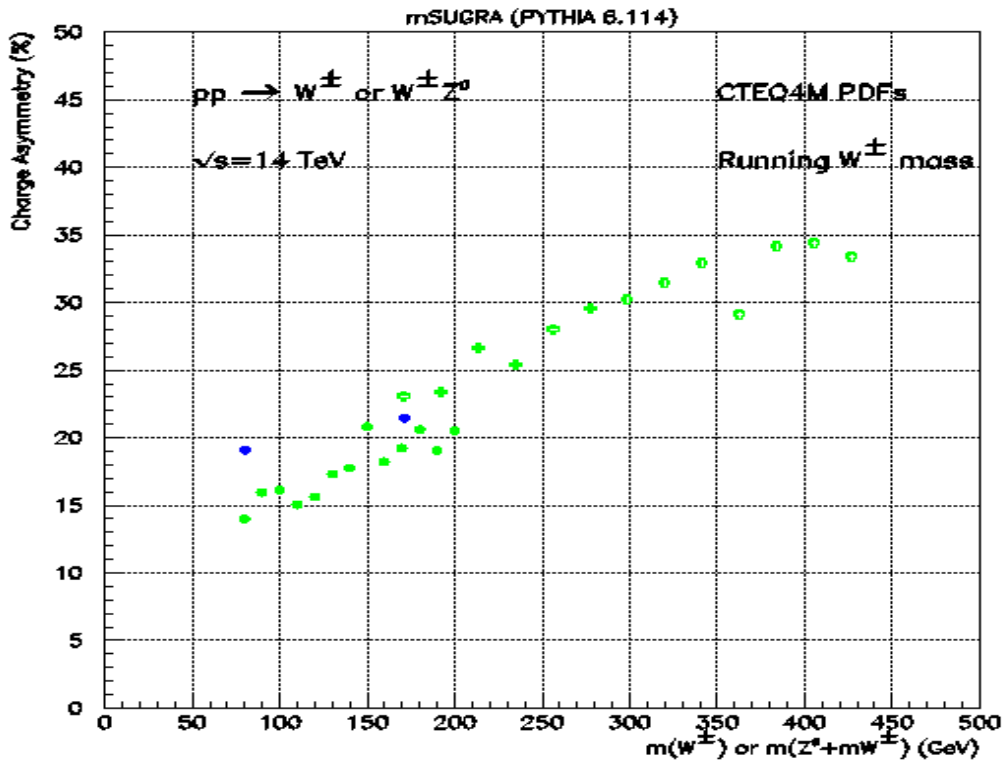}
\vspace*{-3.5cm}
\end{center}
\caption{The charge asymmetry [in \%]  as a function of the mass of the final 
$W$ or $WZ$--system at the LHC with $\sqrt{s}=14$ TeV, using the CTEQ4M parton 
densities.} 
\vspace*{-10.8cm}
\end{figure}

\newpage

In Ref.~\cite{steve}, it has been therefore proposed to apply this method to
the process $pp \ra \chi_2^0 \chi_1^\pm \ra 3l^\pm$+ missing energy. Let
us first recall that the production amplitude is dominated by the $s$--channel
$W$ boson exchange for heavy enough squarks. In this diagram, the 
$u \bar{d}W^+$ vertex depends only on the proton structure function, and
is completely independent of the final SUSY state. However, this vertex
decides of the $W$ boson charge, and thus on the global charge of
the chargino and the trilepton event. One can measure the charge asymmetry 
of the set of events which passes the cuts needed to extract the signal, and
determine the sum of the masses $m_{\chi_2^0}+m_{\chi_1^+}$. In Fig.~23, 
this mass was varied by scanning the MSSM parameter space, and the obtained
asymmetry is shown. As can be seen, the charge asymmetry is always 
proportional to the mass of the final system. \s
 
To confirm the usefulness of this method to measure the $\chi$ masses, 
a detailed Monte-Carlo simulation including the signals and backgrounds
which pass the selection cuts is required. Although not very precise 
[there is a systematic bias from the used structure functions, and there
is still a small dependence on the SUSY parameters due to the small 
interference between the $s$--channel $W$ exchange and the $t$--channel squark
exchange amplitudes], the results are encouraging and the method
deserves further investigations.

\begin{figure}[htb]
%\vspace*{-.05cm}
\epsfxsize=160mm
\epsfysize=160mm
\vspace*{-1.cm}
\begin{center}
\epsffile{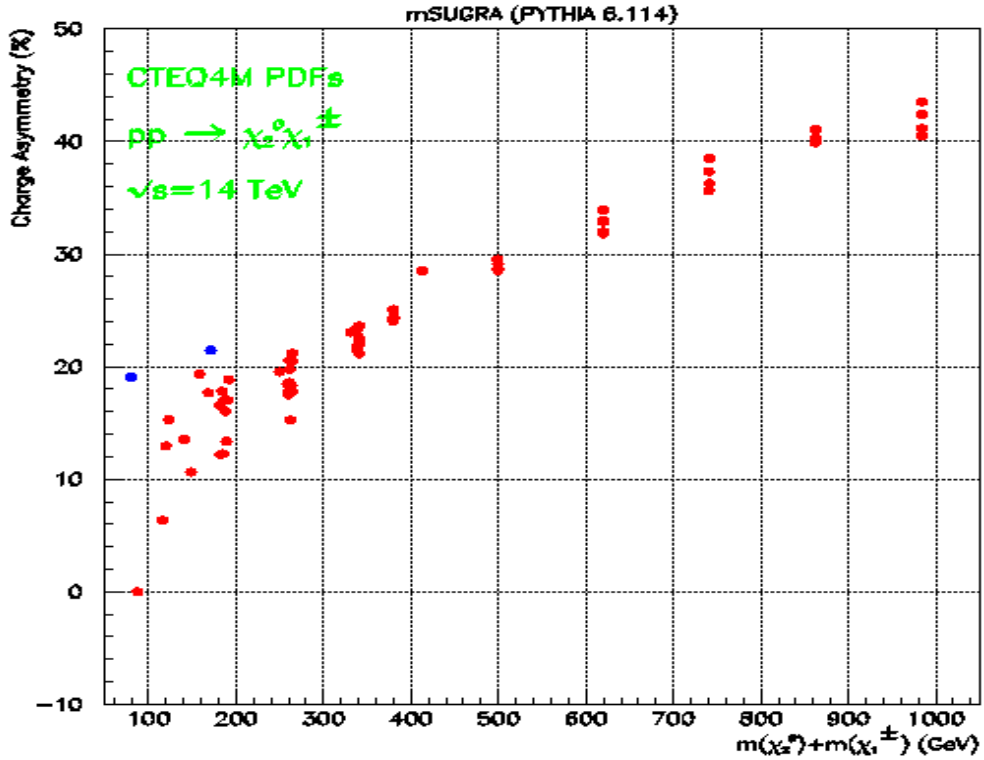}
\vspace*{-3.8cm}
\end{center}
\caption{The charge asymmetry [in \%] a function of $m_{\chi_2^0}+m_{\chi_1^+}$
at the LHC with $\sqrt{s}=14$ TeV, using the CTEQ4M parton densities.}
\vspace*{-10.01cm}
\end{figure}

\newpage

\subsection*{5.4 Two-- and three--body sfermion decays} 

The dominant decay modes of scalar fermions are the two--body decays into
their partner fermions plus neutralinos or charginos:
\beqn 
\tilde{f}_i \ra f \chi_j^0 \ \ ,  \ \ \tilde{f}_i \ra f' \chi_j^\pm
\eeqn
In general at least the decays into the lightest neutralino [which is the
lightest SUSY particle] and chargino is kinematically available. In the
case of squarks, if their masses are larger than the sum of the partner 
quark and the gluino masses, the decay 
\beqn 
\tilde{q} \ra q \tilde{g} \ \  {\rm if}  \ \ m_{\tilde{q}} > m_q
+m_{\tilde{g}} 
\eeqn
largely dominates since it is a strongly interacting process. These decays
are well--known; see for instance Ref.~\cite{HaberKane} for a discussion. \s

In the case of the third generation sfermions, and in particular for stops,
the mixing can be so important that a large splitting between superpartners
is generated; see section 2.4. This opens the possibility of decays of sfermions into 
lighter sfermions and gauge bosons or Higgs bosons: 
\beqn 
\tilde{f}_i &\ra & \tilde{f}'_j + W^\pm  \ , \ \tilde{f}'_j +H^\pm \non \\
           &\ra & \tilde{f}_j + Z \ , \ \tilde{f}_j +h,A,H 
\eeqn
These decays, including the mixing between sfermions have been discussed 
recently; see Ref.~\cite{mixing} for instance. \s  

Finally, the mass of the lightest stop could be smaller than the sum of the 
top and LSP masses, and smaller than the lightest chargino mass. In this case 
the only allowed decay mode will be into a charm quark and the lightest 
neutralino through loops \cite{Hikasa} 
\beqn
\tilde{t}_1 \ra c \chi_1^0 \label{cchi}
\eeqn

\smallskip

There are many situations, however, where these two--body decays of sfermions
are suppressed or kinematically inaccessible. This happens 
for instance for the lightest stop and sbottom squarks in the case of large 
sfermion mixing, and for the scalar partners of the light quarks in models 
where the gaugino mass unification constraint is relaxed, and in gauge 
mediated SUSY breaking models. In this case, various three--body modes might 
become relevant. A list of three--body decays of sfermions which might be 
important includes: \\

$(i)$ Decays of the stops into a $b$--quark, a neutralino and a $W$ or 
$H^+$ boson:
\beqn 
\tilde{t}_{1,2}  \ra   
t^*, \tilde{b}^*, \chi^{+*} \ra  b \; \chi_1^0 \; W^+  \ \ {\rm or} \ \
b \; \chi_1^0 \; H^+  \label{bcw}
\eeqn
The decay $\tilde{t} \ra b \chi_1^0 W^+$ is especially important in the case 
of the lightest stop, when it is lighter than $m_b+m_{\chi^+_1}$ and  
$m_t+m_{\chi_1^0}$. In this case, the lightest stop $\tilde{t}_1$ will decay 
into a charm quark plus the lightest neutralino, eq.~(\ref{cchi}). 
Because it is loop mediated, this decay has a partial decay width that 
is suppressed by four powers of the electroweak coupling and by the CKM angle 
$V_{cb}$, compared  to the usual two--body decay widths. In this case, the 
three--body decay into $b\chi^0W$ with a virtual exchange of a top quark, a 
sbottom and a chargino might become competitive. In fact, the decay is also 
suppressed by two powers of the electroweak coupling as well as by the 
virtuality of the exchanged particles; however, there are some kinematical 
regions where this mode dominates over the $\tilde{t}_1 \ra c\chi_1^0$ decay 
channel. This decay mode has been also discussed in Ref.~\cite{porod}, where 
however only the matrix elements in terms of the four momenta of the particles 
involved have been given.  \s

The decay $\tilde{t} \ra b \chi_1^0 H^+$ is similar to the previous one
with the $W$ boson replaced by the charged Higgs boson $H^+$. In the MSSM
the charged Higgs boson is always heavier than the $W$ and this in principle
disfavors this process kinematically compared to decay into $Wb \chi_1^0$.
However, the couplings of the $H^+$ boson to the third generation quarks and 
squarks can be strongly enhanced, leading to a possible compensation. For 
relatively small masses of the $H^\pm$ boson, the decay branching ratio of 
this channel can be sizeable. \\

$(ii)$ Decays of the heavier stop into the lighter stop or sbottom and 
a fermion pair: 
\beqn
\tilde{t}_2 \ra Z^*, h^*, A^*  \ra \tilde{t}_1 f\bar{f} \ , \ 
\tilde{t}_2 \ra W^*, H^{+*} \ra \tilde{b}  f\bar{f}' 
\eeqn 
This decay occurs when the states $\tilde{t}_1$ or $\tilde{b}$ are rather 
light and there is a splitting between the two stops and/or between 
$\tilde{t}_2$ and the sbottoms, while the mass differences $m_{\tilde{t}_2}
-m_{\tilde{t}_1}$ or $m_{\tilde{t}_2}- m_{\tilde{b}}$ are still smaller than 
$M_Z, M_h, M_A$ or $M_W, M_{H^\pm}$, respectively. When $\tilde{t}_2$ is 
lighter than $m_b 
+ m_{\chi^\pm_1}$ and $m_t+m_{\chi^0}$, the two--body decays [except for the 
loop decay  $\tilde{t} \ra c \chi_1^0$] are forbidden, and these three--body 
modes become relevant. For high values of $\tan \beta$ the corresponding decays 
of the heavier sbottoms [i.e. the stops replaced by the sbottoms in the decays 
above] might also be of some relevance. \\

$(iii$) Decays of squarks into quarks and a stop or a sbottom
via gluino exchange: 
\beqn
\label{bst} 
\tilde{q} \ra  \tilde{g}^*  \ra  q \; t \; \tilde{t} \ \ {\rm or} \ \ 
q \; b \; \tilde{b}
\eeqn
The decay $\tilde{q} \ra q b \tilde{b}$ of a scalar partner of a light 
quark occurs when gluinos are heavier than the squark [but not too much because 
the large virtuality of the exchanged gluino will strongly suppress the 
decay] and when there is a strong mixing in the sbottom sector which makes 
that the lightest sbottom is lighter than all other squarks. One also needs 
that the lightest chargino and neutralinos  are rather heavy and/or of almost 
higgsino type to suppress their couplings to the squarks. However, in
models with the unification of the gaugino masses at the GUT scale, 
the gluino and chargino/neutralino masses are related and for not 
too heavy gluinos [not to suppress the decay width by the $\tilde{g}^*$
virtuality] the decays into the other chargino and/or neutralinos are still 
allowed kinematically and would dominate since these particles would be 
then of the gaugino type. However, relaxing the assumption of gaugino
mass universality, one can find a large area of the parameter space
where the mode eq.~(\ref{bst}) can be sizeable. \s

For small values of $\tan 
\beta$, it is the mixing in the $\tilde{t}$ sector which can be strong and 
$\tilde{t}_1$ possibly lighter than the other squarks; a similar 
situation as previously 
occurs then for the decay $\tilde{q} \ra q t \tilde{t}$ despite 
of the kinematical complication of having a top quark in the decay product 
which requires larger decaying squark masses than in the previous  case. \\

$(iv$) Decays of squarks into quarks and a slepton--lepton pair
via chargino and/or neutralino exchange: 
\beqn
\label{qsl} 
\tilde{q} \ra \chi^{0*} \ , \chi^{\pm *}  \ra q \tilde{l} l
\eeqn
This decay occurs when sleptons are lighter than squarks and the
squark decays into neutralinos and charginos are kinematically shut
or suppressed by the small squark--quark--higgsino couplings. This happens
for instance in gauge mediated supersymmetry breaking models where, first the 
gravitino which couples only weakly to squarks is the LSP and in addition 
there is a hierarchy which makes the squarks heavier than sleptons. 
The next--to--lightest particles could then be the sleptons [and in
particular the stau's because of the mixing for large values of $\tan 
\beta$]. If squarks have masses smaller than the chargino or neutralino
masses so that the two--body decays $\tilde{q} \ra \chi^0 q , \chi^\pm q'$
are kinematically closed, the modes eqs.~(\ref{qsl}) become the dominant
ones. \s

However, even in the ``phenomenological MSSM" discussed before, 
these decays are possibly important if the lightest neutralinos and 
charginos are pure higgsinos [thus suppressing the squark--quark--ino 
couplings and the two--body decays into quarks+inos] and the other charginos 
and neutralinos are not much heavier than the sleptons [not to be hurt by
the virtually of these states]; the relaxation of the gaugino mass
unification assumption will help in enlarging the MSSM parameter range where 
this situation occurs. The three--body decay modes of first/second generation
sleptons into third generation sleptons+leptons have also been discussed in 
Ref.~\cite{kribs} in the context of gauge mediated breaking models.  \\

All these decays are discussed in detail in Ref.~\cite{yann}. The Dalitz 
plot densities for the partial widths d$\Gamma$/(d$x_1$d$x_2$), where 
$x_1$ and $x_2$ are the scaled energies of two final particles, are
given including the full dependence on the final fermion/sfermion
masses and on the couplings [sfermion mixing is included for instance]. 
The two remaining integrations are then performed numerically to obtain
the partial decay widths and the branching ratios. In the cases where 
there is only one massive particle in the final state [such as in case
($iv)$, $(iii)$ for sbottoms and in case $(ii)$ for both stops and
sbottoms] the integration can be done analytically and is given. \s

A detailed discussion of the relative magnitude of the decay modes
$(i)$--$(iv)$ is given in Ref.~\cite{yann}. In addition, the fortran
code calculating the partial widths and branching ratios is available. 
It is based on the subroutines SFERMION, GAUGINO and SUSYCP [which calculate 
the masses and couplings of sfermions, charginos/neutralinos and Higgs bosons
respectively] included in the program SUSPECT discussed in section 3.2. 
Therefore, an interface with SUSPECT is easily possible and will be done in 
a near future. 

\newpage

\subsection*{5.5 Stop and sbottom searches at LEP200} 

As discussed previously, squarks of the third generation, namely the stops 
and the sbottoms, have a special place in the SUSY spectra due to the 
particular Yukawa couplings of their partners, the top and bottom quarks;
stops and sbottoms could even have masses accessible at LEP200. In this
section we discuss the searches of these squarks at LEP200 \cite{LEP200}, 
focusing on the scenarii where R--parity is conserved and in both the 
unconstrained MSSM and the mSUGRA models. The notations and conventions 
are those introduced in Section 3.  

\subsubsection*{5.5.1 Squark production at LEP200}

At LEP200, squarks of the same mass are pair--produced through $\gamma/Z$ 
s--channel exchange. 
One notes that the squark--squark--Z boson coupling can vanish for the special 
value of the mixing angle: $\theta_q= A_q \cos (\sqrt{ e_{\tilde q} s_W^2/
I_3^q})$, which corresponds to 0.98 rad for the stop and 1.17 rad for 
the sbottom. At the Born level, the production cross section is given by 
\cite{refs}:
\beq
\sigma^{\rm Born}={{\pi \alpha^2} \over {s}}\beta^3 \left[e_{\tilde q}^2+
\left({{(v_e^2+a_e^2)v_{\tilde q}^2} \over {16 s_W^4 c_W^4}}s^2 -
{{e_{\tilde q}v_ev_{\tilde q}} \over {2 s_W^2 c_W^2}}s(s-M_Z^2)\right)
{1 \over {(s-M_Z^2)^2+\Gamma_Z^2M_Z^2}}\right]
\eeq
with $v_e=2s^2_W-1/2$, $a_e=-1/2$ and $v_{\tilde q}=2(I_3^q \cos^2\theta_
q-Q_{\tilde q}\sin^2\theta_W)$;  $\beta$ is the squark velocity.
This cross section is minimal for:
\beq
\cos^2\theta_{\rm min}={e_f s_W \over I_3^q} \left[1+\left(1-{{M_Z^2} \over {s}}c_W^2 \right){{L_e+R_e} \over {L_e^2+R_e^2}}\right]
\eeq
with $L_e=s_W^2-1/2$ and $R_e=s_W^2$. The mixing angle values are equal, up 
to a level of 5 $\%$, to those corresponding to a vanishing squark--squark--Z 
boson coupling. \s

The QCD corrections are factorizable and one obtains for pure
gluon exchange and emission in the final state \cite{reft,reft0}:
\beq
\sigma^{\rm QCD}=\sigma^{\rm Born} \left[1+{4 \over 3}{{\alpha_s} \over 
{\pi}}f(\beta)\right] \ \ ; \ \ 
f(\beta)\simeq {{\pi^2} \over {2 \beta}}-{{1+\beta} \over {2}}\left({{\pi^2} \over 2}-3\right)
\eeq
[the expression of $f(\beta)$ is given in the Schwinger approximation which
reproduces the complete formula at the 1.5 $\%$ level]. In the high energy 
limit $\beta \rightarrow 1$, $f(\beta)\;\rightarrow\;3$ leading to a correction 
four times higher than the one corresponding to quark production in the same 
limit. \s

The electromagnetic corrections can be taken into account by convoluting
the cross section with \cite{reft}:
\beq
L_{ee}(x)=\left[\beta_{em}(1-x)^{\beta_{em}-1}(1+{3 \over 4}\beta_{em})-{1 \over 2}\beta_{em}(1+x)
\right]\left[1+\alpha_{em}\left({\pi \over 3}-{1 \over {2\pi}}\right)\right]
\eeq
where $\beta_{em}=2 \alpha_{em}/ \pi \left( {\rm Log} (s/m_e^2)-1\right)$
and $\alpha_{em}$ the QED coupling constant defined at zero--momentum transfer.
One obtains then the total cross section:
\beq
\sigma^{\rm tot} (e^+e^- \rightarrow \tilde{q} \tilde{q}^*)= \int_0^1 L_{ee}(x)
\sigma^{\rm QCD}(xs)dx
\eeq

The various cross sections: $\sigma^{\rm Born}$, $\sigma^{\rm QCD}$, $\sigma^{
\rm QED}$ and the total cross section $\sigma^{\rm tot}$ including all 
corrections  are shown in Fig.~24 at a c.m. energy $\sqrt{s}=172$ GeV in the
case of the lightest stop and sbottom squarks with masses of 80 GeV and
as function of the squark mixing angles. One can see that the QCD corrections, 
which are always positive, increase when $m_{\tilde q} \rightarrow \sqrt{s}/2$,
reaching 60$\%$ for a 80 GeV stop mass. They are constant and equal to 15$\%$ 
far from the production threshold. QED corrections are negative near the 
kinematical threshold and positive when otherwise.
Close to threshold, the QED corrections tend to compensate the QCD corrections
and the result is a total cross section equal at the level of 5$\%$ for 
$m_{\tilde q_1}$=80 GeV to that of the Born approximation.

\begin{figure}[htbp]
\vspace*{-1cm}
\begin{center}
\includegraphics[width=8cm]{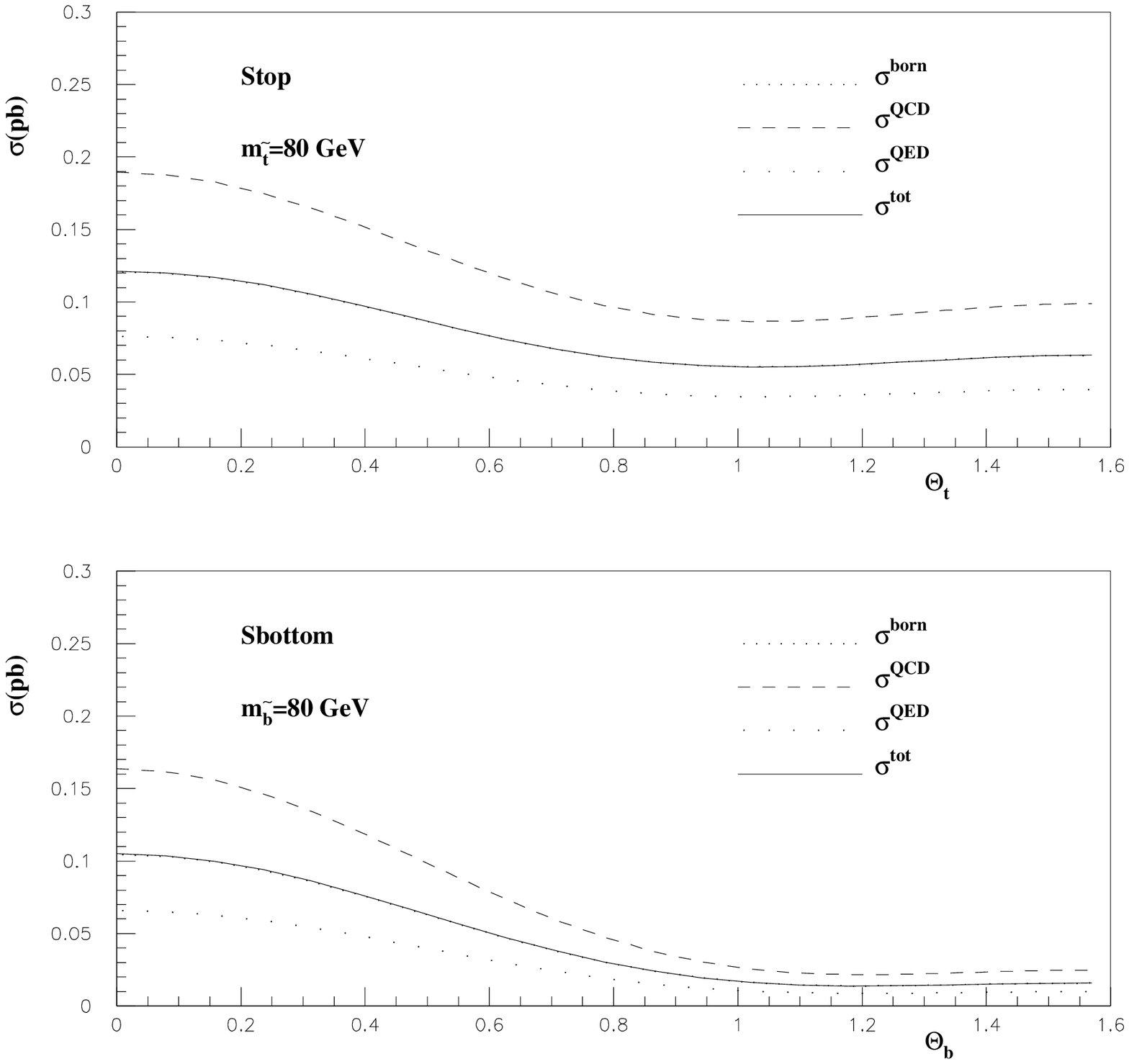}
\vspace*{-1cm}
\end{center}
Figure 24: Stop and sbottom production cross sections at $\sqrt{s}$=172 GeV
as a function of the mixing angles and for a squark mass of 80 GeV.
\end{figure}

\subsubsection*{5.5.2 Squark decays} 

In this section we discuss the decays of the stop and sbottom squarks in
the framework of the uMSSM and mSUGRA. Let us first concentrate on the
case of the lightest stop. According to the discussion in the previous 
section on sfermion decays, there are only a few decay modes relevant for 
stop masses that are kinematically accessible at LEP200, i.e. $m_{\tilde{t}_1}
\lsim 100$ GeV: \s

\nn -- The loop induced flavor changing two body decay mode into 
a charm quark and the lightest neutralino, $\tilde{t}_1 \ra c \chi_{1}^0$,
always occurs since the lightest neutralino is the LSP. The decay width
is given by \cite{Hikasa}: 
\beq
\Gamma(\tilde{t}_1  \rightarrow c \chi_1^0) = {{\alpha} \over {4 s_W^2}}
|\epsilon|^2 m_{\tilde t _1} b_{i1} \left[1-
{{m^2_{\chi^0_i}} \over {m^2_{\tilde t_1}}}\right]^2 
\eeq
where $b_{i1}$ depend on the neutralino parameters $\mu,M_1,M_2,{\rm tan} 
\beta$. The $\epsilon$ parameter takes into account the various possible 
loop contributions and is estimated \cite{Hikasa} to be $\epsilon \sim (1-4)
\times 10^{-4}$. This small value has the consequence that the stop decay 
time is far longer than the strong-interaction time scale, $\tau_{QCD}\sim 
10^{-23}$ s. If this channel dominates, the stop hadronises first before 
decaying. The corresponding decay into charm plus a gluino is ruled out
for stop masses accessible at LEP200, due the experimental bound on the
gluino mass from Tevatron \cite{data}. \s

\nn -- The three--body decays $\tilde{t}_1 \rightarrow b l^+ \tilde \nu$ 
and $\tilde{t}_1 \rightarrow b \tilde l ^+ \nu$ through the exchange
of an off--shell chargino are allowed if $m_{\tilde t_1}>m_{\tilde \nu}+m_b$ 
or $m_{\tilde t_1}> m_{\tilde l}+m_b$. The corresponding widths 
can be found in \cite{Hikasa}. The first of these decays is more favored by
kinematics since the experimental limit \cite{data} on the sneutrino mass, 
$m_{\tilde{\nu}} \gsim 37$ GeV is weaker than the one for charged sleptons, 
$m_{\tilde{l}} \gsim$ 65--85 GeV depending on the flavor. 
The other possible three--body final states are: $\tilde{t}_1 \rightarrow
\;b W^+ \chi_1^0$ and $\tilde{t}_1 \rightarrow\;b H^+\chi_1^0$ through 
exchanges of sbottom, chargino or top quark. Nonetheless, in the MSSM, the 
$H^+$ mass is larger than the $W$ boson mass, and the experimental bound
on the LSP \cite{data} tends to close this channel. \s
 
\nn -- The decay into a bottom and the lightest chargino $\tilde{t}_1 \ra 
b \chi_1^+$ if charginos are lighter than stops. However, since the 
experimental limit on the lightest chargino mass is rather strong
$m_{\chi_1^+} \gsim 90$ GeV \cite{data}, this decay occurs only for stop
masses close to 100 GeV. However, when it occurs, this decay mode largely
dominates since it occurs at the tree--level and is a two--body decay. \s

Thus the most likely decay modes of stops at LEP200 are the $\tilde{t}_1
\ra c\chi_1^0$ and $\tilde{t}_1 \ra b l \tilde{\nu}$. To illustrate the
relative magnitude of these two channels we make the following 
assumptions: $(i$) The lightest neutralino $\chi _1^0$ is the LSP,
($ii$) the lightest stop is lighter than the lightest chargino, and $(iii)$
the lightest stop is heavier than the sleptons. Furthermore, we shall use
the following experimental bounds on the sparticle masses from direct and
indirect searches: a) $m_{\tilde \nu}>$ 37 GeV, b) $m_{\tilde l}>$ 85, 65 
et 67 GeV for selectrons, smuons and staus respectively and c) $m_{\tilde g}>$ 
150 GeV \cite{data}. We vary the parameters $M_2$ and $\mu$ in the range 
[0,1000] GeV and $[-1000,1000]$ GeV, respectively . \s

Fig.~25 shows the accessible region in the plane $(\mu,M_2)$ for which we have
$m_{\chi^+_1}>$ 80 GeV and $m_{\chi_1^0}< m_{\tilde t_1}$= 80 GeV. 
The stop mixing angle has been fixed to $\theta_t$=0 and $\theta_t=\pi/2$. The 
stop partial widths for the two decay modes $\tilde{t}_1 \rightarrow c 
\chi_1^0$ and $\tilde{t}_1 \rightarrow b \tau \tilde \nu$ have been calculated 
for two sneutrino masses: $m_{\tilde \nu}$=42 GeV and 70 GeV. Contours in
the $[M_2, \mu$] plane for the branching ratios of the decay mode $\tilde{t}_1
 \;\rightarrow\;c \chi_1^0$ are shown. One can see that this decay
mode is largely dominating, with a branching ratio larger than 99\%, except 
for small sneutrino masses [which makes the decay mode $\tilde{t}_1 \ra b \tau 
\tilde{\nu}$ more phase--space favored] and small stop mixing angle [which
favors one of the higgsino or wino components of the charginos]. Indeed, for 
$\theta_t=0$, the $\tilde{t}b \chi_1^+$ coupling depends both on the wino and 
higgsino components. Nonetheless, if the chargino is higgsino--like, the 
$\chi_1^+l\tilde{\nu}$ coupling is proportional to the tau Yukawa
coupling which is in that case very small [remember that we are dealing with  
small ${\rm tan} \beta$ values]; this means that for $\theta_t$=0, the decay $\tilde{t}_1 \ra
b l \tilde \nu$ has non negligible values only for a wino--like 
$\chi_1^+$ that is for $|\mu| \gg M_2$ and this is precisely the region
where this decay dominates. On the other hand, for $\theta_t=\pi/2$, 
the coupling stop-bottom-chargino only depends on the higgsino component of
$\chi _1^+$ and the decay $\tilde{t}_1 \rightarrow\;b l \tilde \nu$ is then 
suppressed because the chargino--lepton--sneutrino coupling is in that case 
governed by the lepton Yukawa coupling and is negligible. 

\begin{figure}[htbp]
\vspace*{-1.8cm}
\begin{center}
\includegraphics[width=12cm]{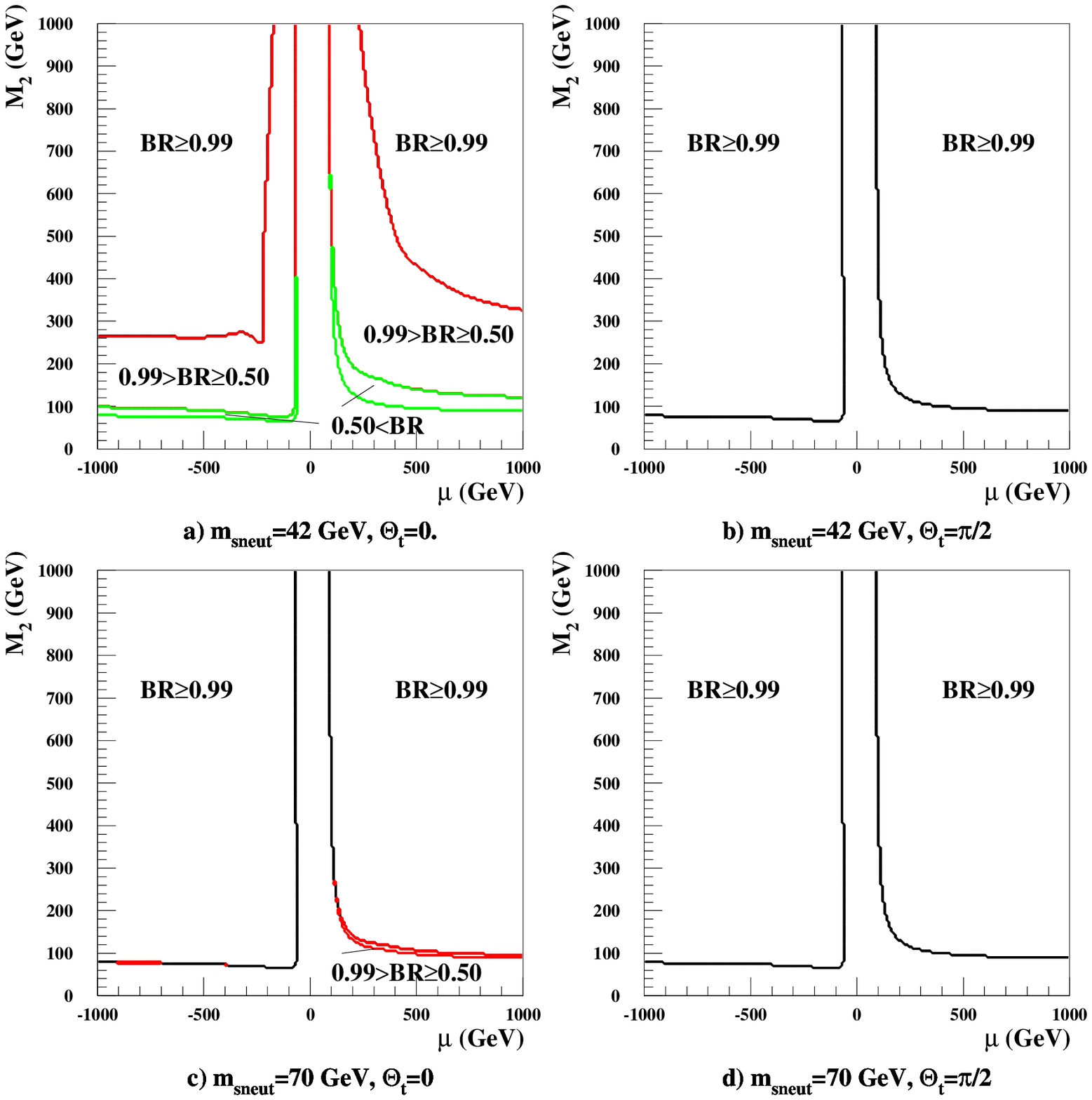}
\end{center}
Figure 25: Contours for the branching ratio $\tilde{t}_1 \ra c \chi_1^0$
in the $(\mu,M_2)$ plane for: ${\rm tan} \beta$=2, $m_{\tilde t_1}$=80 GeV, 
$m_{\chi _1^+}>$ 80 GeV, $m_{\tilde \nu}$=42 or 70 GeV, $\theta_t$=0 
or $\pi/2$.
\end{figure}
The analysis of the stop branching ratio is simpler in mSUGRA due to the
reduced number of free parameters. Let us compare below the two stop decays 
$\tilde{t}_1 \rightarrow \;c \chi_1^0$ and $\tilde{t}_1 \rightarrow\;b l 
\tilde \nu$ in the framework of mSUGRA for the following range of parameters: 
$\alpha_{U}=1/24.3$, $2\leq\tan\beta\leq 8$, $-2.9 \leq A_0\leq 2.9$ [in TeV], 
sign($\mu) = \pm$, $0 \leq m_0 \leq 1000$ and $0 \leq m_{1/2} \leq 1000$. The
regions in the $(m_0,m_{1/2})$ plane for which the following conditions on the
sparticles masses are fulfilled are shown on Fig.~26: $m_{\chi _1^0}<
m_{\tilde t_1} \leq 80$ GeV (25a), $m_{\chi_1^+}>80$ GeV (25b) and 
$m_{\chi_1^0}<m_{\tilde \nu}<m_{\tilde t_1}\leq 80$ GeV (25c). \s

The intersection of the domains in Figs.~26a and 26b gives the accessible 
space for which the decay $\tilde{t}_1 \rightarrow\;c \chi_1^0$ is possible 
with a chargino heavier than the scalar top. One observes that the condition 
$m_{\chi_1^+}>$ 80 GeV strongly reduces the possible space; this is
due to the universality condition in the gaugino sector which prevents a large
mass difference between $\chi_1^0$ and $\chi_1^+$. The intersection 
between the domains in Figs.~26b and 26c is negligible, which implies that
the decay $\tilde{t}_1 \rightarrow\;b l \tilde \nu$ is highly improbable if 
$m_{\chi_1^+}>$ 80 GeV. If the electroweak symmetry breaking conditions 
are imposed, the parameter space in the $(m_0,m_{1/2})$ plane in which this 
decay can occur is strongly reduced as it is observed in Fig.~26. 
In conclusion, the decay $\tilde{t}_1 \;\rightarrow\;c \chi_1^0$ has a 
branching ratio near unity in the constrained MSSM; however the 
corresponding accessible space in the plane $(m_0,m_{1/2})$ is small,
especially if a proper EWSB is required. 

\begin{figure}[htbp]
\vspace*{-1.8cm}
\begin{center}
\includegraphics[width=12cm]{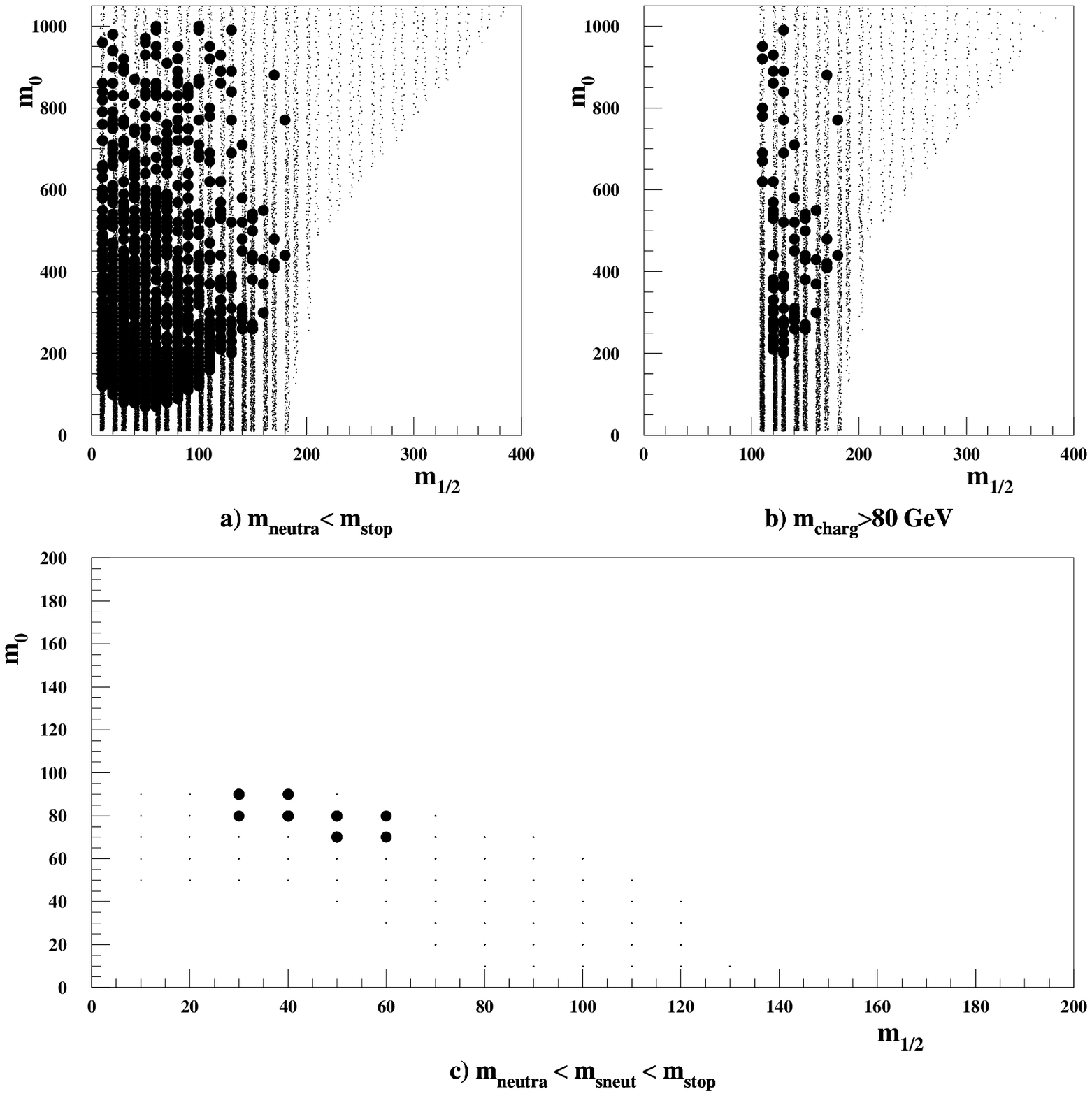}
\end{center}
Figure 26: Accessible domain for the decay $\tilde{t}_1 \ra c \chi_1^0$ in the 
$(m_0,m_{1/2})$ plane for the sets of parameters discussed above, and with
the requirements: $m_{\chi_1^0}< m_{\tilde t_1}$ and $m_{\tilde t_1}
\leq$ 80GeV for a), $m_{\tilde t_1}\leq$80 GeV and $m_{\chi_1^+}>$80 
GeV for b) and $m_{\chi_1^0} \leq m_{\tilde \nu} \leq m_{\tilde t_1}$ 
with $m_{\tilde t_1} \leq$ 80 GeV for c). The points correspond to cases for 
which no correct EWSB is required and the bullets to accessible domains with a 
correct EWSB.
\end{figure}

\smallskip

Finally, let us briefly discuss sbottom decays. The only two accessible
decay modes from sbottoms which can be produced at LEP200, are $\tilde{b}_1
\ra b \chi_{1,2}^0$ [the decay $\tilde b_1 \rightarrow b \tilde g$ is forbidden by Tevatron
results]. The relative magnitudes of these two decays depend on the sbottom mixing angle $\theta_b$
and on $\mu,M_2$ and $\tan \beta$ parameters [we assume here the GUT relation
between $M_1$ and $M_2$].  More precisely, if $M_2 \gg
|\mu|$, $\chi_1^0$ and $\chi_2^0$ are higgsino--like and the couplings $\tilde b_1 b
\chi^0$ do not depend on $\theta_b$ and are similar; the decay $\tilde b_1 \rightarrow b
 \chi_1^0$ then dominates over the decay $\tilde b_1 \rightarrow b \chi_2^0$. 
If $|\mu|\gg M_2$ then $\chi_1^0$ is photino--like and $\chi_2^0$ is zino--like; 
the coupling $\tilde b_1 b \tilde Z$ is minimal for $\theta_b=\pi/ 2$ and in this 
case $\tilde b_1 \rightarrow b \tilde \chi_1^0$ dominates over 
$\tilde b_1 \rightarrow b \chi_2^0$. On the other hand, the coupling
$\tilde b_1 b \tilde Z$ is maximal for $\theta_b=0$ and is larger than the 
$\tilde b_1 b \tilde \gamma$ coupling, leading to a domination of $\tilde b_1 \rightarrow b
\tilde\chi_2^0$ over $\tilde b_1 \rightarrow b \chi_1^0$. The fact that the sbottom
hadronises or not, which is a question of experimental interest when searching for 
sbottom squarks which are mass--degenerate with neutralinos, also depends on $\theta_b$ and 
$\mu,M_2,\tan \beta$ but is complicated by the lack of knowledge of the precise QCD time scale. 
Studies concerning this subject can be found in \cite{LEP200}.

\newpage

\setcounter{equation}{0}
\setcounter{table}{4}
\setcounter{figure}{25}
\renewcommand{\theequation}{6.\arabic{equation}}
\section*{6. Experimental bounds on SUSY Particle Masses} 

\subsection*{6.1. Introduction} 

A wide range of searches for supersymmetric particles are performed at present 
colliders and no deviation with respect to the Standard Model predictions 
was observed yet, unfortunately. Therefore new limits on the masses of
these particles, assuming different models, are set by the various experiments.
In this section we focus on the experimental results obtained both at LEP and 
the Tevatron, interpreted in the constrained MSSM or mSUGRA framework [but for
LEP2, without the constraint of correct radiative electroweak symmetry 
breaking and without assuming any mass unification for Higgs bosons with the 
other scalar particles] assuming that the LSP is the lightest neutralino $\chi^0_{1}$ and 
that R--parity is conserved.
In this model, as discussed in section 2, all SUSY particle masses, their 
couplings and their production cross sections and decay widths are predicted 
in terms of only five free independent parameters: $m_{1/2}, m_0, \mu, {\rm 
tan} \beta$ $m_A$ and $A$; the renormalization group equations are used to determine 
the parameters at low energies. Instead of $m_{1/2}$, the LEP experiments 
usually derive their results as a function of the parameter $M_2$, the SUSY 
breaking mass term associated with the SU(2)$_{\rm L}$ gauge group ($m_{1/2}
\approx M_2/0.81 $). At the Tevatron, the previous assumptions are made in the 
interpretation of the results, with the additional constraint of the radiative 
electroweak symmetry breaking. In this case, $|\mu|$ is no more a free 
parameter and is determined when $m_{0}$, tan$\beta$, $m_{1/2}$ and 
the $Z$ boson mass $M_{Z}$ are fixed; the sign of $\mu$ remains free; 
see section 2. \s

The searches for supersymmetric particles at LEP2 concern sleptons, 
stops, sbottoms, charginos and neutralinos
as detailed in \cite{JeF1}. These various SUSY particles 
decay to SM particles and two LSPs; therefore, SUSY signatures consist of 
some combination of jets or/and leptons and missing energy since the 
LSP escapes detection. The signal topology and the background conditions are 
in practice affected by the SUSY particle and the LSP mass difference 
($\Delta M= m_{\rm SUSY}-m_{\chi^0_1})$ which controls the visible 
energy. In the low  $\Delta M (= 5-10$ GeV) range, the expected topologies 
for the signals are characterized by a low multiplicity and a low visible energy
and the background is dominated by two--photon interactions. For large  
$\Delta M (=50-60$ GeV) values, the signal signatures are very similar to 
those of $W$--pair production.  Since all the background sources are due to 
well calculable processes with reasonable production cross sections  
compared to the signal, most of the decay channels are studied at LEP2. \s

Another important key domain concerns the searches for the MSSM Higgs bosons. 
Presently, only the lighter neutral Higgs bosons $h$ and $A$ can 
be discovered at LEP2 \cite{pat1}, 
since the CP--even Higgs boson $H$ and the
charged Higgs bosons $H^\pm$ are expected to be too heavy.  However searches 
for charged Higgs bosons are also performed at LEP2 in the framework of a 
non--SUSY two--Higgs doublet model. \s

After collecting $ \sim 150 \pb$ at LEP1 ($\rm 1989 \rightarrow 1995 $), 
each experiment (ALEPH, DELPHI, L3, OPAL) has accumulated data at LEP2: 
$\sim 5.5 \pb$ at $ \sqrt{s}=133$ GeV (1995), 
$\sim 10 \pb$  at $ \sqrt{s}=161$ GeV (1996), 
$\sim 10 \pb$  at $ \sqrt{s}=172$ GeV (1996) and  
$\sim 55 \pb$  at $ \sqrt{s}=183$ GeV (1997).
This year, an amount of $150 \pb$ at $\sqrt{s}=189$ GeV (October 1998)
is already recorded by each experiment, but the results including all the 
statistics for 1998 are not yet available; a fraction of the overall 
integrated luminosity is only used corresponding to the statistics 
collected just before the summer conference time  ($\rightarrow$ July 
1998 : $30-40  \pb$). \s

At the Tevatron the main sources for SUSY are squarks and gluinos, 
abundantly produced due to the color factors and the strong coupling
constant. Squarks or gluinos are produced in pair, and decay 
directly or via cascades to at least two LSP's. The classical searches rely 
on large missing transverse energy $(\Ebar)$ caused by the escaping LSPs.
In addition, charginos and neutralinos are searched for via their leptonic 
decay channels by  the two Tevatron experiments CDF and D0. Finally, 
 searches for the MSSM charged Higgs
boson are performed at the Tevatron, and
bounds 
on the charged supersymmetric Higgs boson mass have been set 
by both experiments. Each experiment has collected an integrated 
luminosity  of about 110 pb$^{-1}$ at $\sqrt{s}=1.8$ TeV. \s

Since no evidence for production of supersymmetric particles has been found,
experimental limits on their production cross sections and masses has been 
derived. In what follows, we will pay attention to all the decay channels
used to extract the limits both at LEP2 and Tevatron. The most recent results
have been used, including preliminary results reported at the last 
summer conference in Vancouver and the last LEPC meeting.

\subsection*{6.2 The scalar particle sector}

\subsubsection*{6.2.1 The Higgs bosons}

Due to the expected large mass of the heavier CP even Higgs boson $H$,
only searches for the neutral $h$ and $A$ bosons will be reported in 
what follows. These neutral Higgs bosons can be produced at LEP via two
complementary processes:
\begin{itemize}
\item The Standard Model--like Higgs--strahlung process, $e^+e^-\rightarrow 
hZ$; the cross section for this process is equal to its SM analogue, reduced 
by a factor $\sin^2(\alpha-\beta)$ which means that this process occurs mainly
at low tan$\beta$ values.
\item The associated pair production process, $e^+e^-\rightarrow hA$, 
with a cross section proportional to  $\cos^2(\beta-\alpha)$; this 
process is dominant at large tan$\beta$ values. 
\end{itemize} 

In the first channel, searches are similar to those performed for the SM Higgs 
boson. For tan$\beta >$1 the main decay modes of the $h$ and $A$ bosons are 
into $b\bar{b}$, and to a lesser extent $\tau^+ \tau^-$. Most of the 
experimental analyses required  therefore at least two $b$ quarks in the final 
states, as detailed in Tab.~\ref{tab1}. Clearly, $b$--tagging plays a crucial 
role in all the Higgs searches at LEP; it allows to reach very high 
sensitivities by rejecting most of the $WW$ background. Peculiar decays such 
as $ h\rightarrow AA$ or $h\rightarrow \chi_1^0 \chi_1^0$ have been also 
considered; the sensitivities achieved in the latter case are better
compared to those obtained with the standard decay channels at LEP1
but worse at LEP2. Typical 
efficiencies and background expectations are listed in Tab.~\ref{tab2}, 
where observation and expectation from SM processes agree very well. \s

\begin{figure}[hbtp]
 \begin{center}
  \hspace*{-1.5cm}\mbox{\epsfxsize=10.0cm 
  \epsffile{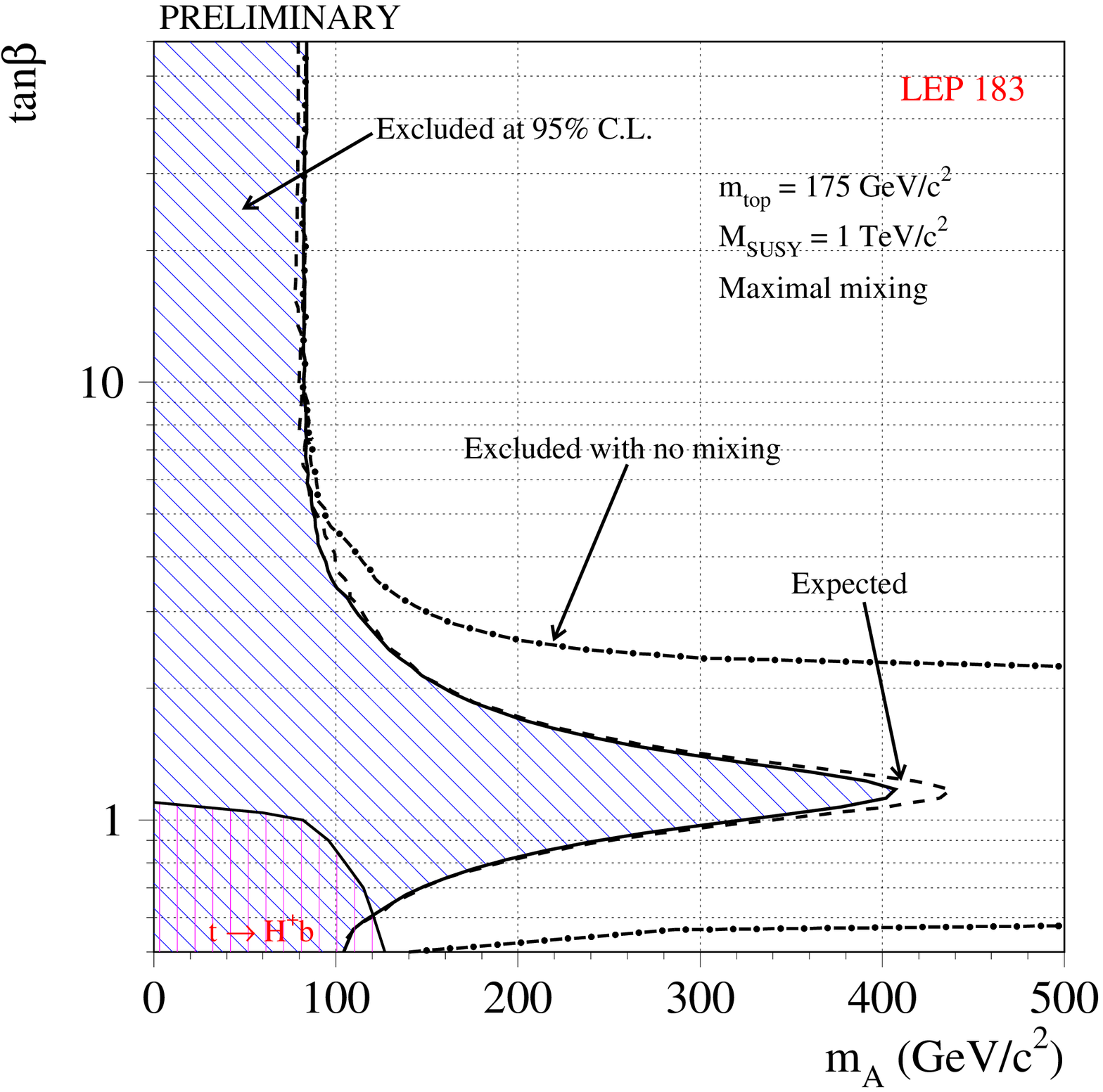}}
  \hspace*{-1.5cm}\mbox{\epsfxsize=10.0cm 
\epsffile{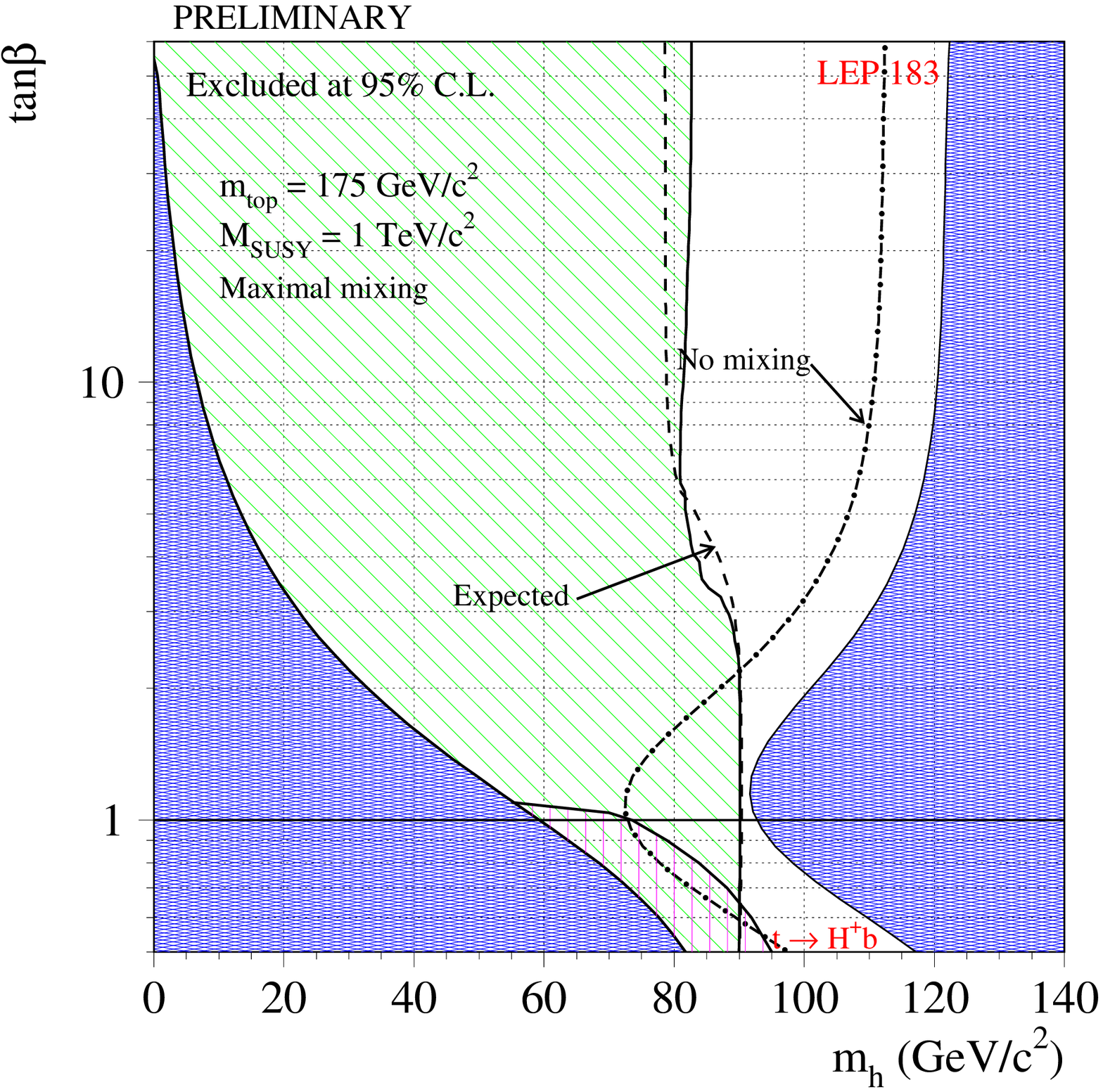}}
  \caption{LEP combined exclusion domains (method D)  
in the plane  $\rm (M_A,\tb)$ (top) or $\rm (M_h,\tb)$ (bottom)
 by the LEP direct 
searches for $\rm e^+e^-\rightarrow$ hZ and hA at center-of-mass energies 
up to 183 GeV assuming either a maximal mixing or no mixing
\cite{h1} and \cite{pat2}.           
          }
  \label{fig:mass_tbmh}
 \end{center}
\end{figure}

Therefore, limits have been extracted
\cite{h2}--\cite{h5}, 
for two ``benchmark" sets of the MSSM 
parameters where $M_{\rm susy}$,  
representing the 
stop mass mean value in Figs.~\ref{fig:mass_tbmh},
is fixed to 1 TeV and $m_{t}= 175$ GeV; $M_A$ varies up to 2 TeV and $\tb$ 
from 0.5 to 50, and either minimal or no stop mixing [$A=0$ TeV, $\mu=-0.1$ 
TeV] or maximal squark mixing [$A=\sqrt{6}$ TeV, $\mu=-0.1$ TeV] is assumed. \s

\begin{table}[htbp]
\begin{center} 
\begin{tabular}{|c|l|l|} 
\hline 
$\sqrt{s}$ (GeV)  &  hA & hZ    \\
\hline
91  & hA $\rm \rightarrow ~q\bar{q}\tau^-\tau^+,\tau^-\tau^+q\bar{q}$ 
  &   Zh$\rm \rightarrow ~\nu\bar{\nu}q\bar{q}$  \\
  &  hA$\rm \rightarrow ~AAA$ & Zh$\rm \rightarrow ~l^-l^+q\bar{q}$  \\
  &  AAA$\rm \rightarrow ~b\bar{b}b\bar{b}b\bar{b}$
  & Zh$\rm \rightarrow ~\tau^-\tau^+q\bar{q}$  \\
  &  & Zh$\rm \rightarrow ~q\bar{q}~\tau^-\tau^+$  \\
  &  &
Zh$\rm \rightarrow ~q\bar{q}h(\rightarrow \chi^0_{1}~\chi^0_{1})$  \\
\hline
136 & hA$\rm \rightarrow ~b\bar{b}b\bar{b}$ & \\
    & hA$\rm \rightarrow ~AAA$ & \\
    & AAA$\rm \rightarrow ~b\bar{b}b\bar{b}b\bar{b}$ & \\
\hline
161,172,183  & hA$\rm \rightarrow ~b\bar{b}b\bar{b}$
   & Zh$\rm \rightarrow ~\nu\bar{\nu}(h\rightarrow  all)$  \\
   & hA$\rm \rightarrow ~b\bar{b}\tau^-\tau^+,\tau^-\tau^+~b\bar{b}$  
& Zh$\rm \rightarrow ~l^-l^+(h\rightarrow  all)$  \\
  &  AAA$\rm \rightarrow ~b\bar{b}b\bar{b}b\bar{b}$
  & Zh$\rm \rightarrow ~\tau^-\tau^+(h\rightarrow  all)$  \\
  &  & Zh$\rm \rightarrow (h\rightarrow  all)~\tau^-\tau^+$  \\
  &  & Zh$\rm \rightarrow ~q\bar{q}b\bar{b}$  \\
  &  &
Zh$\rm \rightarrow ~q\bar{q}$ or $l^-l^ +h(\rightarrow \chi^0_{1}~\chi^0_{1})$  \\
\hline 
\hline 
\end{tabular} 
\caption{Search channels for the neutral Higgs bosons at LEP. 
        }
\label{tab1}
\end{center}
\end{table}
The limits obtained by each experiments within these assumptions are 
listed in Tab.~\ref{tab3}. The combined limit, obtained as described in 
Ref.~\cite{h1} leads to a gain of about 3--4 GeV with respect 
to the individual experiments. 
Four methods for the combination are used,
they all agrees within a spread of $\pm$ 1.9 GeV; conservatively the lower 
limit is quoted in Tab.~\ref{tab3} (method C of Ref.~\cite{h1}), 
while the higher limits (with method D) are given in Figs.~\ref{fig:mass_tbmh}.
These limits are valid for $\tb$ greater than 0.8,
irrespectively of the mixing 
value or of the method used for the combination. For the no mixing scenario, 
the range of $\tb$ between 0.8 and 2.1 is excluded independently of the
method. 
It has to be noticed that the limits have
been improved, for each LEP experiment, 
by almost 4 GeV/c$^2$ with the recent data collected at 189 GeV 
\cite{h0}, reaching therefore
the level of the combined limit at 183 GeV. 
%Although it was pointed out recently in 
%Refs.~\cite{h5},\cite{h20},\cite{h21}, 
%that selected sets of the MSSM parameters  (see 
%Ref.~\cite{h21}) may lead to less stringent exclusions in particular 
%in the case of both light $A$ boson and 
%small bremsstrahlung cross section.] 
It has been pointed out recently in Refs.~\cite{h5},\cite{h20},\cite{h21},     
that selected sets of the MSSM parameters lead to less stringent exclusions,
either because of an unexpectedly small bremsstrahlung cross section or because
of a reduced decay branching ratio into $b\bar{b}$. 
The analysis of Ref.~\cite{h21}
shows however that the probability of occurrence of such sets is extremely
small, typically at the $10^{-3}$ level. The theoretical status of these
pathological parameter sets is currently under investigation. Meanwhile, the
limits obtained in the benchmark cases can be regarded as sufficiently robust
for practical purposes.

\newpage

\begin{table}[htbp]
\begin{center} 
\begin{tabular}{|c|c|c|c|c|} \hline
Selection  & $\epsilon$(\%) for hZ & $\rm N_{signal}^{exp}$ 
             & $\rm N_{Bkg}^{exp}$ & $\rm N^{obs}$  \\
             & $\rm M_h$=80 GeV$/c^2$ &  &  &    \\
\hline 
$ \rm hl^+l^- $ & 78.3 & 1.1 & 2 & 3 \\
$ \rm h\nu\bar{\nu} $ & 20.9 & 0.8 & 0.16 & 0 \\
$ \rm b\bar{b}q\bar{q} $ & 23.6 & 3.6 & 1.4 & 1 \\
$ \rm b\bar{b}\tau^+\tau^- $ & 22.7 & 0.14 & 0.17 & 0 \\
$ \rm \tau^+\tau^-q\bar{q} $ & 9.8  & 0.16 & 0.16 & 0 \\
\hline
Selection  & $\epsilon$(\%) for hA & $\rm N_{signal}^{exp}$ 
             & $\rm N_{Bkg}^{exp}$ & $\rm N^{obs}$  \\

             & $\rm M_h=M_A$=75 GeV$/c^2$ &  &  &    \\
\hline 
$ \rm b\bar{b}b\bar{b} $ & 60.5 & 3.1 & 2.4 & 2 \\
$ \rm b\bar{b}\tau^+\tau^- $ & 28.6 & 0.27 & 0.07 & 0 \\
\hline
\end{tabular}
\caption{
Typical efficiencies and expected/observed number of events
 for the neutral Higgs bosons searches performed by the ALEPH 
experiment \cite{h2} 
at $\sqrt{s}$=183 GeV. }
\label{tab2}
\end{center}
\end{table}
\vspace*{-5mm}
\begin{table}[htbp]
\begin{center} 
\begin{tabular}{|c|c|c|} 
\hline 
     & \multicolumn{2}{|c|}{$\rm M_{h,A}$ lower limit 
 ($\rm GeV/c^2)$}     \\ 
     & \multicolumn{2}{|c|}{up to $\sqrt{s} = 183$ GeV}     \\ 
\cline{2-3}
    & Range of         & lower     \\
    & Individual Limit & LEP combined \\ 
    & (ADLO)           &     \\
\hline
 {\bf h} (Obs.)  & {\bf 70.7--74.4}  & {\bf 78.8} \\ 
 \ \ \ (Exp.) & 67.4--70.3   & 76.3 \\ 
\hline 
 {\bf A} (Obs.)& 71.0--76.1 & 79.1 \\ 
 \ \ \ (Exp.) &    68.4--72.0 & 76.3 \\ 
\hline 
\end{tabular} 
\caption{Individual and LEP combined 
mass limits for the neutral $h$ and $A$ bosons \cite{h1}.
        }
\label{tab3}
\end{center}
\end{table}
At the Tevatron, the most widely studied mechanism for producing neutral 
Higgs bosons is the associated production of the CP--even $h$ boson 
with a $W$ or $Z$ boson. In addition, one has to require the gauge boson
to decay leptonically; this reduces the observable rate by a factor
greater than 75 \%. Therefore a collected luminosity of several fb$^{-1}$ is 
needed to reach a sensitivity competitive with those achieved at LEP.
Recently, in Ref.~\cite{drees}, it has been made use of 
the fact that the $A b\bar{b}$ coupling is proportional to $\tan \beta$, 
i.e. the cross section production grows as $\tan^2 \beta$. Analyzing the 
events from CDF containing  $\tau^-\tau^+ + 2$  jets in the final state 
[used to set limits on third generation leptoquarks], the authors derived 
new bounds in the plane ($M_A,\tb$); in particular if the mass
of the CP--odd Higgs boson is just beyond the present limit, the $\tb$ region 
above 80 is excluded. \s
% \vspace{0.2 cm}

\begin{figure}[hbtp]
 \begin{center}

  \hspace*{-1.5cm}\mbox{\epsfxsize=9.5cm 
  \epsffile{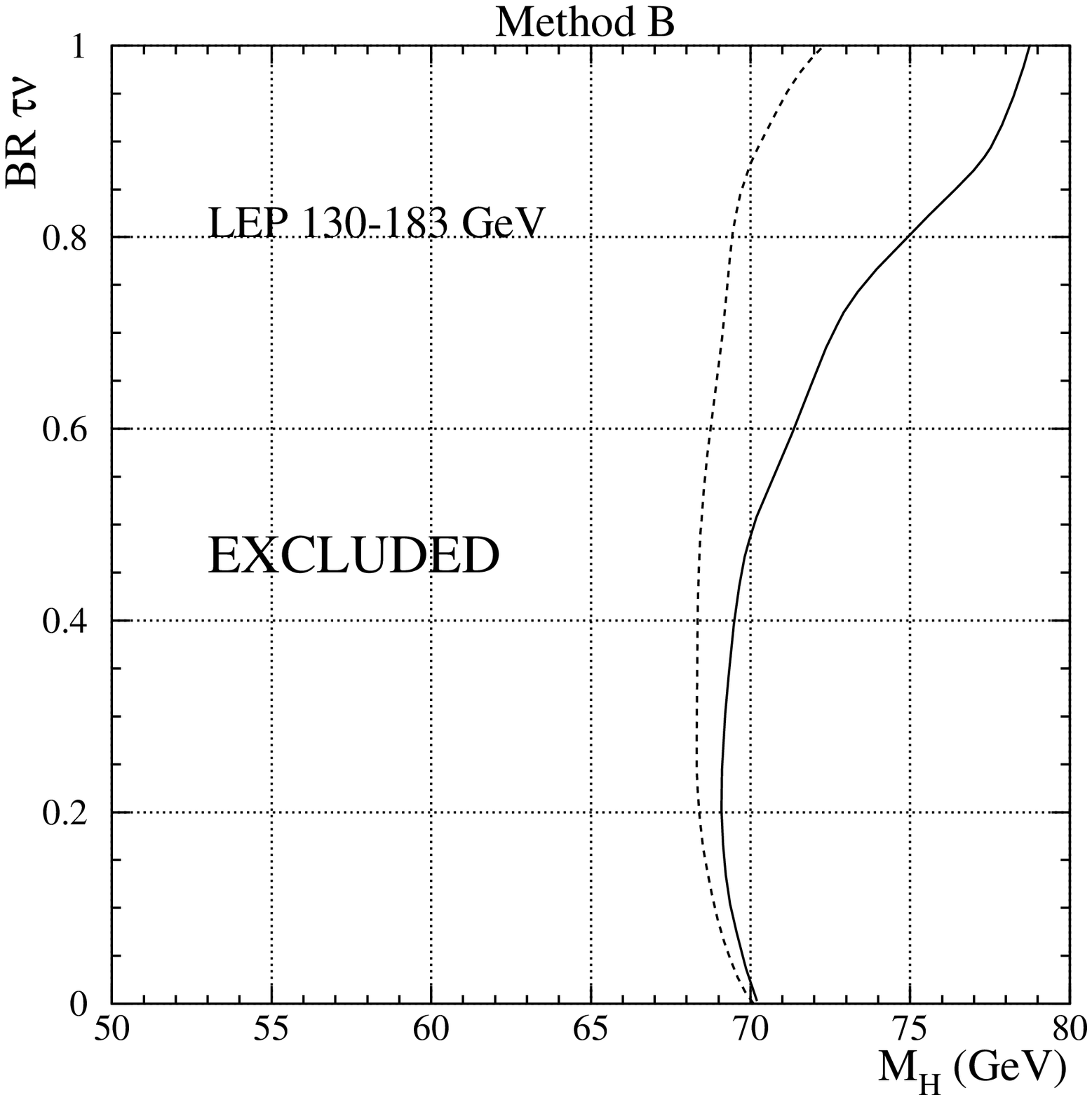}}
  \caption{LEP combined exclusion domains in the plane 
($ M_{H^{\pm}},Br(\rm H^{\pm}\rightarrow \tau \nu_{\tau})$) by the LEP direct 
searches for $\rm e^+e^-\rightarrow H^+H^-$ at center-of-mass energies 
up to 183 GeV \cite{h1}. }
  \label{fig:mchargedH}
 \end{center}
%\end{figure}
%
%%%
%          
%\begin{figure}[hbtp]
 \begin{center} 
 \hspace*{-1.5cm}\mbox{\epsfxsize=9.5cm 
  \epsffile{x5.epsi}}
  \caption{D0 exclusion domains \cite{hc6} in the plane 
($\tan \beta ,M_H^{\pm}$) from top decay.  }
  \label{fig:D0}
 \end{center}
\end{figure}

Let us now turn to charged Higgs bosons. The $H^\pm$ bosons can be produced 
in $e^+e^-$ interactions via the process  $e^+e^- \rightarrow \gamma^*, Z^* 
\rightarrow H^+H^-$ and are expected to decay mainly into the heaviest 
lepton kinematically allowed and its associated neutrino, or into the heaviest 
kinematically allowed quark pair whose decay widths are not Cabibbo 
suppressed; the relative branching ratios are model dependent. Therefore 
searches are performed in the three possible decay modes: $\ee \ra H^+H^- 
\rightarrow \tau^-\bar{\nu_{\tau}} \tau^+\nu_{\tau}$, 
$\tau^+\nu_{\tau}s\bar{c}$ 
and $c\bar{s} s\bar{c}$. 
For each experiment \cite{hc1}--~\cite{hc4} and each decay channel, 
the number of selected events in the data is consistent, 
with the number of expected 
events from SM processes, as can be seen in Tab.~\ref{tab4}. Compared to the 
neutral Higgs boson searches, the sensitivities achieved are lower; this is 
due to the high $WW$ background contamination. 
\begin{table} [htbp]
\begin{center} 
\begin{tabular}{|c|c|c|c|} 
\hline 
 $\rm M_{H^{\pm}}$=60 $\gv$  & $ \tau^+\bar{\nu}_{\tau}\tau^+\nu_{\tau}$
   & $ \tau^+\nu_{\tau}s\bar{c}$  & $c\bar{s}s\bar{c}$ 
   \\
\hline
Efficiency & 24 \% & 42 \%  & 40 \% \\ 
\hline 
Expected events & 9.2 & 30.1 & 99.4 \\ 
\hline 
Observed events & 6   & 28   & 93 \\
\hline 
\end{tabular} 
\caption{Typical efficiencies, number of expected and observed events  
obtained by the L3 experiment in the charged Higgs boson searches
at c.m. energy of 183 GeV \cite{hc3}. }
\label{tab4}
\end{center}
\end{table}
\vspace*{-5mm}
\begin{table} [htbp]
\begin{center} 
\begin{tabular}{|c|c|c|} 
\hline
     & \multicolumn{2}{|c|}{$\rm M_{H^{\pm}}\>$   
Lower limit ($\rm GeV/c^2)$}   \\ 
     & \multicolumn{2}{|c|}{up to $\sqrt{s} = 183$ GeV}     \\ 

\cline{2-3}
      & Range of  & Lower \\ 
      & Individual Limit (ADLO) & LEP combined \\ 

\hline
{\bf H}$^\pm$ (Obs.)& \bf{56.6--59.0} & \bf{68.0} \\ 
\ \ \ \ \ \ (Exp.)& 56.0--62.0 & 69.0 \\ 
\hline 
\end{tabular} 
\caption{Individual and LEP combined 
mass limits for the charged Higgs bosons.
        }
\label{tab5}
\end{center}
\end{table}
The 95 \% C.L. individual limits on the charged Higgs boson mass is shown 
as a function of the branching ratio Br($H^{\pm}\rightarrow \tau^{\pm}
\nu_{\tau})$ in Fig.~\ref{fig:mchargedH}. Combining all LEP results, 
a lower limit on the charged Higgs boson mass of 68 GeV is established at the 
95 \% C.L., assuming just that the sum BR$(H^{+}\rightarrow \tau^+ \nu_{\tau})$ 
+ Br$(H^{+}\rightarrow c \bar{s})$ is equal to one. It represents a 
gain of almost 10 GeV with respect to the individual limits, Tab.~\ref{tab5}.
\s
 
In the MSSM, the charged Higgs bosons are expected to be heavier than the $W$
boson, and  can be searched for in the decay products of the top quarks 
that are produced at the Tevatron. If the charged Higgs boson is light enough,  
the two possible decay channels for the top are either
$t \rightarrow {W^+}b$ or $t \rightarrow {H^+}b$; as an example 
for $M_{H^\pm}=100$ GeV, the decay $t \ra H^+b$ becomes 
dominant for both large ($\gsim 30$) and low ($\lsim 1$) $\tb$ values. 
Moreover for low $\tb$ values $H^+$ decays dominantly 
into $c\bar{s}$ leading to final states containing six jets 
without isolated leptons. The measurement of the top branching 
ratio into $W^+b$, tagged with leptonic events, allows to set 
an upper limit on the branching ratio of the decay $t \ra H^+b$,
assuming an expected $t\bar{t}$ production cross section not larger 
than 5.6 pb; CDF \cite{hc5} and D0 \cite{hc6} set a limit of about 25 \% 
which can be turned into excluded regions in the ($M_{H^\pm},\tb$) plane
as depicted in Fig.~\ref{fig:D0}. For instance, for 
$M_{H^\pm}=100$ GeV, all $\tb$ values below 1.1 and above 
40 are excluded by Tevatron searches \cite{pat2}. These limits may 
be also turned into excluded region in the plane ($M_{h},\tb$) as 
depicted in bottom Fig.~\ref{fig:mass_tbmh} \cite{pat2} and \cite{h1}.
 
\subsubsection*{6.2.2. Scalar leptons}

LEP experiments excluded scalar neutrinos with masses 
up to 43 GeV, from the comparison of the measured $Z$ widths with the SM 
expectation \cite{PDG}. \s

In $e^+e^-$ collisions, the production of scalar muons $\tilde{\mu}$, 
scalar taus $\tilde \tau$ proceeds via $\gamma$ or $Z$ exchange 
in the $s$--channel only, whereas scalar electrons $\tilde e$ can also be 
produced by exchanging neutralinos in the $t$--channel. At the Tevatron, 
charged scalar leptons can be pair produced or produced in association
with a scalar neutrino [for the left handed charged sleptons only]
via the Drell--Yan mechanism. The detection of scalar leptons at the 
Tevatron is difficult due to the high background sources, the low 
production cross sections and the possibly important cascade decay branching 
ratios. Presently, charged scalar leptons can be searched for only at LEP2. \s

The scalar leptons dominantly decay into their SM partners and the 
lightest neutralino $\chi^0_{1}$. If the scalar lepton is not the 
next--to--light SUSY particle (NLSP), cascade decays into  $\chi^0_{2}$ 
or into the lightest chargino $\chi_1^+$ [for $\tilde{l}_L$ only for a pure 
gaugino--type 
chargino] may be possible. The topologies arising from scalar lepton 
production are then usually, acoplanar leptons plus missing energy. 
As already explained in the introduction, the sensitivities for these searches 
depend strongly on the mass difference between the scalar lepton and the 
$\chi^0_{1}$, $\Delta M$. Typical efficiencies and expected events from 
SM processes for different $\Delta M$ ranges are given in Tab.~\ref{tab6}.

\begin{figure}[htbp]
\begin{tabular}{ll}
%\hspace*{-1.5cm} \put(1,268){\bf a)}
\hspace*{.5cm}\epsfig{file=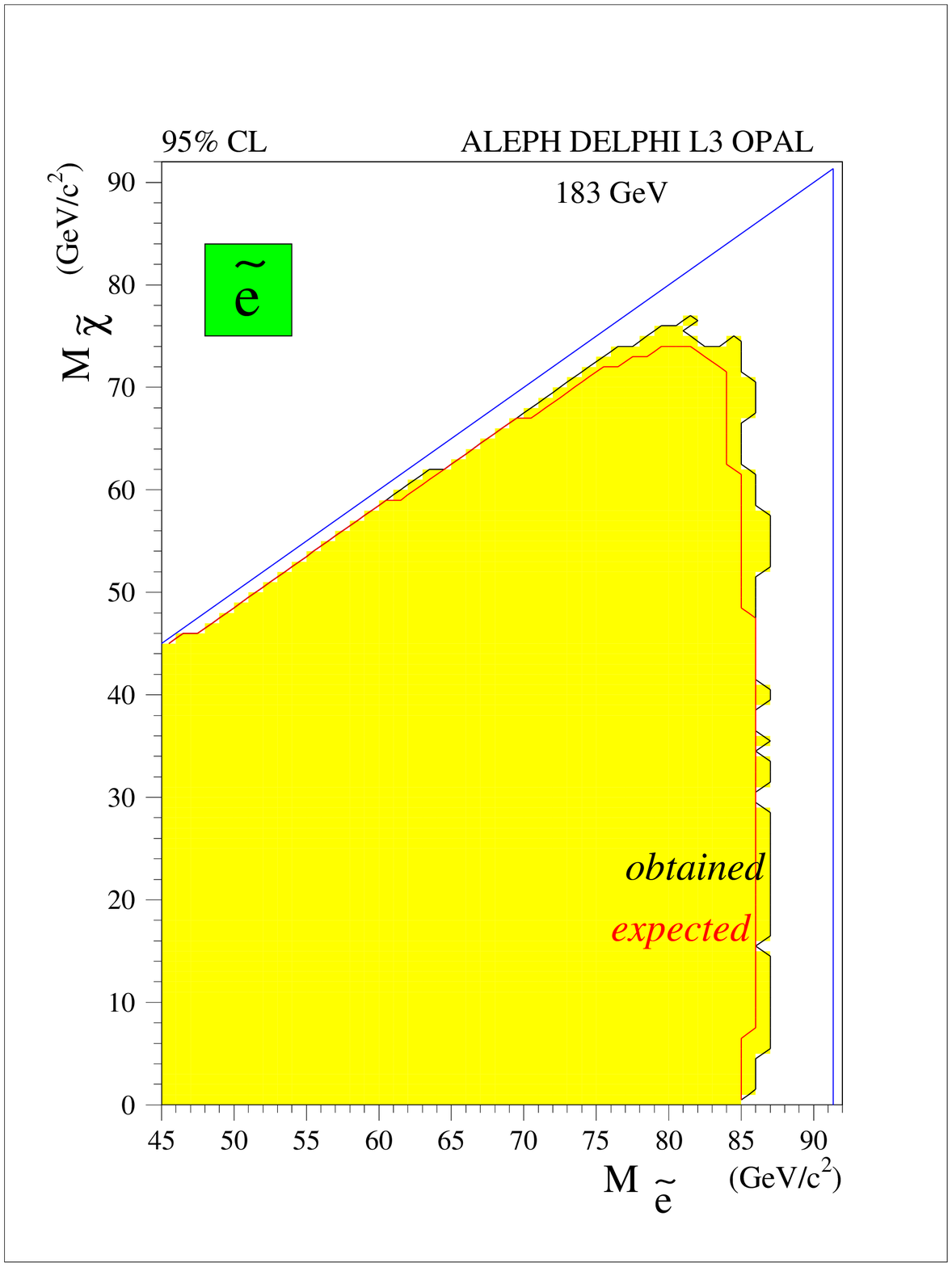,width=6cm}
%\hspace*{2.0cm} \put(60,268){\bf b)}
\hspace*{1.0cm}\epsfig{file=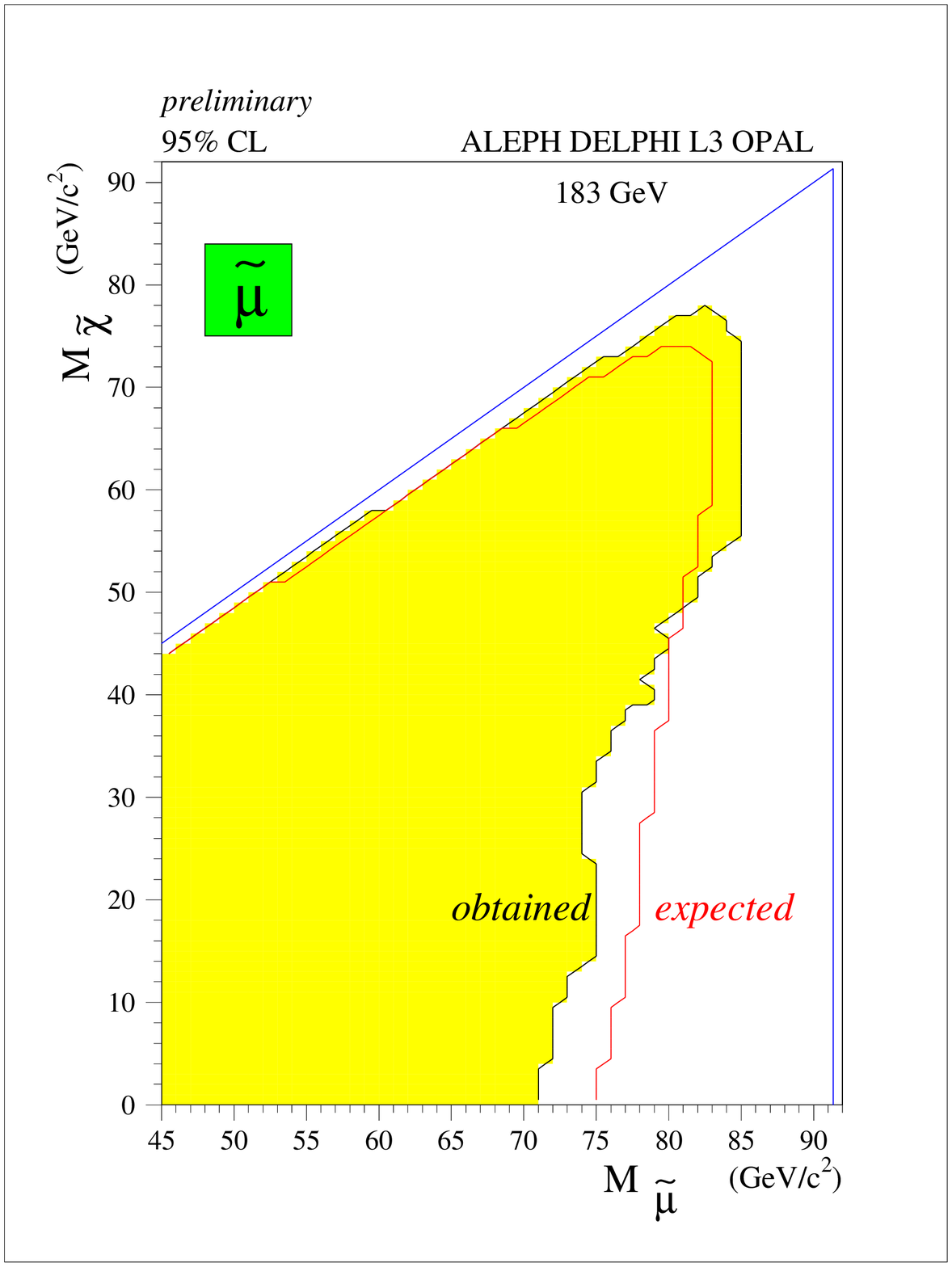,width=6cm}\vspace{0.5 cm}  \\
%\hspace{-1.5cm} \put(1,268){\bf c)}
\hspace*{.5cm}\epsfig{file=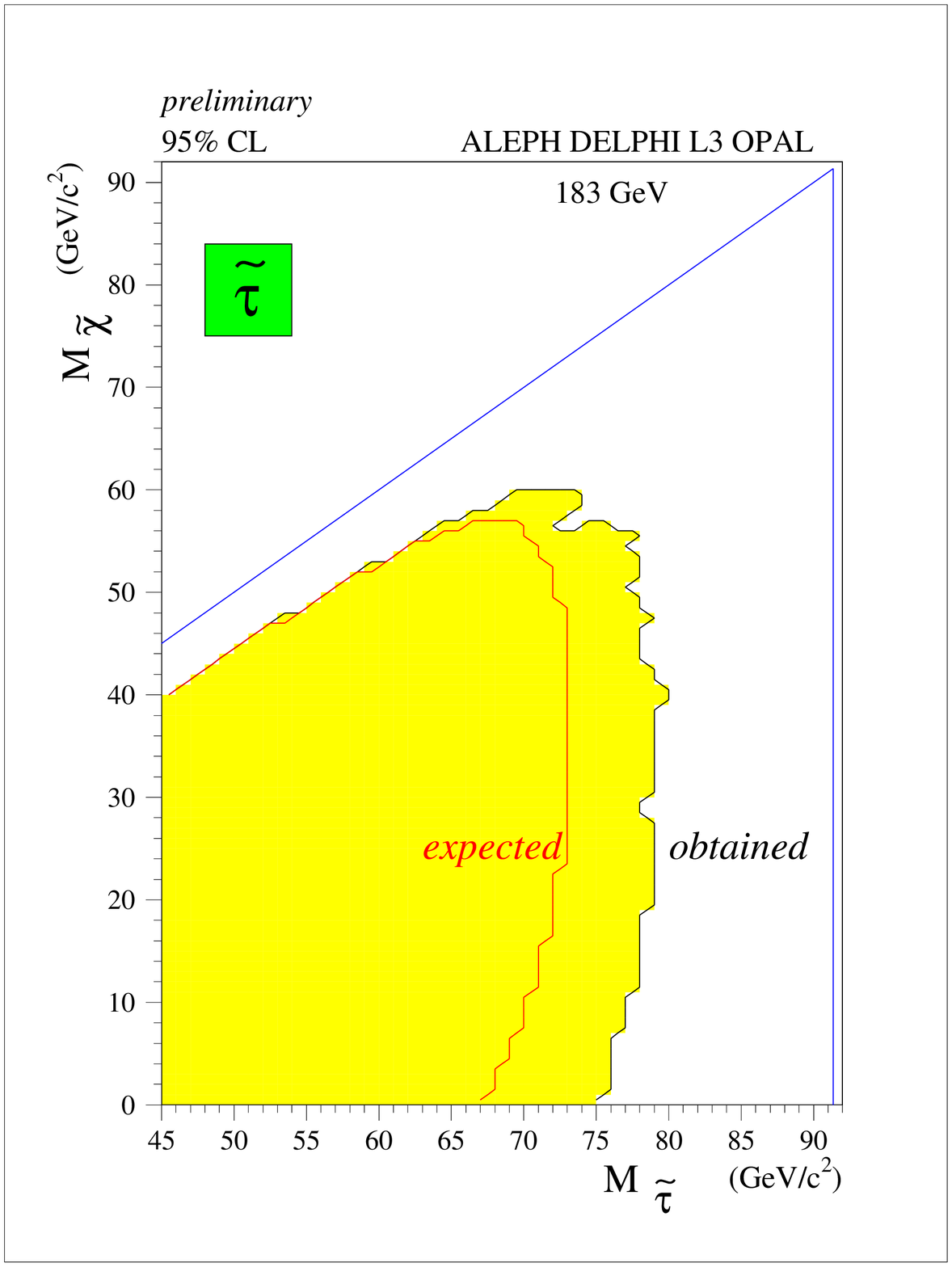,width=6.0cm} 
%\hspace*{3.5cm}\put(100,268){\bf d)}
\hspace*{1.2cm}\epsfig{file=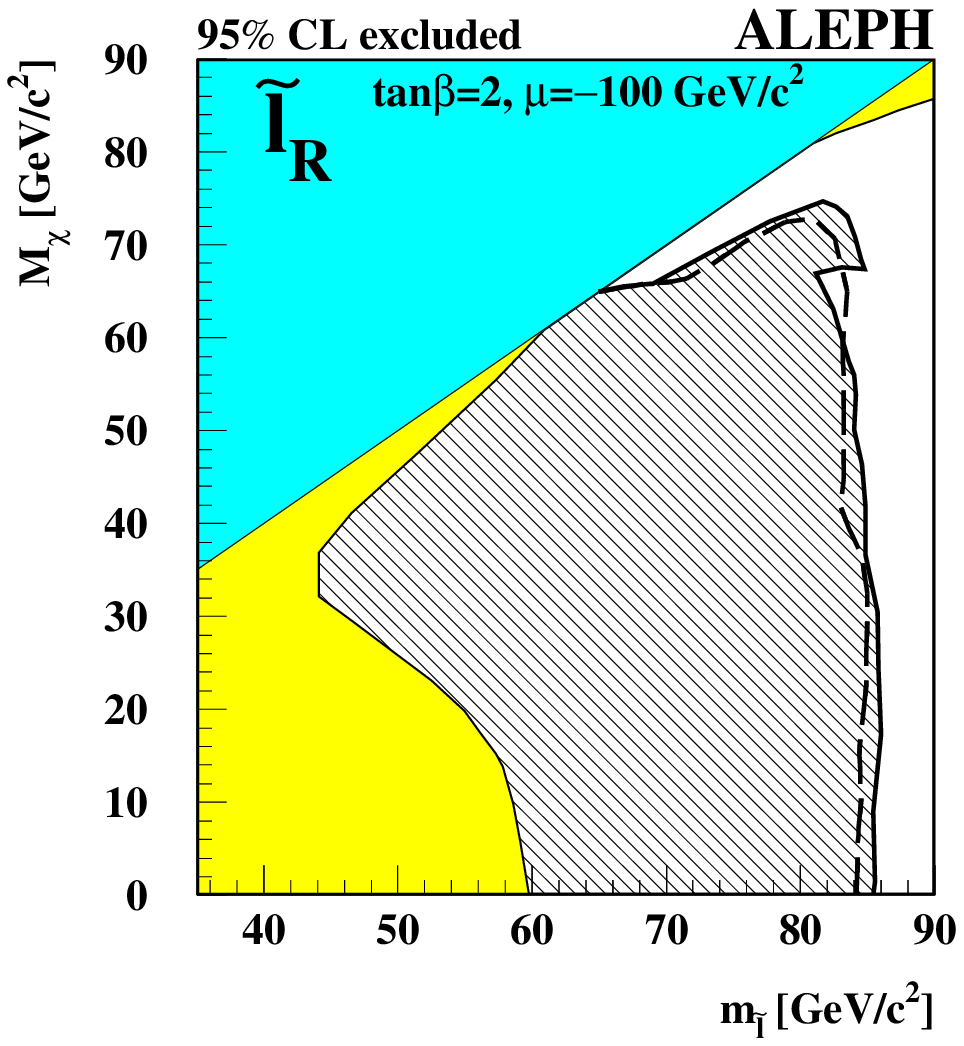,width=8cm}
 \end{tabular}
 \caption{Combined exclusion contours 
          for scalar electrons, muons and 
          taus  \cite{s0} as a function of the LSP mass
          and the limit for sleptons obtained
          by the ALEPH experiment \cite{sl1} when combining all
          selectron and smuon searches up to 183 GeV (bottom--right). 
         }
\label{fig:sleptons}
\end{figure}

\begin{table} [htbp]
\begin{center} 
\begin{tabular}{|c||c|c|c||c|c|c||c|c|c|} \hline
     \multicolumn{10}{|c|}{$m_{\tilde{l}\pm} = 75$ GeV}\\
      \multicolumn{10}{|c|}{$\sqrt{s} = 183$GeV} 
           \\ \cline{1-10} 
{ }  &\multicolumn{3}{|c|}{~~$\tilde{e}_R$~~} 
     &\multicolumn{3}{|c|}{~~$\tilde{\mu}_R$~~ }  
     &\multicolumn{3}{|c|}{~~$\tilde{\tau}_R$~~ } 
            \\   \cline{2-10}
{$\Delta M$ (GeV)}  
& ~~$\epsilon $ (\%) & $N_{\rm sign}^{\rm exp}$ &$N_{\rm back}^{\rm exp}$ 
& ~~$\epsilon $ (\%) & $N_{\rm sign}^{\rm exp}$ &$N_{\rm back}^{\rm exp}$ 
 & ~~$\epsilon $ (\%) & $N_{\rm sign}^{\rm exp}$ &$N_{\rm back}^{\rm exp}$ \\ 
\hline 
0   & 65 & 33.8  & 7.6   & 66  & 4.5  & 6.9  & 43 & 2.9&5.9        \\
15  & 57 & 7.4  & 0.5   & 45   & 4.3  & 0.15 & 38 & 2.4& 3.2  \\
70  & 11 & 0.9  & 0.7   & 13   & 0.9  & 0.9  & 7 & 0.5& 1.5    \\
\hline 
\end{tabular} 
\caption[cascade]{Scalar electron, scalar muon and scalar
tau efficiencies ($\epsilon$) and
           the number of events expected from SM processes
           ($N_{\rm exp}$).  
           Results are obtained at $\sqrt{s} = 183$ GeV
           by the ALEPH experiment \cite{sl1}
           as a function of
           $\Delta M$ for 
           $m_{\tilde{l}\pm}$ = 75 GeV.}
\label{tab6}
\end{center} 
\vspace*{-5mm}
\end{table} 
No excess with respect to the number of events expected from SM processes 
was observed at LEP up to $\sqrt{s}$=183 GeV \cite{sl1}--\cite{sl4}. 
Since in general, the 
cross section for the pair production of $\tilde{l}_R$ is smaller than for 
$\tilde{l}_L$, limits on masses are given by default taking into account 
$\tilde{l}_L \tilde{l}_R$ production only. The expected background from 
$W$--pair production is subtracted for all LEP analyses. All limits are 
derived in the plane ($m_{\chi^0_1}, m_{\tilde{l}_R}$). These
limits depend only slightly on $\mu$ or $\tb$ for $\tilde{\mu}_R$ and 
$\tilde{\tau}_R$, since these parameters may just change their couplings to 
$\chi^0_{2}$ [relevant only for low masses of $\chi^0_{1}$ in the 
region  covered anyway by the $\chi_1^+$ searches] but do not affect their 
cross section for pair--production. For $\tilde{e}_R$, the production cross 
section in the $\chi^0_{1}$ $t$--channel exchange depends on these 
parameters. The combined exclusion contours of all LEP analyses 
\cite{s0} up to 183 GeV are shown 
in Figs.~\ref{fig:sleptons}. The individual limits are derived for different
$\tb$ values 1.4 $\rightarrow$ 2,  and for $\mu=-200$ GeV.  \s

Since we are far from the kinematical limit, more luminosity helps
and a gain ranging from 2 to 10 GeV, depending on the scalar lepton flavor,
is obtained in the combination as shown in Tab.~\ref{tab7}. We should 
mention also that the individual limits obtained recently at $\sqrt{s}$=189 
GeV with an integrated luminosity of 30--40 ${\rm pb}^{-1}$ for each experiment reach the level of the LEP combined [up to 183 GeV] limit \cite{s0}. \s

%\begin{figure}[h]
% \begin{tabular}{ll}
%  \hspace{-1.5cm}
% \mbox{\epsfxsize=8.0cm 
%  \epsffile{sel.eps}}
\begin{table}
\begin{center}
\begin{tabular}{|l|c|l|} 
\hline
     & \multicolumn{2}{|c|}{$\rm m_{\tilde{l}_R}\>$   
lower limit ($\rm GeV/c^2)$}   \\ 
     & \multicolumn{2}{|c|}{up to $\sqrt{s} = 183$GeV}     \\ 

\cline{2-3}
      & Individual Limit (ADLO) & LEP combined \\ 
\hline 

 {\bf $  \tilde{e}_R$} (Obs.) & {\bf 79--83} & {\bf 85 } \\ 
 \ \ \ \ (Exp.)  & 78--83 & 85  \\ 
\hline 
 {\bf $  \tilde{\mu}_R$} (Obs.) & {\bf 55--62} & {\bf 71 } \\ 
 \ \ \ \ (Exp.) & 56--67 & 75 \\ 
\hline 
 {\bf $  \tilde{\tau}_R$} (Obs.)  & {\bf 45--63} & {\bf 75 } \\ 
 \ \ \ \ (Exp.) & 45--56 & 67 \\ 
\hline 
\end{tabular} 
\caption{Individual and LEP combined 
mass limits for $\tilde{e}_R$,
  $\tilde{\mu}_R$, $\tilde{\tau}_R$ up to $\sqrt{s}$ = 183 GeV    
for $\Delta M$ greater than 20 GeV, and assuming $\tb=2$ and $\mu=-200$
GeV for $\tilde{e}_R$. }
\label{tab7}
\end{center}
\vspace*{-7mm}
\end{table}

Assuming a common scalar lepton mass at the GUT scale, 
the relation between the masses of right-- and left--handed 
sleptons can be used to combine 
results of the searches for acoplanar leptons (muons, electrons together)
\cite{sl1} coming from the $\tilde{e}_R \tilde{e}_L$ process when  
the $\tilde{e}_R$ and the $\chi^0_{1}$ are mass degenerate ($\Delta M$ 
below 3 GeV). The result is shown in Fig.~\ref{fig:sleptons}, for $\tb=2$
and $\mu=100$ GeV; this value of $\mu$ minimizes the product of cross section 
times branching fraction for the process $ e^+e^- \rightarrow \tilde{e}_R
\tilde{e}_L$ with $\tilde{e}_L \rightarrow e \chi^0_{1}$ for vanishing 
$\Delta M$. In this case, a scalar lepton mass limit of 65 $\rm GeV/c^2$ is 
set independently of $\Delta M$; this limit holds for higher values of 
$\tb$.

\subsubsection*{6.2.3 Scalar quarks}

The squarks of the two first generations should be abundantly produced at 
the Tevatron which places already limits far above the kinematical reach of 
LEP2. Due to a large Yukawa coupling, the mass of one of the top squarks 
can be significantly smaller compared to those of the other scalar quarks;  
the large mixing between the left and right stops, proportional to $m_t$, 
leads to a large splitting of the two mass eigenstates. The lighter stop 
$\tilde{t}_1$ could be  within the discovery range of LEP2. In addition, for 
large $\tb$ values ($\gsim 30$), the mixing angle in the sbottom sector can 
also be large and the lighter sbottom $\tilde{b}_1$ could also be within the 
reach of LEP2. 
\smallskip

Stop and sbottom production at LEP proceed via $Z/\gamma$ exchange in the 
$s$--channel;
the production cross sections depend on the squark mass and the squark mixing 
angle $\cos\theta_{q}$ and at $\cos\theta_{q}\sim 0.57$(0.39) the stop 
(sbottom) decouples from the $Z$--boson and the cross section is minimal;
see section 5.5. The dominant decay modes of the $\tilde{t}_1$ are expected to 
be $\tilde{t}_1 \rightarrow c\chi^0_{1}$ or $\tilde{t}_1 \rightarrow b 
\tilde{\nu} l^+$; both of these decay modes have been searched for at LEP.
The dominant decay mode of $\tilde{b}_1$ is expected to be  $\tilde{b}_1 
\rightarrow b \chi^0_{1}$. \s

Under the assumption of R--parity conservation, the $\chi^0_{1}$ and the 
$\tilde{\nu}$ are invisible in the detector; thus stop or sbottom pair 
events are characterized by two acoplanar jets or two acoplanar jets plus 
two leptons, with missing energy. When the $\tilde{t}_1 \rightarrow c 
\chi^0_{1}$ 
decay mode is dominant, the corresponding decay width is small enough for 
the top squark to hadronize into a colorless ``stop hadron'' before decaying; 
this feature has been implemented by the LEP experiments in their Monte Carlos.
No excess of events was observed by any of the four LEP experiments
\cite{st1}--\cite{st4}, in the 
searches for stop and sbottom. Typical sensitivities obtained by the OPAL 
\cite{st4} experiment in the various channels, are listed in Tab.~\ref{tab8}. 

\begin{figure}[hbtp]
 \begin{center}  
 \hspace*{-1.5cm}\mbox{\epsfxsize=9.5cm 
  \epsffile{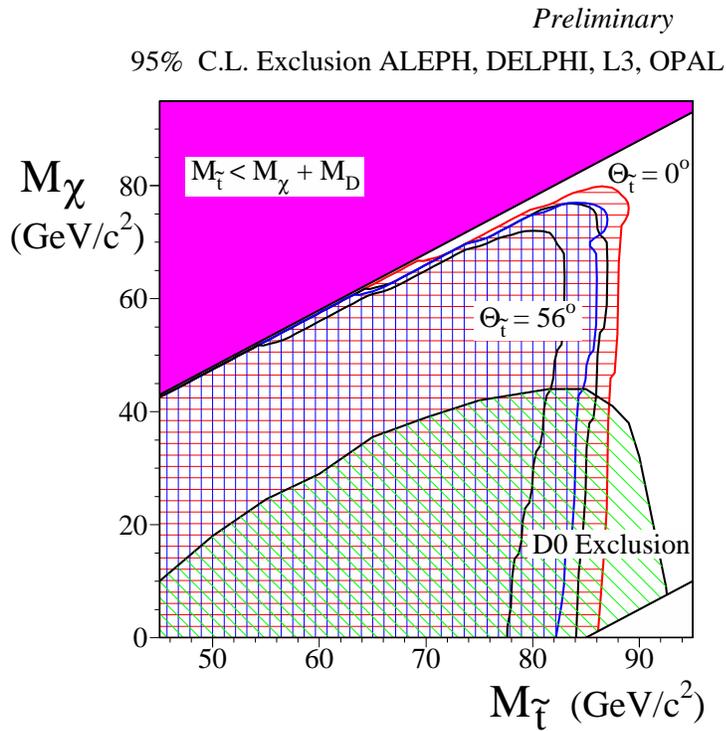}}
 \hspace*{-1.5cm}\mbox{\epsfxsize=10.5cm \epsffile{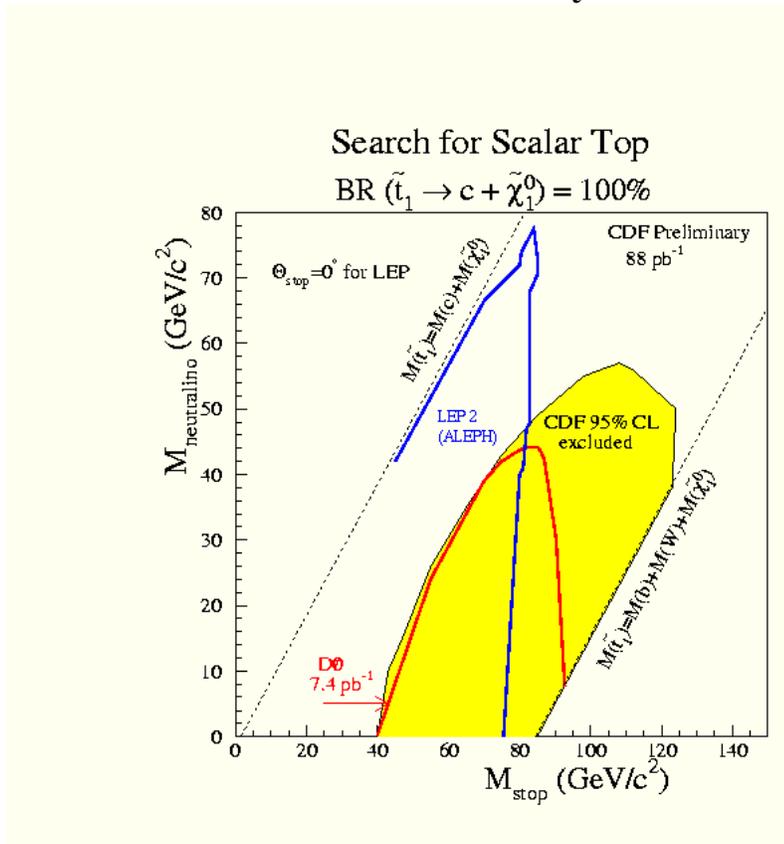}}
  \caption{LEP2 (top) \cite{s0} and CDF (bottom) \cite{st5} exclusion 
domains in the plane ($m_{\chi^0_1},m_{\tilde{t}_1}$) from stop decay 
$\tilde{t}_1 \rightarrow c\chi^0_{1}$ searches.
 }
\label{fig:stop}
\end{center}
\end{figure}

 \begin{table} [htbp]
\begin{center} 
\begin{tabular}{|c||c|c||c|c||c|c||} \hline
     \multicolumn{7}{|c|}{$m_{\tilde{t_1},\tilde{b_1}} = 80$ GeV}\\
      \multicolumn{7}{|c|}{$\sqrt{s} = 183$ GeV} 
           \\ \cline{1-7}
{ }  &\multicolumn{2}{|c|}{$\tilde{t}_1 \rightarrow c\chi^0_{1}$~} 
     &\multicolumn{2}{|c|}{$\tilde{b}_1\rightarrow b \chi^0_{1}$~ }  
     &\multicolumn{2}{|c|}{$\tilde{t}_1\rightarrow bl\tilde{\nu} $~} 
            \\   \cline{2-7}
{~~ $\Delta M$ (GeV)~~}  
& ~~$\epsilon $ (\%)  & $N_{\rm back}^{\rm exp}$ 
& ~~$\epsilon $ (\%)  & $N_{\rm back}^{\rm exp}$ 
 & ~~$\epsilon $ (\%) & $N_{\rm back}^{\rm exp}$ \\ 
\hline 
7   & 23 & 1.97  & 37   & 1.97  & 11.8 &   1.07      \\
15  & 58 & 1.97  & 61   & 1.97  & 64   &   2.05      \\
80  & 37 & 1.97  & 31   & 1.97  & 58   &   2.05      \\
\hline 
\end{tabular} 
\caption[cascade]{$\tilde{t}_1$  and $\tilde{b}_1$ efficiencies ($\epsilon$) and
           the number of events expected from the SM 
           ($N_{\rm exp}$) obtained at $\sqrt{s} = 183$ GeV
           by OPAL \cite{st4}
           as a function of
           $\Delta M$ for 
           $m_{\tilde{q}}$ = 80 GeV.}
\label{tab8}
\end{center}
\end{table}
All LEP results have been combined \cite{s0}
and are summarized  in Tab.~\ref{tab9}
for $\Delta M$ greater than 15 GeV/c$^2$. Since a slight lack of events 
appears for sbottom searches, conservatively, the combination in this channel
is done without any background subtraction. The LEP combined excluded regions 
in the plane ($m_{\chi_1^0}, m_{\tilde{q}}$) are shown in Fig.~\ref{fig:stop}. 
\s

\begin{table}
\begin{center}
\begin{tabular}{|l|c|l|} 
\hline
     & \multicolumn{2}{|c|}{$\rm m_{\tilde{t}_1},m_{\tilde{b}_1}\>$   
lower limits ($\rm GeV/c^2)$}   \\ 
     & \multicolumn{2}{|c|}{up to $\sqrt{s} = 183$GeV}     \\ 
\cline{2-3}
      & Individual Limit (ADLO) & LEP combined \\ 
      &     or AO only: $^*$    & {\bf obs}  \ [exp]             \\     
\hline 
{\bf $\tilde{t}_1\rightarrow c\chi^0_{1}$}, $\rm cos\theta_{t}=1$
& {\bf 80--85} & {\bf 86} (85)  \\ 
{\bf $\tilde{t}_1\rightarrow c\chi^0_{1}$}, $\rm cos\theta_{t}=0.57$
& {\bf 72--81} & {\bf 83} (80)  \\  
\hline  
{\bf $\tilde{t}_1 \rightarrow b\tilde{\nu}l$}, $\rm cos\theta_{t}=1$$^*$ 
& {\bf 84--85} & {\bf 87} (86)  \\ 
{\bf $\tilde{t}_1 \rightarrow b\tilde{\nu}l$}, $\rm cos\theta_{t}=0.57$$^*$ 
& {\bf 80--82} & {\bf 85} (84)  \\  
\hline  
{\bf $\tilde{b}_1\rightarrow b\chi^0_{1}$}, $\rm cos\theta_{b}=1$
& {\bf 78--84} & {\bf 86} (83)  \\ 
{\bf $\tilde{b}_1\rightarrow b\chi^0_{1}$}, $\rm cos\theta_{b}=0.39$
& {\bf 45--68} & {\bf 75} (60)  \\  
\hline 
\end{tabular} 
\caption{Individual and LEP combined 
mass limits for stop and sbottom  
up to $\sqrt{s}= 183$ GeV    
for $\Delta M>15$ GeV, assuming maximal or minimal 
production cross sections. 
}
\label{tab9}
\end{center}
\end{table}
At the Tevatron, top squarks can be directly pair produced or can be produced 
in the top quark decay mode $t \rightarrow \chi^0_{1} \tilde{t}_1$. The stop 
is searched for in the $\tilde{t}_1\rightarrow c\chi^0_{1}$ decay channel but  
also in the mode $\tilde{t}_1\rightarrow b\chi_1^+$ when kinematically 
allowed. Recently, the CDF \cite{st5} 
experiment has reported a higher limit on the $\tilde{t}_1$ mass of about 122 
GeV/c$^2$ for $m_{\chi^0_{1}}$ lower than 50 GeV/c$^2$, the sensitivities in 
the low $\Delta M$ ranges are weak, due to the huge irreducible QCD 
background. The complementary domains excluded both at LEP2 and the Tevatron
are shown in Fig.~\ref{fig:stop}. \s

At hadron colliders, one expects the strongly interacting SUSY particles,
gluinos and squarks, to be produced with the largest cross sections.  
Therefore the search for $\tilde{q} \tilde{q}$, $\tilde{q}\tilde{g}$ and
$\tilde{g}\tilde{g}$ final states has been performed at the Tevatron by both
the CDF \cite{sq1} and D0
\cite{sq2} collaborations. Depending on the mass hierarchy between 
squarks and gluinos, the subsequent decays will be $\tilde{q} \rightarrow q 
\tilde{g}$ or $\tilde{g} \rightarrow q \tilde{q}$; then decays of squarks and 
gluinos proceed according to $\tilde{q}\rightarrow q \chi_1^0$ and $\tilde{g} 
\rightarrow q \bar{q} \chi_1^0$ [the so called direct decays] while for 
sufficiently massive squarks and gluinos, the decays proceed through charginos 
$\chi^\pm_{1,2}$ or heavier neutralinos 
$\chi^{0}_{2,3,4}$, which in turn decay 
to quarks, leptons or neutrinos and one LSP [the so called cascade decays].
In the direct decays scenario, this leads to final states containing a 
significant number of jets and  missing transverse energy carried away by the 
LSPs; it has to be noticed also that gluino decays yield on average one more 
jet than squarks decays. The cascade decays result in the production of a 
larger number of jets, with a reduced but still significant $\Ebar$ and 
occasional production of leptons. The recent dilepton based SUSY search by 
CDF and D0 are sensitive to cascade decays. \s

No excess of data has be found at the two Tevatron experiments
\cite{sq1}--\cite{sq2}; as an 
illustration, the CDF experiment for an integrated luminosity of 19 pb$^{-1}$,
has selected 18 events in the three jets sample and 6 events in the four jets 
sample, these numbers are consistent with estimates of SM processes, 
respectively 33$^{+12}_{-10}~({\rm stat})~^{+19}_{-12}~({\rm syst})$ and  
8$^{+4}_{-3}~({\rm stat})~^{+4}_{-4}~({\rm syst})$. The main background in this 
analysis comes from $W/Z$ + jets processes, top quark production and QCD 
multi--jet production in which the $\Ebar$ originates from mismeasurement. \s
 
\begin{figure}[hbtp!]
 \begin{center}  
\hspace*{-1.5cm}\mbox{\epsfxsize=10.0cm 
  \epsffile{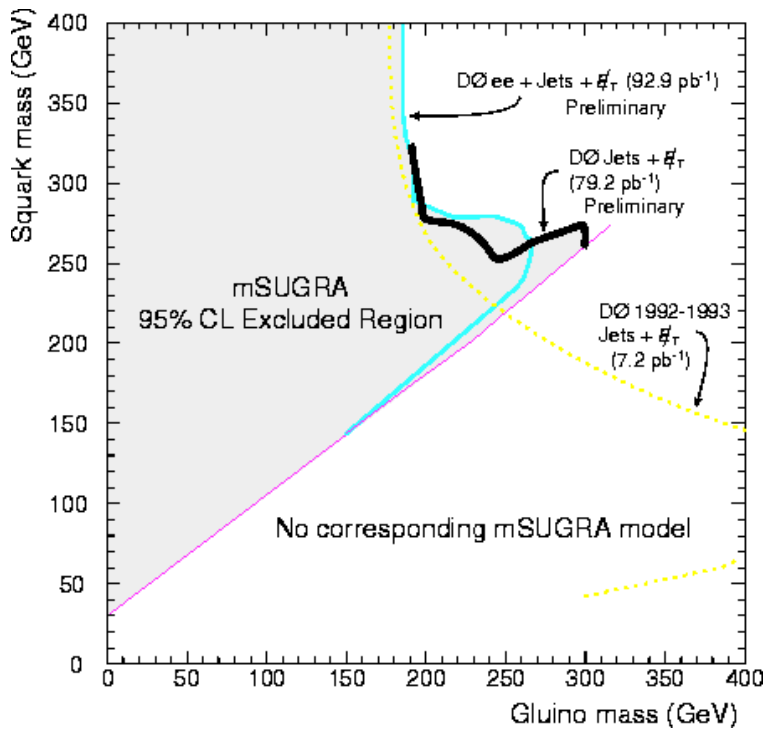}}\vspace*{-7.0 cm}
 \hspace*{-4.3cm}\mbox{\epsfxsize=24.cm\epsffile{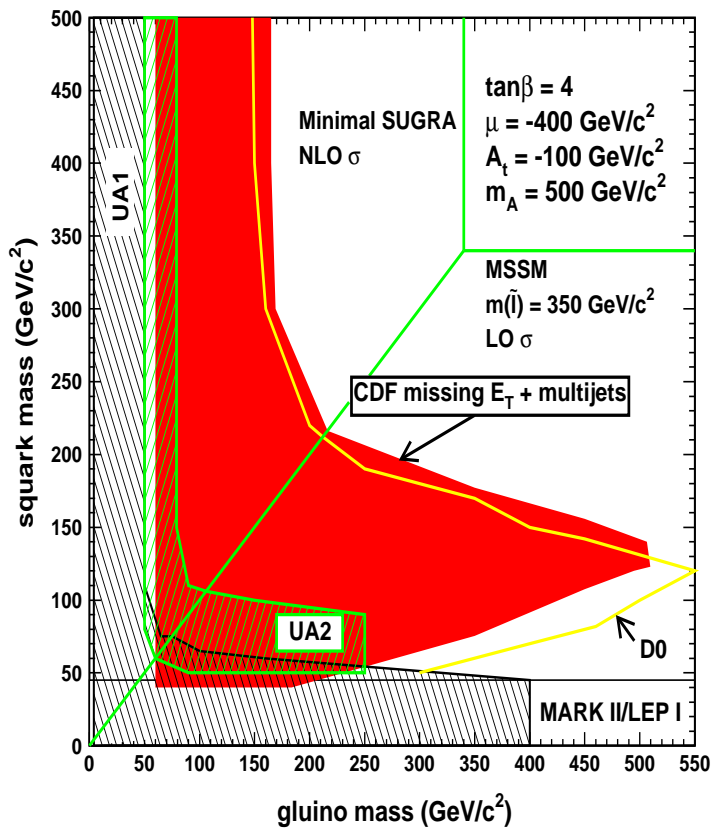}}\vspace*{-13.0 cm}
\caption{D0 (top) \cite{sq2} and CDF (bottom) \cite{sq1} 
exclusion domains in the plane 
($m_{\tilde{g}}, m_{\tilde{q}}$) from the multi--jet plus $\Ebar$ 
or dileptons plus jets plus $\Ebar$ searches at the Tevatron.
}
 \label{fig:squark}
 \end{center}
\end{figure}

These searches yield lower limits on gluino and squark masses and the domains 
excluded by both experiments in the  ($m_{\tilde{g}}, m_{\tilde{q}}$) plane 
are shown in Figs.~\ref{fig:squark}, where it is assumed that all but the top 
squark masses are degenerate. Assuming that squarks are heavier than the 
gluino, a lower limit on the common squark mass equal to 216 and 260 GeV/c$^2$
is set by the CDF and D0 collaborations, respectively. These limits do not 
depend on $\tb$ but hold only for large $\mu$ values [$\mu \gsim 200$ GeV or 
$\mu \lsim -100$ GeV]. A tighter squark mass limit $m_{\tilde{q}}>267$ $\gv$,
using the dilepton channel, is set by D0; in this case however the result is 
more model dependent [charginos and heavy neutralinos have to decay 
leptonically, etc.].

\subsection*{6.3. The gaugino sector}

\subsubsection*{6.3.1 Gluinos}

The gluino mass bounds derived from the $\tilde{q}\tilde{g}$ searches 
previously described, are respectively 173 and 187 GeV/c$^2$ for the CDF 
and D0 collaborations, for any squark mass. When one imposes an equal 
mass for squarks and gluinos, the limits become 216 and 260 $\gv$ respectively,
but are more model dependent. \s

In 1996, two small windows for very light gluinos [below 1.5 $\gv$ and between 
3 and 5 $\gv$] remained unexcluded by any of the existing searches. Such light 
gluinos would modify the decay pattern of the SUSY particles considered; since 
with gluinos lighter than 5 $\gv$, $\Ebar$ becomes small and the usual jets + 
$\Ebar$ signature is no more effective. However, resonant production of squarks
would have a large cross section and, if squarks are not very heavy, broad 
peaks in the dijet mass distributions are expected. Comparison of the observed 
spectrum with theoretical estimates rules out light gluinos if squarks are 
lighter than about 600 $\gv$ \cite{PDG}
. Such light gluinos would also affect  the usual 
phenomenology of QCD at LEP, by changing  the topology of four jet events via
$ g^* \rightarrow \tilde{g}\tilde{g}$ splitting; they would also contribute  
to the running of $\alpha _{s}$ as three additional flavors in leading order, 
up to mass effects. A full four--jet analysis carried out at LEP1 by 
ALEPH \cite{gl2} excludes gluinos with masses smaller than $6.3~\gv$.

\subsubsection*{6.3.2. Charginos and neutralinos}

We now turn to the search for charginos and neutralinos at LEP2. At $e^+e^-$
colliders, charginos $\chi_1^+ \chi_1^-$  (neutralinos $\chi_i^0 \chi_j^0$ 
with ${\small i,j=1, \ldots ,4}$ ordered by their masses) are pair--produced 
via $s$--channel $\gamma/Z$ ($Z$) boson and $t$--channel sneutrino $\tilde{
\nu}$ (selectron $\tilde{e}$) exchange. When the  masses of the sfermions are 
very large, the $\chi_1^\pm$ ($\chi_j^0$ with $j \ge 2$) states decay via an
exchange of a virtual $W$ ($\rm Z$) boson as follows: $\chi_1^\pm \rightarrow 
\chi_1^0 W^* \ra \chi_1^0 f\bar{f}^\prime$ ($\chi^0_{j} \rightarrow \chi_{k}^0 
Z^* \rightarrow \chi_{k}^0 f\bar{f}$ with $k < j$). If the slepton and 
sneutrino masses are comparable to  $M_{W}$ ($M_{Z})$ the charginos 
(next--to--lightest neutralinos) decay via virtual slepton or sneutrino 
exchange and the leptonic branching fraction is enhanced. Finally for slepton 
or sneutrino lighter than the chargino (next to lightest neutralino), 
the decay modes $\chi_1^\pm \rightarrow \tilde{l}^\pm \nu$ or $l^\pm \tilde{
\nu}$ ($\chi^0_{j} \rightarrow \tilde{l}^\pm l^\mp$ or $\tilde{\nu} \nu$) 
become dominant. The radiative decays $\chi_j^0 \rightarrow \chi_{k}^0 \gamma$
are also possible via higher--order diagrams. For all LEP2 studies, 
the gluino is supposed to be heavy, otherwise the chargino decay patterns 
would be different, as reminded in the previous section. \s

The charginos and heavy neutralinos are searched for at LEP2 
\cite{ch1}--\cite{ch4}, in a first time 
within the large $m_0$ scenario, i.e large scalar fermion masses. In this 
case, charginos and neutralinos decay into the LSP accompanied by virtual 
$W$ and $Z$ bosons. The three possible final state topologies arising from 
the $W$ and $Z$ decays are: hadrons plus missing energy, acoplanar lepton 
pairs plus missing energy and mixed final state (hadrons plus leptons) 
plus missing energy. As explained before, signal topologies and the 
associated background sources depend on $\Delta M$. Therefore, for each 
topology, selections were optimized for four different $\Delta M$ ranges: 
the very low (3--5 GeV), the low (5--10 GeV), the medium (20--40 GeV) and 
the large ($\geq$ 50 GeV). In addition, the DELPHI experiment searched for 
topologies arising in the very small $\Delta M$ regime, with decays in 
flight and ISR tags \cite{ch20}. 
Altogether a total of 27 events was observed by the 
four experiments with 25$\pm$5 expected from standard processes with 
efficiencies ranging from 5\% to 65\% depending on the analysis and the 
$\Delta M$ ranges as illustrated in Tab.~\ref{tabch}. 
\begin{table} [htbp]
\begin{center} 
\begin{tabular}{|c|c|c|c|} 
\hline
 \multicolumn{4}{|c|}{$\rm M_{\chi_1^{\pm}}$=91 $\gv$  \ \ \ \     
$\chi_1^\pm \rightarrow \chi_1^0 W^*$ } \\
\hline    
{~~ $\Delta M$ (GeV)~~}  
& ~~$\epsilon$ (\%)  & $N_{\rm back}^{\rm exp}$ &  $N^{\rm obs}$ \\
\hline
5 & 15 & 10.3 $\pm$ 1.1  &     8 \\
20 & 55 & 17.4$\pm$ 1.1  & 18  \\ 
60 & 14   & 17.4$\pm$1.1  & 18\\
\hline 
\end{tabular} 
\caption{Typical efficiencies, number of expected and observed events  
obtained by the DELPHI experiment in the chargino searches
at a c.m. energy of 183 GeV \cite{ch2}. }
\label{tabch}
\end{center}
\end{table}

\smallskip

Depending on the neutralino--chargino field content, one distinguishes the 
following cases for the determination of the lower limits on 
neutralino--chargino masses: \s

\begin{figure}[hbtp!]
\begin{tabular}{ll}
\hspace*{-1.5cm}\mbox{\epsfxsize=8.0cm 
  \epsffile{x14.epsi}}\vspace{-0.0 cm}
\hspace*{0.5cm}\mbox{\epsfxsize=10.0cm 
  \epsffile{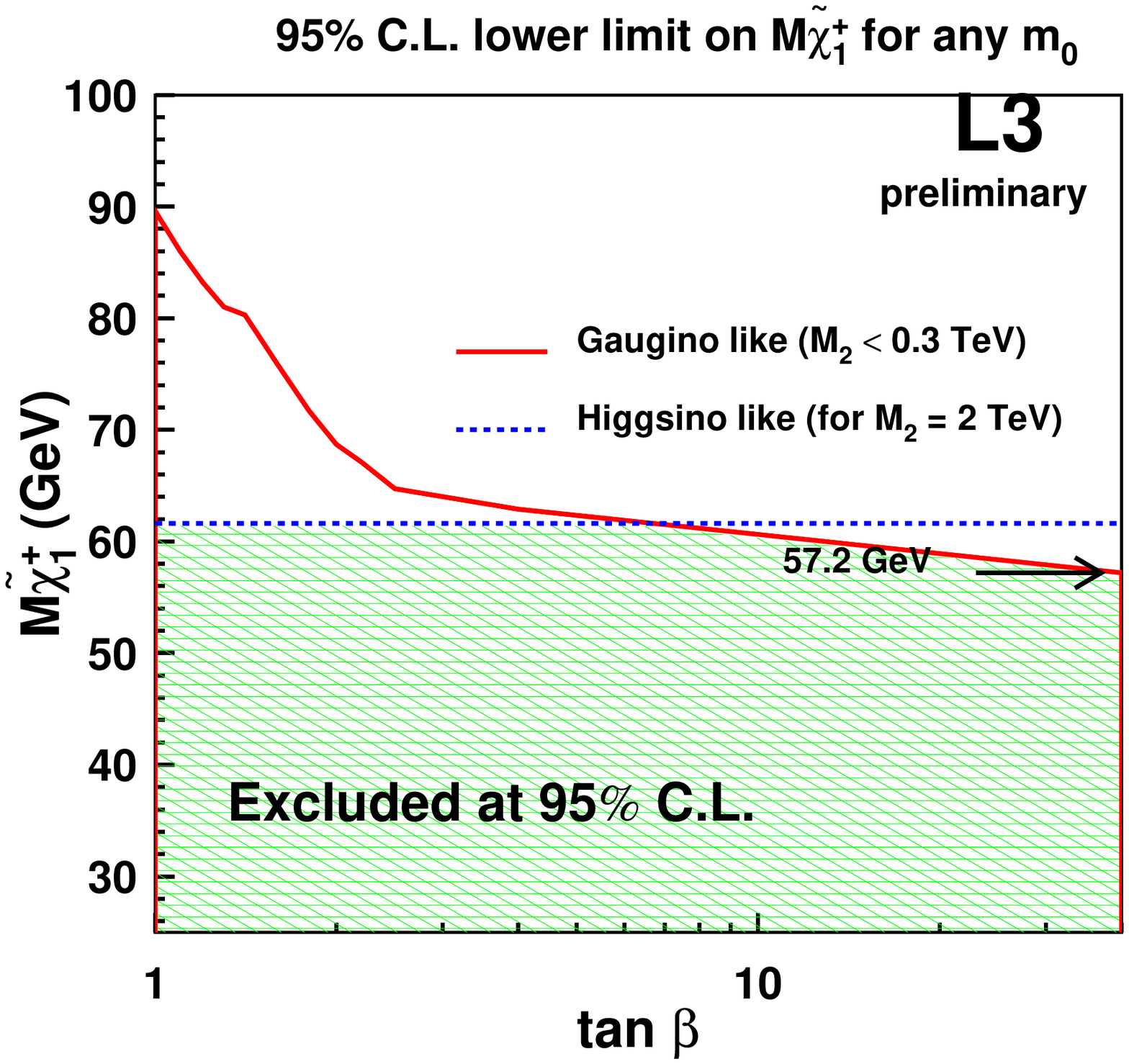}}\vspace{-0.0 cm}
\end{tabular}
  \hspace*{-1.0cm}\mbox{\epsfxsize=17.0cm 
\epsffile{x15.epsi}}
  \caption{Top left: LEP combined \cite{s0} lower limit on the chargino mass   
for different $\Delta M$ values at c.m. energies 
up to 183 GeV. Top right: lower limit on the chargino mass   
for different field content, as a function of $\tb$, by the L3 experiment 
\cite{ch3}.
 Bottom: excluded domains in the plane ($m_{\chi^{\pm}_1},
\Delta M$) in the very low $\Delta M$ range obtained by DELPHI \cite{ch20}
at a c.m. energy up to 183 GeV.                     }
\label{fig:higgsino}
\end{figure}

-- Higgsino--like $\chi_{2}^0$ and $\chi_1^\pm$ ($M_2 \gg |\mu|$):
in this case, the production cross sections do not depend on the scalar lepton
masses, and $\Delta M$ is low and decreases with increasing $M_2$. 
Consequently, the limits on the masses of the next--to--lightest neutralino
and the lightest chargino decrease with $M_2$. The present LEP2 combined 
\cite{s0}
limit on the chargino mass is shown in Fig.~\ref{fig:higgsino} (top left)
turning into a limit on the chargino mass of 63 $\gv$ for $\Delta M 
\geq 3~\gv$. The DELPHI experiment has excluded regions for lower $\Delta M$
ranges as depicted in Fig.~\ref{fig:higgsino}(bottom). Masses lower than 80 $\gv$
are excluded by all LEP experiments for the next--to--lightest neutralino 
$\chi_2^0$ assuming $M_2$ less than 1500 $\gv$. This limit holds 
only for higgsino--like $\chi^0_{2}$, since the coupling to 
the $Z$--boson vanishes for gaugino--like neutralinos.  \s

-- Gaugino--like chargino  ($|\mu| \gg M_2$):
the cross section depends strongly on the scalar neutrino mass. For $50 \leq 
m_{\tilde{\nu}} \leq 80$ $\gv$ the destructive interference term reduces the 
cross section by one order of magnitude compared to what is expected for 
$m_{\tilde{\nu}} \geq 300$ $\gv$. When the two body decay $\chi_1^\pm \ra
l^\pm \tilde{\nu}$ is dominant, the relevant $\Delta M$  becomes equal to 
$ m_{\chi_1^\pm} - m_{\tilde{\nu}}$. Therefore, the limit on the chargino 
mass will depend clearly on the $m_{\tilde{\nu}}$ value. 
For large $m_{\tilde{\nu}}$ values ($\geq$ 300 $\gv$), the situation is the 
ideal one: we benefit there from the best detection sensitivity and the highest 
cross section production. This is the reason why the kinematical limit is 
reached with only a few $\pb$. Up to now, including the recent 189 GeV data, 
LEP experiments exclude a chargino mass below 94.3 $\gv$, irrespective of 
$\tb$. 
Extending the study to all $m_{\tilde{\nu}}$ values  where the chargino 
leptonic decay is enhanced for low slepton masses, dedicated analyses then 
were developed  by the LEP experiments. However, it appears that the chargino 
cannot be directly detected when it is degenerate in mass with the sneutrino
since the soft leptons produced in the final state escape detection. In this
case, the LEP1 limit of 45 $\gv$ remains, that is deduced from the $Z$--boson 
total decay width measurement which is insensitive to the chargino decay
pattern. Recently, this limit has been improved in two independent ways: 
$(i)$ the ALEPH experiment, excludes $m_{\chi_1^\pm}$ below 51 $\gv$ by 
measuring the invisible $W$--boson width, within
some model assumptions  \cite{ch10}; 
and $(ii)$ the L3 experiment \cite{ch3}
exploits the complementarity 
of the scalar lepton searches, excluding chargino masses below 57.2 $\gv$ 
irrespective of $\tb$ and $m_0$, as depicted in Fig.~\ref{fig:higgsino}
(top right).    \s

At the Tevatron, charginos and neutralinos may be produced directly in the 
annihilation process $qq \rightarrow \chi_{2}^0 \chi_1^\pm$ or in the decays of 
heavier squarks as mentioned before. They are searched for in the leptonic 
decays [mainly assuming the decays $\chi_1^\pm \rightarrow l^\pm \nu
\chi_1^0$ and $\chi_2^0 \rightarrow l^+l^- \chi_1^0$]; see section 5.3.
The requirement 
of three leptons in the final state reduces backgrounds to a very small level
but is efficient for the signal only in peculiar cases. Therefore, charginos
and neutralinos may be discovered at the Tevatron above the kinematical reach 
of LEP2, but no robust limits can be derived and the results reported 
today are not competitive with the LEP bounds. \s

\subsubsection*{6.3.3 Neutralino LSP}

Finally, let us discuss the neutralino--LSP mass limits. 
Since the neutralino $\chi_1^0$ is the LSP and escapes detection, direct 
searches for  the LSP neutralino cannot be performed at LEP2. However, 
indirect limits have been derived from the constraints on chargino--neutralino 
and slepton searches. As detailed previously, the limits depend on the 
slepton masses, in particular on $m_{\tilde{\nu}}$. \s

\begin{figure}[hbtp!]
\begin{center}
 \hspace*{-1.5cm}\mbox{\epsfxsize=10.0cm 
  \epsffile{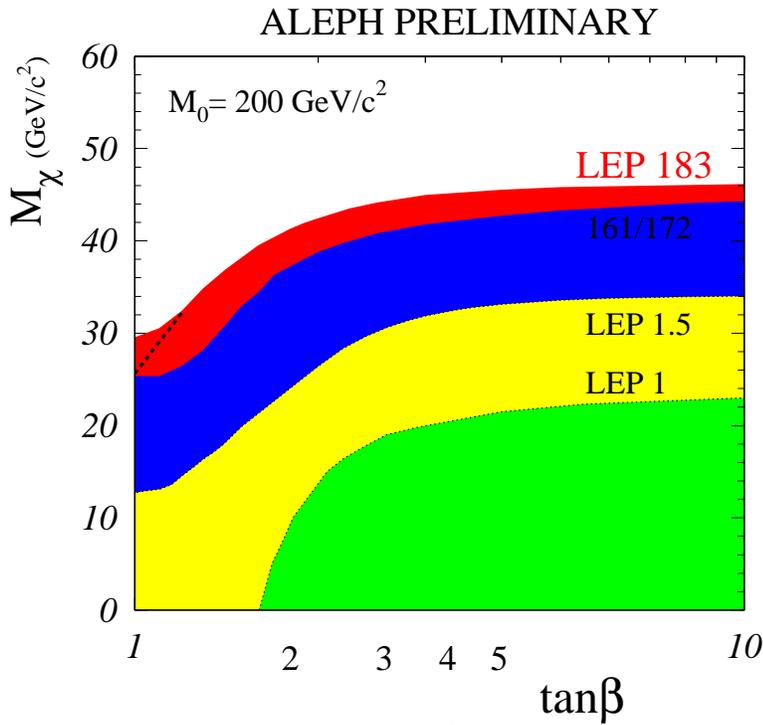}}
  \hspace*{-1.5cm}\mbox{\epsfxsize=10.0cm 
\epsffile{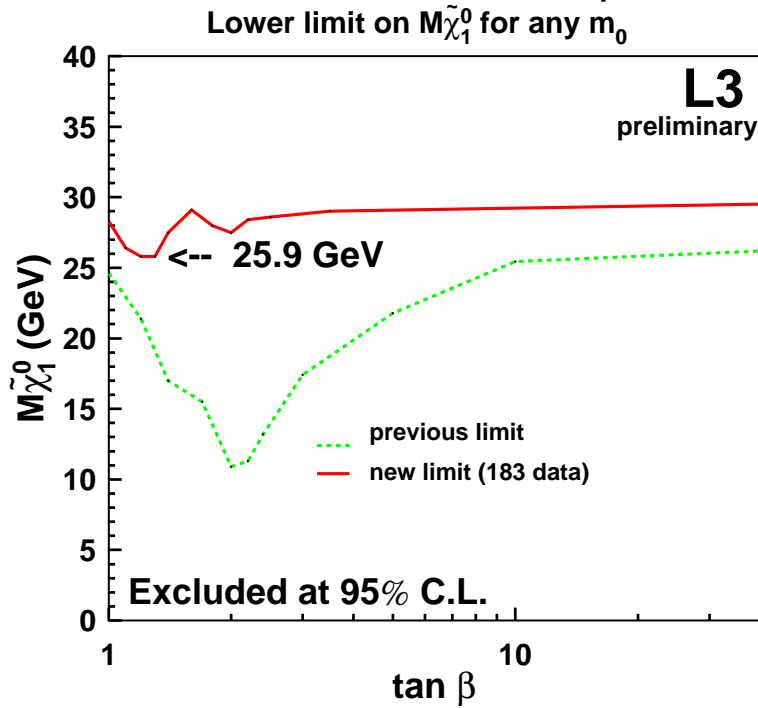}}
  \caption{Top: ALEPH \cite{ch1} lower limit on the LSP mass   
as a function of $\tb$, for large slepton masses,  
at c.m. energies up to 183 GeV. Bottom: 
Lower limit on the LSP mass   
as a function of $\tb$, for any slepton mass 
obtained by L3 \cite{ch3} at c.m. energies 
up to 183 GeV.                     }
\label{fig:LSP}
\end{center}
\end{figure}

-- Large sneutrino mass: For high $\tb$ values, the chargino search provides 
the most stringent bound on $m_{\chi_1^0}$. For $\tb$ values below 2, the 
neutralino searches including the processes $e^+e^- \rightarrow \chi_1^0 
\chi_3^0, \, \chi_1^0 \chi_4^0, \, \chi_2^0 \chi_3^0$ but also the radiative 
decays $ \chi_2^0 \ra \chi_1^0 \gamma$, exclude additional regions of the 
parameter space (mixed region). The LEP1 exclusion limits still play some 
role, due to the fact that the neutralino search is limited in the region 
near the point where the $\chi_1^0$ mass limit is set by the value of the 
coupling to the $Z$--boson rather than by kinematics. The lower limit obtained 
for  $m_{\chi_1^0}$ as a function of $\tb$ and for heavy sleptons is displayed 
in Fig.~\ref{fig:LSP} for LEP center--of--mass energies increasing from 91.5 
up to 183 GeV. A summary of all limits \cite{ch1}--\cite{ch4}
obtained both up to 183 GeV and more 
recently to 189 GeV is listed in Tab.~\ref{tab11}. \s

-- Any sneutrino mass:
For lower slepton masses, as already discussed, the constraints in the 
chargino--neutralino sectors are weaker and the limits on $m_{\chi_1^0}$ 
therefore degradate. Constrains on $m_{\chi_1^0}$ have been obtained 
by all LEP experiments with a systematic scan of the MSSM parameter space, 
as a function of $\tb$ for all $m_0$ values, as depicted in the bottom of
Fig.~\ref{fig:LSP}. For low $\tb$ values ($\leq 1.6)$ the minimum allowed 
value for $m_{\chi_1^0}$ is found in the parameter space where the production 
cross section for charginos is minimal [$m_{\tilde{\nu}} \sim$ 90 $\gv$ in 
the mixed region $\mu \sim -70~ \gv$] 
\cite{ch2}--\cite{ch4} and the heavy neutralino $\chi_{3,4}^0$
are mostly invisible while for higher $\tb$ values ($\geq 2)$ the lower 
$\chi_1^0$ mass limit is found in the parameter space region where the 
chargino and sneutrino mass difference is small and where the 
next--to--lightest neutralinos $\chi_{2}^0$ decay invisibly. The 
exclusion limit came only from slepton searches; it is 
localized in the deep gaugino--like region for charginos ($\mu \leq  -400~\gv$)
\cite{ch1}.
The limits obtained by each experiment are listed in Tab.~\ref{tab11} 
as well as the MSSM parameter ranges used in the scan. 
The most recent limit on the $\tilde{e}$ mass obtained with 189 GeV data
by ALEPH \cite{ch1} allows to set the best absolute limit on the LSP mass of 28 $\gv$, which can be turned into an absolute limit on $M_2$.

\begin{table}
\begin{center}
\begin{tabular}{|l|c|c|} 
\hline
     & \multicolumn{2}{|c|} {$m_{\chi_1^0}>$   ($\rm GeV/c^2)$ }  \\ 
   & up to 183 Gev & up to 189 GeV \\ 
\hline 
%     &                          &                       \\  
Large $m_0, \tb \geq $ 1 &   $\sim$ 30.5 \ \ (ADLO) &  $\sim$ 32.5 \ \ (AO) \\  
\hline 
%&      &                          &                       \\    
Any $ m_0 ,\tb \geq$1  &       &   \\
A \ \ \  $\mu$ up to $-2000$ $\gv$  & 26$^*$    & 28                          \\ 
D \ \ \   $\mu$ up to $-200$ $\gv$  &23.4   &   \\ 
L \ \ \ $\mu$ up to $-500$  $\gv$   & 25.9  & \\ 
O \ \ \ $\mu$ up to $-500$  $\gv$    & 25.4  & \\   
\hline  
\end{tabular} 
\caption{Indirect individual  mass limits for the LSP $\chi_1^0$ 
up to $\sqrt{s}$ = 183, 189 GeV for large $m_0$ or any $m_0$. 
In the latter case, the minimum value
($^{*}$) for ALEPH experiment is found in the deep--gaugino region (up to
 183 GeV data) while other experiments found the minimum in the mixed region.
}
\label{tab11}
\end{center}
\end{table}

\subsection*{6.4 Summary and Bounds on the MSSM parameters}

The summary of the bounds on the SUSY particles obtained at LEP2 and the 
Tevatron are listed in Tab.~\ref{tab13}. We also indicate the parameter
ranges for which these limits hold. 

\smallskip

\begin{table}
\vspace*{-1.2 cm}
\begin{center}
\begin{tabular}{|c|l|c|l|} 
\hline
Particle & Assumptions & Lower limit       & Experimental \\ 
         &             & ($\gv$)              & sources            \\
\hline 
\hline
h        & $\tb \geq$0.8  &  78.8              & LEP2 Comb.  \\
         &              &                    &  \\
  
A        &$\tb \geq$0.8  &  79.1              & LEP2 Comb.    \\
         &   & &     \\ 

H$^{\pm}$  &  
Br($\rm H\rightarrow c\bar{s})$ + Br($\rm H\rightarrow c\bar{s}$)=1
& 68  & LEP2 Comb. \\ 
         &any $\tb$    &  &     \\
%         &                           &   &                  \\   
         & $\tb \leq$1.1 or  $\tb \geq$40       & 100 & CDF/D0           \\
\hline   
\hline
$\tilde{e}_R$ & Br($\tilde{e}_R$ $\rightarrow e \chi_1^0$)=1,
$\Delta$M$\geq 20~\gv$  &  85       & LEP2 Comb.   \\
 & Br($\tilde{e}_R$ $\rightarrow e \chi_1^0$)=1,
$\tb \geq$2  &  65       & LEP2       \\     
         &            &           &     \\     
$\rm \tilde{\mu}_R$  &  BR($\rm \tilde{\mu}_R$ $\rightarrow \mu \chi_1^0$ )=1,
$\Delta$ M $\geq$ 20 $\gv$  &  71       & LEP2 Comb.    \\
 & & & \\
$\rm \tilde{\tau}_R$& BR($\rm \tilde{\tau}_R$ $\rightarrow \tau \chi_1^0$)=1,
$\Delta$ M $\geq 20~\gv$  &  75       & LEP2 Comb.   \\
 & & &  \\
$\tilde{\nu}$    & &  43       & LEP1 -- Z width      \\
 & & &  \\
$\rm \tilde{t}_1$ & BR($\rm \tilde{t}_1$ $\rightarrow  c\chi_1^0$)=1, 
$\Delta$ M $\geq$ 15~$\gv$  &  83       & LEP2 Comb.  \\
 & BR($\rm \tilde{t}_1$ $\rightarrow \nu l\chi_1^0$ )=1, 
$\Delta$ M $\geq$ 15~$\gv$  &  85       & LEP2 Comb.  \\
 & BR($\rm \tilde{t}_1$ $\rightarrow  c\chi_1^0$ )=1,
$ {\rm LEP2;~m_{\chi_1^0}  \leq  50~GeV}$  &  122       & CDF     \\
 & & & \\
$\rm \tilde{b}_1$    & BR($\rm \tilde{b}_1$ $\rightarrow b\chi_1^0$ )=1, 
$\Delta$ M $\geq  15~\gv$  &  75       & LEP2 Comb.  \\
 & & & \\
%          &                           &   &                  \\   
$\rm \tilde{q}$    & $\rm m_{\tilde{q}} \geq m_{\tilde{g}}$
 &  216       & CDF     \\ 
&  $\rm m_{\tilde{q}} \geq m_{\tilde{g}}$  &  260       & D0     \\
\hline
\hline
%          &                           &   &                  \\   
$\rm \tilde{g}$    & $\mu \leq -100$ or $\mu \geq 200~\gv$ 
 &  173       & CDF     \\ 
 & $\mu \leq -100$ or $\mu \geq 200~\gv$ &  187       & D0     \\
          &                           &   &                  \\   
$\chi^{\pm}_{1}$ & Higgsino $\Delta$M $\geq 3~\gv$ & 63 & LEP2 Comb.\\ 
 & Gaugino $\rm m_0$ $\geq 200~ \gv$ & 94.3 & LEP2$^*$ \\ 
    & Gaugino any $\rm m_0$ & 57.2 & LEP2 \\ 
    & & & \\   
$\chi^0_{1}$ & Large $\rm m_0$ &  32.5  &   LEP2$^*$           \\ 
   & Any $\rm m_0$ &  28  &   LEP2$^*$         \\ 
    & & &  \\
\hline 
\end{tabular}
\end{center}
\caption{Summary of lower mass limits obtained at the Tevatron and   
at LEP2 up to $\sqrt{s}$ = 183, 189 GeV [$^*$ at which 
just a small fraction of data has been used]. }
\label{tab13}
\end{table}

\begin{figure}[hbtp!]
\begin{center}
\vspace*{-1.3cm}
 \hspace*{-1.5cm}\mbox{\epsfxsize=10.0cm 
  \epsffile{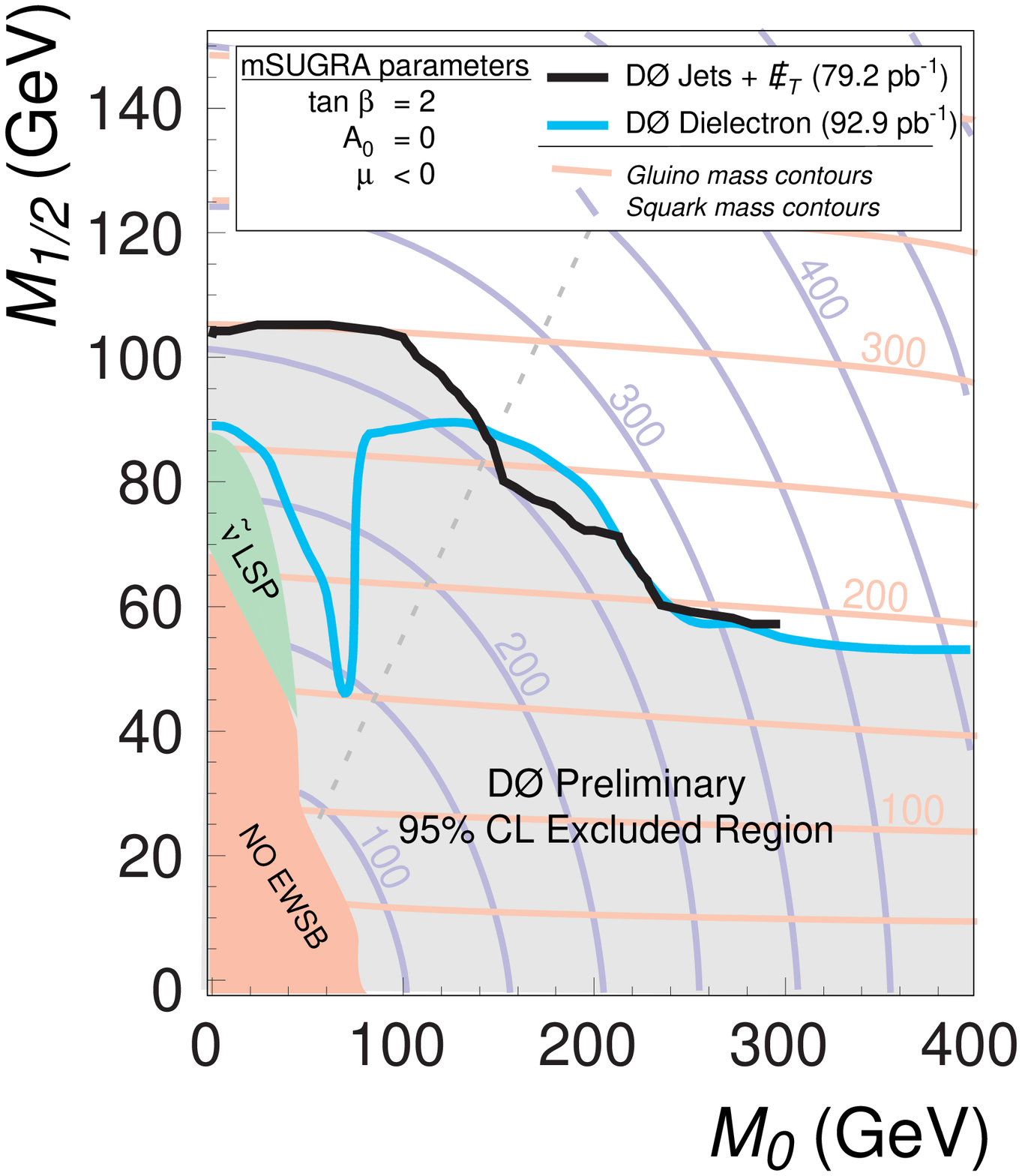}}\vspace{-1.3 cm}
 \hspace*{-1.5cm}\mbox{\epsfxsize=15.0cm 
\epsffile{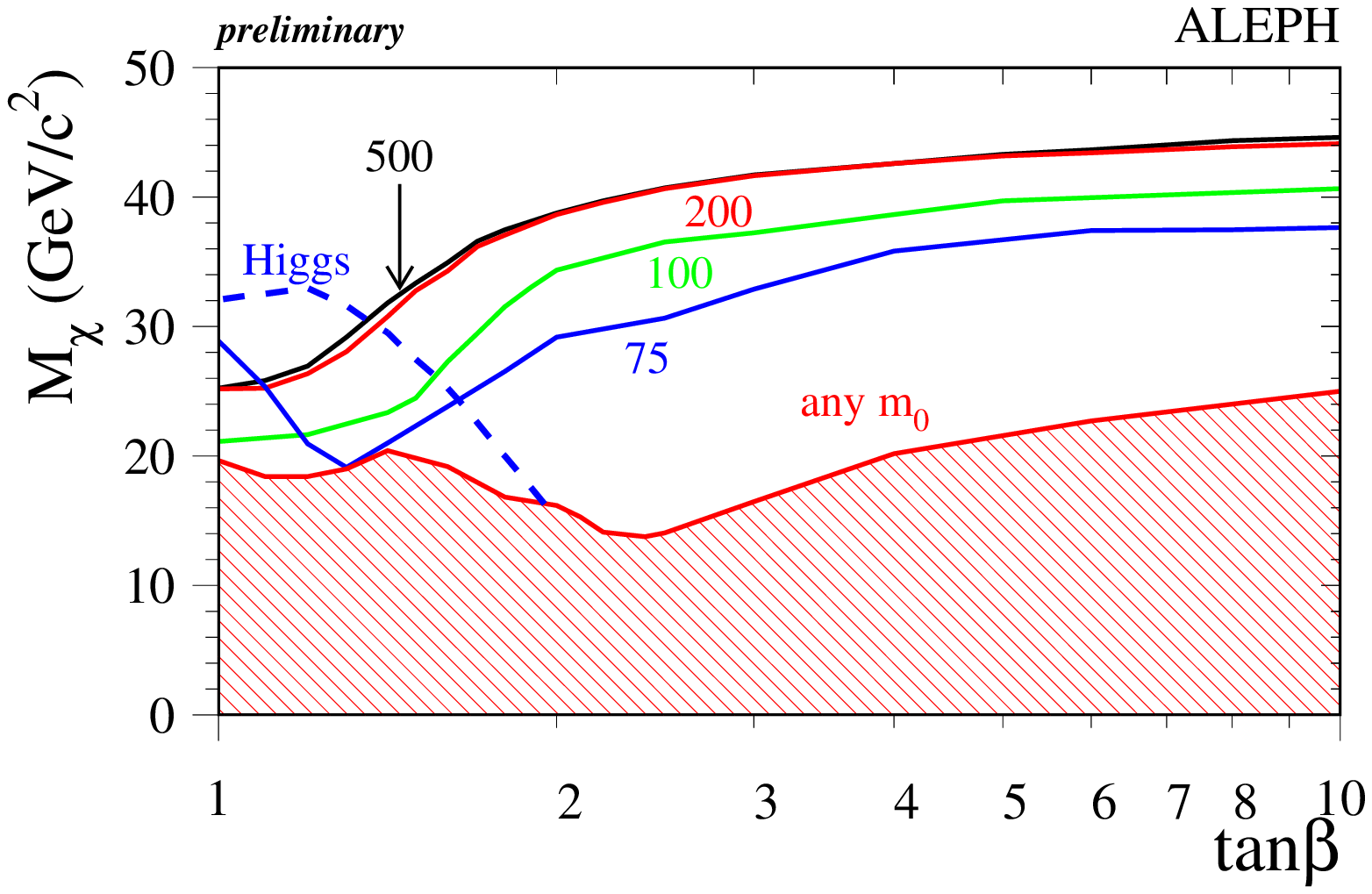}}\vspace{-10 cm}
\caption{Top: 
D0 \cite{sq2} exclusion contours in the plane ($m_0,m_{1/2}$) 
from the multi--jet plus $\Ebar$ or dileptons plus jets plus $\Ebar$ searches.
Bottom: Lower limit on the mass of 
the lightest neutralino as a funtion of $\tb$ for a series of $m_0$ values
by the ALEPH experiment \cite{Laurent} for a c.m. energy up to 172 GeV. 
The dashed curve shows the result coming from the Higgs constraints for 
$m_0$=75 $\gv$ }
\label{fig:dudu}
\end{center}
\end{figure}

\begin{figure}[hbtp!]
\begin{center}
  \hspace*{-1.5cm}\mbox{\epsfxsize=15.0cm \epsffile{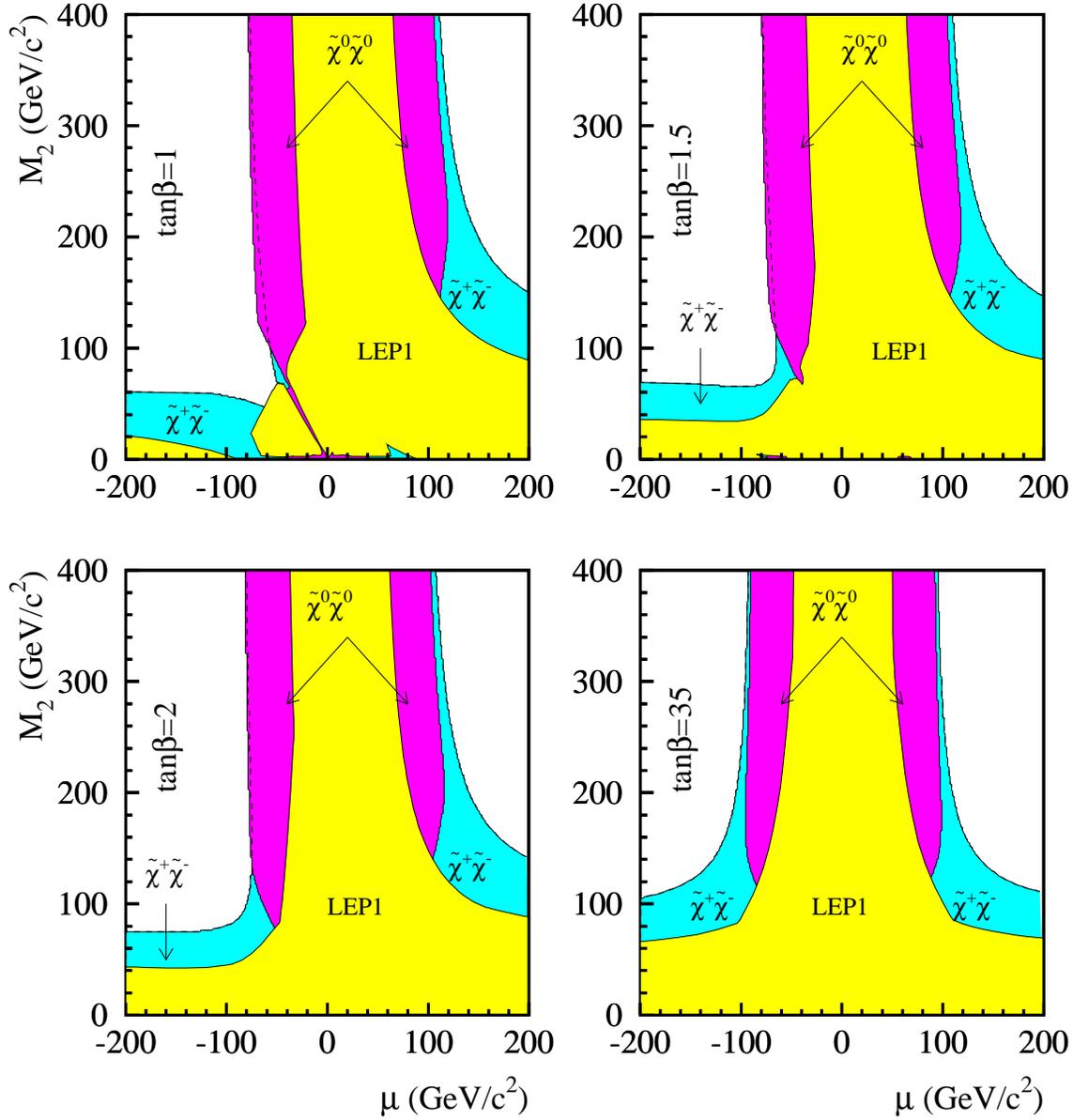}}
  \caption{ Regions in the ($\mu$,$M_2$) plane excluded by DELPHI \cite{ch2}
for different values of $\tb$, assuming large $m_0$. The lightly
shaded areas are those excluded by lower energy LEP1 results.
The intermediate shading shows regions excluded by the chargino search at 
183 GeV. The darkly shaded areas show the regions excluded by the neutralino
searches at these energies.}
\label{fig:mu}
\end{center}
\end{figure}

The limits can be turned into bounds on the cMSSM parameters; 
as an example, at the Tevatron, the excluded region in the ($m_{\tilde{g}},
m_{\tilde{q}}$) plane is translated in the plane 
($m_0, m_{1/2}$) for different 
$\tb$ values, see Fig.~\ref{fig:dudu} (top). 
Similarly, LEP experiments have derived from 
chargino/neutralino searches excluded contours in the plane 
($\mu, M_2$)  for different $m_0$, $\tb$ values
as illustrated on Fig.~\ref{fig:mu}; 
in particular the clear complementarity of the searches for neutralinos and charginos is visible for the low $\tb$ regime. Moreover, 
the combination of the gaugino and the slepton searches allowed to exclude
regions of parameters in the plane ($m_0, m_{1/2}$) and to derive 
an absolute limit on $M_2$ of 47 $\gv$ independently of $m_0$, $\mu$ and 
$\tb$. \s

The low values of the parameter $\tb$ are already constrained, for a given 
set of parameters, by the Higgs boson searches performed at LEP2. The 
combination of results obtained at the Tevatron and at LEP2, 
allow to set more robust and stringent constraints for the low
$\tb$ regime, as already  tested in Ref.~\cite{pat2} and illustrated in 
Fig.~\ref{fig:mass_tbmh}.
 Finally constraints form the Higgs sector and the other 
SUSY particle sector, could be combined to derive improved bounds on the cMSSM 
parameters; a first attempt has been done in Ref.~\cite{Laurent} when 
combining Higgs and SUSY limits obtained at LEP2 for 
a c.m. energy up to 172 GeV. It was shown that the interplay 
between Higgs and stop mass lower limits allows $m_{1/2}$ to be constrained
for low values of $\tb$ and $m_0$: for small (large) mixing, the Higgs
(stop) mass limit provides the most effective constraint. On the bottom of 
Fig.~\ref{fig:dudu},
one can see clearly how the lower limit on the LPS is 
strengthened by taking the Higgs information into account for small 
$\tan \beta$.


\newpage

\end{document}